%% file: dottesi.tex
\documentclass[a4paper,openright,twoside]{report}

\usepackage{listofth}
\usepackage{headerfooter}
\usepackage[dvips,rightbars]{changebar}
\usepackage[english]{babel}
\usepackage{amsbsy}
\usepackage{amsfonts}
\usepackage{amssymb}
\usepackage[mathscr]{eucal}
\usepackage{pstricks}
\usepackage{pst-node}
\usepackage{pst-grad}
\usepackage{pst-coil}
\usepackage{tables}
\input{commands}


\begin{document}

\include{titlepage}
\cleardoublepage

\pagenumbering{roman}
\pageheaderlinetrue
\pageheader{}{Contents.}{}
\pagefooter{}{\roman{page}}{}

\addcontentsline{toc}{chapter}{Contents}
\tableofcontents
\cleardoublepage

\addcontentsline{toc}{chapter}{List of Figures}
\listoffigures
\cleardoublepage

\addcontentsline{toc}{chapter}{List of Tables}
\listoftables
\cleardoublepage

\addcontentsline{toc}{chapter}{List of Definitions}
\listof{Definitions}{defs}
\cleardoublepage

\addcontentsline{toc}{chapter}{List of Propositions}
\listof{Propositions}{props}
\cleardoublepage

\addcontentsline{toc}{chapter}{List of Notations}
\listof{Notations}{nots}

\include{notation}

\cleardoublepage
\baselineskip 15pt
\include{thanks}
\cleardoublepage
\baselineskip 20 pt
\pagefooter{}{[$\, {\mathbb{I}} \, $].\arabic{page}}{}
\pagenumbering{arabic}
\setcounter{page}{1}
\include{intro}
\pagefooter{}{[\arabic{chapter}.\arabic{section}].\arabic{page}}{}

\include{chap01}

\include{chap02}

\include{chap03}

\include{chap04}

\include{chap05}

\include{chap06}

\include{chap07}

\include{chap08}

\include{chap09}

\include{chap10}

\include{concl}

\pagefooter{}{[Appendix~\Alph{chapter}].\arabic{page}}{}

\appendix

\include{appA}

\include{appB}

\include{appC}

\include{appD}




\include{biblio}

\end{document}

%% file: commands.tex
\newcommand{\A}{\mathcal{A}}
\newcommand{\bs}{\boldsymbol}
\newcommand{\Bg}{\boldsymbol{g}}
\newcommand{\Bp}{\boldsymbol{P}}
\newcommand{\Bpi}{\boldsymbol{\pi}}
\newcommand{\Bq}{\boldsymbol{Q}}
\newcommand{\Bs}{\boldsymbol{s}}
\newcommand{\Bsigma}{\boldsymbol{\sigma}}
\newcommand{\Bx}{\boldsymbol{X}}
\newcommand{\Bxi}{\boldsymbol{\xi}}
\newcommand{\Bzeta}{\boldsymbol{\zeta}}
\newcommand{\By}{\boldsymbol{Y}}
\newcommand{\BB}{\boldsymbol{B}}
\newcommand{\BN}{\boldsymbol{N}}
\newcommand{\BT}{\boldsymbol{T}}

\newcommand{\cpu}{m ^{2}}
\newcommand{\dfn}{\stackrel{\mathrm{def.}}{=}}
\newcommand{\Dp}{\boldsymbol{D} ^{(p)}}
\newcommand{\esci}{\! \! \! \! \! \! \! \! \! \! }
\newcommand{\euf}{\mathfrak}

\newcommand{\fact}[1]{\left( #1 \right) !}
\newcommand{\F}{\mathcal{F}}

\newcommand{\hLam}{\hat{\Lambda}}

\newcommand{\hphi}{\hat{\phi}}
\newcommand{\hpsi}{\hat{\psi}}

\newcommand{\hPsi}{\hat{\Psi}}
\newcommand{\hR}{\hat{\mathcal{R}}}
\newcommand{\hx}{\hat{Y}}

\newcommand{\Ham}{{\mathcal{H}}}
\newcommand{\Iuno}{\stackrel{\circ}{\mathbb{I}}{}^{(1)}}
\newcommand{\kawspace}{{{}^{\ast} \mathscr{E} _{\left\{ e _{n} \right\}} \left( \R ^{d} \rightarrow {}^{\ast} \R \right)}}
\newcommand{\Lag}{{\mathcal{L}}}
\newcommand{\mbd}{\mathcal{D} ^{-}}
\newcommand{\mfd}{\mathcal{D} ^{+}}
\newcommand{\Mp}{\mathbb{M} ^{(p+1)}}
\newcommand{\norm}{\left( \oint dl(s) \right) ^{-1}}

\newcommand{\N}{\mathbb{N}}
\newcommand{\Nc}{\mathcal{N}}
\newcommand{\qCq}{\left [ C \right ]}
\newcommand{\rp}{\rho _{p}}
\newcommand{\R}{\mathbb{R}}
\newcommand{\Rns}{{}^{\ast} \R}
\newcommand{\shF}{\check{\mathcal{F}}}
\newcommand{\spbcorr}{$\!\!\!$}
\newcommand{\Sf}{\mathbb{S}}
\newcommand{\Sigmap}{\Sigma ^{(p+1)}}
\newcommand{\Svary}{\Delta S}
\newcommand{\tst}{\left( s \right)}
\newcommand{\T}{\mathbb{T}}
\newcommand{\uespace}{{{}^{\ast} \mathcal{E}}}
\newcommand{\Wsp}{\mathcal{W} ^{(p+1)}}

\newcommand{\Xip}{\Xi ^{(p+1)}}
\newcommand{\beq}{\begin{equation}}
\newcommand{\eeq}{\end{equation}}
\newcommand{\bea}{\begin{eqnarray}}
\newcommand{\eea}{\end{eqnarray}}
\newcommand{\bPPdd}[2]{\bar{P} _{#1 | #2}}

\newcommand{\cexpect}[2]{\mathtt{E} \left \langle #1 \left | #2 \right . \right \rangle}
\newcommand{\cl}[1]{\left[ #1 \right]}
\newcommand{\clue}[2]{\cl{\left\{ #1 \right\} _{#2 \in \N}}}
\newcommand{\cnexpect}[2]{\hat{\mathtt{E}} \left \langle #1 \left | #2 \right . \right \rangle}
\newcommand{\comm}[2]{\left [ #1 , #2 \right ]}
\newcommand{\dete}[1]{\left | #1 \right |}
\newcommand{\equivo}[1]{\stackrel{#1}{\equiv}}
\newcommand{\equo}[1]{\stackrel{#1}{=}}
\newcommand{\expect}[1]{\mathtt{E} \left \langle #1 \right \rangle}
\newcommand{\form}[1]{\bs{#1}}
\newcommand{\funint}[1]{\int \left [ \mathcal{D} #1 \right ]}
\newcommand{\funinte}[3]{\int _{#2} ^{#3} \left [ \mathcal{D} #1 \right ]}
\newcommand{\absv}[1]{\left | #1 \right |}
\newcommand{\multind}[3]{#1 _{#2} \dots #1 _{#3}}
\newcommand{\multint}[4]{\form{d #1} ^{#2 _{#3}} \wedge \dots \wedge \form{d #1} ^{#2 _{#4}}}
\newcommand{\myvary}[1]{\Delta #1}
\newcommand{\nexpect}[1]{\hat{\mathtt{E}} \left \langle #1 \right \rangle}
\newcommand{\norme}[1]{\left( \oint _{#1} dl(s) \right) ^{-1}}
\newcommand{\norbe}[1]{\left( \oint _{#1} d ^{p} \bs{v}(s) \right) ^{-1}}
\newcommand{\nsprob}[1]{{}^{\ast} \mathrm{Prob.} \left\{ #1 \right\}}
\newcommand{\pois}[2]{\left \{ #1 , #2 \right \}}
\newcommand{\prob}[1]{\mathrm{Prob} \left\{ #1 \right\}}
\newcommand{\PPdd}[2]{P _{#1 | #2}}
\newcommand{\PPuu}[2]{P ^{#1 | #2}}
\newcommand{\PPud}[2]{P ^{#1}{}_{| #2}}
\newcommand{\PPdu}[2]{P _{#1}{}^{| #2}}
\newcommand{\qtq}[1]{\left [ #1 \right ]}

\newcommand{\scal}[3]{\left \langle #1 , #2 \right \rangle _{#3}}
\newcommand{\seqn}[2]{\left \{ #1 _{#2} \right\} _{#2 \in \N}}
\newcommand{\sh}[1]{\check{\mathcal{#1}}}
\newcommand{\stand}[1]{st \left( #1 \right)}
\newcommand{\ttt}[1]{\! \left( #1 \right)}
\newcommand{\Tr}[1]{\mathrm{Tr} \left [ #1 \right ]}
\newcommand{\udin}[3]{#1 _{#2} ^{\left( #3 \right)}}
\newcommand{\vect}[1]{\bs{#1}}
\newtheorem{defs}{Definition}[chapter]
\newtheorem{nots}{Notation}[chapter]
\newtheorem{props}{Proposition}[chapter]
\newenvironment{proof}
               {\par
                \addtolength{\parindent}{-1.27 cm} %
                \begin{changebar}\nopagebreak
                {\bf Proof:}\nopagebreak\\
                \rule[2mm]{5cm}{3pt}\nopagebreak
                \par\nopagebreak
                \small
               }
               {\nopagebreak
                \par\nopagebreak
                \rule[2mm]{5cm}{3pt} \hspace{6 cm} $\square$\nopagebreak
                \par\nopagebreak
                \end{changebar}%
                \addtolength{\parindent}{1.27 cm} %
                \normalsize
               }
\newenvironment{start}
               {\par
                \raggedleft
                \begin{minipage}{5cm}
                \raggedleft
                \small \sl
               }
               {
                \end{minipage}
                \par
                \normalsize \rm
                \addtolength{\hsize}{8cm}
               }
\newcommand{\makechaptermyhead}[2]{%
   \cleardoublepage
   \addcontentsline{toc}{chapter}{\protect\numberline{#2}#1}%
   \addtocontents{lof}{\protect\addvspace{10pt}}%
   \addtocontents{lot}{\protect\addvspace{10pt}}%
   \addtocontents{props}{\protect\addvspace{10pt}}%
   \addtocontents{defs}{\protect\addvspace{10pt}}%
   \addtocontents{nots}{\protect\addvspace{10pt}}%
   \thispagestyle{plain}%
   \vspace*{50 pt}%
  {\noindent \raggedright \normalfont
        \huge \bf \space #2 \space #1
        \par\nobreak
        \vskip 40 pt
  }}

%% file: titlepage.tex
\begin{titlepage}
\thispagestyle{empty}
\vspace{.4cm}
\begin{center}
        {\Large UNIVERSIT\`{A} DEGLI STUDI DI TRIESTE}\\[-1.1 cm]
        $$
            \psline(-6,0)(6,0)
        $$
    \\
    \vspace{.1cm}
        {\large \it Dottorato di Ricerca in Fisica}\\
    \vspace{.4cm}
    {\it XI Ciclo}\\
    \vspace{.8 cm}
        {\huge \bf Boundary {\it versus} Bulk}\\[.2cm]
        {\huge \bf Dynamics}\\[.1cm]
        {\small \bf of}\\[.2cm]
        {\huge \bf Extended Objects}\\[.1cm]
        {\small \bf and the}\\[.2cm]
        {\huge \bf Fractal Structure}\\[.1cm]
        {\small \bf of}\\[.2cm]
        {\huge \bf Quantum Spacetime}\\
        \vspace{.4cm}
        {\small \bf NonStandard Treatment}\\
        {\small \bf with}\\
        {\small \bf Detailed Computations}\\
    \vspace{.5 cm}
\end{center}
$$
    \matrix{
            \mathit{Dottorando} \cr
            \mathrm{\bf Dott.\ Stefano\ Ansoldi}
           }
$$
\vspace{1 cm}
$$
\matrix{
        \mathit{Tutore}
        &
        \hspace{1cm}
        &
        \hbox{\it Co--tutore}
        \cr
        \mathrm{\bf Prof.\ Tullio\ Weber}
        &
        &
        \mathrm{\bf Dott.\ Euro\ Spallucci}
        \cr
        \hbox{\sl Universit\`a di Trieste}
        &
        &
        \hbox{\sl Universit\`a di Trieste}}
$$
\vspace{1 cm}
$$
    \matrix{
            \mathit{Coordinatore} \cr
            \mathrm{\bf Prof.\ Paolo\ Schiavon} \cr
            \hbox{\sl Universit\`a di Trieste}
           }
$$
\vspace{1 cm}
$$
    \psline(-6,0)(6,0)
$$
\end{titlepage}
\newpage
\thispagestyle{empty}
\noindent
\cleardoublepage

%% file: notation.tex
\makechaptermyhead{Notations}{I}
\pageheader{}{Notations.}{}

\noindent

\begin{tabular}{ll}
$'$ & derivative with respect to the parameter $s$\\
$'$ & Antisymmetric Derivative on the Boundary\\
$\oint$ & Integral over a manifold without Boundary\\
$[\mathcal{D} \dots]$ & Functional Measure\\
$\dot{\ }$ & Antisymmetric Derivative on the Bulk\\
$\wedge$ & Exterior product of Forms\\
$\partial$ & Boundary\\
$\partial$ & Partial Derivative\\
$\delta$ & Functional Derivative\\
$\delta$ & Variation\\
$\mathcal{W}$ & World Sheet of a String\\
$\Wsp$ & World Hypertube of a $p$-brane\\
$\Xi$ & Parameter Space of a String\\
$\Xip$ & Parameter Space of a $p$-brane\\
$\Sigma$ & Parameter Space of a String\\
$\Sigmap$ & Parameter Space of a $p$-brane\\
$\T$ & Target Space\\
$\Gamma$ & Boundary of the Parameter Space\\
$\gamma$ & Boundary of $\Xi$\\
$\Sf ^{1}$ & Circumpherence \\
$\Sf ^{D}$ & D-dimensional Sphere \\
$C$ & Loop in SpaceTime\\
$\BB$ & Boundary of $\Sigmap$\\
$\Dp$ & $p$-closed Surface in SpaceTime\\
$\Lag$ & Lagrangian Density\\
$\Ham$ & Hamiltonian Density\\
\end{tabular}

\begin{tabular}{ll}
$\forall$ & for all\\
$\exists$ & exists \\
$\in$ & is an element of\\
$\subset$ & is a subset of\\
$\left [ \mathit{expression} \right \rceil _{\dots}$ & {\it expression} evaluated for \dots\\
$\left . \mathit{expression} \right \rceil _{\dots}$ & {\it expression} evaluated for \dots\\
$\left . \right \rceil _{\mathit{condition}}$ & restricted to {\it condition}\\
$\mathbb{M} ^{D}$ & $D$-dimensional Minkowski Space\\
$\dete{\dots}$ & Determinant of \dots\\
$\leftrightarrow$ & Bijection\\
$\square ^{\dots | \dots}$ & $|$ distinguishes internal from spacetime indices\\
$\Delta _{\bs{g}}$ & Laplacian with respect to a metric $\bs{g}$\\
\end{tabular}
\begin{itemize}
    \item Closed indices are summed over.
    \item Vectors tensors and forms, when the indeces are not explicitly written,
    are typed in {\bf boldface}.
    \item Functions have arguments enclosed in round brackets.
    \item Functionals have arguments enclosed in square brackets.
    \item All the quantities for which we give an explicit definition
    are typed in {\sl slanted}, after the definition.
\end{itemize}

%% file: thanks.tex
\makechaptermyhead{Acknowledgments}{I$\!$I}
\pageheader{}{Acknowledgments.}{}

\noindent

I would like to heartily thank Dr. Euro Spallucci for the
valuable and friendly support that he gave me as co-tutor during my Ph.D..

I am also indebted with Ms. Alessandra Richetti and Ms. Rosita Glavina,
System Manager and Secretary at the Department of Theoretical Physics of the
University of Trieste, for their professional
help in a lot of practical problems.

%% file: intro.tex
\pageheader{}{Introduction.}{}

\makechaptermyhead{Introduction}{I$\!$I$\!$I}

\begin{start}
``$\euf{F}$uture?''\\
``Difficult to see.\\
Always in motion is the future.''\\
\end{start}

\noindent{}Subject of this thesis is the study of a closed bosonic string:
the treatment will follow an approach {\it alternative} with
respect to the traditional one and stems from a research line
pursued in by Eguchi.

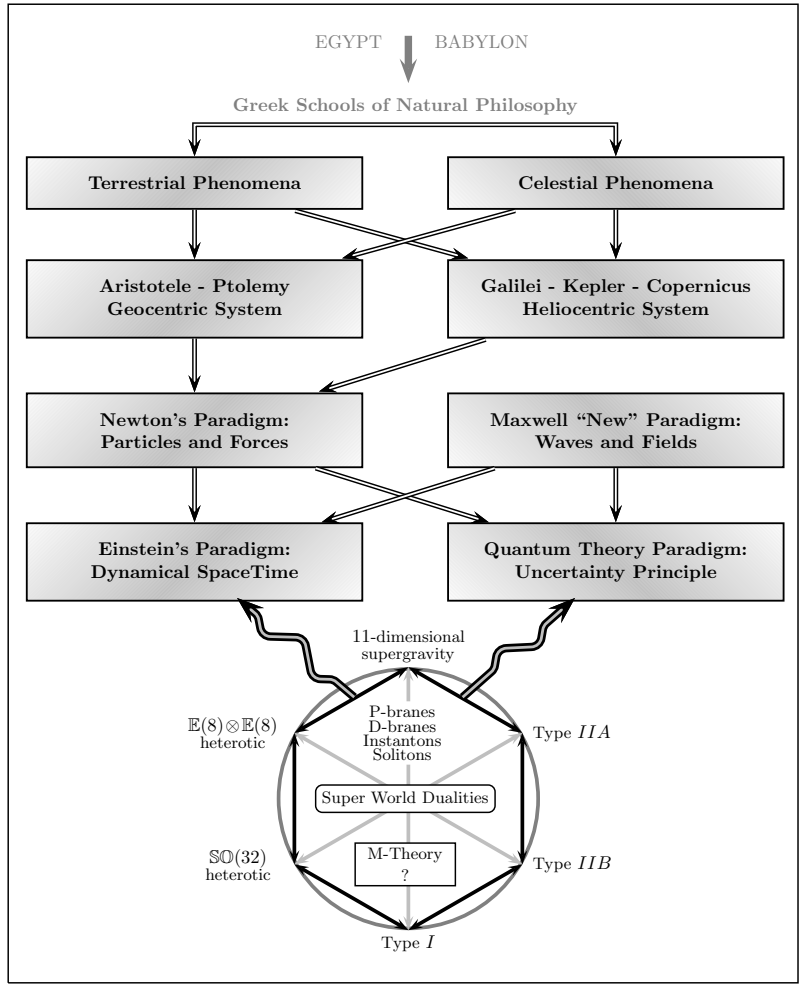
\begin{figure}
    \begin{center}
        \input{fig01}
    \end{center}
    \caption{History of physics shows that conflicting theories eventually
    merge into a broader and deeper synthesis. Will $M$-Theory lead to a UNIQUE
    super synthesis of Quantum Theory, Gravity Theory and supersymmetry?}
\label{fig}
\end{figure}

In particular we will follow an approach
based on an action functional which is
not the  Nambu-Goto proper area of the string world--sheet
but its ``square'', i.e. the Schild action. Starting
from this action we formulate the dynamics of a $1$-(or $p$-)dimensional
extended object following the same procedures and techniques, which
characterize the Quantum Mechanics of point particles. Of course,
since we are dealing with a $1$-dimensional object and not with a
pointlike one, there are some problems in adapting the formalism
and interpreting the results: we tried to solve these problems and
we remarked the relevance  of some results,
which are strictly related to the extended nature of the object
under study.

A second goal that we will pursue in this work is to
give a mathematically consistent formulation of our alternative
description of string dynamics; we will discuss in some depth
the formal ``tricks'' needed to carry on some calculations and the
underlying mathematics. This is our main motivation
in presenting {\it detailed computation,} whenever it is
possible, in order to give the reader a better understanding of the
the results themselves and the way they are derived.

It is also worth pointing out that all the results, that we will
get for the closed string, can be generalized to higher dimensional
closed extended objects ($p$-{\sl{}branes}) without facing  any
further technical problems. This is important, since we are
given the opportunity to focus the simpler
$1$-{\sl{}brane}$\, \equiv \,$string case,
without any loss of generality but with the smaller technical effort.

Shortly presented our goals are the following ones:
\begin{enumerate}
    \item to formulate a Classical Theory of relativistic
    extended objects as a natural generalization of the
    Hamilton--Jacobi Theory
    for point particles;
    \item to quantize the Theory using path--integrals and
    deriving the {\it propagator} as well as the corresponding
    {\it functional wave
    equation}; in other words, we will {\it quantize the Theory in the
    Schr\"odinger representation};
    \item to provide some explicit solutions of the {\it stringy
    Schr\"odinger equation}, with particular emphasis
    on the new characteristic features following from the $1$-dimensional
    extension of the string; these results can also be extended, of course,
    to higher dimensional objects;
    \item to get a better understanding of the general results
    by considering  the particular case of a circular
    string, where, some problems and their solutions are more transparent;
    \item last but not least, to provide an alternative
    interpretation of the
    dynamics of an extended object as a {\it stochastic shape deformation
    process}.
\end{enumerate}
To reach these goals we will use, among others, the following mathematical
tools:
\begin{itemize}
    \item a generalized  Hamilton--Jacobi formulation of the dynamics
    of fields and relativistic  extended objects;
    \item functional integral techniques;
    \item {\it NonStandard Analysis} techniques.
\end{itemize}
In order to provide the reader with an as self contained text
as possible,
the first and the third items will be treated in some detail
in chapters
\ref{1.toeholds}-\ref{2.hamjac} and in appendices
\ref{B.nonstanda}-\ref{C.nonstopro}.

At this stage, we hope that the curious and interested reader would
be asking himself the reason why we propose this {\it non--standard}
formulation for string dynamics. The shortest and probably most general
answer is already present in the very famous book of Green, Schwartz and
Witten.
so that our hope is that a {\it non--standard} approach could be
not an alternative, but a complementary one. With this hope we will
also devote chapter \ref{9.connection} to link our approach with the more
traditional one, which nowadays has a central role in high energy
theoretical physics.  In our opinion, research in Theoretical High
Energy Physics has now the meaning of
investigating about the very nature of
mass and energy, and ultimately about the structure of space and time
themselves.
It may even be argued that the whole history of Physics, to a
large extent, represents the history of the ever changing
notion of space and time in response to our ability to probe
infinitesimally small distance scales as well as larger and
larger cosmological distances.\\
The ``flow chart'' in Figure \ref{fig} summarizes the dialectic
process which has led, through nearly twenty five hundred
years of philosophical speculation and scientific inquiry,
to the current theoretical efforts in search of a
super-synthesis of the two  conflicting paradigms of $20$th
century physics, namely, the Theory of General Relativity and
Quantum Theory. In that Hegelian perspective of the history
of physics, such a super synthesis is regarded by many as the
``holy grail'' of contemporary High Energy Physics. However,
the story of the many efforts towards the formulation of that
synthesis, from supergravity to superbranes, constitutes, in
itself, a fascinating page in the history of theoretical
physics at the threshold of the new millennium. The early
`$80$s excitement about String Theory (``{\it{}The
First String Revolution}'') followed from the prediction that
only the gauge groups $\mathbb{SO} \left( 32 \right)$
and $\mathbb{E} \left( 8 \right) \otimes \mathbb{E} \left( 8 \right)$
provide a quantum mechanically consistent, i.e.,
anomaly free, unified Theory which includes Gravity \cite{gs},
and yet is capable, at least in principle, of
reproducing the standard Electro--Weak Theory below the GUT
scale. However, several fundamental questions were left
unanswered. Perhaps, the most prominent one regards the
choice of the compactification scheme required to bridge the gap between
the multi--dimensional, near--Planckian string--world, and
the low energy, $4$-dimensional universe we live in \cite{nara}.
Some related
problems, such as the vanishing of the cosmological constant (is it
really vanishing, after all?) and the breaking of supersymmetry
were also left without a satisfactory answer. The common feature of all
these unsolved problems is their intrinsically {\it non--perturbative}
character. More or less ten years after the
First String Revolution, the second one, which is still in
progress, has offered a second important clue into the nature
of the superworld. The diagram in Figure \ref{fig} encapsulates the
essential pieces of a vast mosaic out of which the final
Theory of the superworld will eventually emerge. Among those
pieces,  the six surviving viable supermodels known at
present, initially thought to be candidates for the role of a
fundamental Theory of Everything, are now regarded as
different asymptotic realizations, linked by a web of
dualities, of a unique and fundamentally new paradigm of
physics which goes under the name of {\it $M$-Theory}.
The essential components of this
underlying Matrix Theory appear to be string--like objects as
well as other types of extendons, e.g.,
$p$-{\sl{}branes}, $D$-{\sl{}branes}, \dots{}. Moreover, a
new computational approach is taking shape which is based on
the idea of trading off the strongly coupled regime of a
supermodel with the weakly coupled regime of a different
model through a systematic use of dualities.\\
Having said that, the fact remains that $M$-Theory, at
present, is little more than a name for a mysterious
supertheory yet to be fully formulated. In particular, we
have no clue as to what radical modification it will bring to
the notion of spacetime in the short distance regime. In the
meantime, it seems reasonable to attempt to isolate the
essential elements of such non--perturbative approach to the
dynamics of extended objects.

One such approach,
developed over the last few years
\cite{vanzetta}, \cite{noi5}, \cite{noi3},
is a refinement of an early formulation of {\it Quantum String Theory} by
T.Eguchi \cite{egu}, elaborated by following a formal analogy with a
Jacobi--type formulation of the {\it Canonical Quantization of Gravity}.
The relevance of this approach can be traced down to
an intriguing similarity that we can understand between the problem of
quantizing gravity, as described by General Relativity, and that of
quantizing a relativistic string, or any higher dimensional
relativistic extended object. In either case, one can follow
one of two main routes:
\begin{enumerate}
    \item a Quantum Field Theory
    inspired {\it covariant quantization};
    \item a {\it canonical quantization} of the Schr\"odinger type.
\end{enumerate}
The basic idea underlying the covariant approach is to
consider the metric tensor $g _{\mu\nu}(x)$ as an ordinary
matter field and follow the standard quantization procedure, namely,
to Fourier analyze small fluctuation around a classical
background configuration and give the Fourier coefficients the
meaning of creation/annihilation operators of the
gravitational field quanta, the {\it  gravitons}. In the
same fashion, quantization of
the string world--sheet fluctuations leads to a whole
spectrum of particles with different values of mass and spin:
these is a local, short scale and perturbative approach.
Put briefly,
\begin{eqnarray}
&& g_{\mu\nu}(x)=\hbox{background}\,+\,\hbox{``graviton''}\nonumber\\
&& X^{\mu}(\tau,s)=\hbox{zero--mode}\,+\,\hbox{particle spectrum}\nonumber
\quad .
\end{eqnarray}
Against this background, one may elect to forgo the full covariance
of the Quantum Theory of Gravity in favor of the more restricted
symmetry under transformations preserving the
``{\it canonical spacetime slicing}'' into a $1$-parameter family
of spacelike $3$-surfaces. This splitting of space and
time amounts to selecting the spatial components of the
metric, modulo $3$-space reparametrizations,
as the gravitational degrees of freedom to be quantized. This
approach focuses on the quantum mechanical description of the
space itself, rather than the corpuscular content of the
gravitational field. Then, the quantum state of the spatial
$3$-geometry is controlled by the Wheeler--DeWitt equation
\beq
    \left[
        \mathrm{Wheeler-DeWitt\ operator}
    \right]
    \Psi \left[ G _{3} \right]
    =
    0
\label{wd}
\eeq
and the wave functional $\Psi \left[ G _{3} \right]$,
{\it the wave function of the universe},
assigns a probability amplitude to each allowed three
geometry.
We note that in the case of extended objects (strings) the main route
is constituted by the previous one. So to understand in more detail the
differences in the described approaches, let us concentrate for a
while on the case of Gravity. In this case we can understand
that the relation between the two quantization
schemes is akin to the relationship in particle dynamics between
{\it first quantization}, formulated in terms of single particle
wave functions along with the corresponding Schr\"odinger
equation, and the {\it second quantization} expressed in terms of
creation/annihilation operators along
with the corresponding field equations.
Thus, covariant Quantum Gravity is, {\it conceptually}, a second
quantization framework for calculating amplitudes, cross sections,
mean life, etc., for any physical process involving graviton exchange.
Canonical Quantum Gravity, on the other hand, is a Schr\"odinger--type
first quantization framework, which assigns a probability amplitude
for any allowed geometric configuration of three dimensional
physical space. It must be emphasized that there is no immediate
relationship between the graviton field and the wave function
of the universe. Indeed, even if one elevates
$\Psi \left[ G _{3} \right]$ to the role of {\it field operator}, it would
create or destroy  {\it entire $3$-surfaces} instead of single gravitons.
In a more pictorial  language, the
wave functional $\Psi \left[ G_{3} \right]$ becomes a quantum operator
creating/destroying full universes!
Of course, as far as gravity is concerned,
any quantization scheme is affected
by severe problems: perturbative covariant quantization of General
Relativity is not renormalizable, while the
intrinsically non--perturbative Wheeler--DeWitt equation can
be solved only under extreme simplification such as the mini--superspace
approximation; thus probably these shortcomings provided the
impetus toward the formulation of string Theory first,
and Superstring Theory
then, as the only consistent  quantization scheme which accomodates the
graviton in its (second quantized) particle spectrum.
Indeed the key observation that the string position
$X ^{\mu} \! \left( \sigma ^{0} , \sigma ^{1} \right)$ and momentum
$P _{\tau} ^{\mu} \! \left( \sigma ^{0} , \sigma ^{1} \right)$
are dependent on
two variables let to the conclusion that the quantization procedure,
which rises $X ^{\mu}$ and $P ^{\mu}$ to the role of operators, results
in a $2$-dimensional second quantized point particle Quantum Field
Theory \cite{hatfield}. The non--linearity of the Theory is a major
difficulty in this approach since the equations of motions, arising
from the Nambu-Goto action, present intractable computational difficulties.
The way out from this problem is the {\it choice} of a gauge for our
description. In this way constraints are imposed on the {\it gauge freedom}
present in the Theory; their interpretation is then clarified
starting from the Polyakov action and showing how they arise
{\it gauging away} the parameter space metric $g _{ab}$. Canonical
Quantization can then be performed.
Thus, according to the prevalent way of thinking, there is no
compelling reason, nor clear cut procedure to formulate a first quantized
Theory
(i.e. a Quantum Mechanics) of relativistic extended objects. In the case of
strings, this attitude is also deeply rooted in the conventional
interpretation of the world--sheet coordinates
$X ^{\mu} \! \left( \sigma ^{0} , \sigma ^{1} \right)$ as a ``multiplet of
scalar fields'' defined over a $2$-dimensional manifold, $\Sigma$,
covered by the $\ttt{ \sigma ^{0} , \sigma ^{1} }$
coordinate mesh. According to
this point of view, quantizing a relativistic string is formally equivalent
to quantizing a $2$-dimensional Field Theory, bypassing a preliminary
quantum mechanical formulation. However, there are at least two
objections against this kind of reasoning. The first follows from
the analogy between the canonical formulation of General
Relativity and $3$-{\sl{}brane} dynamics, and the second
objection follows from the
``Schr\"odinger representation'' of Quantum Field Theory.
More specifically:
\begin{enumerate}
    \item the Wheeler--DeWitt equation can be interpreted, in a modern
    perspective, as the wave equation for the orbit of a
    relativistic $3$-{\sl{}brane}. In this perspective, then, why not conceive of a
    similar equation for a $1$-{\sl{}brane}?
    \item the functional
    Schr\"odinger representation of Quantum Field Theory assigns a
    probability amplitude to each field configuration over a
    spacelike slice $t = \mathrm{const.}$, and the corresponding wave
    function obeys a functional Schr\"odinger--type equation.
\end{enumerate}
Pushing the above arguments to their natural conclusion, we
are led to entertaining the interesting possibility of
formulating a {\it functional}
Quantum Mechanics for strings and other $p$-{\sl{}branes}. This
approach has received scant attention in the mainstream work
about Quantum String
Theory, presumably because it requires an explicit breaking
of the celebrated reparametrization invariance, which is
the distinctive symmetry of relativistic extended objects.\\
All of the above reasoning leads us to the central question
that we wish to analyze, namely: is there any way to
formulate a reparametrization invariant String Quantum Mechanics?\\
As a matter of fact, a possible answer was suggested by T.Eguchi
as early as $1980$ \cite{egu}, and our own
elaboration of that quantization
scheme \cite{noi} is the topic of this thesis.
We will in particular focus our attention in what follows on the string
as a whole, i.e. as a physical system by itself, restricting our
attention to the case of {\bf closed} strings. Moreover we try to
generalize in a peculiar way what happens for a particle,
i.e, unlike Superstring Theory,
our formulation represents an attempt to construct a
{\it Quantum Mechanical} Theory of (closed) strings
in analogy to the familiar case of point particles.
The ground work of this approach is developed extending
the Hamilton--Jacobi formulation and Feynman's path--integral approach.
Indeed a particle is a $0$-dimensional manifold, and when it evolves
in time it spans (at least classically) a $1$-dimensional manifold,
i.e. its world--line; the time evolution can be described in terms of the
world--line proper length. The way in which we reformulate this observation
for a string is as follows:
{\it a (closed) string is a $1$-dimensional closed manifold, and when
it evolves in ``$\,${\bf{}time}$\,$\footnote{It is worth to use with care
this word!}'' it spans (at least classically) a $2$-dimensional
manifold, i.e. its world--sheet}.
Our implementation of the analogy with the point particle is then
reflected by our unconventional choice of dynamical
variables for the string, namely, the
areas enclosed by the projections of the string loop
onto the coordinate planes and its $2$-form conjugate
momentum. Also central to our own quantum mechanical
approach, is the choice of ``time variable'', which we take
to be the area of the parameter space associated with the
string world--sheet, in analogy to
the point particle case.\\
Thanks to the previous pointed out analogy between our
proposal for string quantization and the canonical quantization of the
Schr\"odinger type for Quantum Gravity, we think that our method
could led to some deeper insight into the problems related to the
structure of spacetime at small scales. Indeed, since the advent of
Quantum Theory and General Relativity, the notion of spacetime as a
preexisting manifold in which physical events take place, is
undergoing a process of radical revision. Thus, reflecting on
those two major revolutions in physics of this century, Edward Witten
writes \cite{witten}, ``{\it Contemporary developments in
theoretical physics suggest that another revolution may be in progress,
through which a new source of ``fuzziness'' may enter physics, and
spacetime itself may be reinterpreted as an approximate, derived
concept}$\,$''. The new source of fuzziness comes from String Theory,
specifically from the introduction of the new fundamental constant,
$\alpha'$, which determines the tension of the string.
Thus, at scales comparable to $(\alpha') ^{1/2}$, spacetime becomes
fuzzy, even in the absence of conventional
quantum effects ($\hbar \simeq 0$).
While the exact nature of this  fuzziness is unclear, it manifests
itself in a new form of Heisenberg's principle, which now depends on both
$\alpha'$ and $\hbar$. Thus, in Witten's words, while ``{\it a proper
theoretical framework for the {\rm [new]}
uncertainty principle has not yet
emerged, {\rm [ \dots ]} the natural framework
of the {\rm [String]} Theory may
eventually prove to be inherently quantum mechanical}$\,$''.\\
The essence of the above remarks, at least in our interpretation, is
that there may exist different degrees of fuzziness in the
{\it making} of spacetime, which set in at various scales of
length, or energy, depending on the nature and resolution of the
Heisenberg microscope used to probe its structure. In other words,
{\it spacetime becomes a sort of dynamical variable, responding to quantum
mechanical resolution just as, in General Relativity, it
responds to mass--energy}. The response of spacetime to mass--energy is
curvature. Its response to resolution seems to be ``fractalization''.\\
Admittedly, in the above discussion, the term ``fuzziness'' is loosely
defined, and the primary aim of chapter \ref{7.fractal} is to suggest a
precise measure of the degree of fuzziness of the quantum mechanical
path of a string. In order to achieve this objective, we need two
things:
\begin{enumerate}
    \item the new form of the uncertainty principle for strings;
    \item the explicit form of the wave--packet for string loops.
\end{enumerate}
Then, we will be able to compute the Hausdorff dimension of a
quantum string and to identify the parameter which controls the
transition from the smooth phase to the fractal phase.\\
It seems worth emphasizing at this introductory stage, before we
embark on a technical discussion, that we are interested primarily in the
analysis of the quantum fluctuations of a string loop. By quantum fluctuations,
we mean a random transition, or quantum jump, between different string
configurations. Since in any such process, the {\it shape} of
the loop changes, we refer to it as a ``shape--shifting'' process. We
find that any such process, random as it is, is subject to an extended
form of the Uncertainty Principle (we call it
{\it Shape Uncertainty Principle}), which forbids the exact,
simultaneous knowledge of the string shape and its area conjugate
momentum. The main consequence of the Shape Uncertainty Principle is the
``fractalization'' of the string orbit in spacetime. The degree of
fuzziness of the string world--sheet will be  measured by its
Hausdorff dimension. We will also try to
quantify the transition from the classical,
or smooth phase, to the quantum, or fractal phase.
A deeper insight into the meaning of the ``fractalization'' process
and its physical interpretation can also be gained in our opinion,
by realizing that the fractal properties of the quantum dynamics of a string
can be interpreted using a stochastic process, which we will call
an {\it Areal Brownian Motion}. This process describes the shape--shifting
process, which is string dynamics, as a fractal modification of the form
of the object due to random fluctuations of its shape.

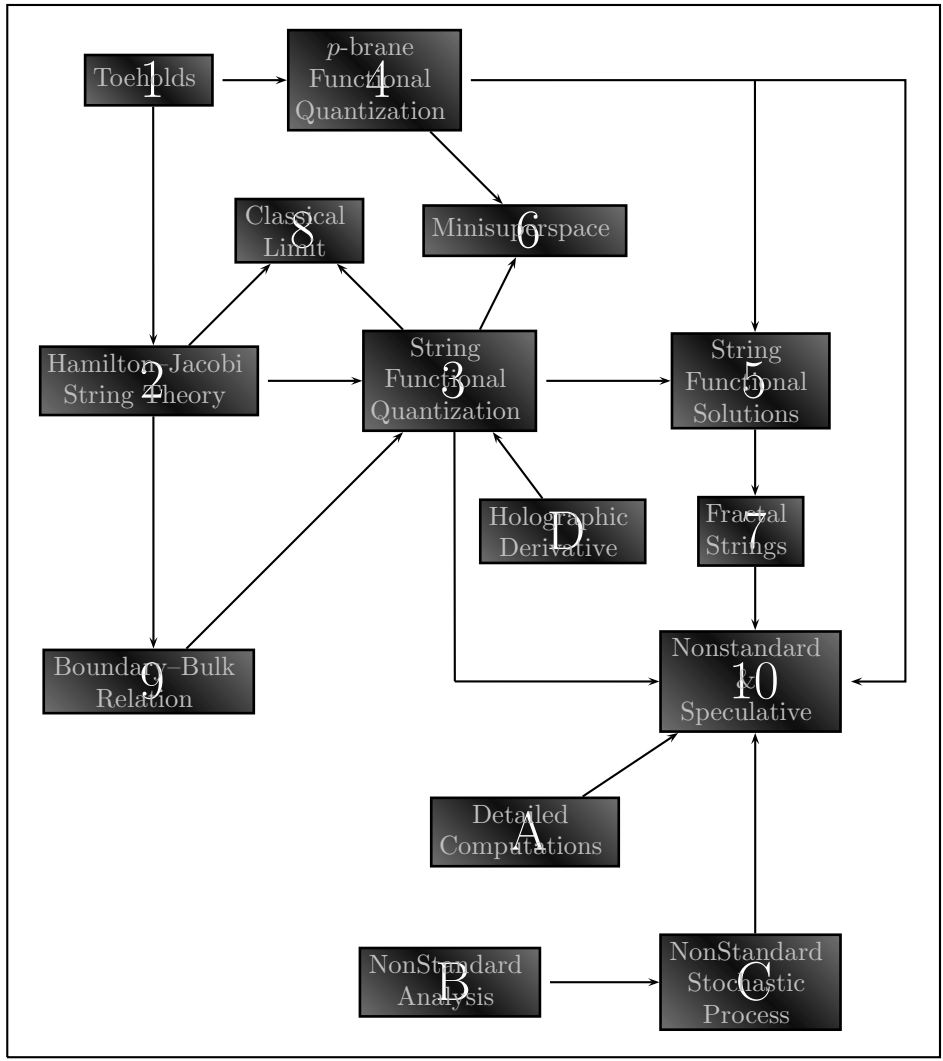
\begin{figure}
    \begin{center}
        \input{fig02}
    \end{center}
    \caption{Logical connection between chapters of the Thesis.}
\label{i.layout}
\end{figure}

Accordingly, the Thesis is structurted as follows\footnote{The
logical interdependence among chapters is shown in figure
\ref{i.layout}.}.

In chapter \ref{1.toeholds} we emphasize the relation between
a closed $1$-dimensional manifold considered as the
only Boundary of a $2$-dimensional simply connected domain (Bulk).
We take here the opportunity to point out the relevance of this passage
from the Bulk to the Boundary defining precisely all the necessary
quantities: this will be relevant in the following chapters. Moreover
a short presentation of the Functional Schr\"odinger Quantization,
that we will take later as a basis for our procedure, is given.

After these introductory, but necessary, topics we address in chapter
\ref{2.hamjac} the problem of describing the Classical Dynamics of
an extended object from the point of view of what we will call its
{\it Boundary Shadow Dynamics}. After the definition of the relevant
quantities we solve this problem using a Functional Hamilton--Jacobi
Formulation. Two different derivations of this result are given, as a
consistency check, and we see how the relevant property of our
formulation turns out to be the fact that it is a Dynamics of {\it Area}.
After the presentation of a classical effective areal equivalent
formulation we also show some possible generalization of the basic concepts
that will be useful later on.

Now that we have a good starting point, namely the Classical Functional
theory, we perform in chapter \ref{3.strfunqua} its quantization: in
particular we will derive the equation for the propagtor that describes
the evolution of a {\it closed bosonic}string as well as the solution to
this equation. Again two different procedures are shown to give the same
result, which reassures us about the correctness of the result. After
an interestiong by--product of our procedure, a functional representation
of the Nambu--Goto string propagatorhus, we turn to the problem of
defining a string quantum state and the corresponding equation for
the probability amplitude, which we will call the
{\it String Functional Schr\"odinger Equation}.

Before proceeding on this road by finding some solutions to this equation,
we present again the procedure, given for the string in the last two
chapters, generalizing it to the case of higher dimensional objects
($p$-branes). This is the topic of chapter (\ref{4.pbrcha}).

Then we proceed in chapter \ref{5.strfunsol} to the derivation
of some solutions to the {\it String Schr\"odinger Equation}, finding
the Plane Wave and the Gaussian Wave Packet. The most interesting properties
of these solutions are also pointed out.

We can so spend chapter \ref{6.mincha} to give a closer look at the
meaning of the results derived before in a very special case, which, by
analogy with similar procedures used in Quantum Gravity, we will call
{\it Minisuperspace Approximation}. Here a drastic reduction of the
number of degrees of freedom of the problem is performed assuming a
circular symmetry for the object. This gives us the opportunity of
making some comments about a very common problem in Quantum Theories,
namely the problem of ordering ambiguities.

Chapter \ref{7.fractal}, which follows,
is a first interesting result that we can deduce
from the work done so far. From the discussion of
the String Functional Schr\"odinger Equation
and its solutions we derive the
Uncertainty Principle for strings as well as its principal
consequence, namely, the fractalization of the string
quantum evolution. This is strongly related to the fractalization
of Quantum SpaceTime.

We then begin in chapter \ref{8.douclalimcha} a process that
will bring the procedure of String Quantization presented in this
Thesis under a new light: here we clarify its relation with
the gauge Theory of the String Geodesic Field, i.e. we show that
in aproperly chosen limit we can recover a Field Theory equivalent
to classical string dynamics.

Then in chapter \ref{9.connection} we show how our formulation can be
given a connection to the usual Polyakov procedure of string quantization:
this is an intersting result that enlightens the properties of
Boundary Dynamics {\it versus} Bulk Dynamics and gives a useful hints
about possible connections with Loop Quantum Gravity and large $N$ QCD.

Finally chapter \ref{10.nonstacha} is devoted to the mathematical fundation
of the {\it Functional Operations} we performed above. Using an interesting
mathematical tool, namely NonStandard Analysis, we show how it is possible
to give rigorous definitions to all the formal steps we took in previous
chapters. We also present two more speculative sections, one about possible
connections with $M$-Theory, the
other about a possible description of the structure of spacetime
in terms of an effective lagrangian based on a covariant,
functional extension of the Ginzburg--Landau model of
superconductivity.

Four appendices follow the main tratment:
\begin{itemize}
    \item in appendix \ref{A.detcal} we present some detailed calculations
    which we skipped in chapter \ref{4.pbrcha};
    \item in appendix \ref{B.nonstanda} we give a brief introduction to
    NonStandard Analysis with the aim of introducing the Ultra Euclidean
    Space;
    \item in appendix \ref{C.nonstopro} we define a Stochastic
    Process on aparticular space defined in terms of NonStandard Analysis;
    \item in appendix \ref{D.looderapp} we give a more detailed discussion
    about Functional and Holographic Derivatives, which are the principal
    quantities in writing our functional equations.
\end{itemize}

Eventought we tried to give an as self contained as possible treatment,
we were not able to cover all the prerequisites in full depth and we
refer the interested reader to the bibliography for further readings.

%% file: fig01.tex
\scalebox{.7}{
\psset{framesep=10pt}
\begin{pspicture}(0.4,0)(15.4,18.6)
\psline(0.4,0)(0.4,18.6)(15.4,18.6)(15.4,0)(0.4,0)
%
%
\pscircle[linewidth=2pt,linecolor=gray](8,3.5){2.5}
\psline[linewidth=2pt,linecolor=lightgray]{<->}(8,6)(8,1)
\rput[c](8,4.75)
                {\psframebox[linestyle=none,linecolor=black,framesep=2pt,fillstyle=solid,fillcolor=white]
                     {\begin{minipage}{2.5cm}
                      \begin{center}
                         {\small P-branes}\\[-1.5mm]
                         {\small D-branes}\\[-1.5mm]
                         {\small Instantons}\\[-1.5mm]
                         {\small Solitons}
                      \end{center}
                      \end{minipage}
                     }
                }
\psline[linewidth=2pt,linecolor=lightgray]{<->}(10.165,4.75)(5.835,2.25)
\psline[linewidth=2pt,linecolor=lightgray]{<->}(10.165,2.25)(5.835,4.75)
\rput[c](8,3.5){\psframebox[fillstyle=solid,fillcolor=white,framesep=3pt,linecolor=black,framearc=.5]
                            {\small Super World Dualities}
               }
\rput[c](8,2.25)
                {\psframebox[framesep=2pt,linecolor=black,fillstyle=solid,fillcolor=white]
                            {
                             \begin{minipage}{1.5cm}
                             \begin{center}
                                 {\small M-Theory}\\
                                 {\small ?}
                             \end{center}
                             \end{minipage}
                            }
                }
\rput[r](5.635,4.75)
         {
          \begin{minipage}{1.7cm}
          \begin{center}
              $\mathbb{E} (8) \otimes \mathbb{E} (8)$\\[-1mm]
              {\small heterotic}
          \end{center}
          \end{minipage}
         }
\rput[r](5.635,2.25)
         {
          \begin{minipage}{1.5cm}
          \begin{center}
              $\mathbb{SO} (32)$\\[-1mm]
              {\small heterotic}
          \end{center}
          \end{minipage}
         }
\rput[t](8,.9){{\small Type} $I$}
\rput[l](10.365,2.25){{\small Type} $IIB$}
\rput[l](10.365,4.75){{\small Type} $IIA$}
\rput[b](8,6.1)
         {
          \begin{minipage}{3cm}
          \begin{center}
              $11${\small{}-dimensional}\\[-1mm]
              {\small supergravity}
          \end{center}
          \end{minipage}
         }
\psline[linewidth=2pt]{<->}(8,6)(10.165,4.75)
\psline[linewidth=2pt]{<->}(10.165,4.75)(10.165,2.25)
\psline[linewidth=2pt]{<->}(10.165,2.25)(8,1)
\psline[linewidth=2pt]{<->}(8,1)(5.835,2.25)
\psline[linewidth=2pt]{<->}(5.835,2.25)(5.835,4.75)
\psline[linewidth=2pt]{<->}(5.835,4.75)(8,6)
\rput[bc](4,8){\rnode[c]{einstein}
                        {
                         \psframebox[fillstyle=gradient,gradbegin=white,gradend=lightgray,gradangle=45,linewidth=1pt,gradmidpoint=.5]
                                    {
                                     \parbox[b]{5.4cm}
                                               {
                                                \centerline{{\bf Einstein's Paradigm:}}
                                                \centerline{{\bf Dynamical SpaceTime}}
                                               }
                                    }
                        }
              }
\rput[bc](12,8){\rnode[c]{quantum}
                        {
                         \psframebox[fillstyle=gradient,gradbegin=white,gradend=lightgray,gradangle=-45,linewidth=1pt,gradmidpoint=.5]
                                    {
                                     \parbox[b]{5.4cm}
                                               {
                                                \centerline{{\bf Quantum Theory Paradigm:}}
                                                \centerline{{\bf Uncertainty Principle}}
                                               }
                                    }
                        }
              }
\rput[bc](4,10.5){\rnode[c]{newton}
                        {
                         \psframebox[fillstyle=gradient,gradbegin=white,gradend=lightgray,gradangle=45,linewidth=1pt,gradmidpoint=.5]
                                    {
                                     \parbox[b]{5.4cm}
                                               {
                                                \centerline{{\bf Newton's Paradigm:}}
                                                \centerline{{\bf Particles and Forces}}
                                               }
                                    }
                        }
              }
\rput[bc](12,10.5){\rnode[c]{maxwell}
                        {
                         \psframebox[fillstyle=gradient,gradbegin=white,gradend=lightgray,gradangle=-45,linewidth=1pt,gradmidpoint=.5]
                                    {
                                     \parbox[b]{5.4cm}
                                               {
                                                \centerline{{\bf Maxwell ``New'' Paradigm:}}
                                                \centerline{{\bf Waves and Fields}}
                                               }
                                    }
                        }
              }
\rput[bc](4,13){\rnode[c]{aristotele}
                        {
                         \psframebox[fillstyle=gradient,gradbegin=white,gradend=lightgray,gradangle=45,linewidth=1pt,gradmidpoint=.5]
                                    {
                                     \parbox[b]{5.4cm}
                                               {
                                                \centerline{{\bf Aristotele - Ptolemy}}
                                                \centerline{{\bf Geocentric System}}
                                               }
                                    }
                        }
              }
\rput[bc](12,13){\rnode[c]{galilei}
                        {
                         \psframebox[fillstyle=gradient,gradbegin=white,gradend=lightgray,gradangle=-45,linewidth=1pt,gradmidpoint=.5]
                                    {
                                     \parbox[b]{5.4cm}
                                               {
                                                \centerline{{\bf Galilei - Kepler - Copernicus}}
                                                \centerline{{\bf Heliocentric System}}
                                               }
                                    }
                        }
              }
\rput[bc](4,15.2){\rnode[c]{terrestrial}
                        {
                         \psframebox[fillstyle=gradient,gradbegin=white,gradend=lightgray,gradangle=45,linewidth=1pt,gradmidpoint=.5]
                                    {
                                     \parbox[b]{5.4cm}
                                               {
                                                \centerline{{\bf Terrestrial Phenomena}}
                                               }
                                    }
                        }
              }
\rput[bc](12,15.2){\rnode[c]{celestial}
                         {
                         \psframebox[fillstyle=gradient,gradbegin=white,gradend=lightgray,gradangle=-45,linewidth=1pt,gradmidpoint=.5]
                                    {
                                     \parbox[b]{5.4cm}
                                               {
                                                \centerline{{\bf Celestial Phenomena}}
                                               }
                                    }
                        }
              }
\rput[tr](7.5,18){{\gray EGYPT}}
\rput[tl](8.5,18){{\gray BABYLON}}
\pnode(6.9175,5.475){circlel}
\pnode(9.0825,5.475){circler}
\psline[linewidth=4pt,linecolor=gray]{->}(8,18)(8,17.1)
\ncangle[armA=6mm,angleA=90,armB=6mm,angleB=90,doubleline=true,doublesep=1pt]{<->}
        {terrestrial}{celestial}\Aput[3pt]{{\gray \bf Greek Schools of Natural Philosophy}}
\ncline[nodesep=.5pt,doubleline=true,doublesep=1pt]{->}{terrestrial}{galilei}
\ncline[doubleline=true,doublesep=1pt]{->}{terrestrial}{aristotele}
\ncline[doubleline=true,doublesep=1pt]{->}{celestial}{galilei}
\ncline[nodesep=.5pt,doubleline=true,doublesep=1pt]{->}{celestial}{aristotele}
\ncline[doubleline=true,doublesep=1pt]{->}{aristotele}{newton}
\ncline[nodesep=.5pt,doubleline=true,doublesep=1pt]{->}{galilei}{newton}
\ncline[doubleline=true,doublesep=1pt]{->}{newton}{einstein}
\ncline[doubleline=true,doublesep=1pt]{->}{newton}{quantum}
\ncline[doubleline=true,doublesep=1pt]{->}{maxwell}{einstein}
\ncline[doubleline=true,doublesep=1pt]{->}{maxwell}{quantum}
\nczigzag[nodesep=-2pt,doubleline=true,doublecolor=lightgray,doublesep=1.5pt,linearc=2mm,coilwidth=5mm,coilheight=2.5,coilaspect=5,linewidth=1.5pt]{<-}{einstein}{circlel}
\nczigzag[nodesep=-2pt,doubleline=true,doublecolor=lightgray,doublesep=1.5pt,linearc=2mm,coilwidth=5mm,coilheight=2.5,coilaspect=5,linewidth=1.5pt]{<-}{quantum}{circler}
\end{pspicture}
}
 

%% file: fig02.tex
\psset{framesep=2pt}
\begin{pspicture}(-1,0)(11.4,14)
\psline(-1,0)(11.4,0)(11.4,14)(-1,14)(-1,0)
%
%
\rput(1,13){\rnode{toeholds}{\psframebox[fillstyle=gradient,gradbegin=gray,gradend=black,gradangle=-45,linewidth=1pt,gradmidpoint=.5]%
                                   {\rput[c](0.8,0.1){{\white \huge 1}}
                                    \begin{minipage}{1.4cm}
                                        \begin{center}
                                            \vspace*{2mm}\hskip 2mm\\[-3mm]
                                            {\lightgray Toeholds}\\[-3.5mm]
                                            \vspace*{1mm}\hskip 1mm
                                        \end{center}
                                    \end{minipage}
                                   }
                            }
           }
\rput(1,9){\rnode{hjst}{\psframebox[fillstyle=gradient,gradbegin=gray,gradend=black,gradangle=-45,linewidth=1pt,gradmidpoint=.5]%
                                   {\rput[c](1.4,0.1){{\white \huge 2}}
                                    \begin{minipage}{2.6cm}
                                        \begin{center}
                                            {\lightgray Hamilton--Jacobi}\\
                                            {\lightgray String Theory}
                                        \end{center}
                                    \end{minipage}
                                    }
                       }
          }
\rput(5,9){\rnode{sfq}{\psframebox[fillstyle=gradient,gradbegin=gray,gradend=black,gradangle=-45,linewidth=1pt,gradmidpoint=.5]%
                                   {\rput[c](1.1,0.1){{\white \huge 3}}
                                    \begin{minipage}{2cm}
                                        \begin{center}
                                            {\lightgray String}\\
                                            {\lightgray Functional}\\
                                            {\lightgray Quantization}
                                        \end{center}
                                    \end{minipage}
                                    }
                       }
          }
\rput(4,13){\rnode{pbrfq}{\psframebox[fillstyle=gradient,gradbegin=gray,gradend=black,gradangle=-45,linewidth=1pt,gradmidpoint=.5]%
                                   {\rput[c](1.1,0.1){{\white \huge 4}}
                                    \begin{minipage}{2cm}
                                        \begin{center}
                                            {$\lightgray{}p$}{\lightgray{}-brane}\\
                                            {\lightgray Functional}\\
                                            {\lightgray Quantization}
                                        \end{center}
                                    \end{minipage}
                                    }
                       }
          }
\rput(9,9){\rnode{sfs}{\psframebox[fillstyle=gradient,gradbegin=gray,gradend=black,gradangle=-45,linewidth=1pt,gradmidpoint=.5]%
                                   {\rput[c](1,0.1){{\white \huge 5}}
                                    \begin{minipage}{1.8cm}
                                        \begin{center}
                                            {\lightgray String}\\
                                            {\lightgray Functional}\\
                                            {\lightgray Solutions}
                                        \end{center}
                                    \end{minipage}
                                    }
                       }
          }
\rput(6,11){\rnode{mini}{\psframebox[fillstyle=gradient,gradbegin=gray,gradend=black,gradangle=-45,linewidth=1pt,gradmidpoint=.5]%
                                   {\rput[c](1.3,0.1){{\white \huge 6}}
                                    \begin{minipage}{2.4cm}
                                        \begin{center}
                                            \vspace*{2mm}\hskip 2mm\\[-3mm]
                                            {\lightgray Minisuperspace}\\[-3.5mm]
                                            \vspace*{1mm}\hskip 1mm
                                        \end{center}
                                    \end{minipage}
                                    }
                       }
          }
\rput(9,7){\rnode{fs}{\psframebox[fillstyle=gradient,gradbegin=gray,gradend=black,gradangle=-45,linewidth=1pt,gradmidpoint=.5]%
                                   {\rput[c](0.65,0.1){{\white \huge 7}}
                                    \begin{minipage}{1.1cm}
                                        \begin{center}
                                            {\lightgray Fractal}\\
                                            {\lightgray Strings}
                                        \end{center}
                                    \end{minipage}
                                    }
                       }
          }
\rput(3,11){\rnode{cl}{\psframebox[fillstyle=gradient,gradbegin=gray,gradend=black,gradangle=-45,linewidth=1pt,gradmidpoint=.5]%
                                   {\rput[c](.8,0.1){{\white \huge 8}}
                                    \begin{minipage}{1.4cm}
                                        \begin{center}
                                            {\lightgray Classical}\\
                                            {\lightgray Limit}
                                        \end{center}
                                    \end{minipage}
                                    }
                       }
          }
\rput(1,5){\rnode{bbr}{\psframebox[fillstyle=gradient,gradbegin=gray,gradend=black,gradangle=-45,linewidth=1pt,gradmidpoint=.5]%
                                   {\rput[c](1.35,0.1){{\white \huge 9}}
                                    \begin{minipage}{2.5cm}
                                        \begin{center}
                                            {\lightgray Boundary--Bulk}\\
                                            {\lightgray Relation}
                                        \end{center}
                                    \end{minipage}
                                    }
                       }
          }
\rput(9,5){\rnode{ns}{\psframebox[fillstyle=gradient,gradbegin=gray,gradend=black,gradangle=-45,linewidth=1pt,gradmidpoint=.5]%
                                   {\rput[c](1.15,0.1){{\white \huge 10}}
                                    \begin{minipage}{2.1cm}
                                        \begin{center}
                                            {\lightgray Nonstandard}\\
                                            {\lightgray \&}\\
                                            {\lightgray Speculative}
                                        \end{center}
                                    \end{minipage}
                                    }
                       }
          }
\rput(6,3){\rnode{dc}{\psframebox[fillstyle=gradient,gradbegin=gray,gradend=black,gradangle=-45,linewidth=1pt,gradmidpoint=.5]%
                                   {\rput[c](1.2,0.1){{\white \huge A}}
                                    \begin{minipage}{2.2cm}
                                        \begin{center}
                                            {\lightgray Detailed}\\
                                            {\lightgray Computations}
                                        \end{center}
                                    \end{minipage}
                                    }
                       }
          }
\rput(5,1){\rnode{nsa}{\psframebox[fillstyle=gradient,gradbegin=gray,gradend=black,gradangle=-45,linewidth=1pt,gradmidpoint=.5]%
                                   {\rput[c](1.15,0.1){{\white \huge B}}
                                    \begin{minipage}{2.1cm}
                                        \begin{center}
                                            {\lightgray NonStandard}\\
                                            {\lightgray Analysis}
                                        \end{center}
                                    \end{minipage}
                                    }
                       }
          }
\rput(9,1){\rnode{nsp}{\psframebox[fillstyle=gradient,gradbegin=gray,gradend=black,gradangle=-45,linewidth=1pt,gradmidpoint=.5]%
                                   {\rput[c](1.15,0.1){{\white \huge C}}
                                    \begin{minipage}{2.1cm}
                                        \begin{center}
                                            {\lightgray NonStandard}\\
                                            {\lightgray Stochastic}\\
                                            {\lightgray Process}
                                        \end{center}
                                    \end{minipage}
                                    }
                       }
          }
\rput(6.5,7){\rnode{hd}{\psframebox[fillstyle=gradient,gradbegin=gray,gradend=black,gradangle=-45,linewidth=1pt,gradmidpoint=.5]%
                                   {\rput[c](1.05,0.1){{\white \huge D}}
                                    \begin{minipage}{1.9cm}
                                        \begin{center}
                                            {\lightgray Holographic}\\
                                            {\lightgray Derivative}
                                        \end{center}
                                    \end{minipage}
                                    }
                       }
          }
\pnode(8.94,13){auxuno}
\pnode(4.95,5){auxdue}
\end{pspicture}
\ncline{->}{toeholds}{pbrfq}
\ncline{->}{toeholds}{hjst}
\ncline{->}{hjst}{cl}
\ncline{->}{hjst}{sfq}
\ncline{->}{hjst}{bbr}
\ncline{->}{sfq}{sfs}
\ncline{->}{sfq}{mini}
\ncline{->}{sfs}{fs}
\ncline{->}{fs}{ns}
\ncline{->}{dc}{ns}
\ncline{->}{nsa}{nsp}
\ncline{->}{nsp}{ns}
\ncline{->}{bbr}{sfq}
\ncline{->}{sfq}{cl}
\ncline{->}{hd}{sfq}
\ncline{->}{pbrfq}{mini}
\ncline{pbrfq}{auxuno}
\ncline{->}{auxuno}{sfs}
\ncline{sfq}{auxdue}
\ncline{->}{auxdue}{ns}
\ncangles[armA=2cm]{->}{auxuno}{ns}
 

%% file: chap01.tex
\pageheader{}{Toeholds.}{}
\chapter{Toeholds}
\label{1.toeholds}

\begin{start}
$\euf{A}$ glowing fire\\
dances in the center\\
of the spartan, low--ceilinged room,\\
creating a kaleidoscope of shadows\\
on the walls.\\
\end{start}

\section{Boundary Dynamics}

In this first chapter we provide an in depth discussion of the main
ideas and motivations underlying the whole Thesis.
In particular, we shall explain how the Quantum Dynamics of a string
(and, more in general, of a $p$-dimensional extended object) can be
viewed as  {\it the Boundary Dynamics} induced by the world--sheet (or,
more in general, by the $(p+1)$-dimensional world--manifold)
quantum vibrations.
This concepts generalizes in a very straightforward way
the quite similar view  of a pointlike particle as the free
end--point of a $1$-dimensional $(p=0)$ world--line.
Now, we have to deal with the problem of a non--local  quantum
mechanics as a consequence of the string spatial extension.
The way, which we consider more appropriate to
approach such a problem, goes through the
relation between the Hamilton--Jacobi\label{1.HamJacrel} formulation
of Classical Mechanics and the associated Quantum formulation.

We think that the above considerations can be taken as a good evidence
of the relevance of the Hamilton--Jacobi Theory in passing from the
classical formulation of a system to the quantum one. This is the reason
why we decided to derive the Dynamics of an extended object starting
from the Hamilton--Jacobi formulation of the Classical Theory and
to devote chapter \ref{2.hamjac} to a detailed exposition of this subject
starting from different viewpoints.

\subsection{``Shadow'' Dynamics}
\label{1.ogielski}

A central point in our treatment is based on the observation that
all physical theories rely on the solution of a (more or less intricate)
system of differential equations (ordinary or partial) encoding the
law of evolution of the system. The set of dynamical equations must be
supplemented with appropriate {\it Boundary} conditions taking into account
the ``action of the {\it environment}$\,$'' on the system under study.
In this kind of description,  Dynamics inside a given region is constrained
by the conditions imposed on the {\it Boundary} of the region itself.
In other world, we have a pure {\it Bulk Dynamics} while the {\it Boundary}
is a {\it non--dynamical} region, where the behavior of the system
is {\it a priori} fixed. In our opinion, {\it Boundary} conditions are often
assigned for the only purpose of convenience without
any further justification.
Against this background, the Hamilton--Jacobi approach ``shifts'' the
physical Dynamics to the {\it Boundary}, and assign a secondary role to
the Dynamics of the {\it Bulk}. A good starting point to illustrate this
different strategy is provided by the Ogielski's formulation
of the closed, bosonic string Dynamics.%
\footnote{Here and in what follows we always
assume to consider systems that admit a Lagrangian Dynamics as well
as an Hamiltonian one.}.
\begin{defs}[Parameter Space]\spbcorr{}.\\
    The \underbar{Parameter Space} is a compact connected
    two dimensional domain
    $\Sigma \in \R ^{2}$ coordinatized by the couple of variables
    $\ttt{\sigma ^{0} , \sigma ^{1}} \equiv \Bsigma$.
\end{defs}
For what concerns Ogielski's treatment we are going to assume a
{\it simply connected} {\sl Parameter Space}. Of course, as we will have
the opportunity to underline in chapter \ref{3.strfunqua},
more complicated
cases can be worth of interest in string Dynamics\footnote{In particular,
a doubly connected space is the most appropriate in treating the free
propagation of a free string between two different configurations.}.
No problems thus arise in the following definition
\begin{defs}[Boundary Space]\spbcorr{}.\\
    The \underbar{Boundary Space} is the Boundary
    $\Gamma \equiv \partial \Sigma$
    of the {\sl Parameter Space}: it is parametrized by a
    parameter $s$, varying in the interval $S \equiv \qtq{s _{0}, s _{1}}$;
    since the Boundary of a Boundary is the empty set,
    $$
        \partial \Gamma = \partial \partial \Sigma = \emptyset
        \quad ,
    $$
    the interval $S$ defined above has its end points identified,
    $$
        s _{0} \equiv s _{1}
        \quad ,
    $$
    i.e. is topologically equivalent to a circle, or
    a $1$-dimensional sphere, $\Sf ^{1}$.
\end{defs}
We note that in more complicate situations, such as a multiply connected
{\sl Parameter Space}, it is useful consider different ways of
parametrizing the {\it Boundary}, usually with a properly chosen set of functions.
Indeed, the {\it Boundary} of a multiply connected space is not always
connected. Thus, it can be useful to
associate different parametrizations to different connected components
of the {\sl Boundary Space}.

We will often use quantities differentiated once with
respect to the parameter $s \in \Sf ^{1}$. Hence, let us introduce
the following notations.
\begin{nots}[First Derivative]\spbcorr{}.
\label{1.firdernot}\\
   We will use a {\it prime} ``$\;'\:$'' to denote the first derivative
   with respect to the unique parameter, say $s$, of a curve: for example
   $$
       \Bsigma ' \ttt{\bar{s}}
       \dfn
       \left . \frac{d \Bsigma \tst}{ds} \right \rceil _{s = \bar{s}}
       =
       \ttt{
           \left .
               \frac{d \sigma ^{0}}{ds}
           \right \rceil _{s = \bar{s}}
           ,
           \left .
               \frac{d \sigma ^{1}}{ds}
           \right \rceil _{s = \bar{s}}
           }
   \quad .
   $$
\end{nots}
Moreover, we are also going to use the more compact
\begin{nots}[Modulus of the Boundary Space Tangent Vector]\spbcorr{}.\\
    The modulus of the vector tangent to the Boundary Space $\Gamma$
    at the point
    $P = \ttt{\sigma ^{0} \ttt{\bar{s}} , \sigma ^{1} \ttt{\bar{s}}}$
    is denoted by
    $$
        \sqrt{\ttt{\vect{\sigma} ' \ttt{\bar{s}}} ^{2}}
        \dfn
        \sqrt{
              \sigma ^{\prime \, i} \ttt{\bar{s}}
              \sigma ' _{i} \ttt{\bar{s}}
             }
        =
        \left .
        \sqrt{
                  \frac{d \sigma ^{i}}{ds} \frac{d \sigma _{i}}{ds}
             }
        \,
        \right \rceil _{s = \bar{s}}
    \quad .
    $$
\end{nots}

Suppose now that the
model we are going to study is defined by the following $2$-form
\beq
    \form{\omega}
    =
    \Lag \ttt{X ^{\mu} , X ^{\mu} {}_{, i} ; \sigma ^{i}}
    d \sigma ^{0} \wedge d \sigma ^{1}
    \quad ,
    \label{1.lagtwoform}
\eeq
where, $\Lag$ is a sufficiently regular function (the {\it Lagrangian
density} of our model),
$X ^{\mu} \equiv X ^{\mu} \ttt{\sigma ^{i}}$ are the fields defined
over  the domain $\Sigma$, and $Y ^{\mu} \equiv Y ^{\mu} \tst$ their values on
the {\it Boundary} given by
$$
    Y ^{\mu} \tst
    \equiv
    X ^{\mu} \ttt{\sigma ^{i} \tst}
    \quad .
$$
The fields $\Bx$ and $\By$ take their values in the below defined target
space.
\begin{defs}[Target Space]\spbcorr{}.\\
    The \underbar{Target Space}, $\T$, is the space in which the fields
    $X ^{\mu}$ (equivalently $Y ^{\mu}$)
    take their values as functions on $\Sigma$.
\end{defs}
In the target space the fields $X ^{\mu} \ttt{\Bsigma}$ define a surface
$\mathcal{W} \dfn X ^{\mu} \ttt{\Sigma}$ bounded by a {\it loop}
$
 C \dfn Y ^{\mu} \ttt{\Gamma}
    =   X ^{\mu} \ttt{\sigma ^{i} \ttt{\Sf ^{1}}}
    =   \partial \mathcal{W}
$.

It is  already clear how the {\it Boundary} of the system
is playing a central role in this description. It is the physical
dynamical object rather than the locus where constraints are imposed
over the classical solutions.
The {\it Boundary} Dynamics will essentially depend
from the chosen Lagrangian density $\Lag$, in equation (\ref{1.lagtwoform}),
for the fields $X ^{\mu}$.
The action corresponding to the Lagrangian  (\ref{1.lagtwoform}) is
\beq
    S = \int _{\Sigma} \form{\omega}
      = \int _{\Sigma}
           \Lag \ttt{X ^{\mu} , X ^{\mu} {}_{, i} ; \sigma ^{i}}
           d \sigma ^{0} \wedge d \sigma ^{1}
    \quad ,
    \label{1.action}
\eeq
Moreover, we can introduce the following quantities:
\begin{defs}[Tangent and Normal Vectors]\spbcorr{}.\\
    The following vectors are the tangent and the normal to the
    Boundary $\Gamma \approx \Sf ^{1}$
    of the {\sl Parameter Space}, and to its image $C$ in the
    {\sl Target Space} $\T$:
    \begin{enumerate}
        \item the tangent unit vector to the ({\sl Parameter Space})
        Boundary
        curve $\Gamma \approx \Sf ^{1}$ at the point defined by the
        value $\bar{s}$ of
        the parameter $s \in \Sf ^{1}$ is
        \beq
            \vect{t} \ttt{\bar{s}}
            =
            \left . \vect{t} \right \rceil _{s = \bar{s}}
            =
            \left .
                \frac{1}{\sqrt{\ttt{\vect{\sigma ' \tst }^{2}}}}
                \frac{d \sigma ^{j} \tst}{d s}
            \right \rceil _{s = \bar{s}}
            \! \! \! \!
            \vect{\partial} _{j}
            =
            \frac{\sigma ^{\prime j} \ttt{\bar{s}}}
                 {\sqrt{\ttt{\vect{\sigma} ' \ttt{\bar{s}}} ^{2}}}
            \vect{\partial} _{j}
        \label{1.tanpar}
        \eeq
        and is called the \underbar{({\sl Parameter Space}) Tangent Unit Vector};
        \item the normalized normal vector to the Boundary
        Space $\Gamma \approx \Sf ^{1}$ at the point
        $\bar{s}$
        \beq
            \vect{n} \ttt{\bar{s}}
            =
            \left . \vect{n} \right \rceil _{s = \bar{s}}
            =
            \left .
                \frac{\epsilon ^{jk}}{\sqrt{\ttt{\vect{\sigma} ' \tst} ^{2}}}
                \frac{d \sigma _{k} \tst}{d s}
            \right \rceil _{s = \bar{s}}
            \! \! \! \!
            \vect{\partial} _{j}
            =
            \frac{\epsilon ^{jk} \sigma ' _{k} \ttt{\bar{s}}}
                 {\sqrt{\ttt{\vect{\sigma} ' \ttt{\bar{s}}} ^{2}}}
            \vect{\partial} _{j}
        \label{1.norpar}
        \eeq
        and is called the \underbar{({\sl Parameter Space}) Normal Vector};
        \item the tangent vector to the {\sl Target Space} Boundary
        curve $C$ at the point  $\bar{s}$
        \beq
            \vect{T} \ttt{\bar{s}}
            =
            \left . \vect{T} \right \rceil _{s = \bar{s}}
            =
            \left .
                \frac{d \sigma ^{j} \tst}{d s}
                Y ^{\mu} {}_{, j} \tst
            \right \rceil _{s = \bar{s}}
            \! \! \! \!
            \vect{\partial} _{\mu}
            =
            \sigma ^{\prime j} \ttt{\bar{s}}
            Y ^{\mu} {}_{, j} \ttt{\bar{s}}
            \vect{\partial} _{\mu}
        \label{1.tantar}
        \eeq
        and is called the \underbar{Linear Velocity Vector} of the
        curve $C$;
        \item the normal vector to the {\sl Target Space} Boundary
        curve $C$ at the point  $\bar{s}$
        \beq
            \vect{N} \ttt{\bar{s}}
            =
            \left . \vect{N} \right \rceil _{s = \bar{s}}
            =
            \epsilon ^{ij}
            \left .
                \frac{d \sigma _{i} \tst}{d s}
                Y ^{\mu} {}_{, j} \tst
            \! \! \! \!
            \right \rceil _{s = \bar{s}}
            \vect{\partial} _{\mu}
            =
            \epsilon ^{ij}
            \sigma ' _{i} \ttt{\bar{s}}
            Y ^{\mu} {}_{, j} \ttt{\bar{s}}
            \vect{\partial} _{\mu}
        \label{1.nortar}
        \eeq
        and is called the \underbar{Acceleration} of the curve $C$.
    \end{enumerate}
\end{defs}
After these definitions, we can prove the following result.
\begin{props}[Boundary Variation in
              $(\vect{t} , \vect{n})$ Coordinates]\spbcorr{}.\\
    A variation of the domain $\Sigma$ near the point
    $\sigma ^{i} \ttt{\bar{s}}$ of the Boundary $\Gamma \approx \Sf ^{1}$
    (in {\sl Parameter Space}) can be decomposed into a tangential
    part, which we denote by $\delta t \ttt{\bar{s}}$,
    and a normal part,
    denoted by $\delta n \ttt{\bar{s}}$, given respectively by:
    \bea
        \delta t \ttt{\bar{s}}
        & = &
        \left .
            \frac{1}{\sqrt{\ttt{\vect{\sigma} ' \tst} ^{2}}}
            \frac{d \sigma ^{j} \tst}{d s}
        \right \rceil _{s = \bar{s}}
        \! \! \! \!
        \delta \sigma _{j} \ttt{\bar{s}}
        \label{1.bauvarresuno}
        \\
        \delta n \ttt{\bar{s}}
        & = &
        \left .
            \frac{\epsilon ^{jk}}
                 {\sqrt{\ttt{\vect{\sigma} ' \tst} ^{2}}}
            \frac{d \sigma _{j} \tst}{d s}
        \right \rceil _{s = \bar{s}}
        \! \! \! \!
        \delta \sigma _{k} \ttt{\bar{s}}
        \quad ,
        \label{1.bauvarresdue}
    \eea
    where $\delta \sigma _{j} \tst$ is the component of the variation
    in the $\vect{\partial} _{j}$ direction.
\end{props}
\begin{proof}
We are going to express the variation
$\vect{\Delta \sigma} \ttt{\bar{s}}$
in the two different basis,
$\ttt{\vect{\partial} _{0} , \vect{\partial} _{1}}$ and
$\ttt{\vect{t} , \vect{n}}$ at the given point
and then we will compare the two results.
Thus first we consider the variation of the Boundary at the
point $\sigma ^{i} \ttt{\bar{s}}$ and decompose it into the basis
formed by the vectors $\vect{\partial} _{i} \: , i = 1 , 2$,
\beq
    \vect{\Delta \sigma} \ttt{\bar{s}}
    =
    \delta \sigma ^{j} \ttt{\bar{s}} \vect{\partial} _{j}
    \quad .
\label{1.bauvaruno}
\eeq
Then we perform the same operation but with respect to
the basis formed by the vectors
(\ref{1.tanpar}-\ref{1.norpar}), thus getting
\bea
    \vect{\Delta \sigma} \ttt{\bar{s}}
    & = &
    \delta n \ttt{\bar{s}} \vect{n} \ttt{\bar{s}}
    +
    \delta t \ttt{\bar{s}} \vect{t} \ttt{\bar{s}}
    \nonumber \\
    & = &
    \delta n \ttt{\bar{s}}
    \frac{\epsilon ^{jk}}{\sqrt{\ttt{\vect{\sigma} ' \ttt{\bar{s}}} ^{2}}}
    \sigma ' _{k} \ttt{\bar{s}}
    \vect{\partial} _{j}
    +
    \delta t \ttt{\bar{s}}
    \frac{1}{\sqrt{\ttt{\vect{\sigma} ' \ttt{\bar{s}}} ^{2}}}
    \sigma ^{\prime j} \ttt{\bar{s}}
    \vect{\partial} _{j}
    \quad ,
    \label{1.bauvardue}
\eea
where we used definitions (\ref{1.tanpar}) and (\ref{1.norpar}) to express
the ({\sl Parameter Space}) {\sl Tangent Unit Vector} and the
({\sl Parameter Space})
{\sl Normal Vector} in terms of the
$\ttt{\vect{\partial} _{0} , \vect{\partial} _{1}}$ basis.
Comparing then equation (\ref{1.bauvaruno})
with equation (\ref{1.bauvardue}) we obtain
\beq
    \delta \sigma ^{j} \ttt{\bar{s}} \vect{\partial} _{j}
    =
    \Delta n \ttt{\bar{s}}
    \frac{\epsilon ^{jk}}{\sqrt{\ttt{\vect{\sigma} ' \ttt{\bar{s}}} ^{2}}}
    \sigma ' _{k} \ttt{\bar{s}}
    \vect{\partial} _{j}
    +
    \Delta t \ttt{\bar{s}}
    \frac{1}{\sqrt{\ttt{\vect{\sigma} ' \ttt{\bar{s}}} ^{2}}}
    \sigma ^{\prime j} \ttt{\bar{s}}
    \vect{\partial} _{j}
    \quad .
    \label{1.bauvarequ}
\eeq
Applying both sides to the one form
$\form{\theta} = \sigma ' _{i} \ttt{\bar{s}} \form{d x} ^{i}$
and suppressing,
for the sake of notational simplicity,  the point $\ttt{\bar{s}}$, we get
\bea
    \delta \sigma ^{j} \sigma ' _{j}
    & =
    \label{1.bauvartre}
    \\
    \delta \sigma ^{j} \sigma ' _{i} \delta ^{i} _{j}
    & =
    \nonumber \\
    \delta \sigma ^{j} \sigma ' _{i}
    \vect{\partial} _{j} \ttt{\form{d x} ^{i}}
    & =
    \nonumber \\
    \delta \sigma ^{j}
    \vect{\partial} _{j} \ttt{\sigma ' _{i} \form{d x} ^{i}}
    & \equo{\footnotemark} &
    \delta n
    \frac{\epsilon ^{jk}}{\sqrt{\ttt{\Bsigma '} ^{2}}}
    \sigma ' _{k}
    \vect{\partial} _{j} \ttt{\sigma ' _{i} \form{d x} ^{i}}
    +
    \delta t
    \frac{\sigma ^{\prime j}}{{\sqrt{\ttt{\Bsigma '} ^{2}}}}
    \vect{\partial} _{j} \ttt{\sigma ' _{i} \form{d x} ^{i}}
    \nonumber \\
    & = &
    \delta n
    \frac{\epsilon ^{jk}}{\sqrt{\ttt{\vect{\sigma} '} ^{2}}}
    \sigma ' _{k}
    \sigma ' _{i}
    \vect{\partial} _{j} \ttt{\form{d x }^{i}}
    +
    \delta t
    \frac{1}{\sqrt{\ttt{\vect{\sigma} '} ^{2}}}
    \sigma ^{\prime j}
    \sigma ' _{i}
    \vect{\partial _{j}} \ttt{\form{d x} ^{i}}
    \nonumber \\
    & = &
    \delta n
    \frac{\epsilon ^{jk}}{\sqrt{\ttt{\vect{\sigma} '} ^{2}}}
    \sigma ' _{k}
    \sigma ' _{i}
    \delta ^{i} _{j}
    +
    \delta t
    \frac{1}{\sqrt{\ttt{\vect{\sigma} '} ^{2}}}
    \sigma ^{\prime j}
    \sigma ' _{i}
    \delta _{j} ^{i}
    \nonumber \\
    & = &
    \delta n
    \frac{\epsilon ^{jk}}{\sqrt{\ttt{\vect{\sigma} '} ^{2}}}
    \sigma ' _{k}
    \sigma ' _{j}
    +
    \delta t
    \frac{1}{\sqrt{\ttt{\vect{\sigma} '} ^{2}}}
    \sigma ^{\prime j}
    \sigma ' _{j}
    \nonumber \\
    & \equo{\footnotemark} &
    0
    +
    \sqrt{\ttt{\vect{\sigma} '} ^{2}}
    \delta t
    \quad ,
    \label{1.bauvarqua}
\eea
\addtocounter{footnote}{-1}
\footnotetext{We use here equation (\ref{1.bauvarequ}).}
\addtocounter{footnote}{1}
\footnotetext{The first term vanishes since it has two symmetric indices
contracted with two antisymmetric ones.}
\noindent so that from the equality of (\ref{1.bauvartre}) and (\ref{1.bauvarqua})
we get exactly result (\ref{1.bauvarresuno}).
Proceeding in the same way but applying (\ref{1.bauvarequ}) to the one form
$
 \form{\tilde{\theta}}
 =
 \epsilon _{lm} \sigma ' _{l} \ttt{\bar{s}} \form{d x} ^{m}$
we get
\bea
    \delta \sigma ^{j} \epsilon _{lj} \sigma ^{\prime l}
    & =
    \\
    \delta \sigma ^{j} \epsilon _{lm} \sigma ^{\prime l} \delta ^{m} _{j}
    & =
    \nonumber \\
    \delta \sigma ^{j} \epsilon _{lm} \sigma ^{\prime l}
    \vect{\partial} _{j} \ttt{\form{d x} ^{m}}
    & =
    \nonumber \\
    \delta \sigma ^{j}
    \vect{\partial} _{j} \ttt{\epsilon _{lm} \sigma ^{\prime l} \form{d x} ^{m}}
    & \equo{\footnotemark} &
    \delta n
    \frac{\epsilon ^{jk}}{\sqrt{\ttt{\vect{\sigma} '} ^{2}}}
    \sigma ' _{k}
    \vect{\partial} _{j} \ttt{\epsilon _{lm} \sigma ^{\prime l} \form{d x} ^{m}}
    +
    \delta t
    \frac{1}{\sqrt{\ttt{\vect{\sigma} '} ^{2}}}
    \sigma ^{\prime j}
    \vect{\partial} _{j} \ttt{\epsilon _{lm} \sigma ^{\prime l} \form{d x} ^{m}}
    \nonumber \\
    & = &
    \delta n
    \frac{\epsilon ^{jk} \epsilon _{lm}}{\sqrt{\ttt{\vect{\sigma} '} ^{2}}}
    \sigma ' _{k}
    \sigma ^{\prime l}
    \vect{\partial} _{j} \ttt{\form{d x ^{m}}}
    +
    \delta t
    \frac{\epsilon _{lm}}{\sqrt{\ttt{\vect{\sigma} '} ^{2}}}
    \sigma ' _{j}
    \sigma ^{\prime l}
    \vect{\partial} _{j} \ttt{\form{d x} ^{m}}
    \nonumber \\
    & = &
    \delta n
    \frac{\epsilon ^{jk} \epsilon _{lm}}{\sqrt{\ttt{\vect{\sigma} '} ^{2}}}
    \sigma ' _{k}
    \sigma ^{\prime l}
    \delta ^{m} _{j}
    +
    \delta t
    \frac{\epsilon ^{lm}}{\sqrt{\ttt{\vect{\sigma} '} ^{2}}}
    \sigma ' _{l}
    \sigma ' _{m}
    \nonumber \\
    & \equo{\footnotemark} &
    \sqrt{\ttt{\vect{\sigma} '} ^{2}} \delta n
    +
    0
\eea
\addtocounter{footnote}{-1}
\footnotetext{This line is equation (\ref{1.bauvarequ})
              applied to the $1$-form $\form{\tilde{\theta}}$.}
\addtocounter{footnote}{1}
\footnotetext{We use the relation
              $$
                  \epsilon ^{m} {}_{k} \epsilon _{lm} = \delta _{kl}
                  \quad ;
              $$
              moreover the second term vanishes as before.}
getting at the end equation (\ref{1.bauvarresdue}).
\end{proof}
We can now define the variation of the oriented area $A$ of the domain
$\Sigma$ at the point $s$ as follows:
\begin{defs}[Parameter Space Area Variation]\spbcorr{}.\\
    The \underbar{{\sl Parameter Space} Area Variation} at the point
    $s$ of the Boundary $\Gamma$ is given by the modulus of the tangent
    vector multiplied by the normal variation
    $$
        \delta A \tst
        =
        \sqrt{\ttt{\Bsigma ' \tst} ^{2}}
        \delta n \tst
    $$
    and can be expressed also as
    $$
        \delta A \tst
        =
        \epsilon ^{ab}
        \sigma ' _{a} \tst
        \delta \sigma _{b} \tst
        \quad .
    $$
\end{defs}
The functional derivative with respect to the
{\sl Parameter Space} area can be defined as
well.
\begin{defs}[Area Funcitonal Derivative]\spbcorr{}.\\
    The functional derivative with respect to a variation of the
    Area in the Parameter Space at the point
    $s \in \Gamma = \partial \Sigma$ is the rescaled normal variation ad
    $s$:
    \beq
        \frac{\delta}{\delta A \tst}
        =
        \frac{1}{\sqrt{\ttt{\Bsigma '} ^{2}}}
        \frac{\delta}{\delta n \tst}
        \quad .
    \eeq
\end{defs}
We observe at this stage that, if the system is invariant under {\it Boundary}
reparametrizations, it is insensitive to the place on the {\it Boundary}
where the Area variation takes place and the {\sl Area Functional
Derivative}
can be traded with an ordinary (partial) derivative.

After this results, we can tackle the problem of what
kind of Dynamics results from the variation of the action
(\ref{1.action}).
We observe that under the term ``variation'', we understand not
only a variation of the fields  defined on the {\it fixed}
domain $\Sigma$, but also a variation of $\Sigma$ itself. As we have already
pointed out, this is  related to the dynamical role
which is  assigned to the {\it Boundary} $\Sf ^{1} \approx \Gamma = \partial \Sigma$
in this formulation: we do not push it
to infinity, or consider it as a non--dynamical, purely geometrical quantity,
on the contrary, the physical, evolving object is the {\it Boundary} itself.
We anticipate that this is a far reaching point of view.
By extremizing the action integral
(\ref{1.action}) we get the following results:
\begin{props}[Normal and Tangential Boundary
              Variation of the Action]\spbcorr{}.\\
    Let $X ^{\mu} \ttt{\sigma _{0} , \sigma _{1}}$ be an extremal
    of the action functional {\rm (\ref{1.action})};
    then the following equations
    for the variation of the action hold:
    \bea
        \frac{\delta S}{\delta n \tst}
        & = &
        \sqrt{\ttt{\vect{\sigma} '} ^{2}}
        \left(
            \frac{\partial \Lag}{\partial N ^{\mu}} N ^{\mu}
            -
            \Lag
        \right)
        \label{1.oginorfunder}
        \\
        \frac{\delta S}{\delta t \tst}
        & = &
        \sqrt{\ttt{\vect{\sigma} '} ^{2}}
        \frac{\partial \Lag}{\partial N ^{\mu}} T ^{\mu}
        \label{1.ogitanfunder}
        \\
        \frac{\delta S}{\delta Y ^{\mu} \tst}
        & = &
        \ttt{\vect{\sigma} '} ^{2}
        \frac{\partial \Lag}{\partial N ^{\mu}}
        \quad ,
        \label{1.ogiNorfunder}
    \eea
    where the quantities $\vect{t}$ and $\vect{n}$
    are defined in equations {\rm (\ref{1.tanpar}-\ref{1.norpar})}
    respectively
    and $T ^{\mu}$ and $N ^{\mu}$ are the components of $\vect{T}$ and
    $\vect{N}$ in equations {\rm (\ref{1.tantar}-\ref{1.nortar})}.
    Moreover $\delta / \ttt{\delta n \tst}$ means that we perform a
    variation of the $\Sigma$ domain at the {\it Boundary} in a direction
    normal to $\Gamma$, $\delta / \ttt{\delta t \tst}$ has the same meaning
    for the tangential direction and $\delta / \ttt{\delta Y ^{\mu} \tst}$
    takes into account variation related to the embedding in the {\sl
    Target Space} $\T$.
\end{props}

It is worthwhile to remark that:
\begin{enumerate}
    \item the vector normal to the {\it Boundary}
    in the {\sl Target Space} $\T$ is not an independent variable;
    it can be eliminated from the functional derivatives
    of the action with respect to
    the normal $\vect{n}$ in {\sl Parameter Space} and the tangent
    $\vect{T}$ in {\sl Target Space}
    (equations (\ref{1.oginorfunder}), (\ref{1.ogiNorfunder}) respectively);
    in this way we combine these two
    equations to get the Hamilton-Jacobi
    equation for the Dynamics of the {\it Boundary};
    \label{1.ogielskiHJ}
    \item the functional derivative
    with respect to the tangent vector $\vect{t}$ in $\Sigma$
    (equation {\ref{1.ogitanfunder}}), gives a
    condition for the action $S$ to be a reparametrization invariant
    functional of  the loop.  The physical meaning is that every
    {\it tangential deformation} of the {\it Boundary} can be absorbed
    in a reparametrization of the loop itself; different
    parametrizations physically correspond to the same loop.
\end{enumerate}

%% file: chap02.tex
\pageheader{}{Hamilton--Jacobi String Theory.}{}
\chapter{Hamilton--Jacobi String Theory}
\label{2.hamjac}

\begin{start}
``$\euf{W}$hat's in there?''\\
``Only what\\
you take with you.''\\
\end{start}

\section{Preliminaries}

In this section we specialise the results already obtained in
subsection \ref{1.ogielski} of chapter
\ref{1.toeholds} to the Dynamics of a closed bosonic string  described
by the following Lagrangian density:
\beq
    \Lag _{\mathrm{Schild}}
    =
    \frac{m ^{2}}{4}
    \dot{X} ^{\mu \nu} \dot{X} _{\mu \nu}
    \quad ,
    \label{2.schild}
\eeq
where
\beq
    \dot{X} ^{\mu \nu}
    =
    \epsilon ^{ab}
    \partial _{\xi ^{a}} X ^{\mu}
    \partial _{\xi ^{b}} X ^{\nu}
    \label{2.dotxmunu}
\eeq
and
\beq
    X ^{\mu}
    =
    X ^{\mu} \ttt{\xi ^{0} , \xi ^{1}}
    \quad .
    \label{2.xmu}
\eeq
The pair of variables $(\xi ^{0} , \xi ^{1})$ ranges in a $2$-dimensional
domain $\Xi$, so that action \ref{1.action} becomes
\beq
    S
    =
    \frac{m ^{2}}{4}
    \int _{\Xi}
           \dot{X} ^{\mu \nu} \ttt{\Bxi}
           \dot{X} _{\mu \nu} \ttt{\Bxi}
           d \xi ^{0} \wedge d \xi ^{1}
    \quad .
    \label{2.action}
\eeq
Our main task is thus to derive the classical Hamilton--Jacobi equation,
reproducing Ogielski's procedure in this particular setting; then
this result will be
used to derive a functional Schr\"o{}dinger equation.
This procedure is quite similar to the one performed to quantize
a free particle.
Even for a single string we need an infinite, continuous, set of indices
labelling the collection of constituent points.
This is the reason why the functional calculus enters  the game.
We would like to remark, once again, the basic difference between
this approach and the traditional way of approaching
string Dynamics: instead of expanding  the string embedding functions in
normal modes to find the excitation spectrum, we  focus on the
{\it wholeness} of the closed string as a geometric object. Some new features
of string Dynamics can be better understood in this way. In particular it
turns out that the Quantum Dynamics of the object is, consistently  described
as a {\sl Shadow Dynamics} of areas parametrized by an
areal time. Before turning to this interesting developments,
it is worth to recall the work by T.Eguchi, who was
the first in suggesting such a non--standard string quantization method,
and give sound motivations for the choice of the Lagrangian density.
This is the subject of the next two sections; after,
we will apply all that to the {\it Functional Schr\"o{}dinger Quantization}
of the bosonic string; the possible relations between this approach and
the more traditional one will be pointed out in chapter \ref{9.connection}.
To connect the mathematical description with the physical interpretation
the following definitions are useful.
\begin{defs}[World--Sheet \& Parametrization]\spbcorr{}.
    \label{2.worldsheet}\\
    Let $\Xi$ be a $2$-dimensional domain with boundary
    $\gamma = \partial\, \Xi$ in the $2$-dimensional
    Minkowski space.
    A \underbar{(Classical)\,\footnotemark
    World--Sheet},
        \footnotetext{\label{2.footuno}We do not make at present
        any assumption about the quantum nature of the string; indeed, as
        it will be clear in the subsequent developments, it is really
        impossible to attribute to the quantum object the same properties
        (smoothness for example) that characterize the classical one.}
    $\mathcal{W}$, is a map of the {\sl Parameter Space} $\Xi$ into the
    $D$-dimensional Minkowski spacetime $\mathbb{M} ^{D}$,
    $$
        \mathcal{W} : \Xi \longrightarrow \mathbb{M} ^{D}
        \quad .
    $$
    If the map $\mathcal{W}$ is described by the embedding
    functions
    $$
        X ^{\mu} : \Xi \longrightarrow \mathbb{M} ^{D}
    $$
    we will refer to $X ^{\mu}$ as a \underbar{Parametrization} of the
    {\sl Classical World--Sheet} $\mathcal{W}$.
\end{defs}
The {\sl World--Sheet} is the {\it Bulk} of the Theory: then a string
is the {\it Boundary} of the {\sl World--Sheet}.
\begin{defs}[Classical Closed Bosonic String \& Parametrization]\spbcorr{}.
    \label{2.string}\\
    Let $\Xi$ be a $2$-dimensional domain with boundary
    $\gamma = \partial \Xi$ in the $2$-dimensional
    Minkowski space and $\mathcal{W}$ the {\sl World--Sheet} defined according to
    the definition {\rm \ref{2.worldsheet}} and parametrized by
    $X ^{\mu} \ttt{\xi ^{0} , \xi ^{1}}$.
    A \underbar{(Classical)\,\footnotemark Bosonic String}
        \footnotetext{\label{2.footdue}Please, see footnote
        \ref{2.footuno} on page \pageref{2.footuno}.}
    is an embedding
    $$
            \tilde{C} :
            \gamma
            \longrightarrow
            \mathbb{M} ^{D}
    $$
    from the boundary of $\Xi$ to the $D$-dimensional Minkowski space.
    Then the embedding functions of the {\sl String}, $Y ^{\mu}$, are
    the embedding functions of the {\sl World--Sheet} $X ^{\mu}$
    restricted to the boundary $\gamma$:
    $$
        Y ^{\mu} \dfn \left. X ^{\mu} \right \rceil _{\gamma}
        :
            \gamma = \partial \Xi
            \longrightarrow
            \tilde{C} \ttt{\gamma} \subset \mathbb{M} ^{D}
        \quad .
    $$
    Moreover, if we consider a one to one and onto map $\xi$, defined as
    $$
       \matrix{
           \xi ^{i} & : & \Sf ^{1} & \longrightarrow & \gamma \cr
                    &   & s
                       & \longrightarrow &
                       \ttt{\xi ^{0} \tst , \xi ^{1} \tst}
                       \quad ,
              }
    $$
    i.e. a parametrization of the boundary $\gamma$, we can
    think a {\sl Classical Closed Bosonic String}
    simply as a map from $\Sf ^{1}$
    to $\mathbb{M} ^{D}$. This definition is clearly diplayed by
   the following notation
    \beq
        Y ^{\mu} \tst = X ^{\mu} \ttt{\xi ^{0} \tst , \xi ^{1} \tst}
        \label{2.baupardef}
    \eeq
    for the parametrization of the boundary. We will also
    call $Y ^{\mu} = Y ^{\mu} \tst$ a \underbar{Parametrization} of
    the {\sl Classsical Closed Bosonic String}.
    Moreover we set
    $$
        C = \tilde{C} \circ \Bxi
        \quad .
    $$
\end{defs}
Of course, at least in the computations related to the Classical
Dynamics of the objects, we always assume that all the defined maps are
sufficiently regular in such a way that all the derivations and integrations
can be unambiguously carried out. The problem of defining the most adequate
functional class for the desciption of the Quantum Dynamics of the objects
will be tackled in a more deep way in chapter \ref{10.nonstacha} and
in two appendices, \ref{B.nonstanda} and \ref{C.nonstopro}.
Indeed this problem is really
a non trivial one, since in view of the peculiar properties of the
Quantum Dynamics it seems reasonable that, in trying to carry the classical
approach toward the quantum one, some regularity requirments imposed
at the classical level should be weakned.

\section{Area Quantization Scheme: Original Formulation}

Eguchi's approach to {\sl String} quantization, which we are going to
briefly review, is an immediate generalization of the
point--particle quantization along the guidelines of the
Feynman--Schwinger method. The essential point is that
reparametrization invariance is not assumed as an original
symmetry of the classical action;  rather, it is a symmetry of the
physical Green functions to be obtained at the very end of the
calculations by means of an appropriate averaging procedure. 
More explicitly, the basic action is not the Nambu--Goto
proper area of the {\sl String} {\sl World--Sheet}, but the ``square'' of
it, i.e. the Schild Lagrangian (\ref{2.schild}).
As discussed in the next section,
the corresponding (Schild) action is invariant under area preserving
transformations only.
Even if this is a restricted symmetry with respect to the full
reparametrization invariance of the Nambu--Goto action, it allows a
non--standard formulation of {\sl String} Dynamics leading to a
new, Jacobi--type, canonical formalism in which {\it the
proper area of the {\sl String} {\sl Parameter Space}
plays the role of evolution parameter}.
In other words, the ``proper time'' appropriate to this problem is neither
the {\sl String} manifold timelike coordinate  $\tau$, nor
the target space the coordinate time $x ^{0}$,
but the {\it invariant} combination of {\sl String} manifold coordinates
provided by
\beq
    A
    =
    \int _{\Xi}
        d \xi ^{0} \wedge d \xi ^{1}
    =
    \frac{1}{2}
    \epsilon _{ab}
    \int _{\Xi}
        d \xi ^{a} \wedge d \xi ^{b}
    \quad .
    \label{2.Adef}
\eeq
Once committed to this unconventional definition of time, the
quantum amplitude for the transition from an initial vanishing
{\sl String} configuration to a final non-vanishing {\sl String} configuration
after a lapse of areal time $A$, is provided by the kernel
$G \qtq{Y \tst ; A}$ which satisfies the following
diffusion--like equation,
or imaginary area Schr\"odinger equation
\beq
    \frac{1}{2}
    \frac{\delta ^{2} K \left [ \By \tst ; A \right ]}
         {
          \delta Y ^{\mu} \tst
          \delta Y _{\mu} \tst
         }
    =
    \frac{\partial K \left [ \By \tst ; A \right ]}{\partial A}
    \quad ,
    \label{2.egufunequ}
\eeq
where, following definition \ref{2.string},
$
    Y ^{\mu} \tst
    =
    X ^{\mu} \left( \xi ^{0} \tst , \xi ^{1} \tst
\right)
$,
the {\sl Parametrization} of the {\sl Classical Closed
Bosonic String}\footnote{We indicate with $C$ the image of the {\sl String}
in spacetime.},
represents the physical {\sl String} coordinate, i.e. the only spacelike boundary
of the {\sl World--Sheet}.
It may be worth to recall that in the Quantum Mechanics
of point particles the
``time'' $t$ is {\it not a measurable quantity but an arbitrary parameter},
since there does not exist a self--adjoint quantum operator with eigenvalues
$t$. Similarly, since there is no self--adjoint operator corresponding to
the {\sl World--Sheet} area, $K \qtq{\By \tst ; A}$
turns out to be explicitly dependent from an arbitrary parameter $A$,
and cannot represent a measurable quantity. However,
the Laplace transformed Green function
is $A$-independent and corresponds to the Feynman propagator
\bea
    G \left [ \By \tst ; E \right ]
    & \equivo{\footnotemark} &
    \int _{0} ^{\infty} d A \,
        K \left [ \By \tst ; A \right ]
        \exp
        \ttt{- E A}
    \nonumber\\
    & = &
    -
    \frac{1}{2 \left( 2 \pi \right) ^{3/2}}
    \int _{0} ^{\infty} \frac{dA}{A ^{3/2}}
        \exp
        \left(
            -
            \frac{\bar{F} \qtq{C}}{2 A}
            -
            \frac{1}{2}
            M ^{2}
            A
        \right)
    \label{2.egupro}
    \\
    \mathrm{where}
    & &
    \bar{F} \qtq{C}
    =
    \frac{1}{4}
    \left(
        F ^{\mu \nu} \qtq{C} \pm {}^{\ast} F ^{\mu\nu} \qtq{C}
    \right) ^{2}
    \\
    \mathrm{and}
    & &
    F ^{\mu \nu} \qtq{C}
    =
    \oint _{C} Y ^{\mu} d Y ^{\nu}
    \quad ,
\eea
where
$F$ stands for the self--dual (anti self--dual) area element.
\footnotetext{Note the naturalness of the inferior bound, namely $0$,
in the area integration: indeed, an area cannot be negative!}

Evidently, this approach is quite different  from the ``normal
mode quantization'' based on the Nambu--Goto action or the
path--integral formulation a la' Polyakov.
Moreover, there are some ambiguities in interpreting the formulae
for the functional equation and the propagator
(\ref{2.egufunequ}-\ref{2.egupro}), because at this
stage it is still not clear how the dependence from the parametrization
of the loop has to be interpreted. If the Green function is independent
from the loop paramter, i.e. it is just a functional of the loop, then the
 right-hand side of equation (\ref{2.egufunequ}) does
not depend on $s$, whereas the left--hand side, does. Instead if the
Green function, or the propagator, explicitly depend on $s$, they cannot
be considered as {\it evolution operators} for the object as a whole,
and this is difficult to reconcile with the invariance under area
preserving transformation of the starting action. Moreover, naively,
we would expect an equation for the kernel independent
from the particular choice of the parametrization, at least of the
loop, because, as already pointed out in the exposition of the previous
chapter, the last one can be considered as the true dynamical
object\footnote{This is true, of course, at the Classical as well as
at the Quantum level.}.
These problems can be solved by a closer analysis of the
the propagator within a formulation derived from the classical
Theory of loop Dynamics. Before facing this task, we would like
to provide a stronger motivation for the choice of the lagrangian
(\ref{2.schild}) we started from.

\section{The Basic Action}
\label{2.basactsec}

As we already anticipated in the previous discussion, our starting
point is somehow unconventional with respect to the well known
formulation of {\sl String} Theory. We do not start  from the Nambu--Goto action:
\beq
    S _{\mathrm{NG}}
    =
    T _{\mathrm{str.}}
    \int _{\Xi}d^2\sigma
        \sqrt{- \frac{1}{2} \dot{X ^{\mu \nu}} \dot{X _{\mu \nu}}}
\quad ,
\label{2.namgotact}
\eeq
where
\beq
    \sqrt{- \frac{1}{2} \dot{X ^{\mu \nu}} \dot{X _{\mu \nu}}}
    \dfn
    \Lag _{\mathrm{NG}}\ ,
\label{2.namgotlag}
\eeq
$\dot{X} ^{\mu \nu} \ttt{\xi ^{0} , \xi ^{1}}$ is defined as in
(\ref{2.dotxmunu}), and $T _{\mathrm{str.}}$ is the {\it string
tension}\footnote{The string tension, $T _{\mathrm{str.}}$.
is exactly the quantity
$$
    \frac{1}{2 \pi \alpha '}
    \quad ,
$$
which appears usually in all the papers on {\sl String} Theory.}

The action (\ref{2.namgotact}) has the remarkable
geometrical meaning of being the proper
area of the {\sl String} {\sl World--Sheet}
in {\sl Target Space}. For this reason, it is
the most natural generalization of the relativistic point particle action,
i.e. the proper length of the particle world--line. The price to pay
for such a well defined geometrical meaning is  the vanishing of the
corresponding canonical Hamiltonian, as a consequence
of the reparametrization invariance of the system. Accordingly,
a first quantization of the system based over a
Schr\"o{}edinger like approach is washed out from the very beginning.
This problem can be successfully skipped by straightforwardly switching
to a second quantized, or field theoretical, framework where the
embedding functions $X^\mu(\sigma^0,\sigma^1)$ are treated as multiplet
of scalar fields living over the $2$-dimensional {\sl String} manifold.
Such a formulation has elevated {\sl String} Theory to the
role of the most valuable candidate to the  Theory of Everything.
Notwithstanding, it seems to us to be worth exploring
different formulations as well, and clarify the eventual relations among
different approaches. The main reason for that stems from the fundamental
developments of the last years, that deeply changed  the general attitude
towards {\sl String} Theory. The basic result was  that the five different
superstring models candidate to the role of ultimate unified Theory were
not really distinct, but related by an intricate web of non--perturbative
duality relations!
Thus, a  non--perturbative formulation of {\sl String} Dynamics is compelling.
Is it possible to
describe the motion of a one--dimensional object without relying on an
expansion into harmonic modes? Is it possible to describe the state
of the system as a whole, without resolving it, from the very
beginning, into a collection of pointlike constituents? And
If the answer is yes, can this new formulation
give at least a slightly better insight toward the non--perturbative
properties that nowadays play such a central role in {\sl String} Theory?

As a personal contribution to  answer, at least partially, these questions we
propose  a  {\it functional generalization of the first quantization procedure
of a point particle through path--integrals methods} \cite{noiAmJ}.
This approach:
\begin{enumerate}
    \item is intrinsically non--perturbative;
    \item encodes the extended nature of
    the physical object into a mathematical formulation in terms of
    ``functionals'' of the {\sl String} configuration,
    instead of ordinary functions.
\end{enumerate}
Of course, we cannot proceed further without a well defined Hamiltonian.
So,  we will spend some time in considering an action functional, different
from (\ref{2.schild}): that is, the one already proposed by Schild in 1977,
the Lagrangian density of which we rewrite as:
$$
    \Lag _{\mathrm{Schild}}
    =
    \frac{m ^{2}}{4}
    \dot{X} ^{\mu \nu} \dot{X} _{\mu \nu}
    \quad .
$$
Before going on, we need some definitions.
\begin{defs}[Holographic Coordinates]\spbcorr{}.
    \label{2.holcordef}\\
    Let $\tilde{C}$ be a {\sl String}, as defined in
    {\rm \ref{2.string}},
    and $Y ^{\mu} \tst$, one of its possible {\sl Parametrizations}.
    The  {\sl String} \underbar{Holographic Coordinates} are
    $$
        Y ^{\mu \nu} \qtq{\gamma}
        =
        \oint _{\gamma} ds
            Y ^{\mu} \ttt{s}
            Y ^{\prime\, \nu} \ttt{s}
        =
        \oint _{C}
            Y ^{\mu}
            d Y ^{\nu}
        \quad .
    $$
\end{defs}
It is possible to see that these are {\it well--defined objects}
for a given {\sl String} thanks to the following proposition.
\begin{props}[Invariance of Holographic Coordinates]\spbcorr{}.\\
    The {\sl Holografic Coordinates} are invariant under
    {\sl String} reparametrization.
\end{props}
\begin{proof}
This result can be understood since the geometrical
interpretation of the {\sl Holographic Coordinates} reveals that
they are nothing but the areas of the projection of the {\sl String}
onto the coordinate planes and these are  invariant quantities.
Moreover, we can also prove in  a few lines this result
observing that if
$$
    R : \Sf ^{1} \longleftrightarrow \Sf ^{1}
$$
with
$$
    \bar{s} = R \tst \in \Sf ^{1} \quad , \qquad \forall s \in \Sf ^{1}
$$
is a reparametrization of the {\sl String}, i.e. a diffeomorphism
of the circle such that
$\Sf ^{1} \approx \bar{\Gamma} = R \ttt{\Gamma}$,
and
$$
        \bar{Y} ^{\mu} \ttt{\bar{s}}
        =
        Y ^{\mu} \ttt{R ^{-1} \tst}
        \qquad
        \mathrm{where}
        \qquad
        \matrix{
                \Sf ^{1}
                &
                \stackrel{R}{\longleftarrow}
                &
                \Sf ^{1}
                &
                \stackrel{Y ^{\mu}}{\longrightarrow}
                &
                \mathbb{M} ^{4}
                \cr
                \bar{s} = R \tst
                &
                \longleftarrow
                &
                s
                &
                \longrightarrow
                &
                Y ^{\mu} \tst
               }
    \quad ,
$$
then we have
\bea
    Y ^{\mu \nu} \qtq{C}
    & = &
    \oint _{\bar{\Gamma} \approx \Sf ^{1}} d \bar{s} \,
        \bar{Y} ^{\mu} \ttt{\bar{s}}
        \frac{d \bar{Y} ^{\nu} \ttt{\bar{s}}}{d \bar{s}}
    \nonumber \\
    & = &
    \oint _{\bar{\Gamma} \approx \Sf ^{1}} d \ttt{R \tst}
        \bar{Y} ^{\mu} \ttt {R \tst}
        \frac{d \bar{Y} ^{\nu} \ttt{R \tst}}{d R \tst}
    \nonumber \\
    & = &
    \oint _{\Gamma \approx \Sf ^{1}} d s \frac{d R \tst}{d s}
        Y ^{\mu} \ttt {R ^{-1} \ttt{R \tst}}
        \frac{d Y ^{\nu} \ttt{R ^{-1} \ttt{R \tst}}}{d s}
        \frac{1}{\displaystyle \frac{d R \tst}{d s}}
    \nonumber \\
    & = &
    \oint _{\Gamma \approx \Sf ^{1}} d s \,
        Y ^{\mu} \tst
        Y ^{\prime \nu} \tst
    \nonumber
    \quad .
\eea
From the right hand sides in the first and the last lines we thus get
the desired result.
\end{proof}

A functional differential operator that will be central in our
description of the {\sl String} Dynamics, will be the functional
derivative with respect to the {\sl Holographic Coordinates}:
Now, we define the following quantity:
\begin{defs}[Holographic Functional Derivative]\spbcorr{}.
\label{2.holfunderdef}\\
    The \underbar{Holographic Functional Derivative} is
    the functional derivative with respect to the tensor
    density corresponding to the
    {\sl Holographic Coordinates}; we will indicate is as
    $$
        \frac{\delta}{\delta Y ^{\mu \nu} \tst}
    $$
    and define it implicitly in terms of the standard functional
    derivative as,
    \beq
        \frac{\delta}{\delta Y ^{\mu} \tst}
        \dfn
        Y ^{\prime \nu}
        \frac{\delta}{\delta Y ^{\mu \nu} \tst}
        \quad .
    \label{2.holfunder}
    \eeq
\end{defs}
More information about this functional operation can be found
in appendix \ref{D.looderapp}.
Here we only note that, while the {\sl Holographic Coordinates}
$Y ^{\mu \nu} \left [ C \right ]$
are {\it functionals} of
the loop $C$, i.e., they contain no reference to a special
point of the loop,
the {\sl holographic Functional Derivative}
$\delta/\delta \sigma ^{\mu \nu} \left( \bar{s} \right)$
operates at the contact point
$\bar{Y} ^{\mu} = Y ^{\mu} \left( \bar{s} \right)$.
Thus, the area derivative
of a functional is no longer a functional.\label{2.aredercom}
Even if the functional
under derivation is reparametrization invariant,  its
area derivative behaves as a scalar density under
redefinition of the loop coordinate.

Moreover since we would like to define the Dynamics of the object we have
actually to concentrate our attention on the {\sl World--Sheet} of the
{\sl String} also, defined as in \ref{2.worldsheet}.
The presence of two parameters $\xi ^{i}$
allows us to define the following dynamical quantities:
\begin{defs}[Local World--Sheet Area Velocity]\spbcorr{}.
    \label{2.locareveldef}\\
    Let $X ^{\mu} \left( \xi ^{i} \right)$ be a {\sl Parametrization}
    of the {\sl String} {\sl World--Sheet}  .
    The \underbar{Local Area Velocity} of the {\sl World--Sheet} is
    $$
    \dot{X} ^{\mu \nu} \left( \xi ^{i} \right)
    \dfn
    \epsilon ^{ab}
    \partial _{a} X ^{\mu} \ttt{\xi ^{i}}
    \partial _{b} X ^{\nu} \ttt{\xi ^{i}}
    \dfn
    \pois{X ^{\mu}}{X ^{\nu}} _{\xi}
    \quad ,
    $$
    where we use the shorthand notation
    $$
        \partial _{a} X ^{\mu} \ttt{\xi ^{i}}
        =
        \frac{\partial X ^{\mu} \ttt{\xi ^{i}}}{\partial \xi ^{a}}
        \quad .
    $$
\end{defs}

A similar quantity can be defined for the {\it Boundary} of the
{\sl World--Sheet} as well:
\begin{defs}[String Area Velocity]\spbcorr{}.
    \label{2.strareveldef}\\
    Let $\dot{X} ^{\mu \nu} \ttt{\xi ^{i}}$ be the
    {\sl Local Area Velocity} of the {\sl World--Sheet}, associated
    with a {\sl String};
    the \underbar{String Area Velocity} is
    \beq
    \dot{X} ^{\mu \nu} \tst
    \dfn
    \left .
        \epsilon ^{ab}
        \partial _{a} X ^{\mu} \left( \xi ^{c} \right)
        \partial _{b} X ^{\nu} \left( \xi ^{c} \right)
    \right \rceil _{\xi ^{c} = \xi ^{c} \tst}
    =
        \pois{X ^{\mu} \left( \xi ^{c} \right)}
             {X ^{\nu} \left( \xi ^{c} \right)} _{\xi ^{c} = \xi ^{c} \tst}
    \quad ,
    \label{2.strareveldefequ}
    \eeq
    i.e. the {\sl Local Area Velocity} of the {\sl World--Sheet}
    computed on the boundary.
\end{defs}
We remark that the {\sl String Area Velocity} implicitly
encodes the information that the {\sl String} is {\it glued} to a
{\sl World--Sheet}. This means that it is {\it impossible} to express
$\dot{X} ^{\mu \nu} \tst$ only in terms of
$Y ^{\mu} \tst$, without any referenced to the {\sl Local Area Velocity} of the
{\sl World--Sheet.} A
{\sl String
} is not just a free fluctuating loop, because it is also the
boundary of a {\sl World--Sheet}. We will have the opportunity in chapter
\ref{9.connection} to spend some more comments about this feature,
enlightening also the relation between loop Dynamics and our formulation
of {\sl String} Dynamics. Momentarily, let us introduce the following notation,
which is useful to distinguish between {\it Bulk} and
{\it Boundary} quantities:
\begin{nots}[String Area Velocity]\spbcorr{}.\\
    We will indicate the {\sl String Area Velocity} with the symbol
    $$
        \dot{Y} ^{\mu \nu} \tst
        \quad .
    $$
\end{nots}
We stress again that this is just a convenient notation to understand
at first sight if we are dealing with a {\it Bulk} quantity or with
a {\it Boundary} one: {\it but$\,$} it is {\it NOT$\,$} possible to compute
$\dot{Y} ^{\mu \nu}$ with equation \ref{2.strareveldefequ} by simply
replacing $X ^{\mu}$ with $Y ^{\mu}$, because the passage to
the boundary is definitely non trivial: namely our {\sl String} is still
attached to its {\sl World--Sheet}!

As already pointed out by Schild \cite{schild} the Lagrangian density
(\ref{2.schild}) has some shortcomings as well as some advantages. The most
relevant problem is that it lacks a full reparametrization invariance, being
invariant only under area preserving transformations, i.e transformations
of the {\sl Parameter Space} of the type
\beq
    \left( \xi ^{0} , \xi  ^{1} \right)
    \longrightarrow
    \left( \bar{\xi} ^{\,0} , \bar{\xi} ^{\,1} \right)
    \quad
    :
    \qquad
    \left |
        \frac{
              \partial
              \ttt{\bar{\xi} ^{\,0} , \bar{\xi} ^{\,1}}
             }
             {
              \partial
              \left( \xi ^{0} , \xi ^{1} \right)
             }
    \right |
    =
    1
    \quad .
\eeq
Neverthless, it is possible to prove by varying the action functional
\beq
    S _{\mathrm{Schild}}
    =
    \int _{\Xi} \Lag _{\mathrm{Schild}} d \xi ^{1} \wedge d \xi ^{2}
    \label{2.schact}
\eeq
that
\begin{props}[Schild String Equations of Motion]\spbcorr{}.\\
    The equations of motion for a {\sl String} (i.e a {\sl String}
    defined by the Schild Lagrangian density) are
    $$
        \epsilon ^{ab} \partial _{a} \left [
                                         \dot{X} _{\mu \nu}
                                         \partial _{b} X ^{\nu}
                                     \right ]
        =
        0
    $$
    and they correspond to motions with constant
    $\ttt{\dot{X} ^{\mu \nu}} ^{2}$.
\end{props}
\begin{proof}
We use the same notation as before, where the {\sl String} is defined
on the {\sl Parameter Space} $\Xi$ and thus the chosen
{\sl Parametrization} for the {\sl World--Sheet}
is denoted by $X \ttt{\Bxi}$. Then the equations of motion for the
{\sl Schild String} are the Euler--Lagrange equations associated with
the Lagrangian Density (\ref{2.schild}). We thus compute
$$
    \frac{\partial \Lag _{\mathrm{Schild}}}{\partial X ^{\alpha}} = 0
$$
and
\bea
    \frac{\partial \Lag _{\mathrm{Schild}}}
         {\partial \ttt{\partial _{a} X ^{\alpha}}}
    & = &
    \frac{\partial \Lag _{\mathrm{Schild}}}
         {\partial \dot{X} ^{\rho \tau}}
    \frac{\partial \dot{X} ^{\rho \tau}}
         {\partial \ttt{\partial _{a} X ^{\alpha}}}
    \nonumber \\
    & = &
    \frac{\partial \Lag _{\mathrm{Schild}}}
         {\partial \dot{X} ^{\rho \tau}}
    \epsilon ^{mn}
    \left[
        \delta _{m} ^{a} \delta ^{\rho} _{\alpha} \partial _{n} X ^{\tau}
        +
        \delta _{n} ^{a} \delta ^{\tau} _{\alpha} \partial _{m} X ^{\rho}
    \right]
    \nonumber \\
    & = &
    \frac{m ^{2}}{4}
    2
    \dot{X} _{\alpha \tau}
    2
    \epsilon ^{an}
    \partial _{n} X ^{\tau}
    \quad .
    \nonumber
\eea
Then
$$
    \partial _{a}
    \left(
        \frac{\partial \Lag _{\mathrm{Schild}}}
             {\partial \ttt{\partial _{a} X ^{\alpha}}}
    \right)
    -
    \frac{\partial \Lag _{\mathrm{Schild}}}{\partial X ^{\alpha}}
    =
    0
$$
is equal to
$$
    \partial _{a}
    \left(
        m ^{2}
        \dot{X} _{\alpha \tau}
        \epsilon ^{an}
        \partial _{n} X ^{\tau}
    \right)
    =
    0
    \quad ,
$$
which is the desired result.
\end{proof}

At the same time with the same procedure, but starting from the
Nambu--Goto Lagrangian density (\ref{2.namgotlag}), we obtain
\begin{props}[Nambu-Goto Equations of Motions]\spbcorr{}.\\
    The equations of motion for a Nambu-Goto {\sl String} are
    $$
        \epsilon ^{ab}
        \partial _{a}
        \left [
            \frac{\dot{X} _{\mu \nu}}
                 {\sqrt{- \frac{1}{2} \dot{X} ^{\mu \nu} \dot{X} _{\mu \nu}}}
            \,
            \partial _{b} X ^{\nu}
        \right ]
        =
        0
        \quad .
    $$
\end{props}
\begin{proof}
The procedure is as before. We now have again
$$
    \frac{\partial \Lag _{\mathrm{NG}}}{\partial X ^{\alpha}} = 0
$$
but
\bea
    \frac{\partial \Lag _{\mathrm{NG}}}
         {\partial \ttt{\partial _{a} X ^{\alpha}}}
    & = &
    \frac{\partial \Lag _{\mathrm{NG}}}
         {\partial \dot{X} ^{\rho \tau}}
    \frac{\partial \dot{X} ^{\rho \tau}}
         {\partial \ttt{\partial _{a} X ^{\alpha}}}
    \nonumber \\
    & = &
    \frac{\partial \Lag _{\mathrm{NG}}}
         {\partial \dot{X} ^{\rho \tau}}
    \epsilon ^{mn}
    \left[
        \delta _{m} ^{a} \delta ^{\rho} _{\alpha} \partial _{n} X ^{\tau}
        +
        \delta _{n} ^{a} \delta ^{\tau} _{\alpha} \partial _{m} X ^{\rho}
    \right]
    \nonumber \\
    & = &
    -
    \frac{1}{2}
    T _{\mathrm{Str.}}
    \frac{2 \dot{X} ^{\alpha \tau}}
         {\sqrt{- \frac{1}{2} \dot{X} ^{\eta \lambda} \dot{X} _{\eta \lambda}}}
    2
    \epsilon ^{an}
    \partial _{n} X ^{\tau}
    \nonumber
    \quad .
\eea
The above equation is meaningful only if
$$
    \dot{X} ^{\eta \lambda} \dot{X} _{\eta \lambda}
    \neq
    0
$$
of course.
Then the Euler--Lagrange equations are
$$
    2
    \,
    T _{\mathrm{Str.}}
    \partial _{a}
    \left(
        \frac{\dot{X} _{\alpha \tau}}
             {\sqrt{- \frac{1}{2} \dot{X} ^{\eta \lambda} \dot{X} _{\eta \lambda}}}
        \epsilon ^{an}
        \partial _{n} X ^{\tau}
    \right)
    =
    0
    \quad ,
$$
as stated in the proposition.
\end{proof}

Thus we can see that
\begin{props}[Schild versus Nambu-Goto Equivalence]\spbcorr{}.\\
    For motions with
    $\ttt{\dot{X} ^{\mu \nu}} ^{2} = \mathrm{const.} \neq 0$
    the Schild {\sl String} has the same Dynamics of the Nambu--Goto {\sl String}.
\end{props}
\begin{proof}
We firstly observe that the following identity holds:
\bea
    \partial _{b}
    \left[
        \frac{
              \partial
              \left(
                    -
                    \frac{1}{4}
                    \dot{X} ^{\eta \lambda}
                    \dot{X} _{\eta \lambda}
              \right)
             }
             {
              \partial _{a} X ^{\mu}
             }
    \right]
    & = &
    -
    \partial _{b}
    \left[
        \frac{1}{4}
        2 \dot{X} _{\eta \lambda}
        \frac{\partial \dot{X} ^{\eta \lambda}}{\partial _{a} X ^{\mu}}
    \right]
    \nonumber \\
    & = &
    -
    \frac{1}{2}
    \partial _{b}
    \left[
        \dot{X} _{\eta \lambda}
        \frac{\partial}{\partial _{a} X ^{\mu}}
        \left(
            \epsilon ^{mn}
            \partial _{m} X ^{\eta}
            \partial _{n} X ^{\lambda}
        \right)
    \right]
    \nonumber \\
    & = &
    -
    \frac{1}{2}
    \partial _{b}
    \left[
        \dot{X} _{\eta \lambda}
        \epsilon ^{mn}
        \left(
            \delta _{n} ^{a} \delta _{\mu} ^{\lambda}
            \partial _{m} X ^{\eta}
            +
            \delta _{m} ^{a} \delta _{\mu} ^{\eta}
            \partial _{n} X ^{\lambda}
        \right)
    \right]
    \nonumber \\
    & = &
    \partial _{b}
    \left[
        \dot{X} _{\mu \lambda}
        \epsilon ^{an}
        \partial _{n} X ^{\lambda}
    \right]
    \quad .
\eea
Then we can use it to see that the equation of motion
for the Schild String
implies
\bea
    0
    & = &
    \partial _{b} X ^{\mu}
    \partial _{a}
    \left [
        \frac{
              \partial
              \left(
                    -
                    \frac{1}{4}
                    \dot{X} ^{\eta \lambda}
                    \dot{X} _{\eta \lambda}
              \right)
             }
             {
              \partial \ttt{\partial _{a} X ^{\mu}}
             }
    \right ]
    \nonumber \\
    & = &
    \partial _{a}
    \left [
        \partial _{b} X ^{\mu}
        \frac{
              \partial
              \left(
                    -
                    \frac{1}{4}
                    \dot{X} ^{\eta \lambda}
                    \dot{X} _{\eta \lambda}
              \right)
             }
             {
              \partial \ttt{\partial _{a} X ^{\mu}}
             }
    \right ]
    -
    \partial ^{2} _{ab} X ^{\mu}
    \left [
        \frac{
              \partial
              \left(
                    -
                    \frac{1}{4}
                    \dot{X} ^{\eta \lambda}
                    \dot{X} _{\eta \lambda}
              \right)
             }
             {
              \partial \ttt{\partial _{a} X ^{\mu}}
             }
    \right ]
    \nonumber \\
    & = &
    \partial _{b}
    \left(
          -
          \frac{1}{2}
          \dot{X} ^{\eta \lambda}
          \dot{X} _{\eta \lambda}
    \right)
    -
    \partial _{b}
    \left(
          -
          \frac{1}{4}
          \dot{X} ^{\eta \lambda}
          \dot{X} _{\eta \lambda}
    \right)
    \nonumber \\
    & = &
    \partial _{b}
    \left(
          -
          \frac{1}{4}
          \dot{X} ^{\eta \lambda}
          \dot{X} _{\eta \lambda}
    \right)
    \quad ;
\eea
then a Schild String motion has
$$
    \dot{X} ^{\eta \lambda}
    \dot{X} _{\eta \lambda}
    =
    \mathrm{const.}
    \quad .
$$
When this condition is satisfied, we see that the Nambu--Goto String
equation is
$$
    2
    T _{\mathrm{Str.}}
    \partial _{a}
    \left(
        \dot{X} _{\alpha \tau}
        \epsilon ^{an}
        \partial _{n} X ^{\tau}
    \right)
    =
    0
    \quad ,
$$
i.e. nothing but the Schild String equation.
This completes the proof.
\end{proof}

The similarity between the Schild {\sl String} and the Nambu--Goto
{\sl String} is hardly
surprising. Indeed, if we consider the Nambu--Goto {\sl String}
as a Classical Field Theory, it can be proved that, being
reparametrization invariant as a dynamical system, if $X ^{\mu}$ are the
coordinate variables, then the Lagrangian does not depend,
besides $X ^{\mu}$,
on all the conjugated momenta, but only on their antisymmetric combination
$$
    \dot{X} ^{\mu \nu}
    =
    \frac{\partial X ^{[ \mu}}{\partial \xi ^{0}}
    \frac{\partial X ^{\nu ]}}{\partial \xi ^{1}}
    =
    \frac{1}{2}
    \epsilon ^{ab}
    \frac{\partial X ^{\mu}}{\partial \xi ^{a}}
    \frac{\partial X ^{\nu}}{\partial \xi ^{b}}
\quad ,
$$
i.e. on the {\sl Local World--Sheet Area Velocity},
constrained in a proper way. Therefore, we can see that in the
Schild Lagrangian density
the main property implied, by reparametrization invariance
in $\Lag _{\mathrm{NG}}$ of equation (\ref{2.namgotlag}), i.e. the
statement about {\it which are the correct} {\bf speed} {\it variables},
is still there. Moreover one thing that we gain with the Schild
formulation is that it allows the description of null {\sl Strings}. The second
one is, of course, that the lack of full reparametrization invariance leads
to a well defined canonical Hamiltonian density. We first give some more
definitions.
\begin{defs}[Bulk Area Momentum]\spbcorr{}.
    \label{2.aremomdef}\\
    The \underbar{Bulk Area momentum} or
    \underbar{World--Sheet Area Momentum} is the momentum
    canonically conjugated to the Local World--Sheet Area Velocity, i.e.
    $$
        P _{\mu \nu} \ttt{\Bxi}
        =
        \frac{\partial \Lag _{\mathrm{Schild}}}
             {\partial \dot{X} ^{\mu \nu} \ttt{\Bxi}}
    \quad .
    $$
\end{defs}
Then, we extend this result from the {\it Bulk} to the {\it Boundary}
as we already did in the case of the {\sl Local World--Sheet Area
Velocity}:
\begin{defs}[Boundary Area Momentum]\spbcorr{}.
    \label{2.bouaremomdef}\\
    The \underbar{Boundary Area Momentum} or
    \underbar{String Area Momentum} is the
    {\sl Bulk Area Momentum} computed on the boundary, i.e.
    $$
        Q _{\mu \nu} \tst
        =
        \left .
            \frac{\partial \Lag}
                 {\partial \dot{X} ^{\mu \nu} \ttt{\Bxi}}
        \right \rceil _{\Bxi = \Bxi \tst}
        =
        \left .
            P _{\mu \nu} \ttt{\Bxi}
        \right \rceil _{\Bxi = \Bxi \tst}
    \quad ,
    $$
    where, $\Bxi = \Bxi \tst$ is a parametrization of the boundary
    $\partial \Xi$ of the domain\footnote{This definition is
    the natural counterpart of definition \ref{2.strareveldef}: again
    we go to the {\it Boundary},
    which is just a curve in the $\Bxi$ variables.
    As the {\sl Parametrization} of the boundary of the {\sl World--Sheet}
    is named with a letter $\By$ which is the next one after the one
    for the {\sl Parametrization} of the {\sl World--Sheet} itself, i.e.
    $\Bx$, in the alphabet, the standard name for the {\sl Boundary
    Area Momentum}, $\Bq$, is the next letter after the one for the
    {\sl Bulk Area Momentum}, namely $\Bp$.} $\Xi$.
\end{defs}
After this digression, which will be useful later on, we go back
to our previous assertion which is the argument of the following
proposition.
\begin{props}[Schild Hamiltonian Density]\spbcorr{}.\\
    The Schild {\sl String}  admits the Hamiltonian density
    \beq
        \Ham _{\mathrm{Schild}}
        =
        \frac{P ^{\mu \nu} \ttt{\Bxi} P _{\mu \nu} \ttt{\Bxi}}{4 m ^{2}}
        \quad ,
        \label{2.schham}
    \eeq
    where $P ^{\mu \nu} = P ^{\mu \nu} \ttt{\xi ^{i}}$ is the
    {\sl Bulk Area Momentum}.
\end{props}
\begin{proof}
Starting from the definition (\ref{2.aremomdef}), and remembering the
Schild Lagrangian Density (\ref{2.schild}), we firstly get
\bea
    P _{\mu \nu} \ttt{\Bxi}
    & = &
    \frac{\partial \Lag _{\mathrm{Schild}}}
         {\partial \dot{X} ^{\mu \nu} \ttt{\Bxi}}
    \nonumber \\
    & = &
    \frac{\partial}
         {\partial \dot{X} ^{\mu \nu} \ttt{\Bxi}}
         \left [
             \frac{m ^{2}}{4}
             \dot{X} ^{\rho \sigma} \ttt{\Bxi}
             \dot{X} _{\rho \sigma} \ttt{\Bxi}
         \right ]
    \nonumber \\
    & = &
    2
    \frac{m ^{2}}{4}
    \dot{X} _{\rho \sigma} \ttt{\Bxi}
    \frac{\partial \dot{X} ^{\rho \sigma} \ttt{\Bxi}}
         {\partial \dot{X} ^{\mu \nu} \ttt{\Bxi}}
    \nonumber \\
    & = &
    \frac{m ^{2}}{2}
    \dot{X} _{\rho \sigma} \ttt{\Bxi}
    \left [
        \delta ^{\mu} _{\rho} \delta ^{\nu} _{\sigma}
        -
        \delta ^{\mu} _{\sigma} \delta ^{\nu} _{\rho}
    \right ]
    \nonumber \\
    & = &
    m ^{2}
    \dot{X} _{\mu \nu} \ttt{\Bxi}
    \quad ,
    \nonumber
\eea
so that
$$
    \dot{X} _{\mu \nu} \ttt{\Bxi}
    =
    \frac{1}{m ^{2}} P _{\mu \nu} \ttt{\Bxi}
    \quad .
$$
Substituting the last equation in the expression for the Hamiltonian,
defined as the Legendre Transform of the Lagrangian (\ref{2.schild}),
\bea
    \Ham _{\mathrm{Schild}}
    & \equo{\footnotemark} &
    \frac{1}{2}
    P ^{\mu \nu} \ttt{\Bxi}
    \dot{X} _{\mu \nu} \ttt{\Bxi}
    -
    \Lag _{\mathrm{Schild}}
    \label{2.schlaglegtra} \\
    & = &
    \frac{1}{2}
    P ^{\mu \nu} \ttt{\Bxi}
    \frac{1}{m ^{2}} P _{\mu \nu} \ttt{\Bxi}
    -
    \frac{m ^{2}}{4}
    \frac{1}{m ^{2}} P _{\mu \nu} \ttt{\Bxi}
    \frac{1}{m ^{2}} P _{\mu \nu} \ttt{\Bxi}
    \nonumber \\
    & = &
    \frac{P ^{\mu \nu} \ttt{\Bxi} P _{\mu \nu} \ttt{\Bxi}}{4 m ^{2}}
    \quad ,
    \nonumber
\eea
we get the desired result.
\end{proof}
\footnotetext{We note that the factor $1/2$ in the Legendre Transform is
there to take into account the antisimmetry of the contracted indices;
in this way, overcounting is avoided. We adhere to the convention
that in the general case the factor is $1/(p!)$
(see also the computations in appendix \ref{A.HamDer}).}

So we see that a quantization procedure starting from the Schild action
can be performed more easily on the same guidelines of the quantization
of a classical particle. In particular we can translate the
Schild Action (\ref{2.schact}) derived from the Lagrangian
(\ref{2.schild}) in the Hamiltonian Formalism.
\begin{props}[Schild Action In Hamiltonian Form]\spbcorr{}.\\
    The Schild Action {\rm (\ref{2.schild})} admits the Hamiltonian
    form
    \beq
        S \qtq{\Bx , \Bp}
        =
        \frac{1}{2}
        \int _{\Xi} d ^{2} \Bxi
        \left [
                P _{\mu \nu} \ttt{\Bxi}
                \dot{X} ^{\mu \nu} \ttt{\Bxi}
                -
                \frac{1}{2 m ^{2}}
                P _{\mu \nu} \ttt{\Bxi}
                P ^{\mu \nu} \ttt{\Bxi}
        \right ]
    \quad .
    \label{2.schhamfun}
    \eeq
\end{props}
\begin{proof}
If we solve for the Lagrangian density in the Legendre Transform
(\ref{2.schlaglegtra}), we substitute the expression
(\ref{2.schham}) for the Hamiltonian and take out a common factor
$1/2$, we obtain
\bea
    \Lag _{\mathrm{Schild}}
    & = &
    \Lag _{\mathrm{Schild}} \ttt{\Bx , \Bp}
    \nonumber \\
    & = &
    \frac{1}{2}
    \left [
            P _{\mu \nu}
            \dot{X} ^{\mu \nu}
            -
            \frac{1}{2 m ^{2}}
            P _{\mu \nu}
            P ^{\mu \nu}
    \right ]
    \nonumber
\eea
for the Lagrangian, which we then put in (\ref{2.schact}).
\end{proof}

Neverthless, we do not like giving up
reparametrization invariance without fighting! We will present in
the following section a possible way out of these seemingly
problematic situation.

\section{Reparametrized Schild Formulation}
\label{2.repschfor}

The central step in our proposal is already present in the following
key remark:
\begin{props}[Schild Lagrangian Variation
              under Reparametrization]\spbcorr{}.
    \label{2.repfirstep}\\
    Under a reparametrization of the domain $\Xi$,
    $$
       \matrix{
          \euf{s} & : & \Sigma & \longrightarrow & \Xi \cr
                  &   & \ttt{\sigma ^{0} , \sigma ^{1}}
                      & \longrightarrow &
                      \ttt{
                           \xi ^{0} \ttt{\sigma ^{0} , \sigma ^{1}}
                           ,
                           \xi ^{1} \ttt{\sigma ^{0} , \sigma ^{1}}
                          }
                      \quad ,
              }
    $$
    the $2$-form
    $$
        \form{\omega}
        =
        \Lag _{\mathrm{Schild}}
            \ttt{
                 X ^{\mu} \ttt{\xi ^{0} , \xi ^{1}}
                 ,
                 \dot X ^{\mu \nu} \ttt{\xi ^{0} , \xi ^{1}}
                }
        \form{d \xi} ^{0} \wedge \form{d \xi} ^{1}
    $$
    transforms as
    $$
        \form{\omega}
        \longrightarrow
        \form{\tilde{\omega}}
    $$
    where,
    $$
        \form{\tilde{\omega}}
        =
        \frac{1}{\dete{\mathcal{J} \ttt{\euf{s}}}}
        \Lag _{\mathrm{Schild}}
            \ttt{
                 \tilde{X} ^{\mu} \ttt{\sigma ^{0} , \sigma ^{1}}
                 ,
                 \dot{\tilde{X}} {}^{\mu \nu} \ttt{\sigma ^{0} , \sigma ^{1}}
                }
        \form{d \sigma} ^{0} \wedge \form{d \sigma} ^{1}
    \quad ,
    $$
    \bea
        \tilde{X} ^{\mu} \ttt{\sigma ^{0} , \sigma ^{1}}
        & = &
        X ^{\mu}
        \ttt{
             \xi ^{0} \ttt{\sigma ^{0} , \sigma ^{1}}
             ,
             \xi ^{1} \ttt{\sigma ^{0} , \sigma ^{1}}
            }
        \quad ,
        \nonumber \\
        \dot{\tilde{X}} {}^{\mu \nu} \ttt{\sigma ^{0} , \sigma ^{1}}
        & = &
        \dot{X} ^{\mu \nu}
        \ttt{
             \xi ^{0} \ttt{\sigma ^{0} , \sigma ^{1}}
             ,
             \xi ^{1} \ttt{\sigma ^{0} , \sigma ^{1}}
            }
    \nonumber
    \eea
    and
    $$
        \mathcal{J} \ttt{\euf{s}}
        \equiv
        \left(
            \matrix{
                    \displaystyle
                    \frac{\partial \xi ^{0}}{\partial \sigma ^{0}}
                    &
                    \displaystyle
                    \frac{\partial \xi ^{0}}{\partial \sigma ^{1}}
                    \cr
                    \cr
                    \displaystyle
                    \frac{\partial \xi ^{1}}{\partial \sigma ^{0}}
                    &
                    \displaystyle
                    \frac{\partial \xi ^{1}}{\partial \sigma ^{1}}
                   }
        \right)
    $$
    is the Jacobean matrix of the transformation $\euf{s}$.
\end{props}
\begin{proof}
The Lagrangian one form $\form{\omega}$
depends only from the {\sl Area Velocity}
$\dot{X} ^{\mu \nu}$. Thus, we  get
\bea
    \dot{\tilde{X}} {}^{\mu \nu}
    & = &
    \epsilon ^{ab}
    \frac{\partial \tilde{X} ^{\mu} \ttt{\Bsigma}}
         {\partial \sigma ^{a}}
    \frac{\partial \tilde{X} ^{\nu} \ttt{\Bsigma}}
         {\partial \sigma ^{b}}
    \nonumber \\
    & = &
    \epsilon ^{ab}
    \frac{\partial X ^{\mu} \ttt{\Bxi \ttt{\Bsigma}}}
         {\partial \sigma ^{a}}
    \frac{\partial X ^{\nu} \ttt{\Bxi \ttt{\Bsigma}}}
         {\partial \sigma ^{b}}
    \nonumber \\
    & = &
    \epsilon ^{ab}
    \frac{\partial X ^{\mu} \ttt{\Bxi}}
         {\partial \xi ^{A}}
    \frac{\partial X ^{\nu} \ttt{\Bxi}}
         {\partial \xi ^{B}}
    \frac{\partial \xi ^{A}}{\partial \sigma ^{a}}
    \frac{\partial \xi ^{B}}{\partial \sigma ^{b}}
    \quad .
    \label{2.areveltra}
\eea
Now, let us consider the quantity
$$
    \epsilon ^{ab}
    \frac{\partial \xi ^{A}}{\partial \sigma ^{a}}
    \frac{\partial \xi ^{B}}{\partial \sigma ^{b}}
    \quad :
$$
this is a totally antisymmetric $2$nd rank tensor in two dimensions,
and thus it must be proportional to the Levi--Civita tensor; then,
\beq
    \epsilon ^{ab}
    \frac{\partial \xi ^{A}}{\partial \sigma ^{a}}
    \frac{\partial \xi ^{B}}{\partial \sigma ^{b}}
    =
    \epsilon ^{AB}
    \kappa \ttt{\Bsigma}
    \quad .
    \label{2.levcivequuno}
\eeq
The expression of $\kappa \ttt{\Bsigma}$ can be deduced,
for example, by looking at the
$a=0$, $b=1$ component of the tensor in the equation above;
then, equation (\ref{2.levcivequuno}) reduces to
\bea
    \kappa \ttt{\Bsigma}
    =
    \epsilon _{01}
    \kappa \ttt{\Bsigma}
    & = &
    \epsilon ^{ab}
    \frac{\partial \xi ^{0}}{\partial \sigma ^{a}}
    \frac{\partial \xi ^{1}}{\partial \sigma ^{b}}
    \nonumber \\
    & = &
    \frac{1}{2}
    \epsilon ^{ab}
    \epsilon _{AB}
    \frac{\partial \xi ^{A}}{\partial \sigma ^{a}}
    \frac{\partial \xi ^{B}}{\partial \sigma ^{b}}
    \nonumber \\
    & = &
    \dete{\mathcal{J} \ttt{\euf{s}}}
    \quad .
\eea
From equation (\ref{2.areveltra}) we thus obtain
\bea
    \dot{\tilde{X}} {}^{\mu \nu}
    & = &
    \dete{\mathcal{J} \ttt{\euf{s}}}
    \epsilon _{AB}
    \frac{\partial X ^{\mu} \ttt{\Bxi}}
         {\partial \xi ^{A}}
    \frac{\partial X ^{\nu} \ttt{\Bxi}}
         {\partial \xi ^{B}}
    \nonumber \\
    & = &
    \dete{\mathcal{J} \ttt{\euf{s}}}
    \dot{X} ^{\mu \nu}
    \nonumber
\eea
or, equivalently
\beq
    \dot{X} ^{\mu \nu}
    =
    \frac{\dot{\tilde{X}} {}^{\mu \nu}}
         {\dete{\mathcal{J} \ttt{\euf{s}}}}
    \quad .
    \label{2.schlagvaruno}
\eeq
At the same time, we have
\bea
    \form{d \xi} ^{0} \wedge \form{d \xi} ^{1}
    & = &
    \frac{1}{2}
    \epsilon _{AB}
    \form{d \xi} ^{A} \wedge \form{d \xi} ^{B}
    \nonumber \\
    & = &
    \frac{1}{2}
    \epsilon _{AB}
    \frac{\partial \xi ^{A}}{\partial \sigma ^{i}}
    \frac{\partial \xi ^{B}}{\partial \sigma ^{j}}
    \form{d \sigma} ^{i} \wedge \form{d \sigma} ^{j}
    \nonumber \\
    & = &
    \frac{1}{2!}
    \epsilon _{AB}
    \frac{\partial \xi ^{A}}{\partial \sigma ^{i}}
    \frac{\partial \xi ^{B}}{\partial \sigma ^{j}}
    \epsilon ^{ij}
    \form{d \sigma} ^{0} \wedge \form{d \sigma} ^{1}
    \nonumber \\
    & = &
    \left [
        \frac{1}{2!}
        \epsilon _{AB}
        \epsilon ^{ij}
        \frac{\partial \xi ^{A}}{\partial \sigma ^{i}}
        \frac{\partial \xi ^{B}}{\partial \sigma ^{j}}
    \right ]
    \form{d \sigma} ^{0} \wedge \form{d \sigma} ^{1}
    \nonumber \\
    & = &
    \dete{
          \frac{\partial \ttt{\xi ^{0} , \xi ^{1}}}
               {\partial \ttt{\sigma ^{0} , \sigma ^{1}}}
         }
    \form{d \sigma} ^{0} \wedge \form{d \sigma} ^{1}
    \nonumber \\
    & = &
    \dete{\mathcal{J} \ttt{\euf{s}}}
    \form{d \sigma} ^{0} \wedge \form{d \sigma} ^{1}
    \quad .
    \label{2.schlagvardue}
\eea
Then, by substituting equations (\ref{2.schlagvaruno}) and
(\ref{2.schlagvardue}) into the Schild Lagrangian $2$-form
$\form{\omega}$, we get
$$
    \tilde{\form{\omega}}
    =
    \frac{\dot{\tilde{X}} {}^{\mu \nu} \dot{\tilde{X}} _{\mu \nu}}
         {\dete{\mathcal{J} \ttt{\euf{s}}}}
    \form{d \sigma} ^{0} \wedge \form{d \sigma} ^{1}
    \quad ,
$$
the desired result.
Incidentally, we observe that if we define in a natural way
$$
    \dot{\xi} ^{AB}
    =
    \epsilon ^{ab}
    \frac{\partial \xi ^{A}}{\partial \sigma ^{a}}
    \frac{\partial \xi ^{B}}{\partial \sigma ^{b}}
$$
then,
\beq
    \dete{\mathcal{J} \ttt{\euf{s}}}
    =
    \frac{1}{2}
    \epsilon _{ab}
    \dot{\xi} ^{ab}
    \quad .
    \label{2.detexp}
\eeq
\end{proof}

This is the key step toward our next proposal:
{\it we can lift the variables $\xi ^{a}$ to the role of dynamical
fields provided we introduce a new pair of  coordinates, and
transforming our original model into a two dimensional field Theory
in six dimensions}. We point out that there is no
more dependence in the $X ^{\mu}$ variables from the $\xi ^{A}$
variables, which are now independent fields: the
{\sl World--Sheet} coordinates, and the new
$\xi ^{M}$ fields are  both
functions of the new parameters\footnote{Please, see notation
(\ref{2.indnot}) below for the change
in the notation of the index.} $\sigma ^{a}$.
\begin{nots}[Indexing of Fields and Parameters]\spbcorr{}.
    \label{2.indnot}\\
    To keep a clear distinction between the new fields $\Bxi$,
    and the new parameters $\sigma ^{a}$, from now on we
    will use capital latin indices for the new fields: $\xi ^{A}$.
    This should also be seen as a clear distinction between
    the previous role of the $\Bxi$ and the present one!
    Moreover, all the quantities already defined on $\Xi$, when referred
    after this point, are to be considered as defined on $\Sigma$
    and thus functions of the $\Bsigma$ variables$\,$\footnote{This is just
    a matter of convention. We are aware that changing the
    name of the variables should not be a problem. However, we believe
    that a good notation can enlighten some important aspects
    of the underlying concepts. Accordingly, from now on we will consistently
    parametrize the {\sl World--Sheet} with $\Bsigma \in \Sigma$.}.
\end{nots}
In light of proposition \ref{2.repfirstep} we  {\it introduce}
the {\it  Reparametrized Schild Lagrangian density}.
\begin{defs}[Reparametrized Schild
             Lagrangian Density]\spbcorr{}.\\
    The \underbar{Reparametrized Schild Lagrangian Density} is
    \beq
        \Lag   =  \Lag ^{\mathrm{rep.}} _{\mathrm{Schild}}
             \dfn \frac{m ^{2}}{2}
                  \frac{
                        \dot{X} ^{\mu \nu} \ttt{\sigma ^{0} , \sigma ^{1}}
                        \dot{X} _{\mu \nu} \ttt{\sigma ^{0} , \sigma ^{1}}
                       }
                       {
                        \epsilon ^{AB}
                        \dot{\xi} _{AB} \ttt{\sigma ^{0} , \sigma ^{1}}
                       }
    \label{2.repschlag}
    \eeq
    so that, the basic dynamical object is now the $2$-form
    \beq
        \form{\Omega} = \frac{m ^{2}}{2}
                        \frac{
                              \dot{X} ^{\mu \nu}
                              \dot{X} _{\mu \nu}
                             }
                             {\epsilon ^{AB} \dot{\xi} _{AB}}
                        \form{d \sigma} ^{0} \wedge \form{d \sigma} ^{1}
        \quad .
    \eeq
\end{defs}
Note that we used here relation (\ref{2.detexp}).
Some clarifications are due at this stage.
\begin{enumerate}
    \item The Theory defined by the previous formulae is
    {\it reparametrization invariant}: thus in this formulation
    we recover the main property of the Nambu--Goto action.
    Neverthless, it is polynomial in the dynamical variables,
    because this $6$-dimensional extension does not treat all the
    fields on the same footing. It is basically different
    from, say, the $6$-dimensional extension that we could obtain
    by simply considering $X ^{\mu}$ as a $6$-dimensional vector, e.g.
    there is no Lorenz symmetry relating the $\xi ^{A}$ and $X ^{\mu}$ fields.
    \item The last, but not the least, we can connect the meaning of the
    additional fields with the discussion about boundary Dynamics we made
    in the previous chapter; we saw that assigning a dynamical role
    to the boundary implies the need to vary the boundary itself. This
    is a non--standard procedure in field Theory.
    In this new framework, we expect to capture the whole boundary Dynamics
    through the variation of the $\xi ^{A}$ fields. These extra fields
    provide a geometrical {\it representation} of the {\sl Parameter Space}
    after an arbitrary reparametrization.
    Thus, they give us the opportunity to build up an
    ``ordinary Field Theory'',
    embodying arbitrary variations of the {\sl Parameter Space} itself.
   \end{enumerate}
Before concluding this section we take the first results
from the {\sl Reparametrized Schild Lagrangian Density},
giving the Hamiltonian formulation of the Theory:
\begin{props}[Conjugated Momenta in Reparametrized Formulation]\spbcorr{}.
    \label{2.repschconmom}\\
    The momenta canonically conjugated to the {\sl Area Velocities}
    are
    \bea
        P ^{\mu \nu}
        & = &
        2 m ^{2}
        \frac{\dot{X} ^{\mu \nu}}{\epsilon _{CD} \dot{\xi} ^{CD}}
        \label{2.Pmunu}
        \\
        \pi ^{AB}
        & = &
        -
        m ^{2}
        \frac{\dot{X} ^{\mu \nu} \dot{X} _{\mu \nu}}
             {\ttt{\epsilon _{CD} \dot{\xi} ^{CD}} ^{2}}
        \epsilon ^{AB}
        =
        -
        m ^{2}
        \frac{P ^{\mu \nu} P _{\mu \nu}}{2}
        \epsilon ^{AB}
        =
        - \epsilon ^{AB}
        \Ham _{\mathrm{Schild}} \ttt{\Bp}
        \quad .
        \label{2.piAB}
    \eea
\end{props}
\begin{proof}
To get the results it is just necessary to comput the derivatives
of the reparametrized Lagrangian Density, with respect to
$\dot{X} ^{\mu \nu}$ and $\dot{\xi} ^{AB}$. We thus have
\bea
    \frac{\partial \Lag}{\partial \dot{X} ^{\mu \nu}}
    & = &
    \frac{m ^{2}}{2}
    \frac{1}{\epsilon ^{AB} \dot{\xi} _{AB}}
    2
    \dot{X} ^{\lambda \eta}
    \frac{\partial \dot{X} ^{\lambda \eta}}{\partial \dot{X} ^{\mu \nu}}
    \nonumber \\
    & = &
    m ^{2}
    \frac{1}{\epsilon ^{AB} \dot{\xi} _{AB}}
    \dot{X} ^{\lambda \eta}
    \left [
        \delta ^{\lambda} _{\mu} \delta ^{\eta} _{\nu}
        -
        \delta ^{\lambda} _{\nu} \delta ^{\eta} _{\mu}
    \right ]
    \nonumber \\
    & = &
    m ^{2}
    \frac{1}{\epsilon ^{AB} \dot{\xi} _{AB}}
    2
    \dot{X} ^{\mu \nu}
    \nonumber \\
    2 & = &
    m ^{2}
    \frac{\dot{X} ^{\mu \nu}}{\epsilon ^{AB} \dot{\xi} _{AB}}
    \nonumber
    \quad .
\eea
In the same way we can obtain the second result:
\bea
    \frac{\partial \Lag}{\partial \dot{\xi} ^{AB}}
    & = &
    \frac{m ^{2}}{2}
    \frac{\dot{X} ^{\mu \nu} \dot{X} _{\mu \nu}}
         {\left( \epsilon ^{EF} \dot{\xi} _{EF} \right) ^{2}}
    \frac{\partial \left( \epsilon _{CD} \dot{\xi} ^{CD} \right)}
         {\partial \dot{\xi} ^{AB}}
    \nonumber \\
    & = &
    \frac{m ^{2}}{2}
    \frac{\dot{X} ^{\mu \nu} \dot{X} _{\mu \nu}}
         {\left( \epsilon ^{EF} \dot{\xi} _{EF} \right) ^{2}}
    \epsilon _{CD}
    \frac{\partial \dot{\xi} ^{CD}}
         {\partial \dot{\xi} ^{AB}}
    \nonumber \\
    & = &
    \frac{m ^{2}}{2}
    \frac{\dot{X} ^{\mu \nu} \dot{X} _{\mu \nu}}
         {\left( \epsilon ^{EF} \dot{\xi} _{EF} \right) ^{2}}
    \epsilon _{CD}
    \left(
        \delta ^{AC} \delta ^{BD}
        -
        \delta ^{AD} \delta ^{BC}
    \right)
    \nonumber \\
    & = &
    m ^{2}
    \frac{\dot{X} ^{\mu \nu} \dot{X} _{\mu \nu}}
         {\left( \epsilon ^{EF} \dot{\xi} _{EF} \right) ^{2}}
    \epsilon _{CD}
    \nonumber \\
    & = &
    \epsilon _{CD}
    \Ham _{\mathrm{Schild}} \ttt{P _{\mu \nu}}
    \quad .
\eea
\end{proof}

According with \cite{kastrup} the $1$-form defining the Hamiltonian
Theory will be
\bea
    \form{\Omega} _{\mathrm{H}}
    & = &
    P _{\mu \nu} \, \form{d X} ^{\mu} \wedge \form{d X} ^{\nu}
    +
    \pi _{AB} \, \form{d \xi} ^{A} \wedge \form{d \xi} ^{B}
    +
    \nonumber \\
    & & \qquad +
    \frac{1}{2}
    N ^{AB}
    \left(
        \pi _{AB} - \epsilon _{AB} \Ham _{\mathrm{Schild}} \ttt{\Bp}
    \right)
    \form{d \sigma} ^{0} \wedge \form{d \sigma} ^{1}
    \quad ,
\label{2.fulrephamfor}
\eea
where $N ^{AB} \equiv N ^{AB} \ttt{\sigma ^{0} , \sigma ^{1}}$ is simply
a Lagrange multiplier enforcing the
relation (\ref{2.piAB}) between the momenta
$P ^{\mu \nu}$ and $\pi ^{AB}$. We also note that the expression
$\Ham _{\mathrm{Schild}} \ttt{\Bp}$ means the quantity which has the
functional dependence of $\Ham _{\mathrm{Schild}}$ but from the variable
which is the conjugated momentum of equation (\ref{2.Pmunu}).

\section{Hamilton--Jacobi Theory}
\label{2.hamjacsec}

In our opinion, as already pointed out on page \pageref{1.HamJacrel},
the Hamilton--Jacobi formulation of the Dynamics of a classical system
is a preferred starting point toward the quantum framework. In this section,
we will derive the Hamilton--Jacobi equation for a closed {\sl String} from
two different perspectives in order to enlighten the properties of the
{\it Boundary} Dynamics,
and to prepare in a solid way the derivation of
a {\it Schr\"o{}dinger like functional equation.}

\subsection{Hamilton--Jacobi Equation: Ogielski Formulation}
\label{2.hamjacsecogi}

As we already pointed out at the end of subsection \ref{1.ogielski}, on
page \pageref{1.ogielskiHJ}, if we are able to eliminate the vector
$\vect{T}$ from the functional derivatives
(\ref{1.oginorfunder}, \ref{1.ogiNorfunder}) of the action (\ref{1.action}),
then we can find a relation between them.  This relation is nothing
but the Hamilton--Jacobi equation. Then, we shall specialise this result to
the Lagrangian density (\ref{2.schild}) for the Schild {\sl String}.
As a preliminary result we can write the {\sl String Area Velocity}
in the basis $\ttt{\BT , \BN}$ given by the vectors tangent to the boundary
and normal to the {\sl World--Sheet} at the boundary.
\begin{props}[String Area Velocity
              in $\ttt{\BT , \BN}$ Coordinates]\spbcorr{}.\\
    The {\sl String Area Velocity} expressed
    in terms of the vector fields
    normal $\vect{N}$ and tangent $\vect{T}$ to the loop $C$,
    representing the {\sl String} in {\sl Target Space}, is
    \beq
        \dot{X} ^{\alpha \beta}
        =
        \frac{1}{\ttt{\Bsigma '} ^{2}}
        \left(
            T ^{\alpha} N ^{\beta}
            -
            T ^{\beta} N ^{\alpha}
        \right)
    \label{2.TNstrarevel}
    \eeq
\end{props}
\begin{proof}
We start from the expression we want to prove and,
remembering the definitions
(\ref{1.tanpar}, \ref{1.norpar}), we can rewrite it as
as
\bea
    \dot{X} ^{\alpha \beta}
    & = &
    \frac{1}{\ttt{\Bsigma '} ^{2}}
    \left(
        T ^{\alpha} N ^{\beta}
        -
        T ^{\beta} N ^{\alpha}
    \right)
    \nonumber \\
    & = &
    \frac{1}{\ttt{\Bsigma '} ^{2}}
    \left(
        \sigma ' _{c} \tst \partial ^{c} X ^{\alpha}
        \epsilon ^{bd}
        \sigma ' _{b} \tst \partial _{d} X ^{\beta}
        -
        \sigma ' _{c} \tst \partial ^{c} X ^{\beta}
        \epsilon ^{bd}
        \sigma ' _{b} \tst \partial _{d} X ^{\alpha}
    \right)
    \nonumber \\
    & = &
    \frac{1}{\ttt{\Bsigma '} ^{2}}
    \left(
        \sigma ' _{c} \tst \partial ^{c} X ^{\alpha}
        \epsilon ^{bd}
        \sigma ' _{b} \tst \partial _{d} X ^{\beta}
        -
        \sigma ' _{d} \tst \partial ^{d} X ^{\beta}
        \epsilon ^{bc}
        \sigma ' _{b} \tst \partial _{c} X ^{\alpha}
    \right)
    \nonumber \\
    & = &
    \frac{1}{\ttt{\Bsigma '} ^{2}}
    \left [
        \sigma ' _{c} \tst
        \epsilon _{bd}
        \sigma ^{\prime b} \tst
        -
        \sigma ' _{d} \tst
        \epsilon _{bc}
        \sigma ^{\prime b} \tst
    \right ]
    \partial ^{c} X ^{\alpha}
    \partial ^{d} X ^{\beta}
    \quad .
\label{2.arevel.0}
\eea
Now, we consider the quantity in square brackets: it is a tensor in
$2$ dimensions with $2$ indices which are skew--symmetric. Thus, it
must be proportional to the Levi--Civita tensor in $2$-dimensions, i.e.
we have
\beq
    \left [
        \sigma ' _{c} \tst
        \epsilon _{bd}
        \sigma ^{\prime b} \tst
        -
        \sigma ' _{d} \tst
        \epsilon _{bc}
        \sigma ^{\prime b} \tst
    \right ]
    =
    k \tst \epsilon _{cd}
    \quad .
\label{2.arevel.1}
\eeq
The expression $k \tst$ can be determined by observing that if $c=0, d=1$, then
we get
$$
    k \tst = \ttt{\sigma ^{\prime \, 0}} ^{2}
             +
             \ttt{\sigma ^{\prime \, 1}} ^{2}
           = \ttt{\Bsigma '} ^{2}
    \quad ;
$$
so, by substituting this result, together with (\ref{2.arevel.1}),
in (\ref{2.arevel.0}) we recover
\bea
    \dot{X} ^{\alpha \beta}
    & = &
    \frac{1}{\Bsigma ^{\prime 2}}
    \Bsigma ^{\prime 2}
    \epsilon _{cd}
    \,
    \partial ^{c} X ^{\alpha}
    \partial ^{d} X ^{\beta}
    \nonumber \\
    & = &
    \epsilon _{cd}
    \,
    \partial ^{c} X ^{\alpha} \ttt{\Bsigma \tst}
    \partial ^{d} X ^{\beta} \ttt{\Bsigma \tst}
    \nonumber \\
    & = &
    \left .
        \epsilon _{cd}
        \,
        \partial ^{c} X ^{\alpha} \ttt{\Bsigma \tst}
        \partial ^{d} X ^{\beta} \ttt{\Bsigma \tst}
    \right \rceil _{\Bsigma = \Bsigma \tst}
    \dfn
    \dot{Y} ^{\alpha \beta} \ttt{s}
    \ ,
\eea
which coincides with the definition \ref{2.strareveldef}\footnote{Please,
remember that with respect to this definition we now have renamed
$\Xi$ with $\Sigma$ and $\Bxi$ with $\Bsigma$.}.
\end{proof}

Again, we remember that $\dot{Y} ^{\alpha \beta} \ttt{s}$ is only a
notational way to make immediatly clear that we have to do with
a {\it Boundary} quantity. By no means it is possible to compute
this result relying only on the {\sl Embedding Function} of the loop $C$ in
{\sl Parameter Space},
because in general there is {\it no} unique {\it normal}
defined for a loop in a space with more than two dimensions. Thus we
see that the Dynamics of the {\sl World--Sheet} is  tightly connected with
the {\it Shadow Dynamics} of the boundary; we shall
comment again about  this remark later on\footnote{The curious reader
could make a quick jump to page \pageref{9.connecbou}, section
\ref{9.connecbou}.}. For the sake of clarity, we also set the following
notation.
\begin{nots}[Linear Velocity Vector]\spbcorr{}.\\
    The {\sl Linear Velocity Vector} of the loop $C$, $\vect{T}$,
    will be indicated with $\By '$ ($Y ^{\prime \mu}$ in components).
\end{nots}

We can now turn to the main result of this section:
\begin{props}[Ogielski Hamilton--Jacobi String Equation]\spbcorr{}.\\
    The Schild {\sl String} Shadow Dynamics is described by the
    following reparametrization invariant Hamilton--Jacobi equation:
    $$
        \frac{\partial S}{\partial A}
        =
        \frac{1}{2 m ^{2}}
        \norme{\Gamma}
        \oint _{\Gamma} \frac{ds}{\sqrt{Y ^{\prime \mu} Y ' _{\mu}}}
            \frac{\delta S}
                 {\delta Y ^{\mu} \tst}
            \frac{\delta S}
                 {\delta Y _{\mu} \tst}
    \quad .
    $$
\end{props}
\begin{proof}
We will use the expression for the {\sl Area Velocity} $\dot{Y} ^{\mu \nu}$
of the {\sl String} on the boundary $\Gamma$ of the domain $\Sigma$,
we derived in the previous proposition.
Starting from the result (\ref{2.TNstrarevel}), we can derive the
following results for the various terms appearing in equations
(\ref{1.oginorfunder}-\ref{1.ogiNorfunder}):
\bea
    \frac{\partial \Lag}{\partial N ^{\mu}}
    & = &
    \frac{m ^{2}}{2}
    \dot{Y} _{\alpha \beta}
    \frac{\partial \dot{Y} ^{\alpha \beta}}{\partial N ^{\mu}}
    \nonumber \\
    & = &
    \frac{m ^{2}}{2}
    \frac{1}{\sqrt{\ttt{\Bsigma '} ^{2}} ^{4}}
    \left(
        T _{\alpha} N _{\beta}
        -
        T _{\beta} N _{\alpha}
    \right)
    \left(
        T ^{\alpha} \delta ^{\beta} _{\mu}
        -
        T ^{\beta} \delta ^{\alpha} _{\mu}
    \right)
    \nonumber \\
    & = &
    \frac{m ^{2}}{2}
    \frac{1}{\sqrt{\ttt{\Bsigma '} ^{2}} ^{4}}
    \left(
        \vect{T} ^{2} N _{\mu}
        -
        T _{\mu} \ttt{\vect{T} \cdot \vect{N}}
        -
        T _{\mu} \ttt{\vect{T} \cdot \vect{N}}
        +
        \vect{T} ^{2} N _{\mu}
    \right)
    \nonumber \\
    & = &
    m ^{2}
    \frac{1}{\sqrt{\ttt{\Bsigma '} ^{2}} ^{4}}
    \left(
        \vect{T} ^{2} N _{\mu}
        -
        T _{\mu} \ttt{\vect{T} \cdot \vect{N}}
    \right)
    \\
    \frac{\partial \Lag}{\partial N ^{\mu}}
    N ^{\mu}
    & = &
    m ^{2}
    \frac{1}{\sqrt{\ttt{\Bsigma '} ^{2}} ^{4}}
    \left(
        \vect{T} ^{2} \vect{N} ^{2}
        -
        \ttt{\vect{T} \cdot \vect{N}} ^{2}
    \right)
    \\
    \frac{\partial \Lag}{\partial N ^{\mu}}
    T ^{\mu}
    & = &
    m ^{2}
    \frac{1}{\sqrt{\ttt{\Bsigma '} ^{2}} ^{4}}
    \left(
        \vect{T} ^{2} \ttt{\vect{T} \cdot \vect{N}}
        -
        \ttt{\vect{T} \cdot \vect{N}} \vect{T} ^{2}
    \right)
    \equiv 0
    \\
    \frac{\partial \Lag}{\partial N ^{\mu}}
    N ^{\mu}
    -
    \Lag
    & = &
    \frac{m ^{2}}{\sqrt{\ttt{\Bsigma '} ^{2}} ^{4}}
    \left(
        \vect{T} ^{2} \vect{N} ^{2}
        -
        \ttt{\vect{T} \cdot \vect{N}} ^{2}
    \right)
    -
    \frac{m ^{2}}{4}
    \frac{1}{\sqrt{\ttt{\Bsigma '} ^{2}} ^{4}}
    \left(
        T ^{\alpha} N ^{\beta}
        -
        T ^{\beta} N ^{\alpha}
    \right)
    \left(
        T _{\alpha} N _{\beta}
        -
        T _{\beta} N _{\alpha}
    \right)
    \nonumber \\
    & = &
    \frac{m ^{2}}{\sqrt{\ttt{\Bsigma '} ^{2}} ^{4}}
    \left(
        \vect{T} ^{2} \vect{N} ^{2}
        -
        \ttt{\vect{T} \cdot \vect{N}} ^{2}
    \right)
    -
    \frac{m ^{2}}{4}
    \frac{1}{\sqrt{\ttt{\Bsigma '} ^{2}} ^{4}}
    2
    \left(
        \vect{T} ^{2} \vect{N} ^{2}
        -
        \ttt{\vect{T} \cdot \vect{N}} ^{2}
    \right)
    \nonumber \\
    & = &
    \frac{m ^{2}}{2}
    \frac{1}{\sqrt{\ttt{\Bsigma '} ^{2}} ^{4}}
    \left(
        \vect{T} ^{2} \vect{N} ^{2}
        -
        \ttt{\vect{T} \cdot \vect{N}} ^{2}
    \right)
    \quad .
\eea
Then, substituting these results,
equations (\ref{1.oginorfunder}-\ref{1.ogiNorfunder}) become:
\bea
    \frac{\delta S}{\delta n \tst}
    & = &
    \frac{m ^{2}}{2}
    \frac{1}{\sqrt{\ttt{\Bsigma '} ^{2}} ^{3/2}}
    \left [
        \vect{T} ^{2} \vect{N} ^{2} - \ttt{\vect{T} \cdot \vect{N}} ^{2}
    \right ]
    \label{2.ogiHJder.1}
    \\
    \frac{\delta S}{\delta Y ^{\mu} \tst}
    & = &
    - m ^{2}
    \frac{1}{\ttt{\Bsigma '} ^{2}}
    \left [
        \vect{T} ^{2} N _{\mu} - \ttt{\vect{T} \cdot \vect{N}} T _{\mu}
    \right ]
    \label{2.ogiHJder.2}
    \\
    \frac{\delta S}{\delta t \tst}
    & = &
    0
    \label{2.ogiHJder.3}
    \quad .
\eea
\label{2.funaredernorareder}
We see, from the last equation, that the model under investigation
is invariant under reparametrization of the loop $\Gamma$ (respectively
$C$) in {\sl Parameter Space} (respectively  {\sl Target Space}),
Then, by combining results (\ref{2.ogiHJder.1}-\ref{2.ogiHJder.2})
it turns out that the local
Hamilton--Jacobi equation satisfied by the system is
$$
    \frac{\partial S}{\partial A}
    =
    \frac{1}{2 m ^{2}}
    \frac{1}{\ttt{\vect{Y} '} ^{2}}
    \frac{\delta S}{\delta Y ^{\mu}}
    \frac{\delta S}{\delta Y _{\mu}}
    \quad .
$$
It has been a common feature of all previous works on this subject to stop
at this stage and to call this one the Hamilton--Jacobi equation for
the {\sl String}.
In our opinion a further step should be carried out to achieve
a reparametrization invariant equation for the Dynamics of the system.
For this reason, we rewrite the result above as
$$
    \sqrt{\ttt{\vect{Y} '} ^{2}}
    \frac{\partial S}{\partial A}
    =
    \frac{1}{2 m ^{2}}
    \frac{1}{\sqrt{\ttt{\vect{Y} '} ^{2}}}
    \frac{\delta S}{\delta Y ^{\mu}}
    \frac{\delta S}{\delta Y _{\mu}}
$$
and  integrate both sides with respect to $s$ to get
\beq
    \frac{1}{2 m ^{2}}
    \norme{\Gamma \approx \Sf ^{1}}
    \oint _{\Gamma \approx \Sf ^{1}} ds
    \frac{1}{\sqrt{\ttt{\vect{Y} '} ^{2}}}
    \frac{\delta S}{\delta Y ^{\mu}}
    \frac{\delta S}{\delta Y _{\mu}}
    =
    \frac{\partial S}{\partial A}
    \quad ,
    \label{2.ogihamjac}
\eeq
which we call the \underbar{{\it String Functional
Hamilton--Jacobi Equation}}.
\end{proof}

\subsection{Hamilton-Jacobi Equation: Reparametrized Formulation}
\label{2.hamjacrepsec}

We are now going to rederive the same result in the reparametrized
framework that we discussed in section \ref{2.repschfor},
i.e. starting from
the Hamiltonian formulation of reparametrized Schild {\sl String} Theory.
This procedure will give us the opportunity to remark some important
features of the model and to address in a more detailed way some key
questions related to the quantization procedure. The first step is to have a
closer look to the classical
equations of motion originating from the action
associated with the canonical $2$-form (\ref{2.fulrephamfor}):
\begin{defs}[Hamiltonian Full Reparametrized Schild Action]\spbcorr{}.\\
    The \underbar{Hamiltonian Full Reparametrized Schild Action} is
    the action associated with the canonical
    $2$-form {\rm (\ref{2.fulrephamfor})}, i.e.
    \bea
        & & \esci \esci
        S \qtq{
               X ^{\mu} ,
               P _{\mu \nu} ,
               \xi ^{A} ,
               \pi _{AB} ,
               N _{AB}
              }
        =
        \nonumber \\
        & & =
        \int _{\mathcal{W}}
        P _{\mu \nu} \form{d X} ^{\mu} \wedge \form{d X} ^{\nu}
        +
        \int _{\Xi}
        \pi _{AB} \form{d \xi} ^{A} \wedge \form{d \xi} ^{B}
        +
        \nonumber \\
        & & \qquad \qquad +
        \frac{1}{2}
        \int _{\Sigma}
        N ^{AB}
        \left(
            \pi _{AB} - \epsilon _{AB} \Ham _{\mathrm{Schild}} \ttt{\Bp}
        \right)
        \form{d \sigma} ^{0} \wedge \form{d \sigma} ^{1}
        \label{2.fulrepact}
        \\
        & & =
        S
        \left [
            X ^{\rho}
            ,
            P _{\sigma \tau}
            ;
            \xi ^{A}
            ,
            \pi _{CD}
        \right ]
        +
        S _{\mathrm{cnstr.}}
        \left [
            X ^{\rho}
            ,
            P ^{\sigma \tau}
            ;
            \xi ^{A}
            ,
            \pi _{CD}
            ;
            N ^{AB}
        \right ]
    \eea
    and its field variables are $\Bx$, $\Bp$, $\Bxi$, $\Bpi$, $\BN$.
\end{defs}
We emphasized in the previous definition that the last term corresponds
to the constraint which must be imposed on the conjugated momenta, i.e.
\beq
    S _{\mathrm{cnstr.}}
    \left [
        X ^{\rho}
        ,
        P ^{\sigma \tau}
        ;
        \xi ^{A}
        ,
        \pi _{CD}
        ;
        N ^{AB}
    \right ]
    =
    \frac{1}{2}
    \int _{\Sigma}
    N ^{AB}
    \left(
        \pi _{AB} - \epsilon _{AB} \Ham _{\mathrm{Schild}} \ttt{\Bp}
    \right)
    \form{d \sigma} ^{0} \wedge \form{d \sigma} ^{1}
    \label{2.conactfun}
    \quad .
\eeq
We stress again, because this is a relevant point, that the dependence in
$\Ham _{\mathrm{Schild}}$ is from the momentum of (\ref{2.Pmunu}).
\begin{props}[Full Reparametrized Theory Equations of Motion]\spbcorr{}.
    \label{2.fullrepequpro}\\
    The equations of motion for the full reparametrized
    Theory {\rm (\ref{2.fulrepact})} are
    \bea
        \dot{X} ^{\mu \nu}
        & = &
        N ^{AB} \epsilon _{AB}
        \frac{P ^{\mu \nu}}{2 m ^{2}}
        \label{2.fulrepeq1}
        \\
        \epsilon ^{ab}
        \partial _{a} P ^{\mu \nu}
        \partial _{b} X ^{\nu}
        & = &
        0
        \label{2.fulrepeq2}
        \\
        \pi _{AB}
        & = &
        \epsilon _{AB}
        \frac{P _{\mu \nu} P ^{\mu \nu}}{4 m ^{2}}
        \label{2.fulrepeq3}
        \\
        \epsilon ^{ab}
        \partial _{a} \pi _{AB}
        \partial _{b} \xi ^{B}
        & = &
        0
        \label{2.fulrepeq4}
        \\
        N ^{AB}
        & = &
        \dot{\xi} ^{AB}
        \quad .
        \label{2.fulrepeq5}
\eea
\end{props}
\begin{proof}
The procedure is as usual to consider a variation of the action
(\ref{2.fulrepact}). Then it is possible to extract the equation
of motion thanks to the fundamental principle of the variational
calculus.
\end{proof}

Now we set up a particular, but by no means restrictive framework
where the result found in the previous section can be generalised
to {\sl Parameter Space} topologies more general than the one
assumed in Ogielski's formulation. Suppose we choose
  a doubly connected {\sl Parameter Space} with the  {\it anulus} topology.
Physically speaking, the boundary corresponding, say,
to the hole can be considered as the initial {\sl String} configuration,
whereas the rest of the boundary of the anulus can be mapped to the
final {\sl String} configuration. We will consider only variations of the
parameter space having the part of the boundary corresponding to the
initial {\sl string} configuration fixed. The rest of the boundary is on the
contrary free and can be varied: it is also parametrized by
$s \in \Sf ^{1} \approx \Gamma$ and $C$ is its image in {\sl Parameter
Space}. Let us  call $C _{0}$ the image in {\sl Parameter Space}
of the initial {\sl String} configuration.
The steps leading to the Hamilton--Jacobi equation are as follows.
First, we remove the Lagrange multiplier from the action using
equation (\ref{2.fulrepeq5}), since it is a non dynamical quantity;
moreover, thanks to equation (\ref{2.fulrepeq3}), we can still
simplify the expression for the action and get the following result:
\bea
    & & \esci \esci
    S _{\mathrm{Red.}}
      \left [
          X ^{\rho} \left( \bs{\sigma} \right)
          ,
          P ^{\sigma \tau} \left( \bs{\sigma} \right)
          ,
          \xi ^{A} \left( \bs{\sigma} \right)
      \right ]
    =
    \nonumber \\
    & & =
    \frac{1}{2}
    \int _{\mathcal{W} \left( \bs{\sigma} \right)}
        P _{\mu \nu} \form{d X} ^{\mu} \wedge \form{d X} ^{\nu}
    +
    \nonumber \\
    & & \qquad \qquad
    -
    \frac{1}{2}
    \epsilon _{AB}
    \int _{\Xi} \form{d \xi} ^{A} \wedge \form{d \xi} ^{B}
        \Ham _{\mathrm{Schild}} \left( P ^{\sigma \tau} \left( \bs{\xi} \right) \right)
    \quad .
    \label{2.redrepact}
\eea
\begin{defs}[Hamiltonian Reduced Reparametrized Schild Action]\spbcorr{}.\\
   We call the expression {\rm (\ref{2.redrepact})} the
   \underbar{Hamiltonian, Reduced, Reparametrized Schild Action},
   because it is the {\sl Hamiltonian  Reparametrized
   Schild Action} with  solved constraints.
\end{defs}
Since we assume the equation of motion for the
{\sl World-Sheet} are satisfied, by varying the action the contribution
from the variation of the {\sl World-Sheet} itself vanishes, and we obtain:
\begin{props}[Restricted Reparametrized Action
              Boundary Variation]\spbcorr{}.\\
    The variation of the reparametrized action due to a variation
    in the boundary $\Gamma \approx \Sf ^{1}$ is
    \beq
        \delta S _{\mathrm{Red.}}
        =
        \int _{\Gamma \approx \Sf ^{1}}
            q _{\mu} \tst \delta Y ^{\mu} \tst ds
        -
        \Ham _{\mathrm{Schild}}
        \delta A
        \quad ,
    \label{2.jacvarpri}
    \eeq
    where
    \bea
        q _{\mu} \tst
        & \dfn &
        \frac{\delta S _{\mathrm{Red.}}}
             {\delta Y ^{\mu} \tst}
        =
        Q _{\mu \nu} \tst Y ^{\prime \nu} \tst
        \label{2.corfunder}
        \\
        A
        & \dfn &
        \frac{1}{2}
        \epsilon _{AB}
        \int _{\Xi}
            d \xi ^{A} \wedge d \xi ^{B}
        \\
        \Ham _{\mathrm{Schild}} \tst
        & = &
        -
        \frac{\delta S_{\mathrm{Red.}}}
             {\delta A \tst}
        \quad ,
        \label{2.arenorder}
        \eea
where $Q _{\mu \nu}$ is the {\sl Boundary Area Momentum} of
definition {\rm (\ref{2.bouaremomdef})}.
\end{props}
This result can be obtained specializing to the particular
case with $p=1$ the computation for the $p$-brane given in
proposition \ref{4.pbrbouvarpro}.
\begin{proof}
Please, see proposition \ref{A.futvarderstrpro}
in appendix \ref{A.FutVarDer} for the detailed computation.
\end{proof}

Thanks to equations
(\ref{2.fulrepeq3}-\ref{2.fulrepeq4}) the Hamiltonian is
constant along a classical trajectory, then we may move the Hamiltonian
outside the area integral and the variation operator.  Moreover,
the functional variation $\delta A$ becomes an ordinary differential
variation $d A$:
\beq
    \Ham _{\mathrm{Schild}}
    \dfn
    E
    \dfn
    - \frac{\partial S _{\mathrm{Red.}}}{\partial A}
    \quad ;
    \label{2.areder}
\eeq
this is the same result obtained in subsection \ref{2.hamjacsecogi}
on page \pageref{2.funaredernorareder} as a direct consequence
of the restricted invariance under loop reparametrizatrions. This result,
is expressed in Ogielski's procedure, as equation (\ref{2.ogiHJder.3}).
Moreover, it shows up in the reparametrized formulation as the constancy
of the Schild Hamiltonian along a classical trajectory, encoded in
equations (\ref{2.fulrepeq3}) and (\ref{2.fulrepeq4}).
Hence, we can see that the Jacobi variational principle in the form of
equation (\ref{2.jacvarpri}) shows that $q _{\mu}$ is conjugated to the
{\sl World--Sheet} boundary variation, while $E \equiv H _{\mathrm{Schild}}$
describes the response of the classical action to an arbitrary area
variation in {\sl Parameter Space}, due to a deformation of the boundary.
Thus if we consider the classical\footnote{As already pointed out
in footnotes \ref{2.footuno} on page \pageref{2.footuno} and
\ref{2.footdue} on page \pageref{2.footdue}, we try to keep a well defined
distinction between quantities related to the classical domain and
quantities associated to the quantum realm.}
Dynamics of the {\sl String} from this point of view,
then $A$ and $X ^{\mu} \tst$ can be interpreted as the classical
``time'' and ``space'' coordinates of the {\sl String} $C$.

Finally we note that in this formulation $E$ can be indentified with
the {\it energy per unit area} associated with
an extremal {\sl World--Sheet} of the action (\ref{2.redrepact}),
while $q _{\mu} \tst$ is the {\it momentum per
unit length} of the {\sl String} {\it loop} $C$. Therefore, the energy--momentum
dispersion relation can be written either as an equation between
{\it densities}
\beq
    \frac{1}{2 m ^{2}}
    q ^{\mu} \tst
    q _{\mu} \tst
    =
    \frac{1}{4 m ^{2}}
    Q _{\mu \nu} \tst
    Q ^{\mu \nu} \tst
    \ttt{\By \tst '} ^{2}
    =
    \ttt{\By \tst '} ^{2}
    E
\label{2.strhamjacpreden}
\eeq
or as an integrated relation
\beq
    \frac{1}{2 m ^{2}}
    \oint _{\Gamma \approx \Sf ^{1}}
    \frac{ds}{\sqrt{\ttt{\By '} ^{2} \tst}}
    q ^{\mu} \tst
    q _{\mu} \tst
    =
    E
    \oint _{\Gamma}
    ds
    \sqrt{\ttt{\By ' \tst} ^{2}}
    \quad .
    \label{2.strhamjacpreint}
\eeq
The above equation once written using
equations (\ref{2.corfunder}, \ref{2.areder}) turns
exactly into (\ref{2.ogihamjac}),
the {\sl Functional Hamilton--Jacobi Equation} for the {\sl String}:
\beq
    -
    \frac{\partial S}{\partial A}
    =
    \frac{1}{2 m ^{2}}
    \norme{\Gamma}
    \oint _{\Gamma} \frac{ds}{\sqrt{Y ^{\prime \mu} Y ' _{\mu}}}
        \frac{\delta S}
             {\delta Y ^{\mu} \tst}
        \frac{\delta S}
             {\delta Y _{\mu} \tst}
    \quad .
    \label{2.ourhamjac}
\eeq
Looking in more detail at this equation, we observe that the
covariant integration over $s$ takes into account all the possible locations
of the  point, along the contour $C$, where the variation can be applied.
But, in this way, every point of $C$ is overcounted a ``number of times''
equal to the {\sl String} proper length. The first factor, in
round parenthesis, is just the {\sl String} proper length and removes such
overcounting. In other words, we sum over all the possible
ways in which one can deform the {\sl String} loop, and then divide by the
total number of them. The net result is that
the left-hand side of equation (\ref{2.ourhamjac}) is insensitive
to the choice of the point where the final {\sl String} $C$ is
deformed. Therefore the right-hand side is a genuine reparametrization
scalar which describes the system's response to the extent of
area variation, irrespective of the way in which the deformation is implemented.
With hindsight, the wave equation proposed in \cite{egu},
\cite{ogi} appears
to be more restrictive than equation (\ref{2.ourhamjac}), in the sense that it
requires the second variation of the line fuctional
to be proportional to $\ttt{\By ' \tst} ^{2}$ at {\it any} point on
the {\sl String} loop, in contrast to equation (\ref{2.ourhamjac}), which represents
an integrated constraint on the {\sl String} as a whole.

\section{Classical Area Effective Formulation}
\label{2.areequide}

It is important to observe that the functional Hamilton--Jacobi equation
(\ref{2.ourhamjac}) can be cast in a form, in which all the derivatives
are with respect to {\it area quantities}. In particular we can trade
the functional derivative with the derivative with respect to the
{\sl Holographic Coordinates}, thanks to equation (\ref{D.funarederrel}).
In particular the Hamilton-Jacobi equation then takes the form
\beq
    -
    \frac{\partial S}{\partial A}
    =
    \frac{1}{4 m ^{2}}
    \norme{\Gamma}
    \oint _{\Gamma} ds \sqrt{Y ^{\prime \mu} Y ' _{\mu}}
        \frac{\delta S}
             {\delta Y ^{\mu \nu} \tst}
        \frac{\delta S}
             {\delta Y _{\mu \nu} \tst}
    \quad .
\label{2.ourhamjacare}
\eeq
Starting from this result we can prove
\begin{props}[Classical Area Hamiltonian Formalism]\spbcorr{}.\\
    The Hamilton--Jacobi equation {\rm (\ref{2.ourhamjacare})}
    is the Hamilton--Jacobi equation associated with the following
    Hamiltonian:
    \beq
        \Ham _{\mathrm{EFF}}
        =
        \frac{1}{4 m ^{2}}
        \norme{\Gamma}
        \oint _{\Gamma} ds \sqrt{\ttt{\By '} ^{2}}
            Q _{\mu \nu} \tst Q ^{\mu \nu} \tst
    \quad ;
    \eeq
    this Hamiltonian can be taken as defining a Theory in which the
    conjugate variables are the {\sl Holographic Coordinates}
    $Y ^{\mu \nu} \qtq{C ; A}$ and the corresponding Conjugate Momentum
    $$
        Q ^{\mu \nu} \qtq{C ; A}
        =
        \frac{1}{2 m ^{2}}
        \norme{\Gamma}
        \oint _{\Gamma} ds \sqrt{\ttt{\By '} ^{2}}
            Q _{\mu \nu} \tst
    \quad ,
    $$
    which is nothing but the ``average'' of the
    {\sl Boundary Area Momentum} taken over the {\sl String}.
    Solving then equation {\rm (\ref{2.ourhamjacare})} is the equivalent to
    find a solution of the following first order system of
    \underbar{Area Hamilton Equations}:
    \beq
        \left\{
            \matrix{
                \frac{\displaystyle d Y ^{\mu \nu} \qtq{C ; A}}
                     {\displaystyle d A}
                =
                Q ^{\mu \nu} \qtq{C ; A}
                \hfill
                \cr
                \cr
                \frac{\displaystyle d Q ^{\mu \nu} \qtq{C ; A}}
                     {\displaystyle d A}
                =
                0
                \hfill
                   }
        \right .
        \quad .
        \label{2.arehameqs}
    \eeq
\end{props}
\begin{proof}
    To get the desired result we have to compute in first istance the
    following functional derivatives:
$$
    \frac{\delta \Ham _{\mathrm{EFF}}}{\delta Q _{\mu \nu} \ttt{t}}
    =
    \frac{1}{2 m ^{2}}
    \norme{\Gamma}
    \sqrt{\ttt{\By \ttt{t}} ^{2}}
    Q _{\mu \nu} \ttt{t}
$$
and
$$
    \frac{\delta \Ham _{\mathrm{EFF}}}{\delta Y _{\mu \nu} \ttt{t}}
    =
    0
    \quad .
$$
As discussed in appendix \ref{D.looderapp} the {\sl Holographic
Derivatives} are explicitly dependent from the loop parameter, $t$ in this
case. To get reparametrization invariant quantities we integrate over
this variable and define
\bea
    \frac{\delta \Ham _{\mathrm{EFF}}}{\delta Q _{\mu \nu} \qCq}
    & \dfn &
    \frac{1}{2 m ^{2}}
    \norme{\Gamma}
    \oint _{\Gamma} d t
    \sqrt{\ttt{\By \ttt{t}} ^{2}}
    Q _{\mu \nu} \ttt{t}
    \dfn
    Q _{\mu \nu} \qtq{C ; A}
    \label{2.hamloomomdef}\\
    \frac{\delta \Ham _{\mathrm{EFF}}}{\delta Y _{\mu \nu} \qCq}
    & \dfn &
    0
    \label{2.hamloocoodef}
    \quad .
\eea
This are well defined {\it loop} quantities, in terms of which we can write
the following Hamilton equations
    \beq
        \left\{
            \matrix{
                \frac{\displaystyle d Y ^{\mu \nu} \qtq{C ; A}}
                     {\displaystyle d A}
                =
                \frac{\displaystyle \delta \Ham}
                     {\displaystyle \delta Q _{\mu \nu} \qCq}
                \hfill
                \cr
                \cr
                \frac{\displaystyle d Q ^{\mu \nu} \qtq{C ; A}}
                     {\displaystyle d A}
                =
                -
                \frac{\displaystyle \delta \Ham}
                     {\displaystyle \delta Y _{\mu \nu} \qCq}
                \hfill
                   }
        \right .
        \quad .
    \eeq
From these equations, after the substitutions of
(\ref{2.hamloomomdef}) and (\ref{2.hamloocoodef}) in the right hand sides,
we get the desired result.
\end{proof}
\begin{props}[Classical Area Newtonian Formalism]\spbcorr{}.\\
    The first order system of {\sl Area Hamilton Equations}
    {\rm (\ref{2.arehameqs})} is equivalent to the following
    second order \underbar{Area Newton Equation}:
    \beq
        \frac{d ^{2} Y ^{\mu \nu} \qtq{C ; A}}{d A ^{2}}
        =
        0
        \quad .
    \label{2.newareequ}
    \eeq
\end{props}
\begin{proof}
Substituting the first equation of the system (\ref{2.arehameqs})
into the second one we obtain the desired result.
\end{proof}
In previous equations the {\sl Area Derivative} $d / \ttt{d A}$
explicitly appears and
to clarify its meaning, as well as other features of this model,
some remarks are worthwhile.

Firstly, we note that our procedure has singled out a very striking
feature of the Dynamics of the {\sl String} described starting from the
Eguchi formulation: {\it Classical String Dynamics can be viewed exactly as
a $2$-dimensional generalisation of particle Dynamics in the sense
that it is a Dynamics of area}. The coordinates of the Theory have length
squared dimensions, as well as the evolution parameter.
Geometrically speaking
we have just {\it raised by one} the particle case, but not simply
by {\it adding an extra parameter.} On the contrary, all
the formulation describes the evolution of some areas (more specifically,
the areas of the projection of the {\sl String} on the coordinate planes)
as their response to the {\sl Parameter Space} (or {\sl World--Sheet})
area variation.

Secondly note that the geometrical description of area naturally
relies on anti--symmetric $2$-forms. These are the natural
generalization of the tangent vector, because they can be
interpreted as the tangent elements to the {\sl World--Sheet} of
the {\sl String}.


As a last observation, we note that it is possible to set up a
statistical formalism for {\sl Strings} ($p$-branes in the general case)
starting from the classical area formalism developed in this section.
In particular we can take as unknown dynamical fields the action
$S \qtq{Y ^{\mu \nu} ; A}$ and the probability density
$P \qtq{Y ^{\mu \nu} ; A}$.
The meaning of this last quantity is that it can be used to obtain
the probability of finding a given loop $C$ that has shadows onto
the coordinate planes described by the {\sl Holographic Coordinates}
$Y ^{\mu \nu} \qtq{C ; A}$.
Then we can take for our fields the following action functional:
\bea
    & & \esci \esci
    \euf{S}
    =
    \int d A
    \funint{C}
        P \qtq{C ; A}
        \left \{
            \frac{1}{4 m ^{2}}
            \norme{\Gamma}
            \cdot
        \right .
    \nonumber \\
    & & \qquad \qquad \qquad
        \left .
            \cdot
            \oint _{\Gamma} ds \sqrt{Y ^{\prime \mu} Y ' _{\mu}}
                \frac{\delta S \qtq{C ; A}}
                     {\delta Y ^{\mu \nu} \tst}
                \frac{\delta S \qtq{C ; A}}
                     {\delta Y _{\mu \nu} \tst}
            +
            \frac{\partial S \qtq{C ; A}}{\partial A}
        \right\}
   \quad .
   \label{2.conequact}
\eea
It can be proved that a variation with respect to $P \qtq{C ; A}$
exactly gives Hamilton--Jacobi equation (\ref{2.ourhamjacare}).
Moreover we can derive a result which will be useful in
chapter \ref{10.nonstacha}.
\begin{props}[Continuity Equation]\spbcorr{}.\\
     The variation of the action {\rm (\ref{2.conequact})}
     with respect to $S \qtq{C ; A}$ results in the
     following equation
     \beq
         \frac{d P \qtq{C ; A}}{d A}
         +
         \frac{1}{4 m ^{2}}
         \norme{\Gamma}
         \oint _{\Gamma} ds \sqrt{{\By '} ^{2}}
             \frac{\delta}
                  {\delta Y ^{\mu \nu} \tst}
             \left [
                P \qtq{C ; A}
                \frac{\delta S \qtq{C ; A}}
                     {\delta Y _{\mu \nu} \tst}
             \right ]
         =
         0
     \quad ,
     \label{2.areconequ}
     \label{8.conequ}
     \eeq
     which we call the \underbar{Continuity Equation for Area Dynamics}.
\end{props}
\begin{proof}
As a preliminary remark we underline that
in this proof we are going to indicate the variation of the
action $S$ with the following notation:
$$
    \Svary \: \mathrm{in\ place\ of} \  \delta S \quad ,
$$
to avoid confusion with the functional derivative symbol $\delta$.
Then we see that a variation of $S$ in (\ref{2.conequact}) yields
\bea
    & & \esci \esci
    \myvary{\euf{S}}
    =
    \int d A
    \funint{C}
        P \qtq{C ; A}
        \left \{
            \frac{1}{2 m ^{2}}
            \norme{\Gamma}
            \cdot
        \right .
    \nonumber \\
    & & \qquad \qquad \qquad
        \left .
            \cdot
            \oint _{\Gamma} ds \sqrt{Y ^{\prime \mu} Y ' _{\mu}}
                \frac{\delta S \qtq{C ; A}}
                     {\delta Y ^{\mu \nu} \tst}
                \frac{\delta \ttt{\Svary \qtq{C ; A}}}
                     {\delta Y _{\mu \nu} \tst}
            +
            \frac{\partial \ttt{\Svary}}{\partial A}
        \right\}
   \quad .
\eea
Integrating by parts and dropping {\sl Boundary} terms as
usual\footnote{This procedure gives no problems with the $dA$ integration;
in principle the meaning of an integration by parts could be much more
troubleful in the functional case where we make the same absumption as
in \cite{}.} gives
\bea
    & & \esci \esci
    \myvary{\euf{S}}
    =
    \int d A
    \funint{C}
        P \qtq{C ; A}
        \left \{
            \frac{1}{2 m ^{2}}
            \norme{\Gamma}
            \cdot
        \right .
    \nonumber \\
    & & \qquad \qquad \qquad
        \left .
            \cdot
            \oint _{\Gamma} ds \sqrt{Y ^{\prime \mu} Y ' _{\mu}}
                \frac{\delta }
                     {\delta Y ^{\mu \nu} \tst}
                \left [
                         P \qtq{C ; A}
                    \frac{\delta \Svary \qtq{C ; A}}
                         {\delta Y _{\mu \nu} \tst}
                \right ]
            +
            \frac{\partial P \qtq{C ; A}}{\partial A}
        \right\}
    \Svary
    \quad ,
\eea
from which the desired result follows.
\end{proof}

\section{Reparametrized Canonical Formulation}
\label{2.repcanforsec}

We will now introduce another equivalent formulation that will give
a deeper insight into further developments and will be useful to understand
the connection between our formulation  of the {\sl String} Dynamics and the more
traditional one. This can be a good opportunity to stress
again some important concepts related to the classical Dynamics
and  a useful premise to the quantization procedure,
to be analysed in detail in the next chapter as well.
We will also take the opportunity to anticipate
some deep results that will become more clear in the following sections.
We start with a definition:
\begin{defs}[Projected World--Sheet Area Momentum]\spbcorr{}.\\
    The {\sl World--Sheet} Area Momentum ({\rm definition \ref{2.aremomdef}})
    projected onto the {\sl World--Sheet} directions,
    $$
        \PPud{m}{\mu}
        \dfn
        P _{\mu \nu} \ttt{\Bsigma}
        \epsilon ^{mn}
        \partial _{n} X ^{\nu} \ttt{\Bsigma}
        \quad ,
    $$
    is the\underbar{Projected World--Sheet Area Momentum}.
\end{defs}
\begin{defs}[Projected World--Sheet Boundary Area Momentum]\spbcorr{}.\\
    The {\sl World--Sheet Boundary Area Momentum} of {\rm definition
    \ref{2.bouaremomdef}}
    projected onto the {\sl World--Sheet}
    directions and calculated on the Boundary,
    $$
        \PPud{m}{\mu}
        \dfn
        \left .
            P _{\mu \nu} \ttt{\Bsigma}
            \epsilon ^{mn}
            \partial _{n} X ^{\nu} \ttt{\Bsigma}
        \right \rceil _{\Bsigma = \Bsigma \tst}
    \quad ,
    $$
    is the \underbar{Projected World--Sheet Boundary Area Momentum}.
\end{defs}
Now it is possible to prove that
\begin{props}[Projected World--Sheet Area Momentum Computation]\spbcorr{}.\\
    The \underbar{Projected World--Sheet Area Momentum} is the momentum
    conjugated to the velocity $\partial _{m} X ^{\nu}$, i.e.
    $$
        \PPud{m}{\mu}
        =
        \frac{\partial \Lag}
             {\partial \ttt{\partial _{n} X ^{\nu}}}
        \quad .
    $$
    Thus, it is the canonical momentum of the {\sl String} {\sl World--Sheet}.
\end{props}
\begin{proof}
The starting point is as natural the Schild Lagrangian
(\ref{2.schild}) and we also remember (\ref{2.dotxmunu}).
Then we have
\bea
    \PPud{m}{\mu}
    & = &
    \frac{\partial \Lag}{\partial \ttt{\partial _{m} X ^{\mu}}}
    \nonumber \\
    & = &
    \frac{m ^{2}}{4}
    2
    \dot{X} _{\rho \tau}
    \frac{\partial X ^{\rho \tau}}{\partial \ttt{\partial _{m} X ^{\mu}}}
    \nonumber \\
    & = &
    \frac{m ^{2}}{2}
    \dot{X} _{\rho \tau}
    \epsilon ^{ab}
    \left(
        \frac{\partial \ttt{\partial _{a} X ^{\rho}}}
             {\partial \ttt{\partial _{m} X ^{\mu}}}
        \partial _{b} X ^{\tau}
        +
        \frac{\partial \ttt{\partial _{b} X ^{\tau}}}
             {\partial \ttt{\partial _{m} X ^{\mu}}}
        \partial _{a} X ^{\rho}
    \right)
    \nonumber \\
    & = &
    \frac{m ^{2}}{2}
    \dot{X} _{\rho \tau}
    \epsilon ^{ab}
    2
    \frac{\partial \ttt{\partial _{a} X ^{\rho}}}
         {\partial \ttt{\partial _{m} X ^{\mu}}}
    \partial _{b} X ^{\tau}
    \nonumber \\
    & = &
    m ^{2}
    \dot{X} _{\rho \tau}
    \epsilon ^{ab}
    \delta _{am} \delta ^{\rho \mu}
    \partial _{b} X ^{\tau}
    \nonumber \\
    & = &
    P _{\mu \nu}
    \epsilon ^{am}
    \partial _{a} X ^{\nu}
    \quad ,
\eea
as required.
\end{proof}

Moreover, the following equality holds.
\begin{props}[Alternative Expression for the Schild Hamiltonian]\spbcorr{}.
    \label{2.althamexp}\\
    The {\sl Schild Hamiltonian} can be written in the following
    equivalent forms
    \beq
        \Ham _{\mathrm{Schild}}
        =
        \frac{P ^{\mu \nu} P _{\mu \nu}}{4 m ^{2}}
        =
        \frac{\PPuu{m}{\mu} \PPdd{m}{\mu}}{2 m ^{2}}
        \label{2.equhamexp}
    \eeq
    and the product of the {\sl Area Velocity} times the
    {\sl World--Sheet Area Momentum} as
    \beq
        P _{\mu \nu} \dot{X} ^{\mu \nu}
        =
        \PPuu{m}{\mu} \partial _{m} X _{\mu}
        \quad .
        \label{2.equmomvelexp}
    \eeq
    All these relations are World--Sheet relations.
\end{props}
\begin{proof}
The first result is just a consequence of ``$\bs{\epsilon}$ tensors
contraction'':
\bea
    \PPuu{m}{\mu} \PPdd{m}{\mu}
    & = &
    \epsilon ^{ma} P ^{\mu \nu} \partial _{a} X _{\nu}
    \epsilon _{mb} P _{\mu \rho} \partial ^{b} X ^{\rho}
    \nonumber \\
    & = &
    P ^{\mu \nu} P _{\mu \rho}
    \partial _{a} X _{\nu} \partial ^{b} X ^{\rho}
    \epsilon ^{ma} \epsilon _{mb}
    \nonumber \\
    & = &
    \frac{1}{2}
    P ^{\mu \nu} P _{\mu \rho}
    \partial _{a} X _{\nu} \partial ^{b} X ^{\rho}
    \delta ^{a} _{b}
    \nonumber \\
    & = &
    \frac{1}{2}
    P ^{\mu \nu} P _{\mu \rho}
    \partial _{a} X _{\nu} \partial ^{a} X ^{\rho}
    \nonumber \\
    & = &
    \frac{1}{2}
    P ^{\mu \nu} P _{\mu \rho}
    \delta ^{\rho} _{\nu}
    \nonumber \\
    & = &
    \frac{1}{2}
    P ^{\mu \nu} P _{\mu \nu}
    \quad .
\eea
Then the result for the Hamiltonian follows. Moreover in the second case
\bea
    \PPuu{m}{\mu} \partial _{m} X _{\mu}
    & = &
    \epsilon ^{mn} P ^{\mu \nu} \partial _{n} X _{\nu}
    \partial _{m} X _{\mu}
    \nonumber \\
    & = &
    P ^{\mu \nu}
    \epsilon ^{mn}
    \partial _{n} X _{\nu}
    \partial _{m} X _{\mu}
    \nonumber \\
    & = &
    P ^{\mu \nu}
    \dot{X} _{\mu \nu}
    \quad .
\eea
\end{proof}

\begin{props}[Bulk--Boundary Interference]\spbcorr{}.\\
    The {\sl Holographic Coordinates} of the {\sl String}
    are the {\sl World--Sheet} integral
    of the {\sl World--Sheet Area Velocity}, i.e.
    $$
        Y ^{\mu \nu} \qtq{C}
        =
        \oint _{C} Y ^{\mu} d Y ^{\nu}
        =
        \int _{\mathcal{W}} d X ^{\mu} \wedge d X ^{\nu}
    \quad ,
    $$
    which  can be written in coordinates
    $$
        Y ^{\mu \nu} \qtq{C}
        =
        \oint _{\Gamma \equiv \Sf ^{1}} ds Y ^{\mu} \tst Y ^{\prime \nu} \tst
        =
        \int _{\Sigma}
            d ^{2} \Bsigma
            \dot{X} ^{\mu \nu} \ttt{\Bsigma}
    \quad .
    $$
\end{props}
\begin{proof}
We prove the result written {\it in an exlpictly choosen coordinate
system}, since the one
in intrinsic forms, simply comes out going to the boundary
$C = \partial \mathcal{W}$.
Thus we have
\bea
    Y ^{\mu \nu} \qtq{C}
    & = &
    \oint _{C} Y ^{\mu} d Y ^{\nu}
    \nonumber \\
    {\scriptstyle 1.}
    & = &
    \oint _{\Gamma \approx \Sf ^{1}} ds \,
        Y ^{\mu} \tst
        Y ^{\prime \nu} \tst
    \nonumber \\
    {\scriptstyle 2.}
    & = &
    \oint _{\Gamma \approx \Sf ^{1}} ds \,
           X ^{\mu} \ttt{\Bsigma \tst}
           \frac{d X ^{\nu} \ttt{\Bsigma \tst}}{d s}
    \nonumber \\
    {\scriptstyle 3.}
    & = &
    \oint _{\partial \Sigma = \Gamma \approx \Sf ^{1}} ds \,
           X ^{\mu} \ttt{\Bsigma \tst}
           \frac{\partial X ^{\nu} \ttt{\Bsigma \tst}}{\partial \sigma ^{a} \tst}
           \frac{d \sigma ^{a}}{d s}
    \nonumber \\
    {\scriptstyle 4.}
    & = &
    \oint _{\partial \Sigma = \Gamma \approx \Sf ^{1}} ds
           \left [
               X ^{\mu} \ttt{\Bsigma \tst}
               \partial _{a} X ^{\nu} \ttt{\Bsigma \tst}
           \right ]
           \frac{d \sigma ^{a} \tst}{d s}
    \nonumber \\
    {\scriptstyle 5.}
    & = &
    \int _{\Sigma} {d ^{2} \Bsigma} \,
           \epsilon ^{ba}
           \partial _{b}
           \left [
               X ^{\mu} \ttt{\Bsigma}
               \partial _{a} X ^{\nu} \ttt{\Bsigma}
           \right ]
    \nonumber \\
    {\scriptstyle 6.}
    & = &
    \int _{\Sigma} {d ^{2} \Bsigma} \,
           \epsilon ^{ab}
           \partial _{a} X ^{\mu} \ttt{\Bsigma}
           \partial _{b} X ^{\nu} \ttt{\Bsigma}
    \nonumber \\
    & = &
    \int _{\Sigma} {d ^{2} \Bsigma} \,
        \dot{X} ^{\mu \nu} \ttt{\Bsigma}
\eea
We used the following properties:
\begin{enumerate}
    \item we first use the boundary parametrization
    $$
        Y ^{\mu} : \Gamma \approx \Sf ^{1} \longrightarrow C
    $$
    for the {\sl String}, \dots
    \item \dots expressing then the {\it Boundary} functions
    as restrictions of the {\it Bulk} ones throgh expression
    (\ref{2.baupardef}), where we remember that after the
    {\it reparametrization procedure} $\Sigma$ substitutes $\Xi$ and thus
    we have $\Bsigma$ in place of $\Bxi$;
    \item we can now use the chain rule in computing the derivatives
    of the {\it Bulk parametrization functions} with respect to the
    boundary parameter and \dots
    \item \dots put in evidence the {\sl Parameter Space} {\it vector}
    $
        \left [
           X ^{\mu} \ttt{\Bsigma \tst}
           \partial _{a} X ^{\nu} \ttt{\Bsigma \tst}
       \right ]
    $,
    of which the present integral is the circulation;
    \item we are now in the position of applying Green's Theorem to
    write the expression as an integration on the {\sl Parameter Space}
    of the $2$-dimensional curl of
    $
        \left [
           X ^{\mu} \ttt{\Bsigma \tst}
           \partial _{a} X ^{\nu} \ttt{\Bsigma \tst}
       \right ]
    $
    \dots
    \item \dots recognising that in computing the $2$-dimensional curl,
    one of the two terms vanishes having two derivative symmetric indices
    contracted with the two totally antisymmetric indices of the
    Levi--Civita tensor.
\end{enumerate}
This completes the proof.
\end{proof}

Having established the results above
(proposition \ref{2.althamexp} in particular),
we can rewrite the action functionals
(\ref{2.fulrepact}) and (\ref{2.schhamfun}) as follows
\begin{props}[Actions in the Reparametrized Mixed Formulation]\spbcorr{}.\\
    In the Reparametrized Mixed Formulation we have the following expressions
    \bea
        S \qtq{\Bx , \Bp , \Bxi}
        & = &
        \frac{1}{2}
        \int _{\Sigma} d ^{2} \Bsigma
        \left [
            \PPuu{m}{\mu}
            \partial _{m} X _{\mu}
            +
            \pi ^{AB}
            \dot{\xi} _{AB}
            +
            N _{AB}
            \left(
                \pi ^{AB}
                -
                \frac{\epsilon ^{AB}}{2 m ^{2}}
                \PPuu{m}{\mu}
                \PPdd{m}{\mu}
            \right)
        \right ]
        \nonumber \\
        S \qtq{\Bx , \Bp}
        & = &
        \frac{1}{2}
        \int _{\Sigma} d ^{2} \Bsigma
        \left [
            \PPuu{m}{\mu}
            \partial _{m} X _{\mu}
            -
            \frac{1}{2 m ^{2}}
                \PPuu{m}{\mu}
                \PPdd{m}{\mu}
        \right ]
        \label{2.mixschhamfun}
    \eea
    for the fully reparametrized action in Hamiltonian form
    {\rm (\ref{2.fulrepact})} and for the
    Hamiltonian formulation of the Schild action {\rm (\ref{2.schhamfun})}
    respectively.
\end{props}
\begin{proof}
The second result immediately comes out from the action (\ref{2.schhamfun})
when we use results (\ref{2.equhamexp}-\ref{2.equmomvelexp}).
The same procedure can then be used to get the first one.
\end{proof}

\section{Covariant Schild Action}
\label{2.covschactsec}

In this section we take the last step in the generalization process
we started in section \ref{2.repschfor} with equation (\ref{2.repschlag}).
We will then see in chapter \ref{9.connection}
how this last step enlightens some very deep
consequences of the formulation we are proposing.
The reader is probably already able to understand what kind of
generalization immediatly comes in mind observing equation
(\ref{2.mixschhamfun}).  In this equation we have a quantity, the
{\sl Area Momentum Projected on the World--Sheet}, which is carrying
a particular indexing, having one, let us call it {\it internal}, index
(the greek {\it one}!) and one {\sl Parameter Space} {\it index} (the
{\it lowercase latin} one). We call the {\it greek} index an internal
one since it just labels a {\it multiplet of fields} defined on
a $2$-dimensional {\sl String} domain. The {\it lowercase latin} index is instead
the index associated with this $2$-dimensional domain, that we
called the {\sl Parameter Space}. When we saturated two of these indices
in the previous computations, we implicitly assumed to close them in
the ``{\it{}flat}\,'' way, but there is no reason to restrict ourself to this
case; in general we can equipe the {\sl Parameter Space} with an arbitrarly
general metric, let us call it $g ^{ab}$. Then the action (\ref{2.mixschhamfun})
turns into the following one.
\begin{defs}[Covariant Schild Action]\spbcorr{}.\\
The covariant version of the Schild Action in Hamiltonian Form is:
\beq
    S _{(\bs{g})} \qtq{\Bx , \Bp , \Bg}
    =
    \int _{\Sigma} d ^{2} \Bsigma \sqrt{\dete{\Bg}} g ^{mn}
        \left [
            \partial _{m} X ^{\mu} \ttt{\Bsigma}
            \PPdd{n}{\mu} \ttt{\Bsigma}
            -
            \frac{1}{2 \cpu}
            \PPdu{m}{\mu} \ttt{\Bsigma}
            \PPdd{n}{\mu} \ttt{\Bsigma}
        \right ]
    \quad ,
\label{2.covschhamact}
\eeq
which we will call the \underbar{Covariant Schild Action}
in Hamiltonian form.
\end{defs}
It can be proved now that
\begin{props}[Equation of Motion from the
              Covariant Schild Action]\spbcorr{}.\\
    The equation of motion for the action
    {\rm (\ref{2.covschhamact})}, $S _{(\bs{g})}$,
    are:
    \bea
        \partial _{m}
        \left [
            \sqrt{\dete{\Bg}}
            \PPuu{m}{\mu}
        \right ]
        & = &
        0
        \label{9.uno}
        \\
        \PPdu{m}{\mu}
        & = &
        \cpu
        \partial _{m}
        X ^{\mu}
        \label{9.due}
        \\
        \PPdd{m}{\mu}
        \left(
            \PPdu{n}{\mu}
            +
            \cpu
            \partial _{n} X ^{\mu}
        \right)
        & = &
        \frac{1}{2}
        g _{mn}
        g ^{ab}
        \PPdd{a}{\alpha}
        \left(
            \PPdu{b}{\alpha}
            +
            \cpu
            \partial _{b} X ^{\alpha}
        \right)
        \label{9.tre}
    \quad .
    \eea
\end{props}
\begin{props}[Solution of the Equation of Motion]\spbcorr{}.
    \label{2.covschsolpro}\\
    The equations {\rm (\ref{9.uno}-\ref{9.tre})} have the following
    solutions:
    \bea
    g _{mn}
    & = &
    \partial _{m} X ^{\mu}
    \partial _{n} X ^{\nu}
    \dfn
    \gamma _{mn} \ttt{\Bx}
    \nonumber \\
    \PPdd{a}{\mu} \ttt{\Bsigma}
    & = &
    \bar{P} _{\mu\nu}
    \epsilon _{a} {}^{n}
    \partial _{n} Y ^{\nu} \ttt{\Bsigma}
    +
    m
    \partial _{a}
    \eta _{\mu} \ttt{\Bsigma}
    \quad .
    \eea
\end{props}
\begin{proof}
Equation (\ref{9.tre}) requires the vanishing of the {\sl String} energy--momentum
tensor $T_{mn}$.
Equation (\ref{9.due}) allows us to solve
equation (\ref{9.tre}) with
respect to the {\sl String} metric:
\beq
    g _{mn}
    =
    \partial _{m} X ^{\mu}
    \partial _{n} X ^{\nu}
    \dfn
    \gamma _{mn} \ttt{\Bx}
    \label{9.gonn}
\eeq
Equations (\ref{9.due}) and (\ref{9.gonn})
show that the on--shell canonical momentum is proportional
to the gradient of the {\sl String} coordinate and the on--shell {\sl String}
metric matches the {\sl World--Sheet} induced metric.
\bea
    P _{\mu\nu}
    & = &
    \bar{P} _{\mu\nu}
    +
    m
    p _{\mu\nu} \ttt{\Bsigma}
    \nonumber \\
    \PPdd{a}{\mu} \ttt{\Bsigma}
    &=&
    \bar{P} _{\mu\nu}
    \epsilon _{a} {}^{n}
    \partial _{n} Y ^{\nu} \ttt{\Bsigma}
    +
    m
    \partial _{a}
    \eta _{\mu} \ttt{\Bsigma}
    \quad ,
    \nonumber \\
    \mathrm{where}
    & &
    \quad
    \epsilon _{mn}
    p _{\mu\nu}
    \partial ^{n} Y ^{\nu}
    =
    \partial _{m}
    \eta _{\mu}
    \quad ,
\label{9.trial}
\eea
$\eta _\mu \ttt{\Bsigma}$ being a $D$-components multiplet of
{\sl World--Sheet}
scalar fields, and $ \bar{P} _{\mu \nu}$ being a constant background over
the {\sl String} manifold, i.e. $\bar{P} _{\mu \nu} $ is the
\underbar{{\it Area Momentum Zero Mode}}:
\beq
    \partial _{m} \bar{P} _{\mu\nu}
    =
    0
    \quad .
\eeq
By averaging the  $\eta$-field over the {\sl String} {\sl World--Sheet},
one can extract its zero frequency component $\bar{\eta} ^{\mu}$:
\bea
    \eta ^{\mu} \ttt{\Bsigma}
    & = &
    \bar{\eta} ^{\mu}
    +
    \tilde{\eta} ^{\mu} \ttt{\Bsigma}
    \quad ,
    \nonumber \\
    \bar{\eta} ^{\mu}
    & \dfn &
    {1 \over \int _{\Sigma} d ^{2} \Bsigma \sqrt{\dete{\bf{g}}}}
    \int _{\Sigma} d ^{2} \Bsigma
    \sqrt{\dete{\bs{g}}}
    \eta ^\mu \ttt{\Bsigma}
    \quad ,
    \nonumber
\eea
where
$\tilde{\eta} ^{\mu} \ttt{\Bsigma}$
describes the {\it Bulk quantum fluctuations},
as measured with respect to the reference value $ \bar{\eta} ^\mu $.
Confining\footnote{By ``{\it confinement}'' we mean that
there is no leakage of the field
current $j_m \dfn \tilde{\eta} ^{\mu} \partial _{m} \tilde{\eta} _{\mu}$
off the {\sl World--Sheet} {\it Boundary}, i.e.
$$
    \oint _{\Gamma} d n ^{a} j _{a}
    =
    0\quad .
$$
}
$\tilde{\eta} ^{\mu} \ttt{\Bsigma}$
to the {\it Bulk} of the {\sl String World--Sheet}
requires appropriate boundary conditions. Accordingly,
we assume that both $ \tilde{\eta} ^{\mu} \ttt{\Bsigma}$ and its
(normal and tangential)
derivatives vanish when restricted on the boundary $\Gamma$:
\begin{eqnarray}
    \left .
        \tilde{\eta} ^{\mu}
    \right \rceil _{\Gamma}
    & = &
    0
    \label{9.b1}
    \\
    \left .
        t ^{m}
        \partial _{m}
        \tilde{\eta} ^{\mu}
    \right \rceil _{\Gamma}
    & = &
    0
    \label{9.b2}\\
    \left .
        n ^{m}
        \partial _{m}
        \tilde{\eta} ^{\mu}
    \right \rceil _{\Gamma}
    & = &
    0
    \quad .
    \label{9.b3}
\end{eqnarray}
\end{proof}
Moreover
\begin{props}[Covariant Schild Action and Nambu--Goto Action]\spbcorr{}.
    \label{2.namgotequpro}\\
    The {\sl Covariant Schild Action}, when computed on shell with respect
    to the {\sl World--Sheet} metric $\Bg$ turns into the Nambu--Goto action
    for the {\sl String}.
\end{props}
\begin{proof}
By inserting the classical
solutions into (\ref{2.covschhamact}) one recovers the Nambu-Goto action.
\end{proof}
With some hindsight, this result follows
from having introduced a non--trivial
metric $g_{mn}(\Bsigma)$ in the {\sl String} manifold. Thus,
the Schild action becomes
diffeormophism invariant as the Nambu--Goto action. Accordingly,
we recovered
the classical equivalence showed in (\ref{2.namgotequpro}).

%% file: chap03.tex
\pageheader{}{String Functional Quantization.}{}
\chapter{String Functional Quantization}
\label{3.strfunqua}

\begin{start}
``$\euf{Y}$ou see,''\\
You can do it.''\\
``I call it luck.''\\
``In my experience,\\
there's no such thing\\
as luck.''\\
\end{start}

\section{Quantum Propagator}
\label{3.strnamgotequsec}

Let us briefly recall the conclusions of section
\ref{2.repschfor} of chapter \ref{2.hamjac}.
After promoting the original pair of {\sl World--Sheet} parameters
to the role of dynamical fields through a {\it Boundary} preserving
transformation, we were able to define the Dynamics for the {\sl String}
as a reparametrization invariant Theory in $6$ dimensions; indeed we had
the usual four spacetime embedding functions
$X ^{\mu} \! \left( \sigma ^{0} , \sigma ^{1} \right)$ as well as the
new fields
$\xi ^{A} \! \left( \sigma ^{0} , \sigma ^{1} \right)$. The corresponding
conjugate momenta, derived from the {\sl Reparametrized Schild Action}
(\ref{2.repschlag}) in proposition \ref{2.repschconmom}, are
the {\sl Bulk Area Momentum}, denoted by
$P _{\mu \nu} \! \left( \sigma ^{0} , \sigma ^{1} \right)$, and
$\pi _{AB} \! \left( \sigma ^{0} , \sigma ^{1} \right)$ respectively:
it is worthwhile to remind that we implemented
the relation between momentum conjugated to $\dot{X} ^{\mu \nu}$
and momentum conjugated to $\dot{\xi} ^{AB}$
(relation (\ref{2.piAB})) in the
Hamiltonian formalism through to the Lagrange multiplier $N ^{AB}$.
This is a straightforward consequence of the reparametrization
invariance. Thus, our preferred starting point to treat
the quantum Dynamics of the system with path--integral techniques
is the Hamiltonian formulation associated to the Hamiltonian $2$-form
(\ref{2.fulrephamfor}), or
equivalently, action (\ref{2.fulrepact}), which we rewrite here for
convenience as
\bea
    & & \esci \esci
    S \qtq{X ^{\mu} , \xi ^{A} ; P _{\mu \nu} , \pi _{AB} ; N ^{AB}}
    =
    \nonumber \\
    & & \qquad \qquad
    =
    S _{P _{\mu \nu}} \qtq{X ^{\mu} , P _{\mu \nu} ; N ^{AB}}
    +
    S ^{(1)} _{\pi _{AB}} \qtq{\pi _{AB} ; N ^{AB}}
    +
    S _{\xi} \qtq{\xi ^{A} , \pi _{AB}}
\quad ,
\eea
where
\bea
    S _{\xi}
    & = &
    \frac{1}{2}
    \int _{\Xi}
    \pi _{AB} \,
    \form{d \xi} ^{A} \wedge \form{d \xi} ^{B}
    \nonumber \\
    S ^{(1)} _{\pi _{AB}}
    & = &
    -
    \frac{1}{2}
    \int _{\Sigma} d ^{2} \bs{\sigma} \,
        N ^{AB} \pi _{AB}
    \label{3.pi1act}
    \\
    S _{P _{\mu \nu}}
    & = &
    \frac{1}{2}
    \left [
        \int _{\mathcal{W}}
        P _{\mu \nu}
        \form{d X} ^{\mu} \wedge \form{d X} ^{\nu}
        +
        \epsilon _{AB}
        \int _{\Sigma} d ^{2} \bs{\sigma}
            N ^{AB} \Ham _{\mathrm{Schild}} \left( P _{\mu \nu} \right)
    \right ]
\quad .
\eea
Let us denote the measure in the functional space of path as
$$
    \left [ \mathcal D \mu \left( \bs{\sigma} \right) \right]
    =
    \left [ \mathcal D X ^{\mu} \left( \bs{\sigma} \right) \right]
    \left [ \mathcal D P _{\mu \nu} \left( \bs{\sigma} \right) \right]
    \left [ \mathcal D \xi ^{A} \left( \bs{\sigma} \right) \right]
    \left [ \mathcal D \pi _{AB} \left( \bs{\sigma} \right) \right]
    \left [ \mathcal D N ^{AB} \left( \bs{\sigma} \right) \right]
$$
Then, the propagator can be expressed by means of the following
path--integral\footnote{
          The normalization constant will be fixed
          after all the functional integrations
          are carried out, by imposing the {\it Boundary}
          condition
          \beq
               \lim _{A \to 0}
                   K \qtq{\By \tst , \By _{0} \tst ; A}
               =
               \delta \qtq{\By \tst , \By _{0} \tst}
               \quad .
          \label{3.boucon}
          \eeq
	  Such a procedure effectively amounts to a
          renormalization of the field--dependent determinants
          produced by gaussian integration.}:
\bea
    K \left [
        Y _{f} ^{\mu} \tst ,
        Y _{i} ^{\mu} \tst ;
        A
      \right ]
    & = &
    \int _{\! \! \! \! \  _{\scriptstyle
                            Y _{i} ^{\mu} \tst \, , \: \zeta _{i} ^{A} \tst}
                           }
         ^{\! \! \! \! \  ^{\scriptstyle
                            Y _{f} ^{\mu} \tst \, , \: \zeta _{f} ^{A} \tst}
                           }
         \! \! \! \! \! \! \! \! \! \! \! \! \! \! \! \! \! \!
         \qtq{\mathcal{D} \mu \ttt{\bs{\sigma}}}
        e ^{
            \frac{i}{\hbar}
            \left\{
                S
                \left [
                    X ^{\rho}
                    ,
                    P _{\sigma \tau}
                    ;
                    \xi ^{A}
                    ,
                    \pi _{CD}
                \right ]
                +
                S _{\mathrm{cnstr.}}
                \left [
                    X ^{\rho}
                    ,
                    P ^{\sigma \tau}
                    ;
                    \xi ^{A}
                    ,
                    \pi _{CD}
                    ;
                    N ^{AB}
                \right ]
            \right\}
           }
    \nonumber \\
    & = &
    \int _{\! \! \! \! \  _{\scriptstyle
                            Y _{i} ^{\mu} \tst \, , \: \zeta _{i} ^{A} \tst}
                           }
         ^{\! \! \! \! \  ^{\scriptstyle
                            Y _{f} ^{\mu} \tst \, , \: \zeta _{f} ^{A} \tst}
                           }
        \! \! \! \! \! \! \! \! \! \! \! \! \! \! \! \! \! \!
        \qtq{\mathcal{D} \mu \left( \bs{\sigma} \right)}
        e ^{
            \frac{i}{\hbar}
            \left [
                S _{\xi}
                +
                S ^{(1)} _{\pi _{AB}}
                +
                S _{P _{\mu \nu}}
            \right ]
           }
    \quad ,
    \label{3.staact}
\eea
where, the initial and final loops are:
\bea
     C _{i} \, & : & \quad Y ^{\mu} _{i} \tst
     \nonumber \\
     C _{f} \, & : & \quad Y ^{\mu} _{f} \tst
     \quad .
     \nonumber
\eea
We recall of course that
\beq
    S _{\mathrm{cnstr.}}
    \left [
        X ^{\rho} 
        ,
        P ^{\sigma \tau} 
        ,
        \xi ^{A} 
        ,
        \pi _{AB} 
        ,
        N ^{AB} 
    \right ]
    =
    -
    \frac{1}{2}
    \int _{\Sigma}
        d ^{2} \bs{\sigma}
        N ^{AB} \left( \bs{\sigma} \right)
        \left [
            \pi _{AB} - \epsilon _{AB} \Ham \left( P _{\mu \nu} \right)
        \right ]
    \quad ,
\eeq
as written in equation (\ref{2.conactfun}).
Now, we are ready to start with the pat--integral computation.
\begin{props}[Integrating out the $\Bxi$ Fields]\spbcorr{}.\\
    Performing the path--integration over the $\Bxi$ fields in
    {\rm (\ref{3.staact})} we get
    \bea
        & & \esci \esci
        \int _{\! \! \! \! \  _{\scriptstyle
                                Y _{i} ^{\mu} \tst \, , \: \zeta _{i} ^{A} \tst}
                               }
             ^{\! \! \! \! \  ^{\scriptstyle
                                Y _{f} ^{\mu} \tst \, , \: \zeta _{f} ^{A} \tst}
                               }
            \! \! \! \! \! \! \! \! \! \! \! \! \! \! \! \! \! \!
            \qtq{\mathcal{D} \mu \left( \bs{\sigma} \right)}
            e ^{
                \frac{i}{\hbar}
                \left [
                    S _{\xi}
                    +
                    S ^{(1)} _{\pi _{AB}}
                    +
                    S _{p _{\mu \nu}}
                \right ]
               }
        =
        \nonumber \\
        & & \qquad
        =
        \funinte{\tilde{\mu} \left( \bs{\sigma} \right)}
                {Y _{i} ^{\mu} \tst}
                {Y _{f} ^{\mu} \tst}
        \delta \left [
                   \epsilon ^{m _{1} m _{2}}
                   \ttt{\partial _{m _{1}} \pi _{a _{1} a _{2}}}
                   \ttt{\partial _{m _{2}} \xi ^{a _{2}}}
               \right ]
            e ^{
                \frac{i}{\hbar}
                \left [
                    S ^{(2)} _{\pi ^{AB}}
                    +
                    S ^{(1)} _{\pi _{AB}}
                    +
                    S _{P _{\mu \nu}}
                \right ]
               }
        \quad ,
	\label{3.firsteact}
    \eea
    where we denote by $\qtq{\mathcal{D} \tilde{\mu} \ttt{\Bsigma}}$
    the remaining measure,
    $$
        \left [ \mathcal D \tilde{\mu} \left( \bs{\sigma} \right) \right]
        =
        \left [ \mathcal D X ^{\mu} \left( \bs{\sigma} \right) \right]
        \left [ \mathcal D P _{\mu \nu} \left( \bs{\sigma} \right) \right]
        \left [ \mathcal D \pi _{AB} \left( \bs{\sigma} \right) \right]
        \left [ \mathcal D N ^{AB} \left( \bs{\sigma} \right) \right]
    $$
    and
    $$
        S ^{(2)} _{\pi ^{AB}}
        =
        \int _{\partial \Xi}
            \form{d \zeta} ^{B}
            \varpi _{AB}
            \zeta ^{A}
        =
        \pi _{AB}
        \int _{\Xi}
            \form{d} \left( \xi ^{A} \form{d \xi} ^{B} \pi _{AB} \right)
        \quad .
    $$
\end{props}
\begin{proof}
To get the desired result we have to prove that,
    \bea
        \funinte{\xi ^{A} \left( \bs{\sigma} \right)}
                {\zeta _{i} ^{A}}
                {\zeta _{f} ^{A}}
            e ^{\frac{i}{\hbar} S _{\xi}}
        & = &
        \delta
        \left [
            \epsilon ^{m _{1} m _{2}}
            \partial _{m _{1}} \pi _{a _{1} a _{2}}
            \partial _{m _{2}} \xi ^{a _{2}}
        \right ]
        e ^{\frac{i}{2 \hbar} S ^{(2)} _{\pi ^{AB}}}
        \\
        S ^{(2)} _{\pi ^{AB}}
        & = &
        \pi _{AB}
        \int _{\Xi \left( \bs{\sigma} \right)}
            \form{d} \left( \xi ^{A} \form{d \xi} ^{B} \right)
    \eea
since these is the only term depending on $\Bxi$ in (\ref{3.staact}).
We now have
\bea
    \funinte{\xi ^{A} \left( \bs{\sigma} \right)}
            {\zeta _{i} ^{A}}
            {\zeta _{f} ^{A}}
        e ^{\frac{i}{\hbar} S _{\xi}}
    & = &
    \funinte{\xi ^{A} \left( \bs{\sigma} \right)}
            {\zeta _{i} ^{A}}
            {\zeta _{f} ^{A}}
    \exp
    \left\{
        \frac{i}{2 \hbar}
        \int _{\Xi}
        \pi _{AB} \,
        d \xi ^{A} \wedge d \xi ^{B}
    \right\}
    \nonumber \\
    & = &
    \funinte{\xi ^{A} \left( \bs{\sigma} \right)}
            {\zeta _{i} ^{A}}
            {\zeta _{f} ^{A}}
    \exp
    \left\{
        \frac{i}{2 \hbar}
        \int _{\Xi}
        \form{d}
        \left(
            \pi _{AB} \,
            \xi ^{A} \form{d \xi} ^{B}
        \right)
        -
        \frac{i}{2 \hbar}
        \int _{\Xi}
        \form{d \pi _{AB}} \wedge \form{d \xi} ^{B}
        \xi ^{A}
    \right\}
    \nonumber \\
    & = &
    \funinte{\xi ^{A} \left( \bs{\sigma} \right)}
            {\zeta _{i} ^{A}}
            {\zeta _{f} ^{A}}
    \exp
    \left\{
        \frac{i}{2 \hbar}
        \int _{\partial \Xi}
            \varpi _{AB} \,
            \zeta ^{A} \form{d \zeta} ^{B}
        -
        \frac{i}{2 \hbar}
        \int _{\Xi}
        \form{d \pi _{AB}} \wedge \form{d \xi} ^{B}
        \xi ^{A}
    \right\}
    \nonumber \\
    & = &
    \exp
    \left\{
        \frac{i}{2 \hbar}
        \int _{\partial \Xi}
            \varpi _{AB} \,
            \zeta ^{A} \form{d \zeta} ^{B}
    \right\}
    \cdot
    \nonumber \\
    & & \qquad \cdot
    \funinte{\xi ^{A} \left( \bs{\sigma} \right)}
            {\zeta _{i} ^{A}}
            {\zeta _{f} ^{A}}
    \exp
    \left\{
        -
        \frac{i}{2 \hbar}
        \int _{\Xi}
        \epsilon ^{m _{1} m _{2}}
        \xi ^{A _{1}}
        \partial _{m _{1}} \pi _{A _{1} A _{2}}
        \partial _{m _{2}} \xi ^{A _{2}}
    \right\}
    \nonumber \\
    & = &
        \delta
        \left [
            \epsilon ^{m _{1} m _{2}}
            \partial _{m _{1}} \pi _{A _{1} A _{2}}
            \partial _{m _{2}} \xi ^{A _{2}}
        \right ]
        e ^{\frac{i}{2 \hbar} S ^{(2)} _{\pi ^{AB}}}
        \nonumber
\eea
as we desired to prove.
\end{proof}

Now we can perform the functional integration over $\pi _{AB}$.
\begin{props}[Integrating out the $\Bpi$ Fields]\spbcorr{}.\\
    Performing the path--integration over the $\Bpi$ fields in
    {\rm (\ref{3.firsteact})} we get
    \bea
        & & \esci \esci
        \int _{\! \! \! \! \  _{\scriptstyle
                                Y _{i} ^{\mu} \tst \, , \: \zeta _{i} ^{A} \tst}
                               }
             ^{\! \! \! \! \  ^{\scriptstyle
                                Y _{f} ^{\mu} \tst \, , \: \zeta _{f} ^{A} \tst}
                               }
            \! \! \! \! \! \! \! \! \! \! \! \! \! \! \! \! \! \!
            \qtq{\mathcal{D} \mu \left( \bs{\sigma} \right)}
            e ^{
                \frac{i}{\hbar}
                \left [
                    S _{\xi}
                    +
                    S ^{(1)} _{\pi _{AB}}
                    +
                    S _{P _{\mu \nu}}
                \right ]
               }
        =
        \nonumber \\
        & & \esci =
        \int _{0} ^{\infty} d E
            e ^{\frac{i}{\hbar} E A}
        \int _{Y _{i} ^{\mu} \tst} ^{Y _{f} ^{\mu} \tst}
            \left [ \mathcal{D} X ^{\mu} \ttt{\Bsigma} \right ]
            \left [ \mathcal{D} N ^{AB} \ttt{\Bsigma} \right ]
            \left [ \mathcal{D} P _{\mu \nu} \ttt{\Bsigma} \right ]
            e ^{
                -
                \frac{i}{\hbar}
                \left [
                    S ^{(2)} _{P _{\mu \nu}}
                    +
                    S ^{(1)} _{N ^{AB}}
                \right ]
               }
        \ ,
    \label{3.secsteact}
    \eea
    where we have
    \beq
        S ^{(1)} _{N _{AB}}
        =
        E
        \int _{\Sigma} d ^{2} \bs{\sigma}
            \epsilon _{AB}
            N ^{AB} \left( \bs{\sigma} \right)
        \label{3.pi1actnew}
        \quad .
    \eeq
\end{props}
\begin{proof}
Of course since the only terms containing $\pi _{AB} \ttt{\Bsigma}$
in equation (\ref{3.firsteact}),
apart from the functional Dirac delta,
are
$S ^{(1)} _{\pi ^{AB}}$
and
$S ^{(2)} _{\pi ^{AB}}$
we have to prove that
\bea
    & & \esci \esci \esci
    \funint{\pi_{AB} \left( \bs{\sigma} \right)}
        \delta
        \left [
            \epsilon ^{m _{1} m _{2}}
            \partial _{m _{1}} \pi _{A _{1} A _{2}}
            \partial _{m _{2}} \xi ^{A _{2}}
        \right ]
        e ^{
            \frac{i}{2 \hbar}
            \left [
                S _{\pi ^{AB}} ^{(1)}
                +
                S _{\pi ^{AB}} ^{(2)}
            \right ]
           }
    =
    \nonumber \\
    & & =
    \int _{0} ^{\infty} d E
        e ^{\frac{i}{\hbar} E A}
        e ^{- \frac{i}{2 \hbar} S ^{(1)} _{N ^{AB}}}
    \quad .
\eea
This result follows since, thanks to the functional Dirac delta
$$
    \delta
    \left [
        \epsilon ^{m _{1} m _{2}}
        \partial _{m _{1}} \pi _{A _{1} A _{2}}
        \partial _{m _{2}} \xi ^{A _{2}}
    \right ]
    \quad ,
$$
the functional integration over $[\mathcal{D} \pi _{AB}]$ is restricted
to the ``classical trajectory'', so that we have (thanks to the classical
Energy balance equation (\ref{2.fulrepeq4}))
\bea
    \pi _{A _{1} A _{2}}
    & = &
    E \epsilon _{A _{1} A _{2}}
    \nonumber \\
    \funint{\pi _{A _{1} A _{2}}}
        \left [ \dots \right ]
    & = &
    \int d E
        \left [ \dots \right ]
    \nonumber
    \quad .
\eea
Substituting the last two expressions gives the desired result.
\end{proof}

Thus we can now write
\beq
    K \left [
          Y ^{\mu} _{i} \ttt{\bs{\sigma}}
          ,
          Y ^{\mu} _{f} \ttt{\bs{\sigma}}
          ;
          A
      \right ]
    =
    \int _{0} ^{\infty} d E
        e ^{\frac{i}{\hbar} E A}
        G \left [
              C _{i}
              ,
              C _{f}
              ;
              E
          \right ]
    \quad ,
\label{3.gretoker}
\eeq
where
\beq
    G \left [
          C _{i}
          ,
          C _{f}
          ;
          E
      \right ]
    =
    \funinte{X ^{\mu} \ttt{\bs{\sigma}}}
            {Y _{i} ^{\mu} \tst}
            {Y _{f} ^{\mu} \tst}
            \left [ \mathcal{D} N ^{AB} \left( \bs{\sigma} \right) \right ]
            \left [ \mathcal{D} P _{\mu \nu} \left( \bs{\sigma} \right) \right ]
        e ^{
            - \frac{i}{\hbar}
            \left [
                S _{p _{\mu \nu}}
                +
                S ^{(1)} _{N ^{AB}}
            \right ]
           }
    \quad .
\eeq
The last step is now the Gaussian Functional Integral over
$P _{\mu \nu}$.
\begin{props}[Integrating out the $\Bp$ Fields]\spbcorr{}.\\
    Performing the Gaussian Functional Integration over $P _{\mu \nu}$
    in equation {\rm (\ref{3.secsteact})} we get
    \bea
        & & \esci
        \int _{\! \! \! \! \  _{\scriptstyle
                                Y _{i} ^{\mu} \tst \, , \: \xi _{i} ^{A} \tst}
                               }
             ^{\! \! \! \! \  ^{\scriptstyle
                                Y _{f} ^{\mu} \tst \, , \: \xi _{f} ^{A} \tst}
                               }
            \! \! \! \! \! \! \! \! \! \! \! \! \! \! \! \! \! \!
            \qtq{\mathcal{D} \mu \left( \bs{\sigma} \right)}
            e ^{
                \frac{i}{\hbar}
                \left [
                    S _{\xi}
                    +
                    S ^{(1)} _{\pi _{AB}}
                    +
                    S _{p _{\mu \nu}}
                \right ]
               }
        =
        \nonumber \\
        & & \qquad
        =
        \int _{0} ^{\infty} d E
            e ^{\frac{i}{\hbar} E A}
        \int _{Y _{i} ^{\mu} \tst} ^{Y _{f} ^{\mu} \tst}
            \left [ \mathcal{D} X ^{\mu} \ttt{\Bsigma} \right ]
            \left [ \mathcal{D} N ^{AB} \ttt{\Bsigma} \right ]
            e ^{
                -
                \frac{i}{\hbar}
                \left [
                    S ^{(1)} _{N ^{AB}}
                    +
                    S ^{(2)} _{N ^{AB}}
                \right ]
               }
    \quad ,
    \label{3.tersteact}
    \eea
    where
    \beq
        S _{N ^{AB}} ^{(2)}
        =
        -
        \int _{\Sigma} d ^{2} \bs{\sigma}
            \frac{m ^{2}}{\epsilon _{AB} N ^{AB}}
            \dot{X} ^{\mu \nu}
            \dot{X} _{\mu \nu}
    \eeq
    \quad .
\end{props}
\begin{proof}
The desired result can be proved if we are able to prove
\beq
    \funint{P _{\mu \nu} \left( \bs{\sigma} \right)}
        e ^{\frac{i}{\hbar} S _{P _{\mu \nu}}}
    =
    e ^{\frac{i}{\hbar} S _{N ^{AB}} ^{(2)}}
    \quad .
\eeq
This result in turn can be derived performing a functional
Gaussian integration. In more detail we have
\bea
    \funint{P _{\mu \nu} \left( \bs{\sigma} \right)}
        e ^{\frac{i}{\hbar} S _{P _{\mu \nu}}}
    & = &
    \funint{P _{\mu \nu} \left( \bs{\sigma} \right)}
    \exp
    \left\{
        \frac{i}{2 \hbar}
        \left [
            \int _{X \left( \bs{\sigma} \right)}
            P _{\mu \nu}
            \form{d X} ^{\mu} \wedge \form{d X} ^{\nu}
            +
        \right .
    \right .
    \nonumber \\
    & &
    \qquad \qquad \qquad \qquad \qquad \qquad +
    \left .
        \left .
            \epsilon _{AB}
            \int _{\mathcal{W}} d ^{2} \bs{\sigma}
                N ^{AB} \Ham _{\mathrm{Schild}} \left( P _{\mu \nu} \right)
        \right ]
    \right\}
    \nonumber \\
    & = &
    \funint{P _{\mu \nu} \left( \bs{\sigma} \right)}
    \exp
    \left\{
        \frac{i}{2 \hbar}
        \left [
            \int _{\Sigma}
            P _{\mu \nu}
            \dot{X} ^{\mu \nu}
            d ^{2} \Bsigma
            +
            \epsilon _{AB}
            \int _{\Sigma} d ^{2} \bs{\sigma}
                N ^{AB} \frac{P _{\mu \nu} P ^{\mu \nu}}{4 m ^{2}}
        \right ]
    \right\}
    \nonumber \\
    & = &
    \funint{P _{\mu \nu} \left( \bs{\sigma} \right)}
    \exp
    \left\{
            \int _{\Sigma} \! \!
            P _{\mu \nu}
            \frac{i \dot{X} ^{\mu \nu}}{2 \hbar}
            d ^{2} \Bsigma
            +
            \int _{\Sigma} \! \! d ^{2} \bs{\sigma}
                P _{\mu \nu}
                \left(
                    \frac{1}{2}
                    \frac{i \epsilon _{AB} N ^{AB}}{4 \hbar m ^{2}}
                \right)
                P ^{\mu \nu}
    \right\}
    \nonumber \\
    & = &
    \exp
    \left\{
        \int _{\Sigma} d ^{2} \bs{\sigma}
            \frac{i \dot{X} ^{\mu \nu}}{2 \hbar}
            \left(
                \frac{1}{2}
                \frac{4 \hbar m ^{2}}{i \epsilon _{AB} N ^{AB}}
            \right)
            \frac{i \dot{X} _{\mu \nu}}{2 \hbar}
    \right\}
    \nonumber \\
    & = &
    \exp
    \left\{
        -
        \frac{i m ^{2}}{2 \hbar}
        \int _{\Sigma} d ^{2} \bs{\sigma}
            \frac{\dot{X} ^{\mu \nu} \dot{X} _{\mu \nu}}
                 {\epsilon _{AB} N ^{AB}}
    \right\}
    \quad .
\eea
\end{proof}

At the end we can thus rewrite
$
    G \left [
          C _{i}
          ,
          C _{f}
          ;
          E
      \right ]
$
as the functional integral over $X ^{\mu} \left( \bs{\sigma} \right)$
and $N ^{AB} \left( \bs{\sigma} \right)$ of the exponential of
$i / \hbar$ times $S _{N ^{AB}} ^{(1)} + S _{N ^{AB}} ^{(2)}$, i.e.
introducing the shorthand notation
$
    N \left( \bs{\sigma} \right)
    =
    \epsilon _{AB}
    N ^{AB} \left( \bs{\sigma} \right)
    /
    2
$,
\beq
    G \left [
          C _{i}
          ,
          C _{f}
          ;
          E
      \right ]
    =
    \funinte{Y ^{\mu} \left( \bs{\sigma} \right)}
            {Y _{i} ^{\mu} \tst}
            {Y _{f} ^{\mu} \tst}
            \left [ \mathcal{D} N ^{AB} \left( \bs{\sigma} \right) \right ]
        \exp
        \left\{
            - \frac{i}{\hbar}
            \int _{\Sigma}
            d ^{2} \bs{\sigma}
            \left [
                -
                \frac{m ^{2}}{4 N}
                \dot{X} ^{\mu \nu}
                \dot{X} _{\mu \nu}
                +
                N
                E
            \right ]
        \right\}
    \label{3.prop}
\eeq
Then
\begin{props}[Schild--Nambu Goto Quantum Equivalence]\spbcorr{}.
\label{3.schnamgotequ}\\
    The saddle point evaluation of the
    {\sl String} propagator {\rm (\ref{3.prop})} is
    \beq
        G \left [ C _{i} , C _{f} ; E \right ]
        \simeq
        \funinte{\Bx \left( \sigma \right)}{C _{i}}{C _{f}}
        \exp
        \left\{
            -
            i
            \sqrt{m ^{2} E }
            \int _\Sigma d ^{2} \sigma
                \sqrt{- \frac{1}{2}\dot{X} ^{\mu \nu} \dot{X} _{\mu \nu}}
        \right\}
        \quad .
    \label{ng}
    \eeq
\end{props}
\begin{proof}
To determine the saddle point we have to solve the following equation
$$
    \frac{d}{d N}
    \left(
        - \frac{m ^{2}}{4 N}
        \dot{X} ^{\mu \nu} \dot{X} _{\mu \nu}
        +
        N
        E
    \right)
    =
    0
    \quad ,
$$
which is
$$
    \frac{m ^{2}}{4 N ^{2}}
    \dot{X} ^{\mu \nu} \dot{X} _{\mu \nu}
    +
    E
    =
    0
$$
The saddle point is given by the following value of $\BN \tst{\Bsigma}$
$$
    \hat{N} \left( \bs{\sigma} \right)
    =
    \sqrt{- \frac{m ^{2} \dot{X} ^{\mu \nu} \dot{X} _{\mu \nu}}{4 E}}
$$
and the substitution of this value in equation
\ref{3.prop} provides us with the result:
\beq
    G \left [
          C _{i}
          ,
          C _{f}
          ;
          E
      \right ]
    =
    \funinte{Y ^{\mu} \left( \bs{\sigma} \right)}
            {Y _{i} ^{\mu} \tst}
            {Y _{f} ^{\mu} \tst}
        e ^{
            - \frac{i}{\hbar}
            \sqrt{m ^{2} E}
            \int _{\Sigma}
            d ^{2} \bs{\sigma}
                \sqrt{- \frac{1}{2} \dot{X} ^{\mu \nu} \dot{X} _{\mu \nu}}
           }
\quad .
\eeq
Then, since $E$ has dimension of inverse length square, in
natural units, we can set the
{\sl String} tension equal to $m ^{2}$, and (\ref{ng}) reproduces
exactly the Nambu--Goto path--integral.
\end{proof}
In terms of the Green function we found above the kernel can be expressed
using equation (\ref{3.gretoker}).

\section{Functional Wave Equation}
\label{3.strfunwvequ}

The purpose of this section is to show how to derive
the functional wave equation for
$K \left[ \By \tst , \By _{0} \tst ; A \right]$
from the corresponding path--integral.
This equation will describe how the {\sl String} responds to a
variation of the final {\it Boundary} $\zeta ^{A} = \zeta ^{A} \tst$, just as the
ordinary Schr\"odinger equation describes a particle reacts
to a shift of the time interval end--point. As we have seen
in section \ref{2.hamjacsec}, the {\sl String} ``natural'' evolution parameter is the
area $A$ of the {\sl Parameter Space}, so that
{\it functional} or {\sl Area Derivatives}
generate ``translations'' in loop space, or {\sl String} deformations in
Minkowski space. Thus, we expect the functional wave
equation to be of order one in $\partial/\partial A$, and of
order two in $\delta/\delta Y ^{\mu} \tst$,
or\footnote{We stress again that {\it functional} and {\sl Area Derivatives}
are related by \cite{migdal}
\beq
    \frac{\delta}{\delta Y ^{\mu} \tst}
    =
    Y ^{\,\prime\,\nu}
    \frac{\delta}{\delta Y ^{\mu \nu}\qtq{C}}
    \quad , \qquad
    Y ^{\,\prime\,\nu}
    \equiv
    \frac{d Y ^{\nu}}{d s}
    \quad .
    \nonumber
\eeq
Therefore, the functional wave equation can be written in
terms of either type of derivative.
Contrary to the statement in ref.\cite{ogi}, area derivatives
are regular even when ordinary functional derivatives are not
\cite{migdal}.}
$\delta/\delta Y ^{\mu\nu} \tst$.\\
The standard procedure to obtain the kernel wave equation
goes through a recurrence relation satisfied by the
discretized version of the Jacobi path--integral
\cite{kawai}, \cite{suga}. However, such a construction is well
defined only when the action is a polynomial in the dynamical
variables. For a non--linear action such
as the Nambu--Goto area functional, a lattice definition of the
path--integral is much less obvious. Moreover, the continuum functional wave
equation is recovered through the highly non--trivial limit of vanishing
lattice spacing \cite{kawai}, \cite{suga}. In any case,
the whole procedure seems disconnected from the classical
approach to {\sl String} Dynamics, whereas we would like to see a logical
continuity between Quantum and Classical Dynamics. Against this
background, it seems useful to offer an {\it alternative}
path--integral derivation of the {\sl String} functional wave equation,
which is deeply rooted in the
Hamiltonian formulation of {\sl String} Dynamics discussed in
subsection \ref{2.hamjacrepsec},
and is basically derived from the same Jacobi
variational principle which we have consistently adopted so far.\\
We first start with a result that will be useful later on
\begin{props}[Kernel Variation]\spbcorr{}.\\
    The kernel variation under infinitesimal deformations of the
    field variables is
    \beq
        \delta K \left[ \By \tst , \By _{0} \tst ; A \right]
        =
        \frac{i}{\hbar}
        \int _{Y ^{\mu} _{0} \tst} ^{Y ^{\mu} \tst}
        \int _{\zeta ^{A} _{0} \tst} ^{\zeta ^{A} \tst}
            [D \mu(\sigma)]
            \ttt{\delta S _{\mathrm{Red.}}}
            \exp \ttt{\frac{i}{\hbar} S _{\mathrm{Red.}}}
    \quad .
    \label{deltak}
    \eeq
\end{props}
\begin{proof}
This result can be derived observing that in the kernel the field
variables are contained only in the action, which in turn is exponentiated:
$$
    K \left[ Y ^{\mu} \tst , Y ^{\mu} _{0} \tst ; A \right]
    =
    \int _{Y ^{\mu} _{0} \tst} ^{Y ^{\mu} \tst}
    \int _{\zeta ^{A}_{0} \tst} ^{\zeta ^{A} \tst}
        [D \mu(\sigma)]
        \exp
        \qtq{\frac{i S  _{\mathrm{Red.}} \qtq{\By , \Bp ; A}}{\hbar}}
    \quad ;
$$
then, applying the usual properties of the exponential, we get
\bea
    \delta K \left[ Y ^{\mu} \tst , Y ^{\mu} _{0} \tst ; A \right]
    & = &
    \delta
    \left[
        \int _{Y ^{\mu} _{0} \tst} ^{Y ^{\mu} \tst}
        \int _{\zeta ^{A} _{0} \tst} ^{\zeta ^{A} \tst}
            [D \mu(\sigma)]
            \exp
            \ttt{\frac{i S  _{\mathrm{Red.}} \qtq{\By , \Bp ; A}}{\hbar}}
    \right]
    \nonumber \\
    & = &
    \int _{Y ^{\mu} _{0} \tst} ^{Y ^{\mu} \tst}
    \int _{\zeta ^{A} _{0} \tst} ^{\zeta ^{A} \tst}
        [D \mu(\sigma)]
        \delta
        \left[
                \exp
                \ttt{\frac{i S _{\mathrm{Red.}} \qtq{\By , \Bp ; A}}{\hbar}}
        \right]
    \nonumber \\
    & = &
    \int _{Y ^{\mu} _{0} \tst} ^{Y ^{\mu} \tst}
    \int _{\zeta ^{A} _{0} \tst} ^{\zeta ^{A} \tst}
        [D \mu(\sigma)]
        \left[
            \frac{i}{\hbar}
            \exp
            \ttt{\frac{i S _{\mathrm{Red.}} \qtq{\By , \Bp ; A}}{\hbar}}
            \delta S \qtq{\By , \Bp ; A}
        \right]
    \quad ,
    \nonumber
\eea
which is the desired result.
\end{proof}
We now observe that only {\it Boundary} variations will contribute to
equation (\ref{deltak}) if we restrict the fields to vary within the family
of classical solutions corresponding to a given initial {\sl String}
configuration, $C _{0} \: : \quad Y _{i} ^{\mu} \tst$. Then
\begin{props}[Kernel Derivatives]\spbcorr{}.\\
The derivatives of the kernel with respect to the dynamical variables,
the {\sl Areal Time} $A$ and the final shape of the loop $Y ^{\mu} \tst$,
are
\bea
    \frac{\partial K \qtq{ Y ^{\mu} \tst , Y ^{\mu} _{0} \tst ; A}}
         {\partial A}
    \! \! \! & = & \! \! \!
    -
    \frac{i E}{\hbar}
    K \qtq{ Y ^{\mu} \tst , Y ^{\mu} _{0} \tst ; A}
    \quad ,
    \label{darea}
    \\
    \frac{\delta K \qtq{ Y ^{\rho} \tst , Y ^{\rho} _{0} \tst ; A}}
         {\delta Y ^{\mu} \tst}
    \! \! \! & = & \! \! \!
    \frac{i}{\hbar}
    \int _{Y ^{\rho} _{0} \tst} ^{Y ^{\rho} \tst}
    \int _{\zeta ^{A} _{0} \tst} ^{\zeta ^{A} \tst}
        \qtq{\mathcal{D} \mu \ttt{\Bsigma}}
        Q _{\mu \nu} Y ^{\prime \, \nu}
    \exp
    \ttt{\frac{i S _{\mathrm{Red.}}}{\hbar}}
    \label{functd}
    \quad .
\eea
\end{props}
\begin{proof}
We are going to use equations (\ref{deltak}) and (\ref{2.strhamjacpreden});
from the last of them we already derived in
subsection (\ref{2.hamjacrepsec}) results
(\ref{2.corfunder}, \ref{2.arenorder}), which we now rewrite for clarity:
\bea
    \frac{\delta S _{\mathrm{Red.}}}
         {\delta Y ^{\mu} \tst}
    & = &
    Q _{\mu \nu} Y ^{\prime \nu}
    \nonumber \\
    \frac{\delta S _{\mathrm{Red.}}}{\delta A}
    =
    \frac{d S _{\mathrm{Red.}}}{d A}
    & = &
    E
    \quad ;
\eea
here, we remembered again that the Hamiltonian $H _{\mathrm{Schild}}$
is constant, $H _{\mathrm{Schild}} \equiv E$, along a classical
trajectory. Thus, from (\ref{deltak})
\bea
    & & \esci \esci
    \frac{\partial K \qtq{ Y ^{\mu} \tst , Y ^{\mu} _{0} \tst ; A}}
         {\partial A}
    =
    \nonumber \\
    & & \qquad
    =
    \frac{i}{\hbar}
    \int _{Y ^{\mu} _{0} \tst} ^{Y ^{\mu} \tst}
    \int _{\zeta ^{A} _{0} \tst} ^{\zeta ^{A} \tst}
        [D \mu(\sigma)]
            \exp
            \ttt{\frac{i S _{\mathrm{Red.}} \qtq{\By , \Bp ; A}}{\hbar}}
            \frac{\delta S _{\mathrm{Red.}} \qtq{\By , \Bp ; A}}{\delta A}
    \nonumber \\
    & & \qquad
    =
    \frac{i}{\hbar}
    \int _{Y ^{\mu} _{0} \tst} ^{Y ^{\mu} \tst}
    \int _{\zeta ^{A} _{0} \tst} ^{\zeta ^{A} \tst}
        [D \mu(\sigma)]
            \exp
            \ttt{\frac{i S _{\mathrm{Red.}} \qtq{\By , \Bp ; A}}{\hbar}}
            \frac{d S _{\mathrm{Red.}} \qtq{\By , \Bp ; A}}{d A}
    \nonumber \\
    & & \qquad
    =
    \frac{i E}{\hbar}
    \int _{Y ^{\mu} _{0} \tst} ^{Y ^{\mu} \tst}
    \int _{\zeta ^{A} _{0} \tst} ^{\zeta ^{A} \tst}
        [D \mu(\sigma)]
            \exp
            \ttt{\frac{i S _{\mathrm{Red.}} \qtq{\By , \Bp ; A}}{\hbar}}
    \nonumber \\
    & & \qquad
    =
    \frac{i E}{\hbar}
    K \qtq{ Y ^{\mu} \tst , Y ^{\mu} _{0} \tst ; A}
    \quad .
\eea
In the same way, we get
\bea
    & & \esci \esci
    \frac{\partial K \qtq{ Y  ^{\rho} \tst , Y _{0}  ^{\rho} \tst ; A}}
         {\partial Y ^{\mu} \tst}
    =
    \nonumber \\
    & & \qquad
    =
    \frac{i}{\hbar}
    \int _{Y ^{\rho} _{0} \tst} ^{Y ^{\rho} \tst}
    \int _{\zeta ^{A} _{0} \tst} ^{\zeta ^{A} \tst}
        [D \mu(\sigma)]
            \exp
            \ttt{\frac{i S _{\mathrm{Red.}} \qtq{\By , \Bp ; A}}{\hbar}}
            \frac{\delta S _{\mathrm{Red.}} \qtq{\By , \Bp ; A}}
                 {\delta Y ^{\mu} \tst}
    \nonumber \\
    & & \qquad
    =
    \frac{i}{\hbar}
    \int _{Y ^{\rho} _{0} \tst} ^{Y ^{\rho} \tst}
    \int _{\zeta ^{A} _{0} \tst} ^{\zeta ^{A} \tst}
        [D \mu(\sigma)]
            \exp
            \ttt{\frac{i S _{\mathrm{Red.}} \qtq{\By , \Bp ; A}}{\hbar}}
            Q _{\mu \nu} \tst Y ^{\prime \nu} \tst
    \quad .
\eea
\end{proof}
After these preliminary steps the main result of this chapter follows:
\begin{props}[String Kernel Functional Schr\"odinger Equation]\spbcorr{}.\\
    The propagation Kernel of the {\sl String} satisfies the following
    Schr\"o{}dinger-like functional equation
    \bea
        & & \esci \esci \esci \esci
        -
        \frac{\hbar ^{2}}{2 m ^{2}}
        \norme{\Gamma \approx \Sf ^{1}}
        \! \! \!
        \oint _{\Gamma \approx \Sf ^{1}}
        \! \!
        \frac{ds}{\sqrt{\ttt{\By '} ^{2}}}
        \frac{\delta ^{2} K \qtq{ \By \tst , \By _{0} \tst ; A}}
             {\delta ^{\mu} Y \tst \delta _{\mu} Y \tst}
        =
        \nonumber \\
        & & \qquad \qquad \qquad \qquad \qquad
        =
        i
        \hbar
        \frac{\partial K \qtq{ \By \tst , \By _{0} \tst ; A}}{\partial A}
    \label{A.kerfunwavequ}
    \quad .
    \eea
\end{props}
\begin{proof}
The proof is a standard procedure, in which the classical Hamilton--Jacobi
equation is assumed as the evolution equation for the mean values of the
quantum operators. Mean values, according to Feynman path--integral
formulation of Quantum Mechanics/Quantum Field Theory can be computed
as functional integrals.
First, we remember result (\ref{functd}),
which is here rewritten for convenience:
\bea
    \frac{\delta}{\delta Y ^{\mu} \tst}
    K \left [
          \By \tst
          ,
          \By _{0} \tst
          ;
          A
      \right ]
    & = &
    \frac{i}{\hbar}
    \int _{\By _{0} \tst} ^{\By \tst}
    \funinte{\mu \tst}
            {\Bzeta _{0} \tst}
            {\Bzeta \tst}
        \frac{\delta S _{\mathrm{Red.}}}{\delta Y ^{\mu} \tst}
        e ^{\frac{i}{\hbar} S _{\mathrm{Red.}}}
    \nonumber \\
    & = &
    \frac{i}{\hbar}
    \int _{\By _{0} \tst} ^{\By \tst}
    \funinte{\mu \tst}
            {\Bzeta _{0} \tst}
            {\Bzeta \tst}
        q _{\mu}
        \exp \ttt{{\frac{i}{\hbar} S _{\mathrm{Red.}}}}
    \label{3.firfunderpro}
    \quad ;
\eea
the definition of equation (\ref{2.corfunder}) has also been used.
We can now take one more functional derivative of the Kernel, again
with respect to $Y ^{\mu}$, to get
\bea
    \frac{\delta ^{2}}
         {\delta Y ^{\mu} \tst \delta Y _{\mu} \tst}
    K \left [
          \By \tst
          ,
          \By _{0} \tst
          ;
          A
      \right ]
    & = &
    -
    \frac{1}{\hbar ^{2}}
    \int _{\By _{0} \tst} ^{\By \tst}
    \funinte{\mu \tst}
            {\Bzeta _{0} \tst}
            {\Bzeta \tst}
        q _{\mu} q ^{\mu}
        \exp \ttt{\frac{i}{\hbar} S _{\mathrm{Red.}}}
    \nonumber \\
    & \equiv &
    -
    \frac{1}{\hbar ^{2}}
    \overline{q ^{\mu} q _{\mu}}
    \label{A.secfundermomave}
    \quad ,
\eea
where we recognize in the right hand side the expectation value
for the square of the {\sl Boundary Area Momentum}.
Moreover we also have equation (\ref{darea}):
\beq
    \frac{\partial}{\partial A}
    K \left [
          \By \tst
          ,
          \By _{0} \tst
          ;
          A
      \right ]
    =
    -
    \frac{i E}{\hbar}
    K \left [
          \By \tst
          ,
          \By _{0} \tst
          ;
          A
      \right ]
    \quad .
    \label{A.kerareder}
\eeq
If we now consider the Hamilton--Jacobi equation (\ref{2.ourhamjac}) we can
interpret it as the evolution equation for the mean values of
the quantum mechanical probabilities, i.e. we can substitute
to $q _{\mu} q ^{\mu}$ the quantum mechanical expectation value
$\overline{q _{\mu} q ^{\mu}}$ so that
\beq
    \frac{1}{2 m ^{2}}
    \oint _{\Gamma \approx \Sf ^{1}} \frac{ds}{\sqrt{\ttt{\By '}^{2}}}
        \overline{q ^{\mu} q _{\mu}}
    =
    E
    \oint _{\Gamma \approx \Sf ^{1}} ds
        \sqrt{\ttt{\By '}^{2}}
\eeq
Using then equations (\ref{A.secfundermomave}, \ref{A.kerareder})
to express the quantities appearing above in terms of the propagator,
we get
\beq
    -
    \frac{\hbar ^{2}}{2 m ^{2}}
    \norm
    \oint _{\Gamma \approx \Sf ^{1}} \frac{ds}{\sqrt{\ttt{\By '}^{2}}}
        \frac{
              \delta ^{2}
              K \left [
                    \By \tst
                    ,
                    \By _{0} \tst
                    ;
                    A
                \right ]
             }
             {\delta Y ^{\mu} \tst \delta Y _{\mu} \tst}
    =
    i
    \hbar
    \frac{\partial}{\partial A}
    K \left [
          \By \tst
          ,
          \By _{0} \tst
          ;
          A
      \right ]
    \quad ,
\eeq
i.e. the equation for the evolution of the kernel of the {\sl String}.
\end{proof}
Thus, $K \left[ \By \tst , \By _{0} \tst ; A \right]$
can be determined either by solving
the functional wave equation (\ref{A.kerfunwavequ}), or by evaluating the
path--integral (\ref{3.staact}).\\
Once equation (\ref{A.kerfunwavequ}) is given, it is straightforward to
show that $G \left[ C , C _{0} ; m ^{2} \right]$ defined in equation
(\ref{3.gretoker}) satisfies the following equation
\bea
    & & \esci \esci
    \left[
        -
        \hbar ^{2}
        \left(
            \int _{0} ^{1} ds
            \sqrt{\ttt{\By '} ^{2}}
        \right) ^{-1}
        \cdot
    \right .
    \nonumber \\
    & & \qquad
    \left .
        \cdot
        \int _{0} ^{1} {ds \over \sqrt{\ttt{\By '} ^{2}}}
            {
             \delta ^{2}
             \over
             \delta Y ^{\mu} \tst
             \delta Y _{\mu} \tst
            }
            +
            m ^{4}
    \right]
    G \qtq{C , C _{0} ; m ^{2}}
    =
    -
    \delta \qtq{ C - C _{0}}
    \quad .
\label{green}
\eea
Therefore, $G \left[ C , C _{0} ; m ^{2} \right]$
can be identified with the Green function for the {\sl String}.

The functional equation (\ref{A.kerfunwavequ}) can be rewritten
in different forms, which are useful to point out some specific
properties. In particular it is natural to ask wether it
is possible to give a formulation in which we have only reference
to {\it Holographic} quantities or not. We thus recall the definition
\ref{2.holfunderdef} as well as the comments following it
on page \pageref{2.aredercom}. In particular we remember that
the {\sl Holographic Functional Derivative} has an explicitly
dependence from the point of the loop where it is calculated.
In order to recover reparametrization invariance, we have
to get rid of the arbitrariness in the choice of the attachment point.
This can be achieved by summing over all its possible
locations along the loop, and then compensating for the
overcounting of the area variation by averaging the result over the proper
length of the loop. This is what we have done in equation
(\ref{A.kerfunwavequ}) and in equation (\ref{2.ourhamjac})
(and we will do the same in (\ref{3.totloomomope})).
The same prescription
enables us to define any other reparametrization invariant
quantity or operator.
Hence, starting with the {\sl Holographic Functional Derivative},
we introduce a simpler but more effective
notation.
\begin{defs}[Loop Derivative]\spbcorr{}.
    The \underbar{Loop Derivative} is the average of the
    {\sl Holographic Derivative} over the {\sl String} loop:
    \beq
        \frac{\delta}{\delta C ^{\mu \nu}}
        \equiv
        \norme{\Gamma}
        \oint _{\Gamma} d l \tst
            \frac{\delta}{\delta Y ^{\mu \nu} \tst}
        \quad .
    \label{3.newnotation}
    \eeq
\end{defs}
This represents a genuine loop operation without
reference to the way in which the loop is parametrized.
We can now cast the functional wave equation for the Kernel
(\ref{A.kerfunwavequ}) in some alternative forms.
\begin{props}[Alternative Forms of
              the Functional Kernel Equation]\spbcorr.{}\\
    The {\sl Functional Wave Equation for the Kernel},
    {\rm (\ref{A.kerfunwavequ})},
    can be cast in the following alternative forms:
    \beq
        -
        \frac{\hbar ^{2}}{2 m ^{2}}
        \norme{\Gamma \approx \Sf ^{1}}
        \! \! \!
        \oint _{\Gamma \approx \Sf ^{1}}
            \! \!
            ds \sqrt{\ttt{\By '} ^{2}}
            \frac{\delta ^{2} K \qtq{\By \tst , \By _{0} \tst ; A}}
                 {\delta \By ^{\mu \nu} \tst \delta \By _{\mu \nu} \tst}
        =
        i
        \hbar
        \frac{\partial K \qtq{ \By \tst , \By _{0} \tst ; A}}{\partial A}
        \label{3.kerfunwavsigequ}
    \eeq
    \beq
        \frac{1}{2}
        \frac{\delta ^{2} K \qtq{ \By \tst , \By _{0} \tst ; A}}
             {\delta C ^{\mu \nu} \delta C _{\mu \nu}}
        =
        i
        \hbar
        \frac{\partial K \qtq{ \By \tst , \By _{0} \tst ; A}}{\partial A}
        \label{3.kerfunwavlopequ}
    \eeq
\end{props}
\begin{proof}
The quickest way to derive this result is to take a look at
equation (\ref{2.strhamjacpreden}). In particular we observe that
the dispersion relation in integrated form can also be written as
$$
    \frac{1}{2 m ^{2}}
    \norme{\Gamma}
    \oint _{\Gamma \approx \Sf ^{1}}
        ds \sqrt{\ttt{\By '} ^{2} \tst}
        Q ^{\mu \nu} \tst
        Q _{\mu \nu} \tst
    =
    E
    \quad .
$$
Of course, as we will stress again on page \pageref{def}
$Q _{\mu \nu} \tst$ is the classical local quantity conjugated
to the operator $i \delta / \delta Y ^{\mu \nu} \tst$, so that,
using a generalized form of the Correpsondence Principle we get exactly
(\ref{3.kerfunwavsigequ}).
Moreover
in the notation (\ref{3.newnotation}), the {\it loop laplacian}
reads
\beq
    \frac{1}{2}
    \frac{\delta ^{2}}
         {
          \delta C ^{\mu \nu}
          \delta C _{\mu \nu}
         }
    \equiv
    \frac{1}{2}
    \norme{\Gamma}
    \oint _{\Gamma} d l \tst
        \frac{\delta ^{2}}
             {
              \delta Y ^{\mu \nu} \tst
              \delta Y _{\mu \nu} \tst
             }
    \quad .
\label{3.lapnewnot}
\eeq
so that from the previous result we recover exactly equation
(\ref{3.kerfunwavlopequ}).
\end{proof}

\section{Computing the Kernel}
\label{5}

\subsection{Integrating the Functional Wave Equation}
\label{5.1}

It is possible to compute
$K \left[ \By \tst , \By _{0} \tst ; A \right]$ exactly
in the ``free'' case because the
Lagrangian corresponding to the Schild Hamiltonian (\ref{2.redrepact}) is
quadratic with respect to the generalized velocities $\dot{X} ^{\mu \nu}$.
Previous experience with this class of Lagrangians
suggests the following ansatz for the {\sl String} quantum kernel.\\[2mm]
{\bf Ansatz (Quantum Kernel)}: {\it The Quantum Kernel of a {\sl String}
has the following functional dependence:}
\beq
    K [\By \tst , \By _{0} \tst ; A]
    =
    {\cal N}
    A ^{\alpha}
    \exp \left(
             \frac{i}{\hbar}
             I [ \By \tst , \By _{0} \tst ; A]
         \right)
    \quad ,
\label{ansatz}
\eeq
{\it where ${\cal N}$ is a normalization constant, and $\alpha$
a real number}.

To solve the problem of determining the form of the propagator is
now equivalent to determine the unknown function
$I \qtq{\By \tst , \By _{0} \tst ; A}$ and the
real number $\alpha$. We make a first step in this direction in the
following
\begin{props}[Amplitude and Phase Equations for the Kernel]\spbcorr{}.
    \label{3.ampphakerequ}\\
    The exponent $\alpha \in \R$ and the phase
    $I[\By , \By _{0} ; A]$ that appear in the ansatz
    {\rm (\ref{ansatz})} satisfy the following two
    independent equations:
    \bea
    {2\alpha m^2\over A}
    &=&
    -\norme{\Gamma}
    \oint _{\Gamma} {ds\over\sqrt{\ttt{\By '} ^{2}}}
    {
     \delta^2 I[\By , \By _{0} ; A]
     \over
     \delta Y^\mu(s) \delta Y_\mu(s)
    }
    \label{ampl}
    \quad ,\\
    2m^2{\partial I[\By , \By _{0} ; A] \over \partial A}
    &=&
    -\norme{\Gamma}
    \oint _{\Gamma} {ds\over\sqrt{\ttt{\By '} ^{2}}}
    {
     \delta I[\By , \By _{0} ; A]
     \over \delta Y^\mu(s)
    }
    {
     \delta I[\By , \By _{0} ; A]
     \over \delta Y_\mu(s)
    }
    \quad .
    \label{phase}
    \eea
\end{props}
\begin{proof}
We substitute the ansatz (\ref{ansatz}) in the functional wave equation
(\ref{A.kerfunwavequ}). Fist we get
\bea
    & & \esci \esci \esci
    -
    \frac{\hbar ^{2}}{2 m ^{2}}
    \norme{\Gamma}
    \oint _{\Gamma}
        \frac{ds}{\sqrt{\ttt{\By ' \tst} ^{2}}}
        \frac{\delta}{\delta Y ^{\mu} \tst}
        \left(
            \frac{i}{\hbar}
            \frac{\delta I}{\delta Y _{\mu} \tst}
            e ^{\frac{i}{\hbar} I}
        \right)
        \mathcal{N}
        A ^{\alpha} =
    \nonumber \\
    & & \qquad \qquad \qquad
    =
    i
    \hbar
    \mathcal{N}
    \frac{A ^{\alpha}}{A}
    e ^{\frac{i}{\hbar} I}
    +
    i
    \hbar
    \mathcal{N}
    A ^{\alpha}
    e ^{\frac{i}{\hbar} I}
    \frac{i}{\hbar}
    \frac{\partial I}{\partial A}
    \quad .
\eea
Rearranging this result, in particular performing the remaining
functional derivative ans dividing by
$\mathcal{N} A ^{\alpha}$we obtain:
\bea
    & & \esci \esci \esci
    -
    \frac{\hbar ^{2}}{2 m ^{2}}
    \norme{\Gamma}
    \oint _{\Gamma}
        \frac{ds}{\sqrt{\ttt{\By ' \tst} ^{2}}}
        \left [
            \left(
                \frac{i}{\hbar}
                \frac{\delta I}{\delta Y _{\mu} \tst}
            \right) ^{2}
            +
            \frac{i}{\hbar}
            \frac{\delta ^{2} I}{\delta Y ^{\mu} \tst \delta Y _{\mu} \tst}
        \right ]
        e ^{\frac{i}{\hbar} I}
        =
    \nonumber \\
    & & \qquad \qquad \qquad
    =
    i
    \hbar
    \frac{1}{A}
    e ^{\frac{i}{\hbar} I}
    +
    i
    \hbar
    \mathcal{N}
    A ^{\alpha}
    e ^{\frac{i}{\hbar} I}
    \frac{i}{\hbar}
    \frac{\partial I}{\partial A}
    \quad .
\eea
The final result then comes out, after dividing by
$\exp \left\{ i I / \hbar \right\}$, by separating the real part
of the resulting equation (which gives (\ref{phase})) from the
imaginary one (which is (\ref{ampl})).
\end{proof}

Comparing equations (\ref{phase}) and (\ref{2.ourhamjac}), we see that
$ I = S _{\mathrm{Red.}} \left[ \By \tst , \By _{0} \tst ; A \right]$
and (\ref{phase}) is just the classical Hamilton--Jacobi equation.
Therefore, the main problem is to determine the form of
$S _{\mathrm{Red.}}$ in the {\sl String} case. We proceed by analogy with the
relativistic point particle case, where $S _{\mathrm{Red.}}$ is a
functional of the world--line length element. In this case we try
with a functional of the natural generalization of that concept,
i.e. the {\sl Holographic Coordinates} of definition (\ref{2.holcordef}).
As a preliminary result we compute some useful derivatives.
\begin{props}[Holographic Derivatives]\spbcorr{}.\\
    The following expressions for
    the first and second functional derivatives of the
    {\sl Holographic Coordinates},
    \bea
    {\delta Y ^{\mu\nu}[C]\over \delta Y^\alpha(s)}&=&
    \delta_\alpha{}^\mu Y^{\,\prime\,\nu}(s)-
    \delta_\alpha{}^\nu Y^{\,\prime\,\mu}(s)
    \label{f1}
    \\
    {\delta^2 Y ^{\mu\nu}[C]\over
    \delta Y^\alpha(s)\delta Y^\beta(u)}&=&\left(\delta_\alpha{}^\mu
    \delta_\beta{}^\nu-\delta_\alpha{}^\nu\delta_\beta{}^\mu\right)
    {d\over ds}\delta(s-u)
    \quad ,
    \label{f2}
    \eea
    hold.
\end{props}
\begin{proof}
We start from the coordinate expression of the {\sl Holographic Coordinates},
$$
    Y ^{\mu \nu} \qCq
    =
    \oint _{\Gamma \approx \Sf ^{1}} du
        Y ^{\mu} \ttt{u}
        \frac{d Y ^{\nu} \ttt{u}}{d u}
    \quad .
$$
Then the results are derived by simply applying the chain rule
for the functional derivative. In particular we have
\bea
    \frac{\delta Y ^{\mu \nu} \qCq}
         {\delta Y ^{\alpha} \tst}
    & = &
    \oint _{\Gamma} d u
        \left [
            \frac{\delta Y ^{\mu} \ttt{u}}{\delta Y ^{\alpha} \tst}
            \frac{d Y ^{\nu} \ttt{u}}{d u}
            +
            Y ^{\mu} \ttt{u}
            \frac{d}{d u}
            \left(
                \frac{\delta Y ^{\nu} \ttt{u}}{\delta Y ^{\alpha} \tst}
            \right)
        \right ]
    \nonumber \\
    & = &
    \oint _{\Gamma} d u
        \left [
            \delta ^{\mu} _{\alpha}
            \delta \ttt{u - s}
            \frac{d Y ^{\nu} \ttt{u}}{d u}
            +
            \delta ^{\nu} _{\alpha}
            Y ^{\mu} \ttt{u}
            \frac{d \delta \ttt{u - s}}{d u}
        \right ]
    \nonumber \\
    & = &
    \oint _{\Gamma} d u
        \left [
            \delta ^{\mu} _{\alpha}
            \delta \ttt{u - s}
            \frac{d Y ^{\nu} \ttt{u}}{d u}
            -
            \delta ^{\nu} _{\alpha}
            \frac{d Y ^{\mu} \ttt{u}}{d u}
            \delta \ttt{u - s}
        \right ]
    \nonumber \\
    & = &
    \delta ^{\mu} _{\alpha}
    Y ^{\prime \nu} \tst
    -
    \delta ^{\nu} _{\alpha}
    Y ^{\prime \mu} \tst
    \quad .
\eea
Then, taking one more functional derivative we obtain
$$
    \frac{\delta ^{2} Y ^{\mu \nu} \qCq}
         {\delta Y ^{\alpha} \tst \delta Y ^{\beta} \ttt{u}}
    =
    \left(
        \delta ^{\mu} _{\alpha}
        \delta ^{\nu} _{\beta}
        -
        \delta ^{\nu} _{\alpha}
        \delta ^{\mu} _{\beta}
    \right)
    \frac{d}{d s} \delta \ttt{s - u}
    \quad .
$$
\end{proof}
We are now ready to determine the complete form of the propagator, which
is done in the following proposition. As in the particle case
we will recognize it as the exponential of the {\it distance squared},
but this time in loop space; here the distance is given by the difference
in the area of the projected shadows onto the coordinate planes, so that
it is useful to use the following notation.
\begin{nots}[Holographic Distance]\spbcorr{}.\\
    We will indicate with $\Sigma ^{\mu \nu} \qtq{C - C _{0}}$
    the area distance between the two loops $C$ and $C _{0}$ in
    the $\mu-\nu$ plane, i.e.
    $$
        \Sigma ^{\mu \nu} \qtq{C - C _{0}}
        =
        Y ^{\mu \nu} \qCq
        -
        Y ^{\mu \nu} \qtq{C _{0}}
        \quad ,
    $$
    and call it the \underbar{Holographic Distance}
    of the considered loops.
\end{nots}
Moreover, we need a suitable definition of the Loop
Space Dirac Delta Function in the {\sl Holographic Coordinates} in order
to fix the normalization constant.
\begin{defs}[Loop Space Dirac Delta Function]\spbcorr{}.
    \label{3.loospadirdelfun}\\
    The \underbar{Loop Space Dirac Delta Function} is defined as
    $$
        \delta[C-C_0]
        \equiv \lim_{\epsilon \to 0}\left({1\over \pi
        \epsilon}\right)^{\ttt{D-1}/2}\exp\left(-{1\over 2\epsilon}
        \Sigma^{\mu\nu}[C-C_0] \Sigma_{\mu\nu}[C-C_0]\right)
        \quad .
    $$
\end{defs}
Then, we can prove
\begin{props}[String Propagation Kernel]\spbcorr{}.\\
    The Propagation Kernel for the {\sl String} is given by
        \beq
        K[\By \tst , \By _{0} \tst ; A]=\left({m^2\over 2i\pi\hbar
        A}\right)^{\ttt{D-1}/2}\exp\left(
        {im^2\over 4\hbar A} \Sigma^{\mu\nu}[C-C_0]
        \Sigma_{\mu\nu}[C-C_0]\right)
        \quad .
        \label{result}
        \eeq
\end{props}
\begin{proof}
Since we have already determined the relation between the phase
of the propagator $I \qtq{Y , Y _{0} ; A}$ and the classical action
$S _{\mathrm{Red.}} \qtq{Y , Y _{0} ; A}$
we introduce the trial solution
\bea
    S _{\mathrm{Red.}}[\By \tst) , \By _{0} \tst ; A]
    &=&
    {\beta\over 4A}
    \left(Y^{\mu\nu}[C]-Y^{\mu\nu}[C_0]\right)
    \left(Y_{\mu\nu}[C]-Y_{\mu\nu}[C_0]\right)
    \nonumber\\
    &\equiv&
    {\beta\over 4A}
    \Sigma^{\mu\nu}[C-C_0]
    \Sigma_{\mu\nu}[C-C_0]
    \quad ,
    \label{trial}
\eea
where, $\beta$ is a second parameter to be fixed by the equations
(\ref{ampl}) and (\ref{phase}). By taking into account (\ref{f1}) and
(\ref{f2}), we find
\beq
    {\delta S _{\mathrm{cl.}}\over\delta Y^\mu(s)}=
    {\beta\over A}\Sigma_{\mu\nu}[C-C_0]Y^{\prime\,\nu}(s)
    \quad .
\eeq
Note that the dependence on the parameter $s$ is only through the
factor $Y ^{\prime\,\nu} \tst$. Then,
\beq
    {\delta^2S _{\mathrm{cl.}} \over \delta Y_\mu(s)\delta Y^\mu(s)}
    =
    {\ttt{D-1}\beta\over A}
    \ttt{\By ' \tst} ^{2}
    \quad .
\eeq
Equations (\ref{phase})and (\ref{ampl}) now give
\beq
    \beta=-{2\alpha\over \ttt{D-1}}m^2\ ,\qquad \alpha=-\frac{\ttt{D-1}}{2}
    \quad .
\eeq
Finally, using definition \ref{3.loospadirdelfun}
the kernel normalization constant is fixed by the {\it Boundary}
condition (\ref{3.boucon}), and we finally obtain the promised
expression of the quantum kernel.
\end{proof}
The above equation, in turn, provides us to the following
representation of the Nambu--Goto propagator.
\begin{props}[Representation of Nambu--Goto String Dynamics]\spbcorr{}.\\
    The Quantum Dynamics of a {\sl String} can be determined thanks to the
    following representation of the Nambu--Goto {\sl String} Propagator
    in terms of the Kernel in the {\sl Holographic Coordinates}:
    \bea
        &&\esci\int_{Y^{\mu}_0 \tst}^{Y ^{\mu} \tst}[DX^\mu(\Bsigma)]
        \exp\left\{ -{im^2\over \hbar}\int_\Sigma
        d^2\Bsigma\sqrt{-{1 \over 2}\dot{X}^{\mu\nu}\dot{X}_{\mu\nu}}\right\}=
        \nonumber\\
        &&=-\int_0^\infty  \! \! \! \! \! dA\, e^{-im^2 A/2\hbar}
        \left({m^2\over 2i\pi\hbar A}\right)^{\ttt{D-1}/2} \! \! \! \! \!
        \exp\left({im^2\over 4\hbar A} \Sigma^{\mu\nu}[C-C_0]
        \Sigma_{\mu\nu}[C-C_0]\right)
        \quad .
        \label{prop}
    \eea
\end{props}
\begin{proof}
From equations (\ref{3.gretoker}), (\ref{ng}) and (\ref{result})
we have
\bea
    & & \esci \esci
    \left(
        \frac{m ^{2}}{2 i \pi \hbar A}
    \right) ^{\ttt{D-1}/2}
    \exp
    \left\{
        \frac{i m ^{2}}{4 \hbar A}
        \Sigma ^{\mu \nu} \qtq{C - C _{0}}
        \Sigma _{\mu \nu} \qtq{C - C _{0}}
    \right\}
    =
    \nonumber \\
    & & \qquad \qquad =
    \int _{0} ^{\infty} d E
        e ^{\frac{i}{\hbar} E A}
        \funinte{X ^{\mu} \ttt{\Bsigma}}
                {Y _{0} ^{\mu} \tst}
                {Y ^{\mu} \tst}
            \exp
            \left\{
                -
                \frac{i m ^{2}}{\hbar}
                \sqrt{
                      -
                      \frac{1}{2}
                      \dot{X} ^{\mu \nu}
                      \dot{X} _{\mu \nu}
                     }
            \right\}
    \quad .
    \nonumber
\eea
From this result we thus get
\bea
    & & \esci \esci
    \funinte{X ^{\mu} \ttt{\Bsigma}}
            {Y _{0} ^{\mu} \tst}
            {Y ^{\mu} \tst}
        \exp
        \left\{
            -
            \frac{i m ^{2}}{\hbar}
            \sqrt{
                  -
                  \frac{1}{2}
                  \dot{X} ^{\mu \nu}
                  \dot{X} _{\mu \nu}
                 }
        \right\}
    =
    \nonumber \\
    & & \qquad \qquad =
    -
    \int _{0} ^{\infty} d A
        e ^{- \frac{i m ^{2} A}{2 \hbar}}
        \left(
            \frac{m ^{2}}{2 i \pi \hbar A}
        \right)
        \exp
        \left\{
            \frac{i m ^{2}}{4 \hbar A}
            \Sigma ^{\mu \nu} \qtq{C - C _{0}}
            \Sigma _{\mu \nu} \qtq{C - C _{0}}
        \right\}
    \quad ,
    \nonumber
\eea
which is the desired result.
\end{proof}
Note that, since no approximation was used to obtain equation
(\ref{prop}), the above representation can also be interpreted as a
new {\it definition} of the
Nambu--Goto path--integral. This definition is based on the classical
Jacobi formulation of {\sl String} Dynamics rather than on the customary
discretization procedure.

\subsection{Integrating the Path--Integral}
\label{5.2}

As a consistency check on the above result, and in order to
clarify some further
properties of the path--integral, it may be useful to offer an
alternative derivation of equation (\ref{prop}) which is
based entirely on the usual gaussian integration technique.
As we have seen in the previous section, the Feynman amplitude
can be written as follows
\bea
&&\esci \esci K[\By \tst , \By _{0} \tst ; A]
=
\funinte{X^\mu(\Bsigma)}{Y^{\mu}_0 \tst}{Y^{\mu} \tst}
\qtq{\mathcal{D}P_{\mu\nu}(\Bsigma)}
\times
\nonumber\\
&&\qquad \qquad \times\exp\left\{
{i\over 2\hbar}\int_{\mathcal{W}}P_{\mu\nu}
\form{dX}^\mu\wedge \form{dX}^\nu
-{i\over 2\hbar}\epsilon_{ab}\int_{\Xi} \form{d\xi}^a\wedge
\form{d\xi}^b H(\Bp)\right\}
\quad .
\label{jak}
\eea
It is possible to evaluate the functional integral (\ref{jak}), without
discretization of the variables; we have just to carefully remember the
meaning of the path--integration in this case, and to distinguish
between {\it Bulk} and {\it Boundary} fields. Moreover, we are going to use the following
\begin{props}[Functional Bulk Integration]\spbcorr{}.\\
The functional integration over the {\it Bulk} coordinates $\Bx$
restricts the functional {\sl Area Momentum} integration to the
classical extremal trajectories of the {\sl String}, namely:
\bea
& & \esci \esci
\funinte{X^\mu(\Bsigma)}{Y^{\mu}_0 \tst}{Y^{\mu} \tst}
\exp
\left\{
{i\over 2\hbar}
\int_{\mathcal{W}}
P_{\mu\nu}\,
\form{dX}^\mu\wedge \form{dX}^\nu
\right\}
=
\nonumber \\
& & \qquad \qquad =
\delta\left[\form{d}\left(P_{\mu\nu}\form{dX}^\nu\right)\right]
\exp\left\{{i\over 2\hbar}
\int_{C}Q_{\mu\nu} Y^\mu \form{dY}^\nu\right\}
\quad .
\eea
\end{props}
\begin{proof}
The proof enlightens a nice feature of the path--integration procedure.
The initial and final values are simply the values of the fields on
the {\it Boundary}, which are held fixed
($Y ^{\mu} \tst$ and $Y ^{\mu} _{0} \tst$). What varies is just
the field on the {\it Bulk} $X ^{\mu} \ttt{\Bsigma}$. We thus firstly
explicitly separate the contribution of the {\it Boundary} from the first
exponent in equation (\ref{jak}) thanks to an integration by parts:
\bea
    \int _{\mathcal{W}}
        P _{\mu \nu}
        \form{d X} ^{\mu} \wedge \form{d X} ^{\nu}
    & = &
    \int _{\mathcal{W}}
        \form{d} \ttt{X ^{\mu} P _{\mu \nu} \form{d X} ^{\nu}}
    -
    \int _{\mathcal{W}}
        X ^{\mu} \form{d} \ttt{P _{\mu \nu} \form{d X} ^{\nu}}
    \nonumber \\
    & = &
    \oint _{C = \partial \mathcal{W}}
        Y ^{\mu} Q _{\mu \nu} \form{d Y} ^{\nu}
    -
    \int _{\mathcal{W}}
        X ^{\mu} \form{d} \ttt{P _{\mu \nu} \form{d X} ^{\nu}}
    \nonumber
\eea
We can then substitute this result into the first factor of (\ref{jak}),
which is the only one depending on $\Bx \ttt{\Bsigma}$:
\bea
&&\esci \esci
\funinte{X^\mu(\Bsigma)}{\By_0 \tst}{\By \tst}
\exp
\left\{
{i\over 2\hbar}
\int_{\mathcal{W}}
P_{\mu\nu}\,
\form{dX} ^{\mu} \wedge \form{dX} ^{\nu}
\right\}
=
\nonumber\\
&&\qquad \qquad=
\exp
\left\{
{i\over 2\hbar}
    \oint _{C}
        Y ^{\mu} Q _{\mu \nu} \form{d Y} ^{\nu}
\right\}
\funinte{X^\mu(\Bsigma)}{\By_0 \tst}{\By \tst}
\exp
\left\{
-
{i\over 2\hbar}
    \int _{\mathcal{W}}
        X ^{\mu} \form{d} \ttt{P _{\mu \nu} \form{d X} ^{\nu}}
\right\}
\nonumber\\
&&\qquad \qquad=
\funinte{X^\mu(\Bsigma)}{\By _0 \tst}{\By \tst}
\exp
\left\{
{i\over 2\hbar}
\left[
    \oint _{C}
        Y ^{\mu} Q _{\mu \nu} \form{d Y} ^{\nu}
    -
    \int _{\mathcal{W}}
        X ^{\mu} \form{d} \ttt{P _{\mu \nu} \form{d X} ^{\nu}}
\right]
\right\}
\nonumber \\
&&\qquad \qquad=\delta
\left[\form{d}\left(P_{\mu\nu}\form{dX}^\nu\right)\right]
\exp\left\{{i\over 2\hbar}
\oint_{C}Q_{\mu\nu} Y^\mu \form{dY}^\nu\right\}
\quad .
\nonumber
\eea
\end{proof}
The functional delta function has support on the classical,
extremal trajectories of the {\sl String}. Therefore, the momentum
integration is restricted
to the classical area--momenta and the residual integration
variables are the components
of the {\sl Boundary Area Momentum}
$Q _{\mu\nu}(s)$ along the {\sl World-Sheet} {\it Boundary}.
Thus we obtain the following result.
\begin{props}[String Propagator from Path--Integration]\spbcorr{}.\\
Performing the functional integrations in {\rm (\ref{jak})}
correctly reproduces the propagator of equation {\rm (\ref{result})}
\end{props}
\begin{proof}
As a matter of fact, {\it Boundary} conditions fix the initial and
final {\sl String} loops $C_0$ and $C$ but not the conjugate momenta.
In analogy to the point particle case, the classical equations of motion on 
the final {\sl World--Sheet} {\it Boundary},
\beq
\left .
\form{d} \left( P_{\mu\nu} \form{dX} ^\nu\right)
\right \rceil _{C}
\quad ,
\eeq
admits a constant area momentum solution
$Q _{\mu\nu} \tst = P _{\mu \nu} \ttt{\Bsigma \tst}$ with
$$
\left .
P_{\mu\nu} \ttt{\Bsigma}
\right \rceil _{\Bsigma = \Bsigma \tst}
T ^{\nu} \tst
=
Q_{\mu\nu} \tst Y^{\prime\,\nu} \tst
=
\mathrm{const.}
\quad .
$$
Hence, the functional integral over the {\sl Boundary Area Momentum}
reduces to a $\ttt{D-1}$-dimensional, generalized, Gaussian integral
\beq
\int [DP_{\mu\nu}(\Bsigma)]
\delta\left[\form{d}\left(P_{\mu\nu}\form{dx}^\nu
\right)\right]
(\dots)=\int [dP_{\mu\nu}](\dots)
\quad .
\eeq
Moreover, the Hamiltonian
is constant over such a classical {\sl World--Sheet} and  can be written
in terms of the {\sl Boundary Area Momentum} $Q _{\mu\nu} \tst$.
In such a way, the path--integral is reduced to the
Gaussian integral over the
$D-1$ components of $P _{\mu\nu}$ which generates normal deformation
of the loop.
\bea
K[\By \tst , Y _{0} \tst ; A]
&=&
{\cal N}\int [dQ_{\rho\tau}] \exp\left\{{i\over \hbar}\left[
Q_{\mu\nu}\int_{C} Y^\mu(s) \form{dY}^\nu -{A\over 4m^2}
Q_{\mu\nu} Q^{\mu\nu}\right]
\right\}
\nonumber\\
&=&{\cal N}\int [dQ_{\rho\tau}] \exp\left\{{i\over \hbar}\left[
{1\over 2} Q_{\mu\nu}\Sigma^{\mu\nu}[C-C_0]
 -{A\over 4m^2} Q_{\mu\nu} Q^{\mu\nu}\right]\right\} .
\label{fine}
\eea
The integral (\ref{fine}) correctly reproduces the expression
(\ref{prop}), i.e.
$$
    K \left[ Y ^{\mu} \tst , Y ^{\mu} _{0} \tst ; A \right]
    =
    \left(
        \frac{m ^{2}}{2 i \pi A}
    \right) ^{\ttt{D-1}/2}
    \exp
    \left\{
        \frac{i m ^{2}}{4 A}
        \left [
            Y ^{\mu \nu} \left [ C \right ]
            -
            Y ^{\mu \nu} \left [ C _{0} \right ]
        \right]
        \left[
            Y _{\mu \nu} \left [ C \right ]
            -
            Y _{\mu \nu} \left [ C _{0} \right ]
        \right]
    \right\}
\quad .
$$
\end{proof}

\section{Summary, Comments, Highlights and More}

Before driving our discussion of the Dynamics of extended objects
toward the more general case of $p$-branes (this is the topic of the next
chapter) we try to summarize what we have done so far, reviewing
briefly our procedure without the bounds that
computations and proofs put on physical intuition\footnote{There
are only two ``{\it{}rigorous}\,'' exception to this claim in this section,
namely proposition \ref{3.funlooschequpro}
and definition \ref{3.totareloomomopedef}.}.

The central point of all our formulation of {\sl String} Dynamics is that
we try to interpret the Quantum Dynamics of a Physical System as
the Dynamics of the {\it Boundary} of what we understand as its
Classical History, the {\it Bulk}.
The Differential Geometric approach to the subject is already in such
an interpretation, and will turn out again and again  in the following
chapters, when we will turn to analyze in more detail the properties of the
Quantum Dynamics. As a result of this geometrical interpretation,
starting from a Classical Hamilton--Jacobi Theory
and developing in this way the connection toward the Quantum Realm,
we see that the correct ``coordinates'' for the description of the
system appear in a natural way. In particular, as we can see from the
expression for the propagator (\ref{result})
the geometric doublet ($Y _{\mu\nu}, A$) represents the correct
set of dynamical variables in our formulation of {\sl String} Quantum Dynamics.
We stress again that this
choice makes it possible to develop a Hamilton--Jacobi Theory
of {\sl String} loops which represents a natural extension of the
familiar formulation of classical and quantum mechanics of
point--particles \cite{noi}, \cite{prop}.
With hindsight, the analogy becomes
transparent when one compares equation \ref{result} with the
amplitude for a relativistic particle of mass $m$ to
propagate from
$x _{0} ^{\mu}$ to $x ^{\mu}$ in a proper-time lapse $T$:
\beq
    K \left( x , x _{0} ; T \right)
    =
    \left(
        \frac{m}{2 \pi T}
    \right) ^{2}
    \exp
    \left [
        \frac{i m}{2 T}
        \left \vert x - x _{0} \right \vert ^{2}
    \right]
    \quad .
\eeq
This comparison suggests the following correspondence between
dynamical variables for particles and {\sl Strings}: a particle position in
$D$-dimensional
spacetime is labeled by $D$ real numbers $x^\mu$ which represent
the projection of the particle position vector along the
coordinate axes. In the {\sl String} case, the conventional\footnote{The
connection between the conventional approach to {\sl String} Theory and
the formulation presented here is deepened in chapter
\ref{9.connection}.} choice is to
consider the position vector for each constituent point, and
then follow their individual dynamical evolution in terms of the
coordinate time $X^0$, or the proper time $\tau$. As a matter of fact,
the canonical {\sl String} quantization is usually implemented in
the proper time gauge $X ^{0} = \tau$. However, this choice explicitly
breaks the reparametrization invariance of the Theory, whereas in
the Hamilton--Jacobi formulation of {\sl String} Dynamics, we have
insisted that reparametrization invariance be manifest at every stage.
The form (\ref{result}) of the {\sl String} propagator reflects that requirement.
Moreover, we called $Y ^{\mu \nu} \qCq$ the
{\sl Holographic Coordinates} of the {\sl String}: they play the
same role as the {\it position vector} in the point-particle
case. Indeed, the six components of $Y ^{\mu \nu} \qCq$ are {\it areas}
and represent the area of the projections
of the loop onto the coordinate planes in spacetime.
Likewise, the reparametrization invariant evolution parameter for the
{\sl String} turns out to be\footnote{neither the coordinate time,
nor the proper time of the constituent points, but} again an {\it area},
i.e. the proper area $A$ of the {\it whole} {\sl String} {\sl Parameter Space}.
Thus, as particle Dynamics is the Dynamics of {\it ``lengths''},
{\sl String} Dynamics turns out to be, in our formulation, a Dynamics
of {\it areas}.
As a matter of fact, the {\sl String}
{\sl World--Sheet} is the spacetime image, through the embedding
$ X ^{\mu} = X ^{\mu} \left( \sigma ^{0} , \sigma ^{1} \right)$,
of a two dimensional manifold $\Sigma$ of coordinates
$\left( \sigma ^{0} , \sigma^1 \right)$.
Thus, just as the proper time $\tau$ is a
measure of the timelike distance between the final and initial
position of a point particle, the proper area $A$ of {\sl Parameter Space}
is a measure of the timelike, or parametric distance between
$C$ and $C_0$, i.e., the final and the initial {\sl String} configurations, or
the {\it Boundary} of the {\sl World--Sheet} in {\sl Target Space}.
In our discussion we always kept fixed the initial loop $C _{0}$,
quantizing then the other free end loop $C$; in this perspective
$$
    \left (
        Y ^{\mu \nu} \qCq
        -
        Y ^{\mu \nu} \qtq{C _{0}}
    \right ) ^{2}
    =
    \left(
        \Sigma ^{\mu \nu} \qtq{C - C _{0}}
    \right) ^{2}
$$
represents the ``spatial distance squared'' between $C$ and
$C _{0}$, and $A$ represents the classical time lapse for the {\sl String} to
change its shape from $C _{0}$ to $C$.\\
The Quantum Formulation of {\sl String} Dynamics based
on the non canonical variables that we carried out through
the evaluation of the {\sl String} kernel can now be completed toward
the derivation of the Schr\"odinger loop equation.
Presently, we are interested in the quantum fluctuations of a
loop. By this, we mean a shape changing transition, and we would like
to assign a probability amplitude to any such
process\footnote{We underline, and after previous comments
this should be deeply motivated, that in order
to do this, we make use of ``areal, or loop, derivatives'', as
described so far and developed, for instance, by Migdal \cite{mig};
as we already pointed out they are
reviewed in appendix \ref{D.looderapp}.}.
Suppose now that the shape of the initial {\sl String} is approximated by
the loop
configuration $C _{0} : \, Y ^{\mu} = Y ^{\mu} _{0} \tst$.
The corresponding ``wave packet'' $\Psi \left [ Y _{0} ; 0 \right ]$
will be concentrated around $C _{0}$. As the areal time increases,
the initial {\sl String} evolves, sweeping a {\sl World--Sheet} which is
the image of a {\sl Parameter Space} of proper
area $A$. Once $C _{0} : \, Y ^{\mu} = Y ^{\mu} _ 0 \tst$
and $A$ are assigned, the final {\sl String}
$Y ^{\mu} = Y ^{\mu} \tst$ can attain any of the
different shapes compatible with the given initial condition and with
the extension of the {\sl World--Sheet Parameter Space}.
Each geometric configuration corresponds to a different ``point'' in
the Holographic Representation of the loop space provided by the
$Y ^{\mu \nu}$ coordinates
\cite{prop}. Then\footnote{Maintaining
the promise we made at the beginning of this section we give an
intuitive treatment in this section, avoiding mathematical problems
which are deferred until chapter \ref{10.nonstacha}.},
$\Psi \left [C ; A \right ]$
{\it will represent the probability amplitude to find a {\sl String} of
shape $C : \, Y ^{\mu} = Y ^{\mu} \left ( s \right)$
as the final {\it Boundary} of the {\sl World--Sheet} of
{\sl Parameter Space} area $A$, originating from $C _{0}$}.
From this vantage
point, the quantum {\sl String} evolution is a random {\it shape--shifting}
process which corresponds, mathematically, to the spreading
of the initial wave packet $\Psi \left [ C _{0} ; 0 \right ]$
throughout loop space. The wave functional $\Psi \left [ C ; A \right ]$
can be obtained by means of the
amplitude (\ref{result}), summing  over all the
initial {\sl String} configurations. This amounts to integrate over all the
allowed loop configurations $Y ^{\mu \nu} \qtq{C _{0}}$:
\bea
    \Psi \left[ C ; A \right]
    & = &
    \sum _{C _{0}}
        K \left [ \By \equiv C , \By _{0} \equiv C _{0} ; A \right]
        \Psi \left[ C _{0} ; 0 \right]
    \nonumber\\
    & = &
    \int _{C _{0}}
        \qtq{\mathcal{D} C _{0}}
        K \left [ \By \equiv C, \By _{0} \equiv C _{0} ; A \right]
        \Psi \left[ C _{0} ; 0 \right]
    \label{3.prowavint}
    \\
    & = &
    \left(
        \frac{m ^{2}}{2 i \pi A}
    \right) ^{3/2}
    \! \! \! \! \! \! \!
    \funint{Y ^{\mu \nu} \qtq{C _{0}}}
        \exp
        \left [
            \frac{i m ^{2}}{4 A}
            \left(
                Y ^{\mu \nu} \qtq{C}
                -
                Y ^{\mu \nu} \qtq{C _{0}}
            \right) ^{2}
        \right]
        \Psi \left[ C _{0} ; 0 \right]
    \quad .
\label{3.prowav}
\eea
Thanks to previous expressions we can see that the {\sl String Functional
Schr\"odinger Equation} for the propagator (\ref{A.kerfunwavequ}) is
equivalent to the following \underbar{Functional
Schr\"odinger Equation} for the wave functional $\Psi \qtq{C ; A}$:
\beq
    -
    \frac{1}{2 m ^{2}}
    \norme{\Gamma \approx \Sf ^{1}}
    \oint _{\Gamma \approx \Sf ^{1}} \frac{d s}{\sqrt{\ttt{\By '} ^{2}}}
        \frac{\delta ^{2} \Psi \left [ C ; A \right]}
             {
              \delta Y ^{\mu} \tst
              \delta Y _{\mu} \tst
             }
    =
    i
    \frac{\partial \Psi \left[ C ; A \right]}
         {\partial A}
    \quad .
\label{3.funlooschequcoo}
\eeq
We turn now to the only proposition of this section.
\begin{props}[String Loop Schr\"odinger Equation]\spbcorr{}.\\
    \label{3.funlooschequpro}
    The {\sl String} Wave Functional $\Psi \qtq{C ; A}$ can be obtained
    by solving the loop Schr\"odinger equation
    \beq
        -
        \frac{1}{4 m ^{2}}
        \norme{\Gamma \approx \Sf ^{1}}
        \oint _{\Gamma \approx \Sf ^{1}} d l \tst
            \frac{\delta ^{2} \Psi \left [ C ; A \right]}
                 {
                  \delta Y ^{\mu \nu} \tst
                  \delta Y _{\mu \nu} \tst
                 }
        =
        i
        \frac{\partial \Psi \left[ C ; A \right]}
             {\partial A}
        \quad .
    \label{3.funlooschequ}
    \eeq
\end{props}
\begin{proof}
To get this equation we take equation \ref{3.kerfunwavsigequ}
multiply both sides by the initial loop wave functional
$\Psi \qtq{C _{0}}$ and functionally integrate over
$\qtq{\mathcal{D} C _{0}}$:
\bea
    & & \esci \esci
    \funint{C _{0}}
        \frac{\hbar ^{2}}{2 m ^{2}}
        \norme{\Gamma}
        \oint _{\Gamma} ds
            \sqrt{\ttt{\By ' \tst} ^{2}}
            \frac{\delta ^{2} K \qtq{Y ^{\mu} , Y _{0} ^{\mu} ; A}}
                 {\delta Y ^{\mu \nu} \tst \delta Y _{\mu \nu} \tst}
            \Psi \qtq{C _{0}}
    =
    \nonumber \\
    & & \qquad \qquad \qquad =
    \funint{C _{0}}
        i \hbar
        \frac{\partial K \qtq{Y ^{\mu} , Y _{0} ^{\mu} ; A}}
             {\partial A}
        \Psi \qtq{C _{0}}
    \quad .
\eea
This equation can be written also as
\bea
    & & \esci \esci \esci
        \frac{\hbar ^{2}}{2 m ^{2}}
        \norme{\Gamma}
        \oint _{\Gamma} ds
            \sqrt{\ttt{\By ' \tst} ^{2}}
            \frac{\delta ^{2}}
                 {\delta Y ^{\mu \nu} \tst \delta Y _{\mu \nu} \tst}
            \left [
                \funint{C _{0}}
                K \qtq{Y ^{\mu} , Y _{0} ^{\mu} ; A}
                \Psi \qtq{C _{0}}
            \right ]
    =
    \nonumber \\
    & & \qquad \qquad \qquad
    \funint{C _{0}}
        i \hbar
        \frac{\partial}{\partial A} \Psi \qtq{C _{0}}
        \left [
            \funint{C _{0}}
            K \qtq{Y ^{\mu} , Y _{0} ^{\mu} ; A}
            \Psi \qtq{C _{0}}
        \right ]
    \quad ,
\eea
where we recognize on both sides that $\Psi \qtq{C ; A}$ appears thanks
to equation (\ref{3.prowavint}); thus we get the desired result.
\end{proof}
We can easily see that Equation (\ref{3.funlooschequ}) is the quantum
transcription, through the Correspondence Principle,
\bea
    H
    & \rightarrow &
    -
    i
    \frac{\partial}{\partial A}
    \label{aham}
    \\
    Q _{\mu \nu} \tst
    & \rightarrow &
    i
    \frac{\delta}{\delta Y ^{\mu\nu} \tst}
    \label{def}
\eea
of the classical relation between the Area Hamiltonian $H$
and the loop momentum density $Q _{\mu \nu} \tst$
\cite{prop}:
\bea
    H \left [ C \right ]
    & = &
    \norme{\Gamma \approx \Sf ^{1}}
    \oint _{\Gamma \approx \Sf ^{1}} d s
        \sqrt{\ttt{\By '} ^{2}}
        \Ham _{\mathrm{Schild.}}
    \nonumber \\
    & = &
    \frac{1}{4 m ^{2}}
    \norme{\Gamma \approx \Sf ^{1}}
    \oint _{\Gamma \approx \Sf ^{1}} d s
        \sqrt{\ttt{\By '} ^{2}}
 	Q _{\mu \nu} \tst
        Q ^{\mu \nu} \tst
    \quad .
\label{ep}
\eea
Once again, we note the analogy
between equation (\ref{ep}) and the
familiar energy momentum relation for a point particle, $H=p ^2 / \ttt{2m}$.
We shall comment on the ``non-relativistic'' form of equation (\ref{ep}) at
the end of this section. Presently, we limit ourselves to
note that the difference between the point--particle case and the
{\sl String} case, stems from the spatial extension of the loop, and is
reflected in equation (\ref{ep}) by the averaging integral of the
momentum squared along the loop itself. Equation (\ref{ep})
represents the total loop energy instead of the energy of a
single constituent {\sl String} bit. Just as the particle linear momentum gives the
direction along which a particle moves and the rate of position change, so the
loop momentum describes the deformation in the loop shape
and the rate of shape change. The corresponding Hamiltonian
describes the energy variation as the loop area varies,
irrespective of the actual point along the loop where the deformation takes
place.\\
Accordingly, the Hamiltonian (\ref{aham}) represents the
{\it generating operator} of the loop area variations, and
the momentum density (\ref{def})  represents the generator of the
deformations in the loop shape at the point $Y ^\mu(s)$.\\
Note that we can define 
(in the same way as we did in equation (\ref{ep}),
passing from the Hamiltonian density $\Ham$ to the integrated quantity $H$)
an integrated {\it Area Loop Momentum}
\beq
    Q ^{\mu \nu} \qtq{C}
    =
    i
    \norme{\Gamma \approx \Sf ^{1}}
    \oint _{\Gamma \approx \Sf ^{1}} d s
    \sqrt{\ttt{\By '} ^{2}}
        Q ^{\mu \nu} \tst
\label{3.totloomom}
\eeq
and through the correspondence (\ref{def}) we obtain the
\begin{defs}[Total Area Loop Momentum Operator]\spbcorr{}.
    \label{3.totareloomomopedef}\\
    The \underbar{Total Area Loop Momentum Operator} obtained according
    to equations {\rm (\ref{3.totloomom})} and {\rm (\ref{def})} is
    \beq
        \euf{Q} ^{\mu \nu} \qtq{C}
        =
        i
        \norme{\Gamma \approx \Sf ^{1}}
        \oint _{\Gamma \approx \Sf ^{1}} d s
            \sqrt{\ttt{\By '} ^{2}}
            \frac{\delta}
                 {\delta Y ^{\mu \nu} \tst}
    \label{3.totloomomope}
    \quad .
    \eeq
Note that we use the name of the image of the loop in {\sl Target Space},
i.e. $C$, to label the {\sl Total Area Loop Momentum Operator} as well as
the {\sl Total Area Loop Momentum}.
\end{defs}
From the above discussion, we are led to deduce that:
\begin{enumerate}
    \item deformations may occur randomly at any point on the loop;
    \item the antisymmetry in the indices $\mu$, $\nu$ guarantees
    that $ Q_{\mu \nu} \tst Y ^{\prime \mu} \tst$
    generates only physical deformations  , i.e. deformations which
    orthogonal to the loop itself
    $$
        Q _{\mu \nu} \tst
        Y ^{\prime \mu} \tst
        Y ^{\prime \nu} \tst
        \equiv
        0
        \quad ;
    $$
    \item a shape changes cost energy because of the {\sl String}
    tension and the addition of a small loop, or ``petal'',
    increases the total length of the {\sl String};
    \item the energy balance condition is provided by equation
    (\ref{ep}) at the classical level and by equation
    (\ref{3.funlooschequ})
    at the quantum level; in both cases the global energy variation
    per unit proper area is obtained by a loop average of the
    double deformation at single point.
\end{enumerate}
The above deductions represent the distinctive features of the
{\sl String} quantum shape--shifting phenomenon.  Moreover it should be
clear now how the described
quantization program is essentially a sort of
quantum mechanics formulated in a space of {\sl String} loops,
i.e., a space in which each point  represents a possible
geometrical configuration of a closed {\sl String}.
We also stress again that all the procedure takes fully advantage from
the enlargement of the canonical phase space
that we obtain promoting the original {\sl World--Sheet} coordinates
$\Bxi$ to the role of {\it dynamical variables$\,$}\footnote{They now
represent fields $\Bxi \left( \Bsigma \right)$
defined over the {\sl String} manifold $\Bsigma$.} and
and introducing $\pi _{AB}$,
the momentum conjugate to $\xi ^{A}$
into the Hamiltonian form of the action\footnote{Please, refer
to table \ref{uno} for a list of all the relevant
dynamical quantities in loop space.}.
\begin{table}
\def\tstrut{\vrule height 5.2ex depth 3.2ex width 0pt}
\begintable
    $
    H \qtq{C}
    =
    \ttt{4 m ^{2} l _{C}} ^{-1}
    \displaystyle
    \oint _{C} d l \tst
        Q _{\mu \nu} \tst
        Q ^{\mu \nu} \tst
    $
    |
    (Schild) Loop Hamiltonian
    \cr
    $
    \Ham _{\mathrm{Schild}} \tst
    =
    \ttt{4 m ^{2} l _{C}} ^{-1}
    Q _{\mu \nu} \tst
    Q ^{\mu \nu} \tst
    $
    |
    (Schild) String Hamiltonian
    \cr
    $
    d l \tst
    \equiv
    \sqrt{\ttt{\By ' \tst} ^{2}}
    Y ^{\prime \, \mu}
    \: , \;
    Y ^{\prime}
    \equiv \frac{d Y ^{\mu} \tst}{d s}
    $
    |
    Loop Invariant Measure
    \cr
    $
    l _{C}
    \dfn
    \displaystyle
    \oint _{C} d l \tst
    $
    |
    Loop Proper Length
    \cr
    $
    Q _{\mu \nu} \tst
    =
    m ^{2}
    \epsilon ^{mn}
    \left .
        \partial _{m} X _{\mu} \ttt{\Bsigma}
        \partial _{n} X _{\nu} \ttt{\Bsigma}
    \right \rceil _{\Bsigma = \Bsigma \tst}
    $
    |
    Area Momentum Density
    \cr
    $
    Q _{\mu \nu} \qtq{C}
    \equiv
    l _{C} ^{-1}
    \displaystyle
    \oint _{C} d l \tst
        Q _{\mu \nu}\tst
    $
    |
    Loop Momentum
\endtable
\def\tstrut{\vrule height 3.1ex depth 1.2ex width 0pt}
\caption{Loop Space functionals and {\it Boundary} fields.}
\label{uno}
\end{table}
This enlargement endows the Schild action
with the full reparametrization invariance under the transformation
$\sigma ^{a} \longrightarrow \tilde{\sigma} ^{a} \left( \Bsigma \right)$,
while {\it preserving the polynomial structure} in the dynamical
variables, which is a necessary condition to solve the
path--integral. In this way we were able to derive without approximations
expressions (\ref{3.gretoker}) and (\ref{result}), which
show the explicit relation between the
{\it fixed area string propagator} $K \left [ \By , \By _{0} ; A \right ]$,
and the {\it fixed ``energy'' string propagator$\,$}\footnote{Or Green
Function}
$G \left [ C , C _{0} ; E \right ]$ without recourse to any {\it ad hoc}
averaging prescription in order to eliminate the $A$ parameter
dependence.
Moreover result (\ref{ng}) allows us to establish the following facts:
\begin{enumerate}
    \item our approach, which directly stems from Eguchi's proposal,
    corresponds to quantizing a {\sl String} by
    keeping fixed the {\it area} of the {\sl String Parameter Space}
    in the path--integral, and then taking the average over the
    {\sl String} tension values;
    \item the Nambu--Goto approach, on the other hand, corresponds
    to  quantizing a {\sl String} by keeping
    fixed the {\sl String} {\it tension} and
    then taking
    the average over the {\sl Parameter Space Areas};
    \item the two quantization schemes are equivalent in the
    saddle point approximation\footnote{This is the content of
    proposition \ref{3.schnamgotequ}.}.
\end{enumerate}
In particular we obtain through equations (\ref{3.gretoker}), (\ref{prop})
a {\it non--perturbative} definition of the Nambu--Goto
propagator (\ref{ng}).

\begin{table}
\def\tstrut{\vrule height 2.6ex depth 1.0ex width 0pt}
\begintable
{\bf Physical Quantity}
\|
\multispan{2}{\hfil Particle \hfil $\longrightarrow$ \hfil String \hfil}
\crthick
\underbar{Object}
\|
Massive Point--Particle
|
\ctr{Non-Vanishing}
\nr
\|
|
\ctr{String Tension}
\cr
\underbar{Mathematical Model}
\|
\ctr{Point $P$}
|
\ctr{SpaceLike Loop $C$}
\nr
\|
\ctr{in $\R^{D}$}
|
\ctr{in $\R^{D+1}$}
\cr
\underbar{Topological Meaning}
\|
\ctr{Boundary (=Endpoint)}
|
\ctr{Boundary}
\nr
\|
\ctr{of a Line}
|
\ctr{of an Open Surface}
\cr
\underbar{Coordinates}
\|
$\left\{ x ^{1} , x ^{2} , x ^{3}\right\}$
|
\ctr{\tstrut Area Element}
\nr
\|
|
\ctr{$Y ^{\mu \nu} \qtq{C} = \oint _{C} Y ^{\mu} d Y ^{\nu}$}
\cr
\underbar{Trajectory}
\|
\ctr{1-parameter family}
|
\ctr{1-parameter family}
\nr
\|
\ctr{of points $\{ \bs{x} (t)\}$}
|
\ctr{of loops $\left\{ Y ^{\mu} \ttt{ s ; A} \right\}$}
\cr
\underbar{Evolution Parameter}
\|
\ctr{``Time'' $t$}
|
\ctr{Area $A$ of the String}
\nr
\|
|
\ctr{Parameter Space}
\cr
\underbar{Translations Generators}
\|
\ctr{Spatial Shifts:}
|
\ctr{Shape Deformations:}
\nr
\|
\ctr{$\frac{\partial}{\partial x ^{i}}$}
|
\ctr{$\frac{\delta}{\delta Y ^{\mu \nu} \tst}$}
\cr
\underbar{Evolution Generator}
\|
\ctr{Time Shifts:}
|
\ctr{Proper Area Variations:}
\nr
\|
\ctr{$\frac{\partial}{\partial t}$}
|
\ctr{$\frac{\partial}{\partial A}$}
\cr
\underbar{Topological Dimension}
\|
\ctr{Particle Trajectory}
|
\ctr{String Trajectory}
\nr
\|
\ctr{$D=1$}
|
\ctr{$D=2$}
\cr
\underbar{Distance}
\|
\ctr{$\left( \vec{\bs{x}} - \vec{\bs{x}} _{0} \right) ^{2}$}
|
\ctr{$\left( Y ^{\mu \nu} \qtq{C} - Y ^{\mu \nu} \qtq{C _{0}} \right)^ {2}$}
\cr
\underbar{Linear Momentum}
\|
\ctr{Rate of Change of}
|
\ctr{Rate of Change of}
\nr
\|
\ctr{Spatial Position}
|
\ctr{String Shape}
\cr
\underbar{Hamiltonian}
\|
\ctr{Time Conjugate}
|
\ctr{Area Conjugate}
\nr
\|
\ctr{Canonical Variable}
|
\ctr{Canonical Variable}
\endtable
\def\tstrut{\vrule height 3.1ex depth 1.2ex width 0pt}
\caption{The Particle/String ``Dictionary''.}
\label{dictionary}
\end{table}

%% file: chap04.tex
\pageheader{}{Generalization: $p$-branes.}{}
\chapter{Generalization: $p$-branes}
\label{4.pbrcha}

\begin{start}
$\euf{O}$k,\\
``it's good to see you''\\
at work again.\\
\end{start}

\section{Reparametrized Schild Action}

The reparametrized formulation we presented in section
(\ref{2.repschfor}) can be formulated also for higher dimensional
objects without stumble into more troubles.
Hence, we start extending some definitions to the case of
a $p$-dimensional, relativistic object in $D$-dimensional space-time:
\begin{defs}[$p$-brane Parameter Space]\spbcorr{}.\\
   The \underbar{Parameter Space} of a $p$-brane is a compact
   connected $(p+1)$-dimensional domain
   $\Xip \subset \Mp$, coordinatized by the
   $(p+1)$-ple of variables
   $\Bxi = \ttt{\xi ^{0} , \dots , \xi ^{p}}$.
\end{defs}

\begin{defs}[$p$-brane Boundary Space]\spbcorr{}.\\
    The \underbar{Boundary Space} of a $p$-brane is the Boundary
    $\BB \dfn \partial \Xip$ of the {\sl Parameter Space}: it is
    parametrized by the $p$-ple
    $\Bs = \ttt{s ^{1} , \dots , s ^{p}}$; we remember that,
    since $\BB$ is already a {\it Boundary}, it has no {\it Boundary}
    $$
        \partial \BB = \emptyset
    \quad .
    $$
\end{defs}

In first instance, we  suppose that our model is defined by the following
$(p+1)$-form
$$
    \form{\omega} ^{(p+1)}
    =
    \Lag \ttt{X ^{\mu} , X ^{\mu} _{,i} ; \xi ^{i}}
    \form{d \xi} ^{0} \wedge \dots \wedge \form{d \xi} ^{p}
    \quad .
$$
We will call $X ^{\mu} \dfn X ^{\mu} \ttt{\xi ^{i}} = X ^{\mu} \ttt{\Bxi}$
the fields on the domain $\Xip$: these are the
{\it World--HyperTube Embedding Functions}.
Then, we will use the following definition
\begin{defs}[World--HyperTube and $p$-brane]\spbcorr{}.\\
    The image of the $p$-brane {\sl Parameter Space} in the
    {\sl Target Space}\footnote{This is the space,
    $\mathbb{M} ^{D}$, where the $p$-brane embedding functions
    take their values.} $\mathbb{T}$ is the \underbar{World--HyperTube},
    $\Wsp$, of the $p$-brane:
    $$
        \Wsp = X ^{\mu} \ttt{\Xip}
        \quad .
    $$
    The \underbar{$p$-brane}, $\Dp$, is the only
    {\it Boundary} of the {\sl World--HyperTube} if this last one
    is simply connected
    with a connected {\it Boundary}. If the {\it Boundary}
    is composed by more than
    one connected component, then we are
    in a multiple $p$-{\sl{}brane} configuration.
    In any case the
    $p$-{\sl{}brane} is a manifold without {\it Boundary}. We will
    call $p$-{\sl{}brane} also the {\sl Boundary Space} $\BB$, by
    extension, in view of the following equalities:
    $$
        X ^{\mu} \ttt{\BB}
        =
        \Dp
        \dfn
        \partial \Wsp
        =
        \partial
        X ^{\mu} \ttt{\Xip}
        \quad .
    $$
\end{defs}
On the $p$-{\sl{}brane}, i.e. on the {\it Boundary} $\BB$, the {\sl World--HyperTube
Embedding Functions} (they are our fields) reduce to
$$
    Y ^{\mu} \ttt{\Bs}
    \dfn
    X ^{\mu} \ttt{\xi ^{i} \ttt{\Bs}}
    \quad ,
$$
which are the \underbar{\it $p$-brane Embedding Functions};
we will always call with different letters fields on the {\it Bulk}
and their values on the {\it Boundary}, as we
already did in the {\sl String} case, to avoid confusion.

It is possible to generalize also the definitions
\ref{2.holcordef} and \ref{2.strareveldef} of
{\sl Holographic Coordinates} and {\sl Area Velocity}, but first we
have to generalize the {\sl Linear Velocity Vector} of a {\sl String},
$C$. For a $p$-{\sl{}brane} we will use the totally antisymmetric
$p$-dimensional volume element at the {\it Boundary} of the $(p+1)$-dimensional
{\sl World--HyperTube}.
\begin{defs}[$p$-brane Tangent Element]\spbcorr{}.
\label{4.pbrtaneledef}\\
    The \underbar{Tangent Element} to the $p$-{\sl{}brane}
    {\sl World--HyperTube} is
    the totally antisymmetric tensor
    \beq
        Y ^{\prime \multind{\mu}{1}{p}}
        =
        \epsilon ^{\multind{A}{1}{p}}
        \frac{\partial Y ^{\mu _{1}}}{\partial \xi ^{A _{1}} }
        \dots
        \frac{\partial Y ^{\mu _{p}}}{\partial \xi ^{A _{p}} }
        \quad .
    \eeq
\end{defs}
Now we can pass directly to the Holographic Coordinates.
\begin{defs}[$p$-brane Holographic Coordinates]\spbcorr{}.
    \label{4.pbrholcoo}\\
    Let $\BB$ be a $p$-{\sl{}brane}, and $Y ^{\mu} \ttt{\Bs}$ one of its
    possible {\sl Parametrizations}.\\
    The \underbar{Holographic Coordinates}
    of the $p$-{\sl{}brane} are
    \bea
        Y ^{\multind{\mu}{0}{p}}
        & \dfn &
        \oint _{\BB}
            d ^{p} \Bs \,
            Y ^{\mu _{0}} \ttt{\Bs}
            Y ^{\prime \multind{\mu}{1}{p}} \ttt{\Bs}
        \nonumber \\
        & = &
        \oint _{\Dp = \partial \Wsp}
            \! \! \! \! \! \! \! \! \! \!
            Y ^{\mu _{0}}
            \multint{Y}{\mu}{1}{p}
        \nonumber \\
        & = &
        \int _{\Wsp}
            \multint{Y}{\mu}{0}{p}
    \quad ,
    \eea
    where $Y ^{\prime \multind{\mu}{1}{p}}$ is the
    $p$-dimensional {\sl Tangent Element}
    to the $p$-{\sl{}brane} of definition (\rm \ref{4.pbrtaneledef}).
\end{defs}
We note incidentally that in this case, even if we are using the
same notation as in the {\sl String} case, the ``$\:'\:$'' symbol has a
more complex meaning than the one assumed in notation \ref{1.firdernot}.
\begin{defs}[Local World--HyperTube Volume Velocity]\spbcorr{}.\\
    Let $X ^{\mu} \ttt{\xi ^{j}}, j = 0 , \dots , p$ be a
    {\sl Parametrization}
    of the {\sl World--HyperTube} of a $p$-{\sl{}brane}.
    The \underbar{Local Volume Velocity} of the {\sl World--HyperTube} is
    $$
        \dot{X} ^{\mu _{0} \dots \mu _{p}} \ttt{\Bxi}
        \dfn
        \epsilon ^{\multind{a}{0}{p}}
        \frac{X ^{\mu _{0}} \ttt{\Bxi}}{\partial \xi ^{a _{0}}}
        \cdot \dots \cdot
        \frac{X ^{\mu _{p}} \ttt{\Bxi}}{\partial \xi ^{a _{p}}}
        \quad .
    $$
\end{defs}
\begin{defs}[$p$-brane Volume Velocity]\spbcorr{}.\\
    Let $\dot{X} ^{\mu _{0} \dots \mu _{p}} \ttt{\Bxi}$ be the
    {\sl Local Volume Velocity} of the {\sl World--HyperTube} $\Wsp$
    associated with a $p$-{\sl{}brane} defined on
    $$
        \BB = \partial \Xip
        \quad ;
    $$
    the \underbar{$p$-brane Volume Velocity} is the {\sl Local
    Volume Velocity} of the {\sl World--HyperTube}
    $\Wsp$ computed on the {\it Boundary}:
    $$
        \dot{X} ^{\mu _{0} \dots \mu _{p}} \ttt{\Bs}
        \dfn
        \left .
            \epsilon ^{\multind{a}{0}{p}}
            \frac{X ^{\mu _{0}} \ttt{\Bxi}}{\partial \xi ^{a _{0}}}
            \cdot \dots \cdot
            \frac{X ^{\mu _{p}} \ttt{\Bxi}}{\partial \xi ^{a _{p}}}
        \right \rceil _{\Bxi = \Bxi \ttt{\Bs}}
    \quad .
    $$
\end{defs}
To have a clearer distinction between {\it Bulk} and {\it Boundary}
quantities we are going to adopt the following convention.
\begin{nots}[$p$-brane Volume Velocity]\spbcorr{}.\\
    We will denote the $p$-{\sl{}brane} {\sl Volume Velocity} with the
    following symbol:
    $$
        \dot{Y} ^{\multind{\mu}{0}{p}} \ttt{\Bs}
        \dfn
        \left .
            \dot{X} ^{\multind{\mu}{0}{p}} \ttt{\Bsigma}
        \right \rceil _{\Bsigma = \Bsigma \ttt{\Bs}}
    $$
\end{nots}
Now, we would like to set up a reparametrization invariant formalism for
a $p$-{\sl{}brane}, but always with a {\it Schild--type} action.
This can be done
as in the {\sl String} case:
the {\it Reparametrized Schild Lagrangian Density}
for a $p$-{\sl{}brane} is the natural generalization of
the Lagrangian (\ref{2.repschlag}). Of course all the procedures
we developed in chapter \ref{2.hamjac}
to motivate this choice are still valid
for higher dimensional objects. Again, we promote the
$\Bxi$ fields to the role of dynamical variables. They explicitly take
into account possible variations of the {\it Boundary} of the {\sl Parameter
Space}, the {\sl Boundary Space} $\BB$. From now on, all the fields,
the new $\Bxi$ as well as the $\Bx$ ones, are defined
on a $(p+1)$-dimensional
manifold that we will call $\Sigmap$, coordinatized by $(p+1)$-variables,
that in all possible cases we will call
$\Bsigma = \ttt{\sigma ^{0} , \dots , \sigma ^{p}}$. All the quantities
defined so far on $\Xip$ are to be interpreted as defined
on $\Sigmap$, since this is now the new name for the {\sl Parameter Space}.
Then, we can give the following
\begin{defs}[$p$-brane Reparametrized Schild Lagrangian Density]\spbcorr{}.\\
The \underbar{Reparametrized Schild Lagrangian Density} for a $p$-{\sl{}brane} is
\beq
    \Lag ^{(p) \mathrm{rep.}} _{\mathrm{Schild}}
    =
    \Lag ^{(p)}
    =
    \frac{\rp}{2 \fact{p+1}}
    \frac{
          \dot{X} ^{\multind{\mu}{0}{p}} \ttt{\Bsigma}
          \dot{X} _{\multind{\mu}{0}{p}} \ttt{\Bsigma}
         }
         {
          \epsilon ^{\multind{A}{0}{p}}
          \dot{\xi} _{\multind{A}{0}{p}} \ttt{\Bsigma}
         }
    \quad .
    \label{4.pbrrepschlag}
\eeq
\end{defs}
Again, we adopt the convention to use {\it uppercase latin indices}
for the new $\xi ^{A} \ttt{\sigma ^{i}}$ fields, and
{\it lowercase latin indices} for the {\sl Parameter Space} variables
$\sigma ^{a}$. Moreover, the quantity $\dot{\Bxi}$ is the natural
generalization of the same quantity already defined for the {\sl String}
(and also for the $\Bx \ttt{\sigma ^{a}}$ fields):
$$
    \dot{\xi} ^{\multind{A}{0}{p}} \ttt{\Bs}
    \dfn
        \epsilon ^{\multind{a}{0}{p}}
        \frac{\partial \xi ^{A _{0}} \ttt{\Bsigma}}{\partial \sigma _{a _{0}}}
        \cdot \dots \cdot
        \frac{\partial \xi ^{A _{p}} \ttt{\Bsigma}}{\partial \sigma ^{a _{p}}}
    \quad .
$$
Now, we can proceed to the generalization of
the Schild action in Hamiltonian form. As we already did
in section (\ref{2.basactsec}) on page \pageref{2.aremomdef}
for the {\sl String}, we firstly give the definition of momenta:
\begin{defs}[$p$-brane Bulk Volume Momentum]\spbcorr{}.
\label{4.pbrwshvolmom}\\
    The \underbar{$p$-brane Bulk Volume Momentum} or
    \underbar{$p$-brane World--HyperTube Volume Momentum} is the momentum
    canonically conjugated to the
    {\sl Local World--HyperTube Volume Velocity}:
    $$
        P _{\multind{\mu}{0}{p}}
        =
        \frac{\partial \Lag ^{p}}
             {\partial \dot{X} ^{\multind{\mu}{0}{p}} \ttt{\Bsigma}}
    \quad .
    $$
\end{defs}
Again, we can extend, or better {\it restrict} (!!!),
this definition from the {\it Bulk} to the {\it Boundary}.
\begin{defs}[$p$-brane Boundary Volume Momentum]\spbcorr{}.
\label{2.pbrbouvolmom}\\
    The \underbar{$p$-brane Boundary Volume Momentum} or
    \underbar{$p$-brane Volume Momentum} is the
    $p$-{\sl{}brane} {\sl Bulk Volume Momentum}
    computed on the {\it Boundary}, i.e.
    $$
        Q _{\multind{\mu}{0}{p}} \ttt{\Bs}
        \dfn
        \left .
            \frac{\partial \Lag ^{p}}
                 {\partial \dot{X} ^{\multind{\mu}{0}{p}} \ttt{\Bsigma}}
        \right \rceil _{\Bsigma = \Bsigma \ttt{\Bs}}
        =
        \left .
            P  _{\multind{\mu}{0}{p}} \ttt{\Bsigma}
        \right \rceil _{\Bsigma = \Bsigma \ttt{\Bs}}
    \quad ,
    $$
    where, $\Bsigma = \Bsigma \ttt{\Bs}$
    is a parametrization of the {\it Boundary}
    $\BB = \partial \Sigmap$ of the domain $\Sigmap$.
\end{defs}
Then, we can express the reparametrized $p$-{\sl{}brane} action as
\begin{defs}[Restricted Reparametrized $p$-brane Action]\spbcorr{}.\\
The \underbar{Restricted Reparametrized Action} for a $p$-{\sl{}brane} is
\bea
    & & \esci \esci
    S \left [ \Bx ( \Bsigma ) , \Bp ( \Bsigma ) , \Bxi ( \Bsigma ) \right ]
    =
    \nonumber \\
    & & =
    \frac{1}{\fact{p+1}}
    \int _{\Wsp}
        P _{\multind{\mu}{0}{p}} \multint{X}{\mu}{0}{p}
    +
    \nonumber \\
    & & \qquad
    -
    \frac{1}{\fact{p+1}}
    \epsilon _{\multind{A}{0}{p}}
    \int _{\Xi}
        \multint{\xi}{A}{0}{p} \Ham \ttt{\Bp}
    \quad ,
\label{4.pbrrepact}
\eea
i.e. the reparametrization of the Hamiltonian form of the Schild Action.
\end{defs}
The Hamiltonian $\Ham ( \Bp )$ is given by\footnote{We set
$\rp = m ^{p+1}$: then in the particle case, $p=0$, $\rho _{1} = m$,
the prefactor becomes correctly $1/ \left( 2m \right )$; for the {\sl String} $p=1$,
$\rho _{1} = m ^{2}$ give $1/ \left( 4 m ^{2} \right)$, and so on.}
\beq
    \Ham \ttt{\Bp}
    =
    \frac{1}{2 \rp \fact{p+1}}
    P _{\multind{\beta}{0}{p}}
    P ^{\multind{\beta}{0}{p}}
\label{pDhamiltonian}
\eeq
and the complete derivation of this result is given in appendix
\ref{A.HamDer}. The Schild Hamiltonian, in the reparametrized formulation
of the model, is proportional to the momenta conjugated to the $\Bxi$ fields;
we stress again, that this is due to the fact that in the reparametrized
formulation the modifications of the {\it Boundary} are encoded in the
variations of the new $\Bxi$ fields. For the sake of  completeness, we write
the expressions for the momenta, which are
\bea
    P _{\multind{\mu}{0}{p}}
    & = &
    \frac{\rp \dot{X} _{\multind{\mu}{0}{p}}}
         {
          \epsilon _{\multind{A}{0}{p}}
          \dot{\xi} ^{\multind{A}{0}{p}}
         }
    \nonumber \\
    \pi ^{\multind{A}{0}{p}}
    & = &
    \epsilon ^{\multind{A}{0}{p}}
    \Ham \ttt{\Bp}
    \label{4.pbrxiconmom}
    \quad .
\eea
In terms of a Lagrange multiplier $N ^{\multind{B}{0}{p}}$ the action
(\ref{4.pbrrepact}) can also be written in terms of the $(p+1)$-form
\bea
    \form{\Omega} ^{\ttt{p+1}}
    & = &
    \frac{1}{\fact{p+1}}
    P _{\multind{\mu}{0}{p}}
    \multint{X}{\mu}{0}{p}
    +
    \nonumber \\
    & & \quad +
    \frac{1}{\fact{p+1}}
    \pi _{\multind{A}{0}{p}}
    \multint{\xi}{A}{0}{p}
    +
    \\
    & & \qquad -
    \frac{1}{\fact{p+1}}
        N ^{\multind{A}{0}{p}}
        \left [
            \pi _{\multind{A}{0}{p}}
            -
            \epsilon _{\multind{A}{0}{p}}
            \Ham ( \Bp )
        \right ]
    \form{d s} ^{0} \wedge \dots \wedge \form{d s} ^{p}
    \nonumber
    \quad ,
\label{4.pbrlagpp1for}
\eea
which is the Hamiltonian form of the reparametrized Lagrangian
(\ref{4.pbrrepschlag}): here the momentum (\ref{4.pbrxiconmom})
enters   explicitly
and, again, the relation between the momenta $\Bp$ and $\Bpi$
is enforced as a constraint by the Lagrange multiplier
$\BN$.

\section{$p$-brane Hamilton Jacobi Theory}
\label{4.pbrhamjacsec}

Having lifted the $p+1$ coordinates $\Bxi$ to the role
of dynamical variables, we achieved a reparametrization invariant
formulation to start from, in order  to derive the functional
Hamilton--Jacobi equation. The procedure is quite similar to the one we already
used in the {\sl String} case. By noting that the embedding functions $\Bx$,
{\it as well as} the $\Bxi$ fields, are now {\it functions of the}
$\Bsigma$'s, we could derive the equation of motion for
the $p$-{\sl{}brane}. However, we prefer to
restrict our attention to the most important of them, which is the one
obtained by variation of the action with respect to the new
$\Bxi$ fields. We emphasize the relevance of this result in
performing the following computation.
\begin{props}[$p$-brane Energy Balance Equation]\spbcorr{}.\\
    The variation of the action {\rm (\ref{4.pbrrepact})}
    with respect to the
    $\Bxi \ttt{\Bsigma}$ fields gives the following equation:
    $$
        \frac{1}{p!}
        \epsilon _{\multind{A}{0}{p}}
        \epsilon ^{\multind{m}{1}{p} m}
            \ttt{\partial _{m _{1}} \xi ^{A _{1}}}
            \cdot \dots \cdot
            \ttt{\partial _{m _{p}} \xi ^{A _{p}}}
            \partial _{m} \Ham \ttt{\Bp}
        =
        0
        \quad .
    $$
\end{props}
\begin{proof}
See appendix \ref{A.EbalDer} for details.
\end{proof}
The result has of course the same meaning as in the {\sl String} case:
it is the energy--balance equations, which again states
that the Hamiltonian is constant along a classical solution.
We can now perform a variation corresponding to a deformation of the
{\it future} {\it Boundary} of the {\sl World--HyperTube} in the functional class
of fields that solve the equation of motion for the {\it Bulk}.
Before that, let us give some  definitions.
\begin{defs}[$p$-brane Projected Boundary Area Momentum]\spbcorr{}.\\
    The \underbar{Projected Boundary Area Momentum}
    of the $p$-{\sl{}brane} is the {\sl Boundary Area Momentum}
    projected in the ``direction'' of the {\sl Tangent Element} to the
    $p$-{\sl{}brane} {\sl World--HyperTube}, i.e.
    \beq
        q _{\mu} \ttt{\Bs}
        =
        Q _{\mu}{}_{\multind{\mu}{1}{p}} \ttt{\Bs}
        Y ^{\prime \multind{\mu}{1}{p}} \ttt{\Bs}
        \quad .
    \label{4.xprimeone}
    \eeq
\end{defs}
As a natural generalization for the
{\it area} of the {\sl Parameter Space} of the {\sl String} we now take the
$(p+1)$-dimensional
{\it HyperVolume} of the {\sl World--HyperTube} of the $p$-{\sl{}brane}, $V$:
\begin{defs}[$p$-brane HyperVolume]\spbcorr{}.\\
    The \underbar{Hypervolume} of a $p$-{\sl{}brane},
    \beq
        V
        \dfn
        \frac{1}{\fact{p+1}}
        \epsilon _{\multind{A}{0}{p}}
        \int _{\Xi ( \Bsigma )}
            \multint{\xi}{A}{0}{p}
        \quad ,
    \eeq
    is the volume associated with its {\sl World--HyperTube}.
\end{defs}
Now, we are ready for the following
\begin{props}[Boundary Variation of the $p$-brane Action]\spbcorr{}.
    \label{4.pbrbouvarpro}\\
    The variation of the action {\rm (\ref{4.pbrrepact})}
    among the functional class
    of field configurations satisfying the classical equation of motion
    for the {\it Bulk} is
    \bea
        \delta S
        & = &
        \frac{1}{p!}
        \int _{\Dp}
            Q _{\multind{\mu}{0}{p}}
            \multint{Y}{\mu}{1}{p}
            \delta Y ^{\mu _{0}} _{(f)}
        -
        \ttt{\delta V} \Ham
        \nonumber \\
        & = &
        \frac{1}{p!}
        \int _{\BB}
            d ^{p} \Bs \,
            q _{\mu} \ttt{\Bs}
            \delta Y ^{\mu} _{(f)} \ttt{\Bs}
        -
        \ttt{\delta V} \Ham
        \quad ,
    \label{4.pbrbouvarres}
    \eea
    where,
    $\delta Y ^{\mu} _{(f)}$  indicates that we performed a variation
    of the embedding functions of the final closed $p$-{\sl{}brane},
    which is the future {\it Boundary} of the considered
    {\sl World--HyperTube}.
\end{props}
\begin{proof}
The detailed computation is performed in appendix \ref{A.FutVarDer}.
\end{proof}

From equation (\ref{4.pbrbouvarres}) in proposition
\ref{4.pbrbouvarpro} we can see that
\bea
    \frac{\delta S}{\delta Y ^{\mu} _{(f)} \ttt{\Bs}}
    & \! \! = \! \! &
    q _{\mu} \ttt{\Bs}
    \label{4.pbrbouvaruno}
    \\
    \Ham \, \equiv \, E
    & \! \! = \! \! &
    - \frac{\delta S}{\delta V}
    \, = \,
    - \frac{\partial S}{\partial V}
    \quad .
    \label{4.pbrbouvardue}
\eea
Moreover, the dispersion relation for the
$p$-{\sl{}brane} has exactly the same form of the one we
already derived for the {\sl String} (compare with equation
(\ref{2.strhamjacpreden}) on page \pageref{2.strhamjacpreden})
\bea
    \frac{1}{2 \rp p!}
    q _{\mu} q ^{\mu}
    & = &
    \frac{1}{2 \rp \fact{p+1}}
    Q _{\mu \multind{\mu}{1}{p}}
    Q ^{\mu \multind{\mu}{1}{p}}
    Y '  _{\multind{\nu}{1}{p}}
    Y ^{\prime \multind{\nu}{1}{p}}
    \nonumber \\
    & = &
    \frac{
          \ttt{\By '} ^{2}
          Q _{\multind{\mu}{0}{p}}
          Q ^{\multind{\mu}{0}{p}}
         }
         {2 \rp \fact{p+1}}
    =
    \ttt{\By '} ^{2}
    E
    \quad .
\eea
We can of course rewrite it in integrated form to regain
reparametrization invariance,
\beq
    \frac{1}{2 \rp p!}
    \oint _{\BB}
        \frac{d ^{p} \Bs}{\sqrt{\ttt{\By '} ^{2}}}
        p _{\mu}
        p ^{\mu}
    =
    E
  \oint _{\BB}
      d ^{p} \Bs
      \sqrt{\ttt{\By '} ^{2}}
  \quad .
\label{4.pbrhamjacpre}
\eeq
Then, we obtain the following result.
\begin{props}[$p$-brane Hamilton--Jacobi Equation]\spbcorr{}.\\
    The Classical Dynamics of a $p$-{\sl{}brane} is regulated by the
    following equation
    \beq
        \frac{1}{2 \rp p!}
        \left(
            \oint _{\BB}
                d ^{p} \Bs
                \sqrt{\ttt{\By '} ^{2}}
        \right) ^{-1}
        \oint _{\BB}
            \frac{d ^{p} \Bs}{\sqrt{\ttt{\By '} ^{2}}}
            \frac{\delta S}{\delta Y ^{\mu} ( \Bs)}
            \frac{\delta S}{\delta Y _{\mu} ( \Bs)}
        =
        -
        \frac{\partial S}{\partial V}
        \quad ,
    \label{4.pbrhamjacres}
    \eeq
    which is nothing but the functional Hamilton--Jacobi equation
    for the $p$-{\sl{}brane}.
\end{props}
\begin{proof}
The hard work has already been done. Thus, we can start  with
equation (\ref{4.pbrhamjacpre})
and use the correspondences (\ref{4.pbrbouvaruno}) and
(\ref{4.pbrbouvardue})
to obtain the functional Hamilton-Jacobi equation for
the $p$-{\sl{}brane} (\ref{4.pbrhamjacpre}).
\end{proof}
Note how in our formulation the equation has the same form as in the {\sl String}
case!!! The only difference is the larger number of indices
carried by the {\sl Holographic Derivative}. Anyway, the physical interpretation
is the same: we have a {\sl Boundary} Dynamics of {\it HyperVolume Elements}
(which are the {\sl Holographic Coordinates})
as {\it shadows} of a {\it Bulk} Dynamics in one more dimension.
We note that equation (\ref{4.pbrhamjacres}) can be expressed
in terms of the $p$-{\sl{}brane Holographic Coordinates} as
\beq
    \frac{1}{2 \rp}
    \left(
        \oint _{\BB}
            d ^{p} \Bs
            \sqrt{\ttt{\By '} ^{2}}
    \right) ^{-1}
    \oint _{\BB}
        d ^{p} \Bs  \sqrt{\ttt{\By '} ^{2}}
        \frac{\delta S}{\delta Y ^{\multind{\mu}{0}{p}} ( \Bs )}
        \frac{\delta S}{\delta Y _{\multind{\mu}{0}{p}} ( \Bs )}
    =
    -
    \frac{\partial S}{\partial V}
    \quad .
\label{4.pbrhamjacreshol}
\eeq

\section{$p$-brane Quantum Dynamics}

\subsection{Equivalence with Nambu--Goto Dynamics}

Having obtained the {\it Functional Hamilton--Jacobi equation} for the
$p$-{\sl{}brane},
we can now turn to the problem of deriving its propagation kernel. As we
have already seen in section \ref{3.strnamgotequsec},
a convenient starting point is the action for the
Reparametrization invariant Hamiltonian Theory with the Lagrange
multiplier $\BN$, which is (\ref{4.pbrlagpp1for}),
\bea
    & & \esci
    S \left [
        \Bx ( \Bsigma ),
        \Bp ( \Bsigma ),
        \Bxi ( \Bsigma ),
        \BN ( \Bsigma );
        V
      \right ]
    =
    \nonumber \\
    & & =
    \frac{1}{\fact{p+1}}
    \int _{\Wsp}
        P _{\multind{\mu}{0}{p}}
        \multint{Y}{\mu}{0}{p}
    +
    \nonumber \\
    & & \quad +
    \frac{1}{\fact{p+1}}
    \int _{\Xip}
        \pi _{\multind{A}{0}{p}}
        \multint{\xi}{A}{0}{p}
    +
    \label{4.pbrfulrepact} \\
    & & \qquad -
    \frac{1}{\fact{p+1}}
    \int _{\Sigmap}
        d ^{2} \Bsigma
        N ^{\multind{A}{0}{p}}
        \left [
            \pi _{\multind{A}{0}{p}}
            -
            \epsilon _{\multind{A}{0}{p}}
            H \ttt{\Bp}
        \right ]
    \quad .
    \nonumber
\eea
An interesting observation about the Lagrange multiplier $\BN$ can
be reported at this stage:
\begin{props}[Meaning of the Lagrange Multiplier $\BN$]\spbcorr{}.\\
    The Lagrange multiplier $\BN$ is the Levi--Civita tensor
    in the $\Bxi$
    variables\,\footnote{Which now, we remember, are fields!}.
\end{props}
\begin{proof}
This result can be obtained varying the action (\ref{4.pbrfulrepact}) with 
respect to the $\Bpi$ field momentum, the one conjugated to $\Bxi$. Then, we get
 the following result (which is explicitly derived in section \ref{A.LMulDer})
for $N ^{\multind{A}{0}{p}}$,
\beq
    N ^{\multind{A}{0}{p}} \ttt{\Bsigma}
    =
    \dot{\xi} ^{\multind{A}{0}{p}}
    =
    \epsilon ^{\multind{m}{0}{p}}
    \ttt{\partial _{m _{0}} \xi ^{A _{0}}}
    \cdot
    \dots
    \cdot
    \ttt{\partial _{m _{p}} \xi ^{A _{p}}}
    \quad ,
\label{4.LMulEqo}
\eeq
which is the desired result.
\end{proof}

Now, we can find for the $p$-{\sl{}brane} the result corresponding to
proposition \ref{3.schnamgotequ}.
\begin{props}[$p$-brane Schild--Nambu Goto Quantum Equivalence]\spbcorr{}.\\
    The quantum propagation of a reparametrized Schild $p$-{\sl{}brane} in the
    saddle point approximation and with energy
    $$
        E = \frac{\rp}{2}
    $$
    is equivalent to the quantum propagation of a Nambu--Goto $p$-{\sl{}brane}.
\end{props}
\begin{proof}
To get the desired result we will compute the
propagation kernel
$K \qtq{\By \ttt{\Bs} , \By _{0} \ttt{\Bs} ; V}$,
starting from the {\it Schild Reparametrized Formulation}
expressed in the action (\ref{4.pbrfulrepact}) by means of the following
path--integral
\bea
    & & \esci \esci
    K
    \left [
        \By \ttt{\hat{\Bs}} ,
        \By _{0} \ttt{\hat{\Bs}} ;
        V
    \right ]
    =
    \int _{\By  _{0} (\Bs)} ^{\By (\Bs)}
    \int _{\Bzeta  _{0} (\Bs)} ^{\Bzeta (\Bs)}
        [\mathcal{D} \Bx ( \Bsigma )]
        [\mathcal{D} \Bxi ( \Bsigma )]
        [\mathcal{D} \Bp ( \Bsigma )]
        [\mathcal{D} \Bpi ( \Bsigma )]
        [\mathcal{D} \BN ( \Bsigma )]
    \cdot
    \nonumber \\
    & & \qquad \qquad \qquad \qquad \qquad \qquad \qquad \qquad \cdot
    \exp
    \left\{
        \frac{i}{\hbar}
        S \left [
            \Bx ( \Bsigma ),
            \Bp ( \Bsigma ),
            \Bxi ( \Bsigma ),
            \BN ( \Bsigma );
            V
          \right ]
    \right\}
    \label{4.kerderintuno}
    =
    \nonumber \\
    & = &
    \int _{\By _{0} (\Bs)} ^{\By (\Bs)}
    \int _{\Bzeta _{0} (\Bs)} ^{\Bzeta (\Bs)}
        [\mathcal{D} \Bx ( \Bsigma )]
        [\mathcal{D} \Bxi ( \Bsigma )]
        [\mathcal{D} \Bp ( \Bsigma )]
        [\mathcal{D} \Bpi ( \Bsigma )]
        [\mathcal{D} \BN ( \Bsigma )]
    \cdot
    \nonumber \\
    & & \qquad
    \cdot
    \exp
    \left\{
    \frac{i}{\fact{p+1}}
    \int _{\Wsp}
        P _{\multind{\mu}{1}{p+1}}
        \multint{x}{\mu}{1}{p+1}
    +
    \right .
    \nonumber \\
    & & \qquad \qquad \quad +
    \frac{i}{\fact{p+1}}
    \int _{\Xip}
        \pi _{\multind{A}{1}{p+1}}
        \multint{\xi}{A}{1}{p+1}
    +
    \nonumber \\
    & & \qquad \qquad \quad -
    \left .
    \frac{i}{\fact{p+1}}
    \int _{\Sigmap}
        d ^{p+1} \Bsigma
        N ^{\multind{A}{1}{p+1}}
        \left [
            \pi _{\multind{A}{1}{p+1}}
            -
            \epsilon _{\multind{A}{1}{p+1}}
            \Ham ( \Bp )
        \right ]
    \right\}
\eea
through the following four steps:
\begin{enumerate}
    \item integrate out the $\Bxi ( \Bsigma )$ fields using the energy
    balance equation and obtaining a 
    functional Dirac Delta, which meets the requirement that the
    momentum $\Bpi ( \Bsigma ) $ satisfies the classical equation
    of motion;
    \bea
    	& & \esci \esci
    	K
    	\left [
    	    \By ( \Bs ),
    	    \By _{0} ( \Bs );
    	    V
    	\right ]
    	=
    	\nonumber \\
    	& = &
    	\int _{\By _{0} (\Bs)} ^{\By (\Bs)}
    	    [\mathcal{D} \Bx ( \Bsigma )]
    	    [\mathcal{D} \Bp ( \Bsigma )]
    	    [\mathcal{D} \Bpi ( \Bsigma )]
    	    [\mathcal{D} \BN ( \Bsigma )]
    	\cdot
    	\nonumber \\
    	& & \quad
    	\cdot
        \delta
        \left [
            \epsilon ^{\multind{m}{1}{p+1}}
            \ttt{\partial _{m _{p+1}} \pi _{\multind{a}{1}{p+1}}}
            \cdot
            \ttt{\partial _{m _{1}} \xi ^{a _{1}}}
            \cdot
            \dots
            \cdot
            \ttt{\partial _{m _{p}} \xi ^{a _{p}}}
        \right ]
    	\cdot
    	\nonumber \\
    	& & \quad
    	\cdot
    	\exp
    	\left\{
    	\frac{i}{\fact{p+1}}
    	\int _{\Wsp}
    	    P _{\multind{\mu}{1}{p+1}}
    	    \multint{X}{\mu}{1}{p+1}
    	+
    	\right .
    	\nonumber \\
    	& & \quad \qquad \quad +
            \frac{i}{\fact{p+1}}
                \int _{\Xip}
                \form{d} \left(
                      \xi ^{a _{p+1}}
                      \pi _{\multind{a}{1}{p+1}}
                      \multint{\xi}{a}{1}{p}
                  \right)
    	+
    	\nonumber \\
    	& & \quad \qquad \quad -
    	\left .
    	\frac{1}{\fact{p+1}}
    	\int _{\Sigmap}
    	    d ^{p+1} \Bsigma
    	    N ^{\multind{a}{i}{p+1}}
    	    \left [
    		\pi _{\multind{a}{1}{p+1}}
    		-
    		\epsilon _{\multind{a}{1}{p+1}}
    		\Ham ( \Bp )
    	    \right ]
    	\right\}
    \quad ;
    \eea
    \item take advantage of this to integrate out the momentum
    $\Bpi ( \Bsigma ) $ also;
    \bea
    	& & \esci
    	K
    	\left [
    	    \By ( \Bs ),
    	    \By _{0} ( \Bs );
    	    V
    	\right ]
    	=
    	\nonumber \\
    	& = &
    	\int _{0} ^{\infty}
    	    d E
            e ^{iEV}
    	\int _{\By _{0} ( \Bs )} ^{\By ( \Bs )}
    	    [\mathcal{D} \Bx ( \Bsigma )]
    	    [\mathcal{D} \Bp ( \Bsigma )]
    	    [\mathcal{D} \BN ( \Bsigma )]
    	\cdot
    	\nonumber \\
    	& & \qquad
    	\cdot
    	\exp
    	\left\{
    	\frac{i}{\fact{p+1}}
    	\int _{\Wsp}
    	    P _{\multind{\mu}{1}{p+1}}
    	    \multint{X}{\mu}{1}{p+1}
    	+
    	\right .
    	\nonumber \\
    	& & \qquad \qquad \quad
            -
            \frac{i}{\fact{p+1}}
        \left .
                \epsilon _{\multind{a}{1}{p+1}}
                \int _{\Sigmap}
        	    d ^{p+1} \Bsigma
        	    N ^{\multind{a}{1}{p+1}}
    	            \left [
    		        E
    		        -
        		\Ham ( \Bp )
    	            \right ]
    	\right.
        \Bigr\}
    \quad ;
    \eea
    \item remembering now the form of the Hamiltonian $\Ham ( \Bp )$ and
    defining
    \beq
        N ( \Bsigma )
        =
        \frac{1}{\fact{p+1}}
        \epsilon _{\multind{A}{1}{p+1}}
        N ^{\multind{A}{1}{p+1}}
        \quad ,
    \eeq
    we can cast the last expression into the following one,
    \bea
    	& & \esci
    	K
    	\left [
    	    \By ( \Bs ),
    	    \By _{0} ( \Bs );
    	    V
    	\right ]
    	=
    	\nonumber \\
    	& = &
    	\int _{0} ^{\infty}
    	    d E
            e ^{iEV}
    	\int _{\By _{0} (\Bs)} ^{\By ( \Bs )}
    	    [\mathcal{D} \Bx ( \Bsigma )]
    	    [\mathcal{D} \Bp ( \Bsigma )]
    	    [\mathcal{D} \BN ( \Bsigma )]
    	\cdot
    	\nonumber \\
    	& & \qquad
    	\cdot
    	\exp
    	\left\{
        i
    	\int _{\Sigmap}
    	    d ^{p+1} \Bsigma
    	    E
    	    N
    	\right\}
    	\cdot
    	\nonumber \\
    	& & \qquad \cdot
    	\exp
    	\left\{
        i
    	\int _{\Sigmap}
    	    d ^{p+1} \Bsigma
    	    \left [
                \frac{N}{2 \rp \fact{p+1}}
                P _{\multind{\mu}{1}{p+1}}
                P ^{\multind{\mu}{1}{p+1}}
                +
            \right .
        \right.
    	\nonumber \\
    	& & \qquad \qquad \qquad \qquad \qquad \qquad \qquad \qquad +
        \left .
            \left .
        	P _{\multind{\mu}{1}{p+1}}
    	        \frac{\dot{X} ^{\multind{\mu}{1}{p+1}}}{\fact{p+1}}
             \right ]
        \right\}
    \quad ,
    \eea
    where, we recognize a {\it functional gaussian integral} in
    $\Bp ( \Bsigma ) $ which gives
    \bea
    	& & \esci
    	K
    	\left [
    	    \By ( \Bs ),
    	    \By _{0} ( \Bs  );
    	    V
    	\right ]
    	=
    	\nonumber \\
    	& = &
    	\int _{0} ^{\infty}
    	    d E
            e ^{iEV}
    	\int _{\By _{0} ( \Bs )} ^{\By ( \Bs )}
    	    [\mathcal{D} \Bx ( \Bsigma )]
    	    [\mathcal{D} \BN ( \Bsigma )]
    	\cdot
    	\nonumber \\
    	& & \qquad
    	\cdot
    	\exp
    	\left\{
    	-
    	i
        \int _{\Sigmap}
    	    d ^{p+1} \Bsigma
    	    \left [
        	-
                \frac{\rp}{2 N \fact{p+1}}
                \dot{X} _{\multind{\mu}{1}{p+1}}
                \dot{X} ^{\multind{\mu}{1}{p+1}}
                +
                N
                E
            \right ]
        \right\}
    \quad ;
    \label{4.befsadpoieva}
    \eea
    \item \dots calculate the result around the saddle point
    $
        \bar{\BN} ( \Bsigma )
        =
        \sqrt{- \rp \left( \dot{\Bx} \right) ^{2} / \qtq{2 E \fact{p+1}}}
    $
    to get rid of the $\BN ( \Bsigma )$
    dependence in the result, obtaining
    \bea
        & & \esci \esci
    	K
    	\left [
    	    \By ( \Bs ),
    	    \By _{0} ( \Bs  );
    	    V
    	\right ]
    	=
    	\int _{0} ^{\infty}
    	    d E
            e ^{iEV}
            \int _{\By _{0} ( \Bs )} ^{\By ( \Bs)}
                [\mathcal{D} \Bx ( \Bsigma )]
                \cdot
        \nonumber \\
        & & \qquad \qquad \qquad \cdot
                \exp
                \left\{
                    -
                    i
                    \left( 2 E \rp \right) ^{\frac{1}{2}}
                    \int _{\Sigmap}
                        d ^{p+1} \Bsigma
                        \sqrt{
                              -
                              \frac{\rp}{\fact{p+1}}
                              \ttt{\dot{\Bx}} ^{2}
                             }
                \right\}
        \quad .
    \label{4.proexp}
    \eea
\end{enumerate}
\end{proof}

Once the propagator is known, we get
\begin{props}[$p$-brane Green Function]\spbcorr{}.\\
    The $p$-{\sl{}brane} Green Function in the saddle point approximation
    is
    $$
    G [\BB , \BB _{0} ; E ]
    =
    \int _{\By _{0} ( \Bs )} ^{\By ( \Bs)}
        [\mathcal{D} \Bx ( \Bsigma )]
        \exp
        \left\{
            -
    	    i
    	    \left( 2 E \rp \right) ^{\frac{1}{2}}
    	    \int _{\Sigmap}
    	        d ^{p+1} \Bsigma
    	        \sqrt{
                      -
                      \frac{\rp}{\fact{p+1}}
                      \ttt{\dot{\Bx}} ^{2}
    	             }
        \right\}
        \quad .
    $$
\end{props}
\begin{proof}
As a matter of fact, there is nothing to prove: thanks to the relation
\beq
    K
    \left [
	\By ( \Bs ),
	\By _{0} ( \Bs );
	V
    \right ]
    =
    \int _{0} ^{\infty}
    d E
    e ^{iEV}
    G [\BB , \BB _{0} ; E ]
\eeq
we have just to read the correct result in equation (\ref{4.proexp}),
which gives the desired result,
\end{proof}

\subsection{Functional Schr\"o{}dinger Equation and $p$-brane}

The next step is to find the functional wave equation for the kernel
$K
 \left [
     \By ( \Bs ),
     \By _{0} ( \Bs );
     V
 \right ]$, a task which we will perform by means of a
{\it path--integral} technique.
\begin{props}[$p$-brane Variation of the Kernel]\spbcorr{}.\\
The variation of the kernel under an
infinitesimal variation of the fields is
\bea
    & & \esci
    \delta
    K
    \left [
    	\By ( \Bs ),
    	\By _{0} ( \Bs );
    	V
    \right ]
    =
    \nonumber \\
    & & =
    \frac{i}{\hbar}
    \int _{\By ( \Bs _{0})} ^{\By ( \Bs)}
    \int _{\Bzeta ( \Bs _{0})} ^{\Bzeta ( \Bs)}
        [\mathcal{D} \mu ( \Bsigma )]
        \delta
        S
        \left [
            \Bx ,
            \Bp ,
            \Bxi ,
            \BN ;
            V
        \right ]
        \exp
        \left\{
            \frac{i}{\hbar}
            S \left [
                  \Bx ,
                  \Bp ,
                  \Bxi ,
                  \BN ;
                  V
              \right ]
        \right\}
\quad ,
\label{KernelVar}
\eea
where, we collected all the functional
integrations in $[ \mathcal{D} \mu ( \Bsigma )]$,
$$
    [\mathcal{D} \mu (\Bsigma )]
    =
    [\mathcal{D} \Bx ( \Bsigma )]
    [\mathcal{D} \Bxi ( \Bsigma )]
    [\mathcal{D} \Bp ( \Bsigma )]
    [\mathcal{D} \Bpi ( \Bsigma )]
    [\mathcal{D} \BN ( \Bsigma )]
    \quad .
$$
\end{props}
\begin{proof}
We start from the expression for the Kernel in equation
(\ref{4.kerderintuno}), noting that a variation only affects
the exponential, i.e. the action in the exponent. The result then
follows by the chain rule applied to that exponential, i.e. we use
$$
    \delta \left( e ^{\frac{i}{\hbar} S} \right)
    =
    \frac{i}{\hbar}
    e ^{\frac{i}{\hbar} S}
    \ttt{\delta S}
\quad .
$$
\end{proof}
Now we remember that we again restrict the field variation within the family of
classical  solutions; this is exactly the same computation we
performed in appendix \ref{A.FutVarDer}, from which we keep the result of
formula (\ref{A.classvar}):
\beq
    \delta S _{\mathrm{cl.}} \left [ \BB ; V \right ]
    =
    \frac{1}{p!}
    \int _{\partial \Wsp}
              p _{\multind{\mu}{1}{p+1}}
              \multint{Y}{\mu}{2}{p+1}
              \delta Y ^{\mu _{1}}
    -
    E \delta V
\label{ActionVar}
\eeq
where, we only substituted $E$ for $\Ham$.
The resulting equations for the kernel, derived in appendix
\ref{A.kerwavder} are
\bea
    \frac{
          \partial K
              \left [
                  \BB ( \Bs ),
                  \BB _{0} ( \Bs );
                  V
              \right ]
         }
         {\partial V}
    & = &
    -
    \frac{i E}{\hbar}
    K
        \left [
            \BB ( \Bs ),
            \BB _{0} ( \Bs );
            V
        \right ]
    \\
    \frac{
          \delta K
              \left [
                  \BB ( \Bs ),
                  \BB _{0} ( \Bs );
                  V
              \right ]
         }
         {\delta Y ^{\mu} ( \Bs )}
    & = &
    \frac{i}{\hbar}
    \int _{\By _{0} ( \Bs )} ^{\By ( \Bs )}
    \int _{\Bzeta _{0} ( \Bs )} ^{\Bzeta ( \Bs )}
        \left [ \mathcal{D} \mu ( \Bsigma ) \right ]
        P _{ \mu \multind{\mu}{1}{p}}
        Y ^{\prime \multind{\mu}{1}{p}}
        \exp \left( \frac{i S}{\hbar} \right)
\eea
and we recall the comment which follows formula
(\ref{A.xprimedef}) about the
definition of $\By '$.
By comparing the last two equations with the functional Jacobi
equation for the $p$-{\sl{}brane} (\ref{4.pbrhamjacres}),
we obtain the functional
wave equation for the kernel (see appendix \ref{A.kerfunwavder}):
\bea
    & &
    - \frac{\hbar ^{2}}{2 \rp \fact{p}}
    \left(
        \oint _{\BB}
            d ^{p} \Bs
            \sqrt{\ttt{\By '} ^{2}}
    \right) ^{-1}
    \cdot
    \nonumber \\
    & & \qquad \qquad \cdot
    \oint _{\BB}
        \frac{d ^{p} \Bs}{\sqrt{\ttt{\By '} ^{2}}}
        \frac{
              \delta ^{2}
              K
              \left [
                  \BB ( \Bs ),
                  \BB _{0} ( \Bs );
                  V
              \right ]
             }
             {\delta Y ^{\mu} ( \Bs ) \delta Y _{\mu} ( \Bs )}
             =
             i \hbar
             \frac{
                   \partial
                   K
                   \left [
                       \BB ( \Bs ),
                       \BB _{0} ( \Bs );
                       V
                   \right ]
                  }
                  {\partial V}
\quad ,
\label{kerfunwav}
\eea
By Fourier transforming, we get
$$
    \left [
 	- \frac{\hbar ^{2}}{p!}
 	\left(
 	    \oint _{\BB}
 		d ^{p} \Bs
 		\sqrt{\ttt{\By '} ^{2}}
 	\right) ^{-1}
        \! \! \! \! \!
 	\oint _{\BB} \! \!
 	    \frac{d ^{p} \Bs}{\sqrt{Y ^{ \prime 2}}}
 	    \frac{\delta ^{2}}
 		 {\delta Y ^{\mu} ( \Bs ) \delta Y _{\mu} ( \Bs )}
    +
    \rp ^{2}
    \right ]
    G
    \left [
    	\BB ( \Bs ),
    	\BB _{0} ( \Bs );
        V
    \right ]
    =
    -
    \delta
    \left [
        \BB - \BB _{0}
    \right ]
$$
and we can  identify $G \left [ \BB ( \Bs ), \BB _{0} ( \Bs ); V \right ] $
with the Green function for the $p$-{\sl{}brane}.

\subsection{$p$-brane Quantum Propagator}

The kernel wave equation can be solved then through the following
ansatz:\\[2mm]
{\bf Ansatz ($p$-brane Quantum Kernel)}: {\it The Quantum Kernel of a
$p$-{\sl{}brane} has the following functional dependence:}
\beq
    K
    \left [
    	\BB ( \Bs ) ,
    	\BB _{0} ( \Bs ) ;
    	V
    \right ]
    =
    {\cal N}
    V ^{\alpha}
    \exp
    \left\{
            \frac{i}{\hbar}
        I
        \left [
            \BB ( \Bs ) ,
            \BB _{0} ( \Bs ) ;
            V
        \right ]
    \right\}
    \quad ,
\label{4.pbrans}
\eeq
{\it where ${\cal N}$ is the normalization constant, and $\alpha$
a real number}.
\begin{props}[$p$-brane Amplitude and Phase Kernel Equations]\spbcorr{}.\\
The exponent $\alpha$ and the phase $I \qtq{\By , \By _{0} ; V}$
satisfy two equations, namely
\bea
    2 \rp \fact{p}
    \frac{\alpha}{V}
    \! \! \! \!
    & = &
    \! \! \! \!
    -
    \left(
        \oint _{\BB}
            d ^{p} \Bs
            \sqrt{\By ^{\prime 2}}
    \right) ^{-1}
    \! \!
    \oint _{\BB}
        \frac{d ^{p} \Bs}{\sqrt{Y ^{\prime 2}}}
        \frac{
              \delta ^{2}
              I
              \left [
                  \BB, 
                  \BB _{0}, 
                  V
              \right ]
             }
             {\delta Y ^{\mu} ( \Bs ) \delta Y _{\mu} ( \Bs )}
    \label{trial1}
    \\
    2 \rp \fact{p}
    \frac{
          \partial I
          \left [
              \BB, 
              \BB _{0}, 
              V
          \right ]}{\partial V}
    \! \! \! \!
    & = &
    \! \! \! \!
    -
    \left(
        \oint _{\BB}
            d ^{p} \Bs
            \sqrt{\ttt{\By '} ^{2}}
    \right) ^{-1}
    \! \!
    \oint _{\BB}
        \frac{d ^{p} \Bs}{\sqrt{\ttt{\By '} ^{2}}}
        \cdot
    \nonumber \\
    & & \qquad \qquad
    \cdot
        \frac{
              \delta
              I
              \left [
                  \BB ,
                  \BB _{0} ,
                  V
              \right ]
             }
             {\delta Y ^{\mu} ( \Bs )}
        \frac{
              \delta
              	  I
              	  \left [
              	      \BB ,
              	      \BB _{0} ,
              	      V
              	  \right ]
             }
             {\delta Y _{\mu} ( \Bs )}
\label{trial2}
\quad ,
\eea
in exact analogy with the {\sl String} case
(see {\rm proposition \ref{3.ampphakerequ}}.)
\end{props}
\begin{proof}
Following the same procedure as in chapter
\ref{3.strfunqua},  we first inserti the ansatz (\ref{4.pbrans})
into (\ref{kerfunwav}).
The results for the second functional derivative,
after the calculation of
$$
    \frac{\delta K}{\delta Y ^{\mu} \ttt{t}}
    =
    \mathcal{N}
    \frac{i}{\hbar}
    V ^{\alpha}
    e ^{\frac{i}{\hbar} I}
    \frac{\delta I}{\delta Y ^{\mu} \ttt{t}}
    \quad ,
$$
is
$$
    \frac{\delta ^{2} K}{\delta Y ^{\mu} \ttt{t} \delta Y _{\mu} \ttt{t}}
    =
    \mathcal{N}
    V ^{\alpha}
    e ^{\frac{i}{\hbar} I}
    \left [
        - \frac{1}{\hbar ^{2}}
        \left(
            \frac{\delta I}{\delta Y ^{\mu} \ttt{t}}
        \right) ^{2}
        +
        \frac{i}{\hbar}
        \frac{\delta ^{2} I}{\delta Y ^{\mu} \ttt{t} \delta Y _{\mu} \ttt{t}}
    \right ]
    \quad .
$$
Moreover the first derivative with respect to $V$ is
$$
    \frac{\partial K}{\partial V}
    =
    \mathcal{N}
    V ^{\alpha}
    e ^{\frac{i}{\hbar} I}
    \left(
        \frac{\alpha}{V}
        +
        \frac{i}{\hbar}
        \frac{\partial I}{\partial V}
    \right)
    \quad .
$$
Substituting the last two equations in (\ref{kerfunwav})
and separating the real and imaginary part, we obtain the desired result.
\end{proof}
Now, using the formulae in appendix \ref{A.detcal}, the following
result can be proved.
\begin{props}[$p$-brane Kernel and Green Function]\spbcorr{}.\\
The Propagation Kernel and the Green Function of a $p$-{\sl{}brane} are given
by
\beq
    K[\By(s),\By_0(s);V]
    =
    \left({\rp\over 2i\pi\hbar V}\right) ^{\ttt{D-p}/2}
    \exp
    \left(
        {i\rp\over 4\hbar V}
        \Sigma^{\mu\nu}[\BB-\BB_0]
        \Sigma_{\mu\nu}[\BB-\BB_0]
    \right)
    \quad .
    \label{4.pbrprores}
\eeq
\beq
    G \qtq{\BB , \BB _{0} ; V}
    =
    \int _{0} ^{\infty}
        dV
        e ^{-i \rp V / 2 \hbar}
        \exp
        \left(
            {i\rp\over 4\hbar V}
            \Sigma^{\mu\nu}[\BB-\BB_0]
            \Sigma_{\mu\nu}[\BB-\BB_0]
        \right)
    \label{4.pbrgreres}
\eeq
respectively.
\end{props}
\begin{proof}
As a first step,
from the comparison of equations (\ref{trial2}) and (\ref{4.pbrhamjacres})
we understand that the phase
$I
 \left [
     \BB ,
     \BB _{0} ,
     V
 \right ]
$
is nothing but the classical action
\beq
    I
    \left [
        \By ,
        \By _{0} ,
        V
    \right ]
    =
    S _{\mathrm{cl.}}
    \left [
        \By ,
        \By _{0} ,
        V
    \right ]
    \quad ,
\label{4.phaequact}
\eeq
which we shall evaluate in analogy with the {\sl String} case.
The generalization of the {\sl String} {\sl Holographic
Coordinates$\,$}\footnote{Which in turn are the generalization
of the point particle line element.} are of course the
$p$-{\sl{}brane} {\sl Holographic Coordinates} of definition
\ref{4.pbrholcoo}, $Y ^{\multind{\mu}{1}{p+1}} [ \BB ]$,
which we rewrite here for convenience:
\begin{equation}
    Y ^{\multind{\mu}{1}{p+1}} [ \BB ]
    \equiv
    \oint _{\Dp}
        Y ^{\mu _{1}} \multint{Y}{\mu}{2}{p+1}
    =
    \oint _{\BB}
        d ^{p } \Bs \,
        Y ^{\mu _{1}} ( \Bs )
        Y ^{\prime \multind{\mu}{2}{p+1}}
\quad .
\label{hypel}
\end{equation}
We quote the results, calculated in appendix
\ref{A.funcdersig}, for the first
and second functional derivatives of the
$p$-{\sl{}brane} {\sl Holographic Coordinates}, which are:
\begin{eqnarray}
    \frac{\delta Y ^{\multind{\mu}{1}{p+1}} \left [ \BB \right ]}
         {\delta Y ^{\alpha} (\bar{\Bs})}
    & = &
    \delta ^{\mu _{1}} _{\alpha}
    Y ^{\prime \multind{\mu}{2}{p+1}} (\bar{\Bs})
    -
    \sum _{i} ^{2,p+1}
        \delta ^{\mu _{i}} _{\alpha}
        Y   ^{
              \prime
              \multind{\mu}{2}{i-1}
              \check{\mu _{i}}
              \mu _{1}
              \multind{\mu}{i+1}{p+1}
             }
    \label{fundersig1}
    \\
    \frac{\delta Y ^{\multind{\mu}{1}{p+1}} \left [ \BB \right ]}
         {\delta Y ^{\alpha} (\bar{\Bs}) \delta Y ^{\beta} (\tilde{\Bs})}
    & = &
    \sum _{j} ^{2,p+1}
        \delta ^{\mu _{1} \mu _{j}} _{\alpha \beta}
        \epsilon ^{\multind{a}{2}{p+1}}
        \left(
            \partial _{a _{2}} Y ^{\mu _{2}}
        \right)
        \cdot
        \dots
        \cdot
        \check{
               \left(
                   \partial _{a _{j}} Y ^{\mu _{j}}
               \right)
               }
        \cdot
        \dots
        \cdot
        \left(
            \partial _{a _{p+1}} Y ^{\mu _{p+1}}
        \right)
    +
    \nonumber \\
    & &
    -
    \sum _{i,j \atop i \neq j} ^{2,p+1}
        \delta ^{\mu _{i}} _{\alpha}
        \delta ^{\mu _{j}} _{\beta}
        \epsilon ^{\multind{a}{2}{i} \dots \multind{a}{j}{p+1}}
    \cdot
        \left(
            \partial _{a _{2}} Y ^{\mu _{2}}
        \right)
        \cdot
        \dots
    \cdot
    \nonumber \\
    & & \qquad \qquad \cdot
        \dots
        \cdot
        \left(
            \partial _{a _{i}} Y ^{\mu _{1}}
        \right)
        \cdot
        \dots
        \cdot
        \left(
            \partial _{a _{p+1}} Y ^{\mu _{p+1}}
        \right)
        \partial _{a _{j}} \delta \left( \bar{\Bs} - \tilde{\Bs} \right)
\quad .
\label{fundersig2}
\end{eqnarray}
Then we can write our ``guess'' for the action, or equivalently,
for the phase (\ref{4.phaequact}):
\begin{eqnarray}
    S _{\mathrm{cl.}} \left [ \BB ( \Bs ), \BB _{0} ( \Bs ) ; V \right ]
    & = &
    \frac{\beta}{2 \left( p + 1 \right) V}
    \left(
        Y ^{\multind{\mu}{1}{p+1}} [ \BB ]
        -
        Y ^{\multind{\mu}{1}{p+1}} [ \BB _{0}]
    \right)
    \cdot
    \nonumber \\
    & & \qquad \qquad \qquad \qquad \cdot
    \left(
        Y _{\multind{\mu}{1}{p+1}} [ \BB ]
        -
        Y _{\multind{\mu}{1}{p+1}} [ \BB _{0}]
    \right)
    \\
    & \equiv &
    \frac{\beta}{2 \left( p + 1 \right) V}
    \Sigma ^{\multind{\mu}{1}{p+1}} [ \BB - \BB _{0}]
    \Sigma _{\multind{\mu}{1}{p+1}} [ \BB - \BB _{0}]
    \label{4.claactgue}
\quad ,
\label{claact}
\end{eqnarray}
where
\beq
    \Sigma ^{\multind{\mu}{1}{p+1}}
    =
    Y ^{\multind{\mu}{1}{p+1}} [ \BB ]
    -
    Y ^{\multind{\mu}{1}{p+1}} [ \BB _{0}]
    \quad .
    \label{4.sigdef}
\eeq
Starting from (\ref{4.claactgue}) we compute in appendix
(\ref{A.funderclaact}) the following equations
expressing the functional derivatives of the classical action:
\begin{eqnarray}
    \frac{\delta S _{\mathrm{cl.}}}{\delta Y ^{\mu _{1}}(\Bs)}
    & = &
    \frac{\beta}{V}
    \Sigma ^{\multind{\mu}{1}{p+1}} [ \BB - \BB _{0}]
    Y ' _{\multind{\mu}{2}{p+1}}
    \label{funderact1}
    \\
    \frac{\delta ^{2} S _{cl.}}
         {\delta Y ^{\mu _{1}}(\Bs) \delta Y _{\mu _{1}}(\Bs)}
    & = &
    \left( D - p \right)
    \frac{\beta}{V}
    \ttt{\By ' ( \Bs )} ^{2}
    \label{fundeeract2}
\quad .
\end{eqnarray}
Then, by taking into account (\ref{4.phaequact}), and
substituting into equations (\ref{trial1}, \ref{trial2}),
we obtain the solutions
\begin{eqnarray}
    \alpha
    & = &
    -
    \frac{D - p}{2 \left( p + 1 \right)}
    \\
    \beta
    & = &
    \rp \fact{p}
    \quad .
\end{eqnarray}
If we define
$$
    \delta \left [ \BB - \BB _{0} \right ]
    =
    \lim _{\epsilon \to 0}
        \left(
            \frac{1}{\pi \epsilon}
        \right) ^{\left( D - p \right) / 2}
        \exp
        \left(
            -
            \frac{1}{\fact{p+1} \epsilon}
            \Sigma ^{\multind{\mu}{1}{p+1}} [ \BB - \BB _{0}]
            \Sigma _{\multind{\mu}{1}{p+1}} [ \BB - \BB _{0}]
        \right)
$$
then, the kernel turns out to be
\beq
    K[\By(s),\By_0(s);V]
    =
    \left({\rp\over 2i\pi\hbar V}\right) ^{\ttt{D - p}/2}
    \exp
    \left(
        {i\rp\over 4\hbar V}
        \Sigma^{\multind{\mu}{1}{p+1}}[\BB-\BB_0]
        \Sigma_{\multind{\mu}{1}{p+1}}[\BB-\BB_0]
    \right)\ ,
    \quad
\eeq
and the corresponding Green Function is
\beq
    G \qtq{\BB , \BB _{0} ; V}
    =
    \int _{0} ^{\infty}
        dV
        e ^{-i \rp V / 2 \hbar}
        \exp
        \left(
            {i\rp\over 4\hbar V}
            \Sigma^{\multind{\mu}{1}{p+1}}[\BB-\BB_0]
            \Sigma_{\multind{\mu}{1}{p+1}}[\BB-\BB_0]
        \right)
    \quad .
\eeq
These are the desired results.
\end{proof}
As a consistency check, it is possible to derive the same quantities by applying
gaussian integration techniques to
\begin{equation}
    K \left [
          \By , \By _{0} ; V
      \right ]
    =
    \int _{\By _{0} ( \Bs )}  ^{\By ( \Bs )}
        [\mathcal{D} \Bx ^{\mu} ( \Bsigma ) ]
        [\mathcal{D} P _{\mu \nu} ( \Bsigma ) ]
        \exp \left\{
                    i S \left[
                               \Bx ( \Bsigma ) ,
                               \Bp ( \Bsigma ) ;
                               \Bxi ( \Bsigma )
                        \right]
             \right\}
    \quad ,
\end{equation}
which is the usual path--integral expression for the kernel with
$S \left[ \By ( \Bsigma ) , \Bp ( \Bsigma ) ; \Bxi ( \Bsigma ) \right]$
given by formula (\ref{4.pbrrepact}).

%% file: chap05.tex
\pageheader{}{String Functional Solutions.}{}
\chapter{String Functional Solutions}
\label{5.strfunsol}

\begin{start}
``$\euf{W}$ell,\\
see what you can do with'' it.\\
\end{start}

After the higher dimensional digression of the previous chapter
we return to our main task:
the next step in our program to set up a (spacetime covariant)
functional quantum mechanics of closed {\sl Strings}, is to find the
basic solutions of equation (\ref{3.funlooschequ}). As in the quantum
mechanics of point particles we can factorize explicitly the $A$
dependence, which gives nothing but an exponential factor. We note that
from now on we will work in $4$-dimensional spacetime: anyway,
no complications arise in keeping an arbitrary dimension.
\begin{props}[Stationary States Schr\"odinger Equation]\spbcorr{}.\\
    A solution of the wave equation {\rm (\ref{3.funlooschequ})}
    can be always written
    in the form
    $$
        \Psi \qtq{C ; A}
        =
        \Phi \qtq{C}
        e ^{i E A / \hbar}
    $$
    provided that the functional of loops $\Phi \qtq{C}$
    satisfies
    \beq
        -
        \frac{\hbar ^{2}}{4 m ^{2}}
        \norme{\Gamma}
        \oint _{\Gamma} \sqrt{\ttt{\By ' \tst} ^{2}} ds
        \frac{\delta ^{2} \Psi \qtq{C}}
             {
              \delta Y ^{\mu \nu} \tst
              \delta Y _{\mu \nu} \tst
             }
        =
        E
        \Psi \qtq{C}
        \quad ,
    \label{5.stastefunwavequ}
    \eeq
    the \underbar{Stationary States Schr\"odinger Equation},
    which reads, in the more compact notation of equation
    {\rm (\ref{3.newnotation})},
    $$
        -
        \frac{\hbar ^{2}}{4 m ^{2}}
        \frac{\delta ^{2} \Psi \qtq{C}}
             {
              \delta C ^{\mu \nu}
              \delta C _{\mu \nu}
             }
        =
        E
        \Psi \qtq{C}
        \quad .
    $$
\end{props}
\begin{proof}
We compute
\bea
    \frac{\partial \Psi \qtq{C ; A}}{\partial A}
    & = &
    \Phi \qCq
    \frac{\partial \left( e ^{- \frac{i}{\hbar} E A} \right)}{\partial A}
    \nonumber \\
    & = &
    \Phi \qCq
    (- \frac{i E}{\hbar})
    e ^{- \frac{i}{\hbar} E A}
    \label{5.staareder}
    \\
    & = &
    - \frac{i}{\hbar} E
    \Psi \qtq{C ; A}
    \quad ;
\eea
substituting then (\ref{5.staareder}) in (\ref{3.funlooschequ})
and simplifing by $\exp \left( - i E A / \hbar \right)$ we obtain
(\ref{5.stastefunwavequ}) (or its compact form using
(\ref{3.lapnewnot}) also).
\end{proof}

\section{Plane Wave Solution}

Now we can turn to the task of determining some particular
solutions of equation (\ref{3.funlooschequ}), or better of its stationary
version (\ref{5.stastefunwavequ}).
\begin{props}[Plane Wave Solution]\spbcorr{}.\\
    The Plane Wave
    \bea
        \Phi _{\mathrm{p}} \qtq{C}
        & = &
        {\Nc}
        \exp \left\{
                 \frac{i}{2}
                 \oint _{C}
                     Q _{\mu \nu}
                     Y ^{\mu} d Y ^{\nu}
             \right\}
        \nonumber\\
        & = &
        {\Nc}
        \exp \left\{
                 \frac{i}{2}
                 \oint _{\Gamma \equiv \Sf ^{1}} ds\,
                     Q _{\mu \nu} \tst
                     Y ^{\mu} \tst
                     \frac{d Y ^{\nu} \tst}
                          {d s}
             \right\}
    \label{5.plawavcooexp}
    \eea
    is a solution of equation {\rm (\ref{5.stastefunwavequ})}
    if the {\sl Boundary Area Momentum} $Q ^{\mu \nu} \tst$
    satisfies the classical dispersion relation. $\Nc$ is a normalization
    constant.
\end{props}
\begin{proof}
We need to evaluate the second functional variation of $\Phi _{\mathrm{p}}
\qtq{C}$
corresponding to the addition of two petals to $C$, say at the point
$\bar{Y} ^{\mu} = Y ^{\mu} \ttt{\bar{s}}$.
The first variation of $\Psi _{\mathrm{p}} \qtq{C}$
can be obtained from Eq.(\ref{5.plawavcooexp})
\beq
    \delta _{\mathrm{p}} \Phi \qtq{C}
    =
    \frac{1}{2}
    Q _{\mu \nu} \ttt{\bar{s}}
    \delta Y ^{\mu \nu} \ttt{\bar{s}}
    \Phi _{\mathrm{p}} \qtq{C}
\quad .
\label{5.plawavfirvar}
\eeq
Reopening the loop $C$ at the same contact point and adding a second
infinitesimal loop, we arrive at the second variation of
$ \Phi _{\mathrm{p}} \qtq{C}$,
\beq
    \delta ^{2}
    \Phi _{\mathrm{p}} \qtq{C}
    =
    \frac{1}{4}
    Q _{\mu \nu} \ttt{\bar{s}}
    Q _{\rho \tau} \ttt{\bar{s}}
    \delta Y ^{\mu \nu} \ttt{\bar{s}}
    \delta Y ^{\rho\tau} \ttt{\bar{s}}
    \Phi _{\mathrm{p}} \qtq{C}
\quad .
\label{varii}
\eeq
Then, Eq.(\ref{5.plawavcooexp}) solves equation (\ref{3.funlooschequ}) if
\beq
    E
    =
    \frac{1}{4 m ^{2}}
    \norme{\Gamma}
    \oint _{\Gamma} d l \left( s \right)
	Q _{\mu \nu} \left( s \right)
        Q ^{\mu \nu} \left( s \right)
\quad ,
\eeq
which, as claimed in the proposition, is exactly
the classical dispersion relation between {\sl String} energy
and momentum.
\end{proof}

\begin{props}[Boundary Momentum Operator Eigenstate]\spbcorr{}.\\
    The wave functional {\rm (\ref{5.plawavcooexp})} represents an
    eigenstate of the loop total momentum operator with
    eigenvalue $\bar{Q} ^{\mu \nu} = \expect{Q ^{\mu \nu} \qtq{C}}$,
    i.e. the momentum average of the loop:
    \beq
        \bar{Q} ^{\mu \nu} \qtq{C}
        \dfn
        \expect{Q ^{\mu \nu} \qtq{C}}
        =
        \norme{\Gamma}
        \oint _{\Gamma} d l \tst Q _{\mu \nu} \tst
    \quad .
    \label{5.avemom}
    \eeq
\end{props}
\begin{proof}
We have just to apply the total loop momentum operator
(\ref{3.totloomomope}) to our plane wave
$$
    \Phi  _{\mathrm{p}} \qtq{C}
    =
    {\Nc}
    \exp \left\{
             \frac{i}{2}
             \oint _{C}
                 Q _{\mu \nu}
                 Y ^{\mu} d Y ^{\nu}
         \right\}
$$
to get
\bea
    i
    \norme{\Gamma}
    \oint _{\Gamma} d l \tst
        \frac{\delta \Phi _{\mathrm{p}} \qtq{C}}
             {\delta \sigma ^{\mu \nu} \tst}
    & = &
    \left\{
        \norme{\Gamma}
        \oint _{\Gamma} d l \tst
	    Q _{\mu \nu} \tst
    \right\}
    \Phi _{\mathrm{p}} \qtq{C}
    \nonumber \\
    & = &
    \bar{Q} ^{\mu \nu} \qtq{C}
    \Phi _{\mathrm{p}} \qtq{C}
\label{ploop} ,
\eea
where, $\bar{Q} ^{\mu \nu} \qtq{C}$ is exactly the momentum
$Q ^{\mu \nu} \tst$ averaged over the loop, as defined in equation
(\ref{5.avemom}) above.
\end{proof}

Having determined a set of solutions to the stationary part of equation
(\ref{3.funlooschequ}), the complete solutions describing the
Quantum evolution
from an initial state
$
    \Psi \qtq{C _{0}, 0}
    =
    \Phi _{\mathrm{p}} \qtq{C _{0}}
$
($\Phi _{\mathrm{p}} \qtq{C _{0}}$ being
the Plane Wave solution determined above)
to a final state $\Psi \qtq{C ; A}$ can be obtained
by means of the amplitude (\ref{result}), using equation (\ref{3.prowav})
\bea
    \Psi \left[ C ; A \right]
    & = &
    \sum _{C _{0}}
        K \left [ C , C _{0} ; A \right]
        \Psi \left[ C _{0} ; 0 \right]
    \nonumber\\
    & = &
    \left(
        \frac{m ^{2}}{2 i \pi A}
    \right) ^{3/2}
    \funint{Y ^{\alpha \beta} \ttt{C _{0}}}
    \exp
    \left [
        \frac{i m ^{2}}{4 A}
        \left(
            Y ^{\mu \nu} \qtq{C}
            -
            Y ^{\mu \nu} \qtq{C}
        \right) ^{2}
    \right ]
    \Phi _{\mathrm{p}} \qtq{C _{0}}
    \nonumber\\
    & = &
    \frac{1}{\left( 2 \pi \right) ^{3/2}}
    \exp
    \frac{i}{2}
    \left [
        \oint _{C}
            Q _{\mu \nu}
            Y ^{\mu}
            d Y ^{\nu}
        -
        \frac{A}{2 m ^{2}}
        \norme{\Gamma}
        \oint _{\Gamma} d l \left( s \right)
            Q _{\mu \nu}
            Q ^{\mu \nu}
    \right]
    \nonumber\\
    & = &
    \frac{1}{\left( 2 \pi \right) ^{3/2}}
    \exp
    \frac{i}{2}
    \left[
        \oint _{C}
            Q _{\mu\nu}
            Y ^{\mu}
            d Y ^{\nu}
            -
            E
            A
    \right]
    \quad .
\label{5.plawavfulsol}
\eea
As one would expect, the solution (\ref{5.plawavfulsol}) represents a
``monochromatic {\sl String} wave train'' extending all over loop space.

\section{Gaussian Loop Wave--Packet}

The quantum state represented by equation (\ref{5.plawavfulsol})
is completely de--localized in loop space, which means that all the {\sl String}
 shapes are equally probable, or that the {\sl String} has no definite shape at all.
Thus, even though the wave functional
(\ref{5.plawavfulsol}) is a solution of
equation (\ref{3.funlooschequ}), it does not have an immediate physical
interpretation: at most, it can be used to describe a flux in
loop space rather than to describe a single physical object.
Physically acceptable one--{\sl{}String} states are obtained by a suitable
superposition of ``elementary'' plane wave solutions. The quantum
state closest to a classical {\sl String} will be described by a
{\it Gaussian Wave--Packet}.
\begin{props}[Gaussian Wave--Packet Solution]\spbcorr{}.\\
    The functional
    \beq
        \Phi _{\mathrm{G}} \qtq{C}
        =
        \left [
            \frac{1}{2 \pi \left ( \Delta \sigma \right) ^{2}}
        \right] ^{3/4}
        \! \! \! \! \! \! \!
        \exp
        \left(
            \frac{i}{2}
            \oint _{C}
                Y ^{\mu}
                d Y ^{\nu}
                Q _{\mu \nu}
        \right)
        \exp
        \left[
            -
            \frac{1}{4 \left( \Delta \sigma \right) ^{2}}
            \left(
                \oint _{C}
                    Y ^{\mu}
                    d Y ^{\nu}
            \right) ^{2}
        \right]
    \quad ,
    \label{5.gauloowav}
    \eeq
    is a gaussian loop wave function solution of the Functional
    Stationary Scr\"oedinger equation {\rm (\ref{3.funlooschequ})}.
    Its areal evolution spreads
    throughout loop space and its width $\Delta \sigma$ broadens
    as
    $$
        \Delta \sigma \ttt{A}
        =
        \Delta \sigma
        \left [
            1 + \frac{A ^{2}}{4 m ^{4} \ttt{\Delta \sigma} ^{4}}
        \right ] ^{1/2}
    $$
\end{props}
\begin{proof}
By inserting Eq.(\ref{5.gauloowav}) into Eq.(\ref{3.prowav}), and
integrating out $Y ^{\mu \nu} \qCq$, we find
\bea
    & & \esci \esci
    \Psi \left [ C ; A \right ]
    =
    \left[
        \frac{1}{2 \pi \left( \Delta \sigma \right) ^{2}}
    \right ] ^{3/4}
    \frac{1}
         {
          \left(
             1
             +
             i
             A / m ^ {2}
             \left( \Delta \sigma \right) ^{2}
          \right) ^{3/2}
         }
         \exp
         \left\{
             \frac{1}
                  {
                   \left(
                       1
                       +
                       i
                       A / m ^ {2}
                       \left( \Delta \sigma \right) ^{2}
                   \right)
                  }
        \right .
        \cdot
        \nonumber\\
        & & \cdot
        \left .
            \left [
                -
                \frac{1}{4 \left( \Delta \sigma \right) ^{2}}
                Y ^{\mu \nu} \left [ C \right ]
                Y _{\mu \nu} \left [ C \right ]
                +
                \frac{i}{2}
                \oint _{C}
                    Y ^{\mu}
                    d Y ^{\nu}
                    Q _{\mu \nu} \left( x \right)
                -
                \frac{i A}{4 m ^{2}}
                Q _{\mu \nu} \left [ C \right ]
                Q ^{\mu \nu} \left [ C \right ]
            \right ]
        \right\}
        \quad .
\label{5.gauloowavfin}
\eea
The wave functional represented by equation (\ref{5.gauloowavfin})
spreads throughout loop space in conformity with the laws of quantum
mechanics. In particular, the center of the Wave--Packet moves
according to the stationary phase principle, i.e.
\beq
    Y ^{\mu \nu} \left[ C \right ]
    -
    \frac{A}{m ^{2}}
    Q ^{\mu \nu} \left [ C \right ]
    =
    0
\eeq
and the width broadens as $A$ increases
\beq
    \Delta \sigma
    \left( A \right)
    =
    \Delta \sigma
    \left(
        1
        +
        A ^{2} / 4 m ^{4}
        \left( \Delta \sigma \right) ^{4}
    \right) ^{1/2}
\quad .
\eeq
\end{proof}

Some comment are in order to clarify the physical meaning of the various
quantities. $\Delta \sigma$ represents the width,
or position uncertainty in loop space,
corresponding to an uncertainty in the physical shape of the loop.
Thus, as discussed previously, $A$ represents a measure of the ``timelike''
distance between the initial and final {\sl String} loop. Then,
$m ^{2} \left( \Delta \sigma \right) ^{2}$
represent the wavepacket mean life.
As long as $A \ll m ^{2} \left( \Delta \sigma \right) ^{2}$,
the wavepacket maintains its original width $\Delta\sigma$.
However, as  $A$ increases with respect to
$m ^{2} \left( \Delta \sigma \right) ^{2}$, the
wavepacket becomes  broader and the initial {\sl String}
shape decays in the background space.
Notice that, for a sharp  initial Wave--Packets, i.e., for
$\left( \Delta \sigma \right) \ll 2 \pi \alpha' $,
the shape shifting process is more ``rapid''
than for larger Wave--Packets. Hence, {\sl Strings} with a
well defined initial shape will sink faster into the sea of
quantum fluctuations than broadly defined {\sl String} loops.

%% file: chap06.tex
\pageheader{}{``Minisuperspace''.}{}
\chapter{``Minisuperspace''}
\label{6.mincha}

\begin{start}
``$\euf{I}$can't even see.\\
How am I supposed to \dots ''\\
``Your eyes deceive you.\\
Don't trust them.''\\
\end{start}

\section{Preliminaries}

We are now going to give a brief exposition of a very particular
setting in which it is possible to analize a specific realization
of the concepts associated to the {\sl Functional Schr\"odinger
Equation} that we derived in chapters \ref{3.strfunqua} and
\ref{4.pbrcha} for the {\sl String} and the $p$-{\sl{}brane}
respectively. The reason
that motivates such an approach is only slightly related to the
clarification of the concepts we already explained in previous
chapters. On the contrary we can see how we can use them, althought
in a very indirect way, to tackle a completely different problem,
which often appears in quantizing geometric theories, namely the
problem of ordering ambiguities. The reason why the treatment we gave
in previous sections can be helpful in this situation is that we
have a problem, even if with  an infinte number of degrees
of freedom, expressed in different formulations. In particular we have
seen that the functional equation can be considered in both, form
(\ref{3.funlooschequcoo}) as well as form
(\ref{3.funlooschequ}). These equations
are again two different ways of writing the equation of motion for
a system which involves an infinte number of degrees of freedom: this
situation is somewhat far beyond the possibilities of any possible
exact computation of the solutions in very general case. Moreover,
as we alread showed in chapter \ref{5.strfunsol}, even the simplest
possible solutions to the {\sl Functional Schr\"odinger
Equation} for the {\sl String} involve some functional steps that
can offer difficulties in grasping the physical meaning of the computations
and obscure some points, which are instead worth to be pointed out.
In this chapter we will thus pass from the general formulation, to a
more particular one, that we will call, by analogy to what is often done
in General Relativity to tackle similar problems, the
{\it Minisuperspace Approximation}. The next section presents the
{\sl String} case, whereas the $p$-{\sl{}brane} one is exposed in the one after
the next.

\section{The String}

We limit ourself to the case with in which we have a
{\it circular} {\sl String} in $2+1$ dimensions; moreover,
since we want to have the simplest possible case, we take the circular
{\sl String} lying in the $1-2$ plane, i.e. without extension in the
$0$ direction. A possible parametrization for the image of the
{\sl Boundary Space} in the {\sl Target Space} $\T$ is thus:
\beq
    Y ^{\mu} (s) = \left(
                         0 ,
                         R \cos \left( 2 \pi s \right) ,
                         R \sin \left( 2 \pi s \right)
                   \right)
\quad ,
\eeq
where $s \in \Sf ^{1}$ is the parameter labeling different points of the
{\sl String}. For convenience we use the representation of $\Sf ^{1}$ consisting
of the closed interval $\qtq{0 , 1} \in \R$ with end point identified.
\begin{nots}[Closed Unit Interval with End Points Identified]\spbcorr{}.\\
    The closed unit real interval with end points identified is denoted by
    \beq
        \Iuno
        \dfn
        \qtq{0 , 1 ^{\equiv 0}}
        \quad .
    \eeq
\end{nots}
It is then possible to derive the expression for
the {\sl Target Space} tangent vector to
the loop $C = Y ^{\mu} \ttt{\Iuno}$.
\begin{props}[Minisuperspace Linear Velocity Vector]\spbcorr{}.\\
    The {\sl Linear Velocity Vector} to the {\sl String} $C$
    is given by
    \beq
        Y^{\prime \mu} \tst = \frac{\partial Y ^{\mu}}{\partial s}
                       = \left(
                               0
                               ,
                               -
                               2 \pi R \sin \left( 2 \pi s \right)
                               ,
                               2 \pi R  \cos \left( 2 \pi s \right)
                         \right)
    \eeq
    and the only nonvanishing {\sl Holographic Coordinate} is
    $Y _{12} \qCq = \pi R ^{2}$.
    Moreover we also quote the result
    \beq
        \ttt{\By '} ^{2}
        =
        4
        \pi ^{2}
        R ^{2}
        \quad ,
    \eeq
    so that the normalization is
    $$
        \norme{\Iuno} = 2 \pi R
        \quad .
    $$
\end{props}
\begin{proof}
Directly from the definition, we can calculate
\beq
    Y ^{\mu \nu} = \oint _{\Iuno} Y ^{\mu} \form{d Y} ^{\nu}
\quad :
\eeq
it is a $3 \times 3$ matrix, totally antisymmetric, so that it has
$3$ independent entries. Two of them vanish because the {\sl String} lies
in the $1-2$ plane, so it does not extend in the $0$ one:
\bea
    Y _{10} & = & \oint _{\Iuno} Y _{1} \form{d Y} _{0}
    \nonumber \\
                 & = & \int _{0} ^{1} ds Y _{1} Y ' _{0}
    \nonumber \\
                 & = & 0
    \\
    Y _{20} & = & \oint _{\Iuno} Y _{2} \form{d Y} _{0}
    \nonumber \\
                 & = & \int _{0} ^{1} ds Y _{2} Y ' _{0}
    \nonumber \\
                 & = & 0
    \quad .
\eea
The only nonzero component is $Y _{12}$
\bea
    Y _{12} & = & \oint _{\Iuno} Y _{1} d Y _{2}
    \nonumber \\
                 & = & \int _{0} ^{1} ds Y _{1} Y ' _{2}
    \nonumber \\
                 & = & \int _{0} ^{1} ds 2 \pi R ^{2}
                                           \cos ^{2} \left( 2 \pi s \right)
    \nonumber \\
                 & = & \pi R ^{2}
\quad .
\eea
Then
the other two results follow quickly: we first compute
the modulus of the vector $Y ^{\prime \mu}$
$$
    Y ^{\prime \mu} \tst
    Y ^{\prime} _{\mu} \tst
    =
    0
    +
    \ttt{2 \pi R} ^{2}
    \sin ^{2} \ttt{2 \pi s}
    +
    \ttt{2 \pi R} ^{2}
    \cos ^{2} \ttt{2 \pi s}
    =
    \ttt{2 \pi R} ^{2}
    =
    4 \pi ^{2} R ^{2}
$$

and then its square root:
$$
    \sqrt{\ttt{\By ' \tst} ^{2}}
    =
    \sqrt{
          Y ^{\prime \mu} \tst
          Y ^{\prime} _{\mu} \tst
         }
    =
    2 \pi R
    \quad .
$$
The integration
over the loop parameter is trivial because the chosen parametrization
implies no parameter dependence in the integrand, so that
$$
    \int _{\Iuno} ds
        \sqrt{\ttt{\By ' \tst} ^{2}}
    =
    \int _{0} ^{1} ds
        2 \pi R
    =
    2 \pi R
    \quad .
$$
\end{proof}
The {\sl Holographic Derivative} can also
be calculated using the last result.
\begin{props}[Minisuperspace Holographic Derivative]\spbcorr{}.\\
The only non--vanishing {\sl Holographic Derivative}
is the one with respect
to $Y _{12}$:
\beq
    \frac{\delta}{\delta Y ^{12}}
    =
    \frac{1}{2 \pi R} \frac{d}{d R}
\quad .
\eeq
\end{props}
\begin{proof}
The result follows by translating the results about the {\sl Holographic
Derivative} in infinitesimal form. The only non--vanishing is the one
related to $Y ^{12}$ so that we first get
$$
    d Y ^{12} = \delta Y ^{12} \tst = 2 \pi R dR
    \quad ,
$$
where thanks to the circular symmetry there is no dependence
from the loop parameter $s$ anymore. Then the functional derivative
is just an ordinary derivative with respect to $R$:
\beq
    \frac{\delta}{\delta Y ^{12} \tst} =
    \frac{d}{d \left( \pi R ^{2} \right)} =
    \frac{1}{2 \pi R} \frac{d}{d R}
\quad ;
\eeq
this is the desired result.
\end{proof}
We can now remember relation (\ref{2.holfunder})
between {\sl Holographic Derivatives}
and {\sl Ordinary Functional Derivatives}:
\beq
    \frac{\delta}{\delta Y ^{\mu} (s)} \sim
    Y ^{\prime \nu} (s) \frac{\delta}{\delta Y ^{\mu \nu} (s)}
    \quad .
\eeq

\begin{props}[Ordinary Functional Derivatives]\spbcorr{}.\\
The {\sl Ordinary Functional Derivatives} with respect to the loop shape
$Y ^{\mu}$ are
\bea
    \frac{\delta}{\delta Y ^{1} (s)} & \sim &
           \cos \left( 2 \pi s \right) \frac{d}{d R}
    \\
    \frac{\delta}{\delta Y ^{2} (s)} & \sim &
           \sin \left( 2 \pi s \right) \frac{d}{d R}
    \\
    \frac{\delta}{\delta Y ^{0} (s)} & \sim &
           0
\quad ,
\eea
where the symbol $\sim$ is used since the equality holds only for
reparametrization invariant functional in an integrated
way$\,$\footnotemark{}.
\end{props}
\footnotetext{Please see appendix
\ref{D.looderapp} for a more detailed explanation.}
\begin{proof}
We compute these quantities using relation
(\ref{D.funarederrel}) and the expressions for
the {\sl Holographic Derivative} that we derived in the
proposition above. Then we have
\bea
    \frac{\delta}{\delta Y ^{1} (s)} & = &
            Y ^{\prime 0} (s) \frac{\delta}{\delta Y ^{10} (s)} +
            Y ^{\prime 1} (s) \frac{\delta}{\delta Y ^{11} (s)} +
            Y ^{\prime 2} (s) \frac{\delta}{\delta Y ^{12} (s)}
    \nonumber \\
                                     & = &
            0 +
            0 +
            R \left(
                    \sin \left( 2 \pi s \right)
              \right) '
              \frac{1}{2 \pi R}
              \frac{d}{d R}
    \nonumber \\
                                     & = &
           \cos \left( 2 \pi s \right) \frac{d}{d R}
\quad ;
\\
    \frac{\delta}{\delta Y ^{2} (s)} & = &
            Y ^{\prime 0} (s) \frac{\delta}{\delta Y ^{20} (s)} +
            Y ^{\prime 1} (s) \frac{\delta}{\delta Y ^{21} (s)} +
            Y ^{\prime 2} (s) \frac{\delta}{\delta Y ^{22} (s)}
    \nonumber \\
                                     & = &
            \sin \left( 2 \pi s \right) \frac{d}{d R}
\quad ;
\\
    \frac{\delta}{\delta Y ^{0} (s)} & = &
            Y ^{\prime 0} (s) \frac{\delta}{\delta Y ^{00} (s)} +
            Y ^{\prime 1} (s) \frac{\delta}{\delta Y ^{01} (s)} +
            Y ^{\prime 2} (s) \frac{\delta}{\delta Y ^{02} (s)} =
            0
\quad .
\eea
These are the desired results.
\end{proof}
As we will see in the sequel, the functional derivatives are in some
sense misleading objects. As explained in appendix \ref{D.looderapp}
this objects are not reparametrization invariant: in our opinion
reparametrization invariance is an important property that should
be preserved by any type of physically meaningful operator. Thus we
can trace to the non invariance of the functional derivatives the
following (and {\bf negative}) proposition.
\begin{props}[Ordering Problems]\spbcorr{}.\\
The functional laplacian computed in terms of the
{\sl Holographic Derivatives}
is not consistent with the same operator computed in terms of the
{\sl Ordinary Functional Derivatives}.
\end{props}
\begin{proof}
We first calculate the second order operator which is the
functional laplacian, in terms of the
functional derivatives; we find
\bea
    \frac{\delta ^{2}}{\delta Y ^{\mu} (s) \delta Y _{\mu} (s)}
    & = &
        \frac{\delta ^{2}}{\delta Y ^{0} (s) \delta Y _{0} (s)} +
        \frac{\delta ^{2}}{\delta Y ^{1} (s) \delta Y _{1} (s)} +
        \frac{\delta ^{2}}{\delta Y ^{2} (s) \delta Y _{2} (s)}
    \nonumber \\
    & = &
        \cos \left( 2 \pi s \right)
        \frac{d}{dR} \left(
                           \cos \left( 2 \pi s \right) \frac{d}{d R}
                     \right)
      + \sin \left( 2 \pi s \right)
        \frac{d}{dR} \left(
                           \sin \left( 2 \pi s \right) \frac{d}{d R}
                     \right)
    \nonumber \\
    & = &
    \frac{d ^{2}}{d R ^{2}}
    \quad ,
    \label{6.ordprofirres}
\eea
so that the corresponding integral operator is
\beq
    \int _{0} ^{1} \frac{ds}{\sqrt{\ttt{\By '} ^{2}}}
                   \frac{\delta ^{2}}{\delta Y ^{\nu} (s) \delta Y _{\nu} (s)}
    =
    \int _{0} ^{1} \frac{ds}{2 \pi R} \frac{d ^{2}}{d R ^{2}}
    =
    \frac{ds}{2 \pi R} \frac{d ^{2}}{d R ^{2}}
\quad .
\eeq
Now we will compute the same quantity using
the {\sl Holographic Derivatives}; note that since we have
relation (\ref{2.holfunder}) between functional
and holographic derivatives, it might be that ordering ambiguities could
arise in defining this second oreder operator, because of the
$Y ^{\prime \mu} \tst$ factor. We have
\bea
    \frac{1}{2} \int _{0} ^{1} ds
                    \sqrt{\ttt{\By '} ^{2}}
                    \frac{\delta ^{2}}
                         {\delta Y ^{\nu \tau} (s) \delta Y _{\nu \tau} (s)}
    & = &
    \frac{1}{2} \int _{0} ^{1} ds
                    \sqrt{\ttt{\By '} ^{2}}
                    2
                    \frac{1}{2 \pi R}
                    \frac{d}{d R}
                    \left(
                          \frac{1}{2 \pi R} \frac{d}{d R}
                    \right)
    \nonumber \\
    & = & \int _{0} ^{1} \frac{ds}{2 \pi} \frac{d}{d R}
                         \left( \frac{1}{R} \frac{d}{d R} \right)
    \nonumber \\
    & = & \frac{1}{2 \pi} \frac{d}{d R}
          \left( \frac{1}{R} \frac{d}{d R} \right)
\quad .
\eea
This result is in disagreement with (\ref{6.ordprofirres}),
because we have
$$
    \int _{0} ^{1} \frac{ds}{\sqrt{\ttt{\By '} ^{2}}}
    \frac{\delta ^{2}}{\delta Y ^{\mu} \tst \delta Y _{\mu} \tst}
    =
    \frac{1}{2}
    \int _{0} ^{1} ds \sqrt{\ttt{\By '} ^{2}}
    \frac{\delta ^{2}}{\delta Y ^{\mu \nu} \tst \delta Y _{\mu \nu} \tst}
$$
but of course
$$
    \frac{ds}{2 \pi R} \frac{d ^{2}}{d R ^{2}}
    \neq
    \frac{1}{2 \pi} \frac{d}{d R}
    \left( \frac{1}{R} \frac{d}{d R} \right)
\quad .
$$
\end{proof}
Some light into this problem can be given by the observation \cite{hoso}
that the functional differential operator does not transform
{\it covariantly} under a reparametrization of the loop. We thus make
the following absumption:\\[2mm]
{\bf Axiom (Quantization Procedure)}: {\it When we quantize a Theory
having some invariance$\,$\footnote{Reparametrization invariance is the
interesting case for the Theory of Extended Objects and General Relativity}
we have to promote to the role of Quantum Operators only quantities
that respect the invariance of the Theory and transform covariantly}.

This can be euristically understood, since good quantum numbers are
to be associated with properties that do not change under a symmetry
transformation, and so must then be the corresponding operators. In the
case of a symmetry under reparametrization we can easily construct
invariant functional quantities from covariant ones, simply by integration
with the natural ``volume element'':
$$
    \oint _{\Gamma} ds \sqrt{\ttt{\By '} ^{2}}
    \quad ;
$$
the analogy with Quantum Gravity is transparent at this stage.

Going back to our particular case we see that a functional differential
operator that transforms {\it covariantly}, is
\beq
    \frac{\Delta}{\Delta Y ^{\mu} \tst}
    =
    \frac{1}{\sqrt{\ttt{\By '} ^{2}}}
    \frac{\delta}{\delta Y ^{\mu} \tst}
    \quad .
    \label{6.covopedef}
\eeq
Moreover we now observe that all the procedure we performed in chapter
\ref{3.strfunqua} to get the functional equation (\ref{3.funlooschequcoo})
can be repeated using the operator of equation (\ref{6.covopedef})
and getting the following result.
\begin{props}[Functional Schr\"odinger Equation: Covariant Formalism]\spbcorr{}.\\
    In terms of the functional differential operator
    {\rm (\ref{6.covopedef})} the functional
    equation {\rm (\ref{3.funlooschequcoo})}
    can be rewritten as
    \beq
        -
        \frac{1}{2 m ^{2}}
        \norme{\Gamma \approx \Sf ^{1}}
        \oint _{\Gamma \approx \Sf ^{1}} d s \sqrt{\ttt{\By '} ^{2}}
            \frac{\Delta ^{2} \Psi \left [ C ; A \right]}
                 {
                  \Delta Y ^{\mu} \left( s \right)
                  \Delta Y _{\mu} \left( s \right)
                 }
        =
        i
        \frac{\partial \Psi \left[ C ; A \right]}
             {\partial A}
        \quad .
    \label{3.funloocovschequ}
    \eeq
\end{props}
\begin{proof}
To derive the desired result we recall the classical dispersion relation
written in integrated form (\ref{2.strhamjacpreint}). If we define
$$
    \tilde{q} ^{\mu} \tst
    =
    \frac{q ^{\mu} \tst}{\sqrt{{\By ' \tst} ^{2}}}
    =
    \frac{1}{\sqrt{\ttt{\By '} ^{2} \tst}}
    \frac{\delta S _{\mathrm{Red.}}}{\delta Y ^{\mu} \tst}
    =
    \frac{\Delta S _{\mathrm{Red.}}}{\Delta Y ^{\mu} \tst}
$$
i.e. a sort of normalized {\sl Boundary Linear Momentum} that equation
can be rewritten as
\beq
    \frac{1}{2 m ^{2}}
    \norme{\Gamma}
    \oint _{\Gamma \approx \Sf ^{1}} ds
        \sqrt{\ttt{\By '} ^{2} \tst}
        \tilde{q} ^{\mu} \tst
        \tilde{q} _{\mu} \tst
    =
    E
    \quad .
    \label{6.enebalequ}
\eeq
Then the first and second functional derivatives of the kernel
(cf. equations (\ref{3.firfunderpro}) and (\ref{A.secfundermomave}))
expressed in terms of the covariant functional derivative are
\bea
    \frac{\Delta}{\Delta Y ^{\mu} \tst}
    K \left [
          \By \tst
          ,
          \By _{0} \tst
          ;
          A
      \right ]
    & = &
    \frac{i}{\hbar}
    \int _{\By _{0} \tst} ^{\By \tst}
    \funinte{\mu \tst}
            {\Bzeta _{0} \tst}
            {\Bzeta \tst}
        \frac{q _{\mu}}{\sqrt{\ttt{\By ' \tst} ^{2}}}
        \exp \ttt{{\frac{i}{\hbar} S _{\mathrm{Red.}}}}
    \nonumber \\
    & = &
    \frac{i}{\hbar}
    \int _{\By _{0} \tst} ^{\By \tst}
    \funinte{\mu \tst}
            {\Bzeta _{0} \tst}
            {\Bzeta \tst}
        \tilde{q} _{\mu}
        \exp \ttt{{\frac{i}{\hbar} S _{\mathrm{Red.}}}}
    \\
    \frac{\Delta ^{2}}
         {\Delta Y ^{\mu} \tst \Delta Y _{\mu} \tst}
    K \left [
          \By \tst
          ,
          \By _{0} \tst
          ;
          A
      \right ]
    & = &
    -
    \frac{1}{\hbar ^{2}}
    \int _{\By _{0} \tst} ^{\By \tst}
    \funinte{\mu \tst}
            {\Bzeta _{0} \tst}
            {\Bzeta \tst}
        \tilde{q} _{\mu} \tilde{q} ^{\mu}
        \exp \ttt{\frac{i}{\hbar} S _{\mathrm{Red.}}}
    \nonumber \\
    & \equiv &
    -
    \frac{1}{\hbar ^{2}}
    \overline{\tilde{q} ^{\mu} \tilde{q} _{\mu}}
    \label{6.tiqtiqseccovfun}
    \quad .
\eea
Now the last form of the Classical Dispersion Relation, (\ref{6.enebalequ}),
can be interpreted as the evolution equation for the mean values of
quantum operators, so that substituting in it (\ref{6.tiqtiqseccovfun})
as well as (\ref{2.areder}), which is unchanged,
we get
\beq
    -
    \frac{1}{2 m ^{2}}
    \norme{\Gamma \approx \Sf ^{1}}
    \oint _{\Gamma \approx \Sf ^{1}} d s \sqrt{\ttt{\By '} ^{2}}
        \frac{\Delta ^{2} \Psi \left [ C ; A \right]}
             {
              \Delta Y ^{\mu} \left( s \right)
              \Delta Y _{\mu} \left( s \right)
             }
    =
    i
    \frac{\partial \Psi \left[ C ; A \right]}
         {\partial A}
    \quad ,
\eeq
the desired result.
\end{proof}
Moreover using this form of the equation all ordering ambiguities disappear
since
\begin{props}[Solution of Ordering Ambiguities]\spbcorr{}.\\
    The operator
    $$
        \frac{\Delta}{\Delta Y ^{\mu} \tst}
    $$
    in the minisuperspace approximation is given by
    $$
        \frac{1}{2 \pi R} \frac{d}{d R}
    $$
    so that the second order operator
    $$
        \frac{1}{2 m ^{2}}
        \norme{\Iuno}
        \oint _{\Iuno} d s \sqrt{\ttt{\By '} ^{2}}
            \frac{\Delta ^{2} \Psi \left [ C ; A \right]}
                 {
                  \Delta Y ^{\mu} \left( s \right)
                  \Delta Y _{\mu} \left( s \right)
                 }
    $$
    coincides with
    $$
        \frac{1}{2 m ^{2}}
        \norme{\Iuno}
        \oint _{\Iuno} d s \sqrt{\ttt{\By '} ^{2}}
            \frac{\delta ^{2} \Psi \left [ C ; A \right]}
                 {
                  \delta Y ^{\mu \nu} \left( s \right)
                  \delta Y _{\mu \nu} \left( s \right)
                 }
        \quad .
    $$
\end{props}
\begin{proof}
Since we have
$$
    \frac{\Delta}{\delta Y ^{\mu} \tst}
    =
    \left(
        0
        ,
        \frac{\cos \ttt{2 \pi s}}{2 \pi R}
        \frac{d}{d R}
        ,
        \frac{\sin \ttt{2 \pi s}}{2 \pi R}
        \frac{d}{d R}
    \right)
$$
then we can compute
\bea
    \frac{\Delta ^{2}}{\Delta Y ^{\mu} \tst \Delta Y _{\mu} \tst}
    & = &
    \frac{\Delta ^{2}}{\Delta Y ^{0} \tst \Delta Y _{0} \tst}
    +
    \frac{\Delta ^{2}}{\Delta Y ^{1} \tst \Delta Y _{1} \tst}
    +
    \frac{\Delta ^{2}}{\Delta Y ^{2} \tst \Delta Y _{2} \tst}
    \nonumber \\
    & = &
    0
    +
    \cos ^{2} \ttt{2 \pi s}
    \frac{1}{2 \pi R}
    \frac{d}{d R}
    \left(
        \frac{1}{2 \pi R}
        \frac{d}{d R}
    \right)
    +
    \sin ^{2} \ttt{2 \pi s}
    \frac{1}{2 \pi R}
    \frac{d}{d R}
    \left(
        \frac{1}{2 \pi R}
        \frac{d}{d R}
    \right)
    \nonumber \\
    & = &
    \frac{1}{4 \pi ^{2} R}
    \frac{d}{d R}
    \left(
        \frac{1}{R}
        \frac{d}{d R}
    \right)
    \nonumber
    \quad .
\eea
At the same time we have
\beq
    \frac{\delta ^{2}}{\delta Y ^{\mu \nu} \tst \Delta Y _{\mu \nu} \tst}
    =
    \frac{1}{4 \pi ^{2} R}
    \frac{d}{d R}
    \left(
        \frac{1}{R}
        \frac{d}{d R}
    \right)
    \label{6.minholderlap}
\eeq
and comparing the last two results we conclude the proof.
\end{proof}
Now we have a completely non--ambiguous expression for
the functional Laplacian, i.e. the fundamental quantity in determining
the functional wave equation for the {\sl String} in the minisuperspace
approximation. We also hope, at least, to have brought to the light
some problems that can be present in the functional formulation
and that make it absolutely non-trivial. We can now conclude this
section:
\begin{defs}[Minisuperspace Stationary Schr\"odinger Equation]\spbcorr{}.\\
    The minisuperspace approximation to the {\sl Stationary
    States Functional Wave Equation}
    for a circular {\sl String} lying in the $1-2$ plane of the $2+1$
    dimensional {\sl Target Space} is the following differential
    equation:
    \beq
        \frac{1}{2 m ^{2}}
        \frac{1}{2 \pi R} \frac{d}{d R}
        \left(
            \frac{1}{2 \pi R} \frac{d \Phi \ttt{R}}{d R}
        \right)
        =
        E
        \Phi \ttt{R}
        \quad .
    \label{6.minstrequ}
    \eeq
\end{defs}
\begin{proof}
This result simply follows substituting, the result for the
{\sl Holographic Derivative} (say, equation (\ref{6.minholderlap})) into
the {\sl Stationary States Schr\"odinger Equation}
(\ref{5.stastefunwavequ}). Of course now the loop $C$ is described
only by its radius $R$ so that functionals of $C$ become ordinary
functions of the real variable $R$ in the same way as {\sl Functional}
and {\sl Holographic Derivative} converted to ordinary derivatives
with respect to $R$.
\end{proof}

\section{The $p$-brane}
\label{6.minpbrequsec}

We will here shortly give the equation corresponding to (\ref{6.minstrequ})
in the more general case of a $p$-{\sl{}brane} living in a $D + 1 > p+1$
dimensional {\sl Target Space}. The main reason for this short digression
is to translate the abstract formulation given in the previous chapters,
and its very simple representation presented in the section above,
into a slightly more complicated case, in which we can see general
ideas at work but not in the simplest possible case. To avoid spending
too much paper for this part, we rely on a geometrical reinterpretation
of the results of previous section. In particular the circular
{\sl String} is a $1$-sphere, $\Sf ^{1}$, and the normalization integral is
the circumpherence, i.e. the {\it surface of the} $1$-sphere. Moreover
since the {\sl String} is in $1-2$ plane, then the areas of its projections
onto the $0-1$ and $0-2$ planes are vanishing, as well as the corresponding
{\sl Holographic Coordinates}, $Y ^{01}$ and $Y ^{02}$.
On the contrary $Y ^{12}$ is the area of the shadow on the $1-2$ plane
of the of the circle, i.e. a disk, so that we obtained the natural
result
$$
    Y ^{12} = \pi R ^{2}
    \quad .
$$
We now turn these considerations into the more general case of a
$p$-{\sl{}brane} in $(D+1)$-dimensions.
\begin{props}[Hyperspherical $p$-brane]\spbcorr{}.\\
    Let us consider an Hyperspherical $p$-{\sl{}brane}, i.e. a
    $p$-{\sl{}brane}
    which is a $p$-sphere, $\Sf ^{p}$, contained in a
    $(p+1)$-dimensional Hypersubspace cooresponding to the
    $1$, $2$, \dots , $p+1$ cartesian coordinates,
    $X ^{(1)}$ , $X ^{(2)}$, \dots{}, $X ^{(p+1)}$. Let us call
    $R$ its radius; the {\sl Stationary States Functional Schr\"odinger
    Equation} for this extended object is given by
    $$
        \frac{\hbar ^{2} p}{2 \rp}
        \left [
            \frac{\Gamma \ttt{\frac{p+1}{2}}}
                 {\ttt{2 \pi} ^{\frac{p+1}{2}}}
        \right ]
        \frac{1}{R ^{p}}
        \frac{d}{d R}
        \left(
            \frac{1}{R ^{p}}
            \frac{d \Psi ^{(\mathrm{p})}}{d R}
        \right)
        =
        E
        \Psi ^{(\mathrm{p})}
        \quad .
    $$
\end{props}
\begin{proof}
We will derive previous equation using the natural generalizations
of the concept of circunference of a circle and area of a disk. In
particular in the case of a $p$-{\sl{}brane} the normalization factor
$$
    \oint _{\Sf ^{p}} d ^{p} \Bs
        \sqrt{\ttt{\By ' \ttt{\Bs}} ^{2}}
$$
is the generalization of the length of a one sphere,
$\Sf ^{1}$: thus it is
the hypersurface of a $p$ sphere, $\Sf ^{p}$, i.e.
\beq
    \oint _{\Sf ^{p}} d ^{p} \Bs
        \sqrt{\ttt{\By ' \ttt{\Bs}} ^{2}}
    =
    \frac{\ttt{2 \pi} ^{\frac{p+1}{2}}}
         {\Gamma \ttt{\frac{p+1}{2}}}
    R ^{p}
    \quad .
    \label{6.pbrminnor}
\eeq
We can now safely turn to the holographic coordinates; since we are
considering our $p$-{\sl{}brane} contained in the hyperplane defined by
the coordinates with indices $1$, \dots{}, $p+1$, and the
{\sl Holographic Coordinates} have $p+1$ totally antisymmetric
indices varying in the range $1$, \dots{}, $D$, we have that the only
non-zero components are those indexed by a permutation of
$1$, \dots{}, $p+1$; there are $\fact{p+1}$ possible combination of the
indices and thus $\fact{p+1}$ non vanishing $Y ^{\multind{\mu}{0}{p}}$.
Moreover their value is just the volume of $\Sf ^{p}$, i.e.
we get for all the nonvanishing {\sl Holographic Coordinates}:
$$
    Y ^{\gamma \ttt{1} \dots \gamma \ttt{p+1}}
    =
    \frac{\ttt{p+1} \ttt{2 \pi} ^{\frac{p+1}{2}}}
         {\Gamma \ttt{\frac{p+1}{2}}}
    R ^{p+1}
    \quad , \qquad
    \forall \gamma \in \euf{S} _{p+1}
    \quad ,
$$
$\euf{S}$ being the symmetric group of order $p+1$.
From the above we directly get the expression for the
nonvanishing {\sl Hypervolume Derivatives}, which are
\beq
    \frac{\delta}
         {\delta Y ^{\gamma \ttt{1} \dots \gamma \ttt{p+1}}}
    =
    \frac{\Gamma \ttt{\frac{p+1}{2}}}
         {\ttt{2 \pi} ^{\frac{p+1}{2}} R ^{p}}
    \frac{d}{d R}
    \quad , \qquad
    \forall \gamma \in \euf{S} _{p+1}
    \quad .
    \label{6.pbrminareder}
\eeq
Substituting then equations (\ref{6.pbrminareder}) and (\ref{6.pbrminnor})
into the stationary equation derived from (\ref{4.pbrhamjacreshol}),
we get the desired result.
\end{proof}

%% file: chap07.tex
\pageheader{}{Fractal Strings.}{}
\chapter{Fractal Strings}
\label{7.fractal}

\begin{start}
``$\euf{Y}$ou're going to find\\
that many of the truths we cling\\
do depend greatly\\
on our own point of view.''\\
\end{start}

\section{The Shape Uncertainty Principle}

In this section we discuss the new form that the Uncertainty
Principle takes in the functional Theory of {\sl String} loops.
\begin{nots}[Fourier Transform Related Width]\spbcorr{}.
\label{7.dePdesrelnot}
    The width in the Momentum Representation $\Delta Q$ is related to the
    width in {\sl Holographic Coordinates} Representation by
    \beq
        \Delta Q = \frac{1}{2 \Delta \sigma}
        \quad ,
        \label{7.dePdesrel}
    \eeq
    qhich with our convention is the usual relation between (Functional)
    Fourier transformed quantities.
\end{nots}

The Uncertainty Principle in Quantum Mechanics gives a definite
lower bound to the accuracy about the simultaneous
knowledge of the position and momentum of a pointlike
particle. In our description the role of the position
is played by the {\sl Holographic Coordinates} of the loop,
$Y ^{\mu \nu} \qtq{C}$, and the role of the momentum is played by
the {\sl Boundary Area Momentum} $Q ^{\mu \nu} \qtq{C}$. We first prove
a useful
\begin{props}[Holographic/Area Functional Fourier
              Transform]\spbcorr{}.\\
    Given the Gaussian Wave Packet {\rm (\ref{5.gauloowav})}
    in the ``Holographic Coordinates Representation'', which is
    \beq
        \Phi _{\mathrm{G}} \qtq{C}
        =
        \left [
            \frac{1}{2 \pi \left ( \Delta \sigma \right) ^{2}}
        \right] ^{3/4}
        \exp
        \left(
            \frac{i}{2}
            \oint _{C}
                Y ^{\mu}
                d Y ^{\nu}
                Q _{\mu \nu}
        \right)
        \exp
        \left[
            -
            \frac{1}{4 \left( \Delta \sigma \right) ^{2}}
            \left(
                \oint _{C}
                    Y ^{\mu}
                    d Y ^{\nu}
            \right) ^{2}
        \right]
    \quad ,
    \label{7.gauloowav}
    \eeq
    its functional Fourier transform, i.e. the corresponding wave functional
    in the ``Area Momentum Representation'' is
    \beq
        \tilde{\Phi} _{\mathrm{G}} \qtq{Q}
        =
        \frac{1}
             {
              \left [
                  \pi \ttt{\Delta Q} ^{2}
              \right] ^{3/4}
             }
             \exp
             \left [
                 -
                 \frac{1}
                      {4
                       \ttt{\Delta Q} ^{2}
                      }
                 \norm
                 \oint _{C} d l \tst
                 \left(
                     Q _{\mu \nu} \left( s \right)
                     -
                     K _{\mu \nu}
                 \right) ^{2}
             \right]
    \quad ,
    \label{7.gauloowavmomrep}
    \eeq
where the relation between $\Delta \sigma$ and $\Delta Q$
is given in notation {\rm \ref{7.dePdesrelnot}}.
\end{props}
\begin{proof}
The Functional Fourier Transform of expression (\ref{7.gauloowav})
is
\bea
    \tilde{\Phi} _{\mathrm{G}} \qtq{C}
    & \! \! \! = \! \! \! &
    \frac{1}{\ttt{2 \pi} ^{3/2}}
    \funint{Y ^{\mu \nu} \qtq{C}}
        \Phi _{\mathrm{G}} \qtq{C}
        \exp \left [
                 -
                 \frac{i}{2}
                 \oint _{C}
                     Y ^{\mu}
                     d Y ^{\nu}
                     K _{\mu \nu}
             \right ]
    \nonumber \\
    & \! \! \! = \! \! \! &
    \frac{1}{\ttt{2 \pi} ^{3/2}}
    \left [
        \frac{1}{2 \pi \ttt{\Delta \sigma} ^{2}}
    \right ] ^{3/4}
    \funint{Y ^{\mu \nu} \qtq{C}}
        \exp \left [
                 \frac{i}{2}
                 \oint _{C}
                     Y ^{\mu}
                     d Y ^{\nu}
                     \ttt{Q _{\mu \nu} - K _{\mu \nu}}
             \right ]
    \cdot
    \nonumber \\
    & & \qquad \qquad \cdot
        \exp \left\{
                 -
                 \frac{\ttt{\Delta \sigma} ^{2}}{4}
                 \left [
                     \oint _{C}
                         Y ^{\mu} d Y ^{\nu}
                 \right ] ^{2}
             \right\}
    \nonumber
\eea
and the final result can be computed performing the path integration:
\bea
    & & \esci \esci
    \frac{1}{\ttt{2 \pi} ^{3/2}}
    \left [
        \frac{1}{2 \pi \ttt{\Delta \sigma} ^{2}}
    \right ] ^{3/4}
    \funint{Y ^{\mu \nu} \qtq{C}}
        \exp \left [
                 \frac{i}{2}
                 \oint _{C}
                     Y ^{\mu}
                     d Y ^{\nu}
                     \ttt{Q _{\mu \nu} - K _{\mu \nu}}
             \right ]
    \cdot
    \nonumber \\
    & & \qquad \qquad \cdot
        \exp \left\{
                 -
                 \frac{\ttt{\Delta \sigma} ^{2}}{4}
                 \left [
                     \oint _{C}
                         Y ^{\mu} d Y ^{\nu}
                 \right ] ^{2}
             \right\}
    \nonumber \\
    & & \qquad =
    \frac{1}{\ttt{2 \pi} ^{3/2}}
    \left [
        \frac{1}{2 \pi \ttt{\Delta \sigma} ^{2}}
    \right ] ^{3/4}
    \funint{Y ^{\mu \nu} \qtq{C}}
    \cdot
    \nonumber \\
    & & \qquad \qquad \cdot
        \exp \left\{
                 \oint _{C} dl(s)
                         Y ^{\mu}
                         Y ^{ \prime \nu}
                        \frac{i \ttt{Q _{\mu \nu} - K _{\mu \nu}}}{2}
                 -
                 \frac{1}{2}
                 \frac{\ttt{\Delta \sigma} ^{2}}{2}
                 \left [
                     \oint _{C} dl(s)
                         Y ^{\mu} Y ^{\prime \nu}
                 \right ] ^{2}
             \right\}
    \nonumber \\
    & & \qquad =
    \frac{1}{\ttt{2 \pi} ^{3/2}}
    \ttt{\frac{\ttt{\Delta \sigma} ^{2}}{2}} ^{3/2}
    \left [
        \frac{1}{2 \pi \ttt{\Delta \sigma} ^{2}}
    \right ] ^{3/4}
    \cdot
    \nonumber \\
    & & \qquad \qquad \cdot
    \funint{Y ^{\mu \nu} \qtq{C}}
        \exp \left\{
                 -
                 \frac{1}{2}
                 \frac{2}{\ttt{\Delta \sigma} ^{2}}
                 \oint _{C} dl(s)
                     \left [
                        \frac{i \ttt{Q _{\mu \nu} - K _{\mu \nu}}}{2}
                     \right ] ^{2}
             \right\}
    +
    \nonumber \\
    & & \qquad =
    \frac{1}{\ttt{2 \pi} ^{3/2}}
    \left [
        \frac{1}{2 \pi \ttt{\Delta \sigma} ^{2}}
    \right ] ^{3/4}
    \! \! \! \!
    \exp \left\{
             \norme{\Gamma}
	     \frac{\ttt{\Delta \sigma} ^{2}}{4}
	     \left [
		 \oint _{C} d l \tst
		     \ttt{Q _{\mu \nu} - K _{\mu \nu}} ^{2}
	     \right ] ^{2}
	 \right\}
    \quad .
\eea
This is again a Gaussian wavepacket, whose ``center of mass''
moves in loop space with a momentum $K _{\mu \nu}$.
\end{proof}
We are now ready to prove the following results:
\begin{props}[Holographic/Area Expectations:
              Gaussian Wavepacket]\spbcorr{}.\\
    For a Gaussian wavepacket of the form {\rm (\ref{5.gauloowav})} in the
    ``Holographic Coordinates Representation'' the expectation values
    of the ``position in loop space'' and of the ``position squared
    in loop space'' are:
    \bea
        \expect{Y ^{\mu \nu} \qtq{C}}
        & = &
        0
        \label{7.sigexp}
        \\
        \expect{Y ^{\mu \nu} \qtq{C} Y _{\mu \nu} \qtq{C}}
        & = &
        3 \ttt{\Delta \sigma} ^{2}
        =
        \frac{3}{2 \ttt{\Delta Q} ^{2}}
        \quad .
        \label{7.sigsigexp}
    \eea
    Moreover the corresponding quantities for the momentum are
    \bea
        \expect{Q ^{\mu \nu} \qtq{C}}
        & = &
        K _{\mu \nu} \qtq{C}
        \label{7.momexp}
        \\
        \expect{Q ^{\mu \nu} \qtq{C} Q _{\mu \nu} \qtq{C}}
        & = &
        \frac{1}{2} K ^{\mu \nu} K _{\mu \nu}
        +
        6 \ttt{\Delta Q} ^{2}
        \quad .
        \label{7.mommomexp}
    \eea
\end{props}
\begin{proof}
To compute this expectation values we have to compute the following
functional integrals respectively,
\bea
    \expect{Y ^{\mu \nu} \qtq{C}}
    & = &
    \funint{Y ^{\lambda \rho} \tst}
        Y ^{\mu \nu} \qtq{C}
        \bigl| \Psi \qtq{C ; A} \bigr| ^{2}
    \\
    \expect{Y ^{\mu \nu} \qtq{C} Y _{\mu \nu} \qtq{C}}
    & = &
    \funint{Y ^{\lambda \rho} \tst}
        Y ^{\mu \nu} \qtq{C}
        Y _{\mu \nu} \qtq{C}
        \bigl| \Psi \qtq{C ; A} \bigr| ^{2}
    \quad ;
\eea
since the dependence from the area $A$ in $\Psi \qtq{C ; A}$ resides
in the exponential, it goes away computing the modulus square so that
at the end we get
\bea
    \expect{Y ^{\mu \nu} \qtq{C}}
    & = &
    \funint{Y ^{\lambda \rho} \tst}
        Y ^{\mu \nu} \qtq{C}
        \bigl | \Phi _{G} \qtq{C} \bigr | ^{2}
    \\
    \expect{Y ^{\mu \nu} \qtq{C} Y _{\mu \nu} \qtq{C}}
    & = &
    \funint{Y ^{\lambda \rho} \tst}
        Y ^{\mu \nu} \qtq{C}
        Y _{\mu \nu} \qtq{C}
        \bigl | \Phi _{G} \qtq{C} \bigr | ^{2}
    \quad ,
\eea
where we can see from (\ref{7.gauloowav}) that
the loop probability density still has a Gaussian form
\bea
    \bigl \vert
        \Phi _{\mathrm{G}} \qtq{C}
    \bigr \vert ^{2}
    & = &
    \left [
        \frac{ \left( \Delta Q \right) ^{2}}{2 \pi}
    \right ] ^{3/2}
    \exp
    \left [
        -
        \frac{\left( \Delta Q \right) ^{2}}{4}
        Y ^{\mu \nu} \left [ C \right ]
        Y _{\mu \nu} \left [ C \right ]
    \right ]
    \nonumber\\
    & \equiv &
    \left [
        \frac{1}{4 \pi \left( \Delta \sigma \right) ^{2}}
    \right ] ^{3/2}
    \exp
    \left[
        -
        \frac{1}{4 \left( \Delta \sigma \right) ^{2}}
        Y ^{\mu \nu} \left [ C \right ]
        Y _{\mu \nu} \left [ C \right ]
    \right]
\label{densc}
\eea
centered around the vanishing loop with a dispersion given by
$\Delta \sigma$.
Now after integratinig by parts, we can use the known result for
functional gaussian integration to get the desired expressions.
In particular
\bea
    \expect{Y ^{\mu \nu} \qtq{C}}
    & = &
    \funint{Y ^{\lambda \rho} \tst}
        Y ^{\mu \nu} \qtq{C}
        \bigl | \Phi _{G} \qtq{C} \bigr | ^{2}
    \nonumber \\
    & = &
    \funint{Y ^{\lambda \rho} \tst}
        Y ^{\mu \nu} \qtq{C}
        \ttt{\frac{1}{4 \pi \ttt{\Delta \sigma} ^{2}}} ^{3/2}
        \exp
        \left\{
            -
            \frac{1}{4 \ttt{\Delta \sigma} ^{2}}
            Y ^{\mu \nu} \qCq
            Y _{\mu \nu} \qCq
        \right\}
    \nonumber \\
    & = &
    0
\eea
since the integrand is odd. With the same procedure, but a bit of more
effort we also have
\beq
    \expect{Y ^{\mu \nu} \qtq{C} Y _{\mu \nu} \qtq{C}}
    =
    \funint{Y ^{\lambda \rho} \tst}
        Y ^{\mu \nu} \qtq{C}
        Y _{\mu \nu} \qtq{C}
        \bigl | \Phi _{G} \qtq{C} \bigr | ^{2}
    \nonumber \\
    =
    3 \ttt{\Delta \sigma} ^{2}
    \quad .
\eeq
The same procedure can be followed for the expectation values related
to the {\sl Boundary Area Momentum}.
\end{proof}

Let us introduce the following notation:
\begin{nots}[String Mean Square Deviations
             in Position and Momentum]\spbcorr{}.\\
    $\Delta \Sigma _{Y} ^{2}$ is the
    \underbar{Mean Square Deviation in the Position
    of the String}$\,$\footnote{Or, equivalently,
    the \underbar{String Position Uncertainty}.}.\\
    $\Delta \Sigma _{Q} ^{2}$ is the
    \underbar{Mean Square Deviation in the Momentum
    of the String}$\,$\footnote{Or, equivalently,
    the \underbar{String Momentum Uncertainty}.}.
\end{nots}
The following proposition is now just a matter of applying
previous results.
\begin{props}[Shape Uncertainty Principle]\spbcorr{}.\\
    The {\sl Quantum Shadow Dynamics} of the {\sl String}
    is such that it is impossible
    to have at the Areal Time $A$ an exact knowledge of both the
    {\sl Holographic Position} of the object and of its conjugated
    {\sl Boundary Area Momentum}; in
    particular
    \beq
        \Delta \Sigma _{Y}
        \Delta \Sigma _{Q}
        =
        \frac{3}{\sqrt{2}}
        \quad , \qquad
        \left( \hbar = 1 \quad \hbox{units} \right)
        \quad .
    \label{7.heiuncpri}
    \eeq
\end{props}
\begin{proof}
From the definition of {\it mean square deviation} of a
stochastic variable $Z$ and the equality
\beq
    \mathtt{Var} Z
    =
    \expect{\ttt{Z - \expect{Z}} ^{2}}
    =
    \expect{Z} ^{2} - \expect{Z ^{2}}
\eeq
we can just substitute results (\ref{7.sigexp}-\ref{7.sigsigexp})
in the following equation
$$
    \Delta \Sigma _{Y} ^{2}
    =
    \frac{1}{2}
    \expect{Y ^{\mu \nu} \qtq{C} Y _{\mu \nu} \qtq{C}}
    -
    \expect{Y ^{\mu \nu} \qtq{C}}
    \expect{Y _{\mu \nu} \qtq{C}}
$$
to get
\beq
    \Delta \Sigma _{Y} ^{2}
    =
    \frac{3}{2 \ttt{\Delta Q} ^{2}}
    \quad .
\label{7.sigsqudev}
\eeq
In the same way starting from (\ref{7.momexp}-\ref{7.mommomexp}) and
$$
    \Delta \Sigma _{Q} ^{2}
    =
    \frac{1}{2}
    \expect{Q ^{\mu \nu} \qtq{C} Q _{\mu \nu} \qtq{C}}
    -
    \expect{Q ^{\mu \nu} \qtq{C}}
    \expect{Q _{\mu \nu} \qtq{C}}
$$
the {\sl String} momentum {\it mean square deviation}, or
{\it uncertainty} squared, is
\beq
    \Delta \Sigma _{Q} ^{2}
    =
    3
    \left( \Delta Q \right) ^{2}
    \quad .
\label{7.momsqudev}
\eeq
Then, comparing equation (\ref{7.sigsqudev}) with equation (\ref{7.momsqudev}),
we find that the uncertainties are related by
$$
    \Delta \Sigma _{Y}
    \Delta \Sigma _{Q}
    =
    \frac{3}{\sqrt{2}}
    \quad .
$$
\end{proof}

Equation (\ref{7.heiuncpri}) represents the new form that the Heisenberg
Principle takes when {\sl String} Quantum Mechanics is formulated in terms
of diffusion in loop space, or quantum shape shifting.
Just as a pointlike particle cannot have a definite position in space
and a definite linear momentum at the same time, a physical
{\sl String} cannot have a definite shape and a
definite rate of shape changing at a given areal time.
In other words, a {\sl String} loop cannot be totally at rest neither in
physical nor in loop space: it is subject to a zero--point motion
characterized by
\bea
    \left \langle
        Q _{\mu \nu} \left [ C \right ]
    \right \rangle
    & = &
    0
    \\
    \left [
        \frac{1}{2}
        \left \langle
            Q _{\mu \nu} \left [ C \right ]
            Q ^{\mu \nu} \left [ C \right ]
	\right \rangle
    \right ] ^{1/2}
    =
    \sqrt{3}
    \left( \Delta Q \right)
    & = &
    \Delta \Sigma _{Q}
    \quad .
\eea
In such a state a physical {\sl String} undergoes a {\it zero--point
shape shifting}, and the loop momentum attains its minimum
value compatible with an area resolution $\Delta\sigma$.\\
To keep ourselves as close as possible to Heisenberg's seminal
idea, we interpret the lack of a definite shape
as follows:  as we increase the resolution of the ``microscope''
used to probe the structure of the {\sl String}, more and more quantum petals
will appear along the loop. The picture emerging out of this is
that of a classical line turning into a {\it fractal} object as
we move from the classical domain of physics to the quantum
realm of quantum fluctuations. If so,  two questions immediately
arise:
\begin{enumerate}
    \item a classical bosonic {\sl String} is a closed line of
    topological dimension one. Its spacetime image consists of a
    smooth, timelike, two--dimensional world--sheet. Then, if a quantum
    {\sl String} is a fractal object, which Hausdorff
    dimension should be assigned to it?
    \item Is there any {\it critical scale} characterizing the
    classical--to--fractal geometrical transition?
\end{enumerate}
These two questions will be addressed in the next section.


\section{Fractal Strings}

In the path--integral formulation of quantum mechanics, Feynman
and Hibbs noted that the trajectory of a particle is continuous but
nowhere differentiable. We extend this result to the quantum
mechanical path of a relativistic {\sl String} and find that the
``trajectory'', in this case, is a fractal surface with Hausdorff
dimension three.  Depending on the resolution of the detecting
apparatus, the extra dimension is perceived as ``fuzziness'' of the
{\sl String} {\sl World--Sheet}. We give an interpretation of this phenomenon
in terms of a new form of the uncertainty principle for {\sl Strings} and
study the transition from the smooth to the fractal phase.

As pointed out in appendix \ref{D.looderapp}
the functional area derivative introduces a non differentiable process.
Non--differentiability is the hallmark of fractal objects.
Thus, anticipating one of our results, quantum loop fluctuations,
interpreted as singular shape--changing transitions resulting
from ``petal addition'', are responsible for the fractalization
of the {\sl String}.
Evidently, in order to give substance to this idea, we must
formulate the shape uncertainty principle for loops, and the
centerpiece of this whole discussion is again the loop wave functional
$\Psi \left [ C ; A \right ]$, whose precise meaning we already discussed.

\subsection{The Hausdorf Dimension of a Quantum String}

One of the major achievements of Feynman's formulation of quantum
mechanics was to restore the particle's trajectory concept  at the
quantum level. However, the dominant contribution to the
``sum over histories'' is provided by trajectories which are nowhere
differentiable \cite{fh}. Non differentiability is the hallmark of
fractal lines. In fact, it seems that Feynman and Hibbs were aware that the
quantum mechanical path of a particle is inherently fractal.
This idea was revisited and further explored by Abbott and
Wise in the case of a free non--relativistic particle \cite{abbw},
and the extension to relativistic particles was carried
out by several authors, but without a general
agreement \cite{dhaus}, \cite{nott}.
This is one of the reasons for setting up a quantum
mechanical, rather than a field theoretical, framework, even for
relativistic objects. It enables us to adapt the Abbott--Wise
discussion to the {\sl String} case, with the following basic substitution:
the point particle, erratically moving through euclidean
space, is replaced by the {\sl String} configuration whose representative
point randomly drifts through loop space. We have shown in the
previous chapters  that ``flow of time'' for particles is
replaced by area variations for {\sl Strings}. Hence, the image of
an abstract linear ``trajectory'' connecting the two ``points'' $C _{0}$
and $C$ in a lapse of time $A$, corresponds, physically, to a family
of closed lines stacked into a two-dimensional surface of
proper area $A$. Here is where the quantum mechanical aspect
of our approach and our choice of dynamical variables seem to
have a distinct advantage over the more conventional
relativistic description of {\sl String} Dynamics. The conventional
picture of a {\sl String World--Sheet}, consisting of a
collection of world--lines associated with each constituent point, is
replaced by a {\sl World--Sheet} ``foliation'' consisting of a
stack of closed
lines  labeled by the internal parameter $A$. In other words, we
interpret the {\sl String} {\sl World--Sheet} as a sequence of ``snapshots''
of single closed lines ordered with respect to the area of the
{\sl Parameter Space} associated with the {\sl World--Sheet}
formed by them. Then, the randomness of the ``motion'' of a point in
loop space is a reflection of the non-differentiability of the
{\sl String} {\sl World--Sheet}, which, in turn, is due to zero--point quantum
fluctuations: the random addition of petals to each loop results in a
fuzziness, or graininess of the world surface, by which the {\sl String} stack
acquires an effective thickness. One can expect that this
graininess becomes apparent only when one can resolve the
surface small irregularities.\\
The technical discussion on which this picture is based, follows
closely the analysis by Abbott and Wise \cite{abbw}. Thus, let us
divide the {\sl String} internal coordinates domain, which is the
{\sl Parameter Space}, into
$N$ strips of area $\Delta A$. Accordingly, the {\sl String} stack is
approximated by the discrete set of the $N + 1$ loops,
$$
    Y _{(n)} ^{\mu} \tst
    =
    Y ^{\mu} \ttt{s ; n \Delta A}
    \quad , \qquad
    n = 0 , 1 , \dots , N
    \quad .
$$
Suppose now that we take a snapshot of each one of them and measure their
``area in {\sl Parameter Space}''.
If the cross section of the emulsion grains is $\Delta \sigma$, then we have
an area indeterminacy greater or equal to $\Delta \sigma$.
Then, the total area of the surface subtended by the last and
first loop in the stack will be given by
\beq
    \left \langle S \right \rangle
    =
    N
    \left \langle \Delta S \right \rangle
\label{smedio}
\eeq
where, $\langle \Delta S \rangle$ is the average area variation
in the interval $\Delta A$,
\beq
    \left \langle \Delta S \right \rangle
    \equiv
    \left[
	\funint{Y ^{\lambda \rho}}
            Y ^{\mu \nu} \left [ C \right ]
            Y _{\mu \nu} \left [ C \right ]
	\bigl \vert
            \Psi \left [ C ; \Delta A \right ]
        \bigr \vert ^{2}
    \right]^{1/2} .
\eeq
The finite resolution in $Y ^{\mu \nu}$ is properly taken into
account by choosing for $\Psi \qtq{C ; \Delta A}$ a gaussian
wave functional of the type (\ref{5.gauloowav}).
After these preliminary remarks we can give a central definition
for the contents of this chapter.
\begin{defs}[$D$--measure]\spbcorr{}.\\
    The \underbar{$D$--measure} $\mathcal{S} _{D}$ of the
    previously considered stack of fluctuating loops is
    \beq
        \mathcal{S} _{D}
        \dfn
        N
        \left \langle \Delta S \right \rangle
        \left( \Delta \sigma \right) ^{D - 2}
    \eeq
    where $D \in \R$ and we call $\Delta \sigma$ the
    \underbar{resolution}
    of the measure\footnote{Since we have in mind physical application
    of this concept we are willing to leave this ambiguous joke between
    different meanings of the word {\it measure}, in particular
    between the more rigorous mathematical one and the more ``phenomenological''
    one.}.
\end{defs}
Of course, not all values of $D$ give equally useful $D$-measures;
in particular we can single out a particular value of $D$ thanks to
the following definition.
\begin{defs}[Hausdorff Dimension \& Hausdorff Measure]\spbcorr{}.\\
    The \underbar{Hausdorff Dimension} of the stack of fluctuating
    loops is the value $D = D _{H}$ such that its $D _{H}$-measure
    is independent from the resolution $\Delta \sigma$. Then the
    $D _{H}$-measure is called the \underbar{Hausdorff Measure}
    of the stack and denoted with ${\mathcal{S}} _{H}$
\end{defs}
Now, we have at our disposal all the tools that we need to determine the
{\it Hausdorff Dimension} associated to the loop {\sl Shadow Dynamics}.
In particular, we can show that
\begin{props}[Hausdorff Dimension of the Quantum Dynamic Process]\spbcorr{}.\\
    The {\sl Quantum Shadow Dynamics} is characterized by
    the Hausdorff dimension $D _{H} = 3$.
\end{props}
\begin{proof}
To get the desired result we represent the
quantum state of the {\sl String} by the loop functional
\beq
    \Psi \qtq{C}
    =
    \frac{1}{\left( 2 \pi \right) ^{3/2}}
    \funint{P _{\mu \nu} \tst}
	\tilde{\Phi} _{\mathrm{G}} \qtq{P}
	\exp
        \left\{
           \frac{i}{2}
           \oint _{C}
               Q _{\mu \nu}
               Y ^{\mu}
               \form{d Y} ^{\nu}
        \right\}
\eeq
where $\tilde{\Phi} _{\mathrm{G}} \ttt{P}$
is a gaussian momentum distribution centered
around a vanishing {\sl String} average momentum $K _{\mu\nu} = 0$,
i.e. we consider a free loop subject only to zero-point fluctuations.
Then,
\beq
    \left \langle \Delta S \right \rangle
    \propto
    \frac{\Delta A}{4 m^{2} \Delta \sigma}
    \sqrt{
          1
          +
          \left(
              \frac{4 m ^{2} \left( \Delta \sigma \right) ^{2}}
                           {\Delta A}
          \right) ^{2}
         }
    \quad ,
\eeq
and to determine the {\sl String} fractal dimension we keep
$\Delta A$ fixed and take the limit
$\Delta \sigma \rightarrow 0$:
\beq
    \mathcal{S} _{H}
    \approx
    \frac{N \Delta A}{4 m ^{2} \Delta \sigma}
    \left(\Delta \sigma \right) ^{D _{H} -2} .
\eeq
Hence, in order to eliminate the dependence on
$\Delta \sigma$, $\displaystyle{D _{H} = 3}$.
\end{proof}

As in the point particle case, quantum
fluctuations increase by one unit the dimension of the
{\sl String} classical path. As discussed in the following subsection,
the appearance of an extra dimension is perceived as
fuzziness of the {\sl String} manifold and the next question to be
addressed is which parameter, in our quantum mechanical approach,
controls the transition from classical to fractal geometry.

\subsection{Classical--to--Fractal Geometric Transition}

The role of area resolution, as pointed out in the previous section,
leads us to search for a {\it critical area} characterizing the transition
from Classical to Fractal Geometry of the {\sl String} stack.
We can define
\begin{defs}[DeBroglie Area of a String]\spbcorr{}.\\
    The \underbar{DeBroglie Area}, $\Lambda _{DB}$,
    of a {\sl String} is the critical
    area characterizing the transition between the Classical and
    the Quantum behaviour of the {\sl String}.
\end{defs}
Now, we would like to understand the features related to such a
transition. In particular we will characterize in terms of the
{\sl Hausdorff Dimension}  the transition, so that
the distinction between the Classical and the Quantum behavior will
be associated to the distinction between the standard and the fractal
geometry of the {\sl String} stack. This is the content of the following
\begin{props}[Classical--to--Fractal Transition]\spbcorr{}.\\
    Let us consider a {\sl String} with a non--vanishing average momentum
    $K _{\mu \nu}$ along the loop $C$. The classical  behavior of
    the {\sl String} is characterized by the Hausdorff Dimension
    $D _{H} = 2$ for the stack of {\sl Strings} that represents its
    classical evolution,  whereas the quantum behavior
    is characterized by the Hausdorff Dimension $D _{H} = 3$. Moreover,
    the De Broglie Area of the {\sl String} can be identified with the
    quantity
    $$
        \left [
            \frac{1}{2}
            K _{\mu \nu} \qCq
            K ^{\mu \nu} \qCq
        \right ] ^{1/2}
    $$
\end{props}
\begin{proof}
Since we consider the case in which the {\sl String} possesses a non
vanishing average momentum we may represent it with a
Gaussian wave packet of the form (\ref{5.gauloowavfin})
remembering that $K _{\mu \nu}$ is constant along the loop.
The $A$--dependent wave functional, in the
Holographic Coordinate Representation, corresponding to an initial
wave packet of the form (\ref{5.gauloowavfin}) is thus
\bea
    & & \esci \esci
    \Psi _K \left [ C ; A \right ]
    =
    \frac{
          \left[
              \left( \Delta \sigma \right) ^{2}
              /
              \left( 2 \pi \right)
          \right ] ^{3/4}
         }
         {
   	  \left[
              \left( \Delta \sigma \right) ^{2}
              +
              i
              A
              /
              \left( 2 m^2 \right)
          \right ] ^{3/2}
         }
    \cdot
    \nonumber \\
    & & \quad \quad \cdot
    \exp
    \left\{
        -
        \frac{
              \left(
                  Y ^{\mu \nu}
                  Y _{\mu \nu}
	          -
                  2
                  i
                  \left( \Delta \sigma \right)
                  Y ^{\mu \nu}
                  K _{\mu \nu}
                  +
                  i
                  A
                  \left( \Delta \sigma \right) ^{2}
                  K _{\mu \nu}
                  K ^{\mu \nu}
                  /
                  \left( 4 m ^{2} \right)
	      \right)
             }
             {
	      2
              \left [
                  \left( \Delta \sigma \right) ^{2}
                  +
                  i
                  A
                  /
                  \left ( 2 m ^{2} \right)
              \right ]
             }
    \right\}
    \quad ,
\eea
where we have used equation (\ref{7.dePdesrel}) to exchange
$\Delta Q$ with $\Delta \sigma$.
The corresponding probability density ``evolves'' as follows
\beq
    \bigl \vert \Psi _{K} \left [ C ; A \right ] \bigr \vert ^{2}
    =
    \frac{\left( 2 \pi \right) ^{-3/2}}
         {
          \left[
              \left( \delta \sigma \right) ^{2}
              +
              A ^{2}
              /
              \left(
                  4
                  \left( \delta \sigma \right) ^{2}
                  m ^{4}
              \right)
          \right] ^{3/2}
         }
    \exp
    \left\{
        -
        \frac{
              \left [
                  Y ^{\mu \nu} \left [C \right ]
                  -
	          A
                  K ^{\mu \nu}
                  /
                  \left( 2 \left( \delta \sigma \right) m ^{2} \right)
              \right ] ^{2}
             }
             {
	      \left[
                  \left( \delta \sigma \right) ^{2}
                  +
                  A ^{2}
                  /
                  \left(
                      4 \left( \delta \sigma \right) ^{2} m^4
                  \right)
              \right]
             }
    \right\}
    \quad .
\label{gaussdens}
\eeq
Therefore, the average area variation
$\left \langle \Delta S \right \rangle$,
when the loop wave packet drifts with a momentum
$K _{\mu \nu} \left [ C \right ]$, is
\beq
    \left \langle \Delta S \right \rangle
    \equiv
    \left[
	\funint{Y ^{\lambda \rho}}
            Y ^{\mu \nu} \left [ C \right ]
            Y _{\mu \nu} \left [ C \right ]
	    \bigl \vert
                \Psi _{K} \left [ C ; \Delta A \right ]
            \bigr \vert ^{2}
    \right] ^{1/2}
\quad .
\label{dsk}
\eeq
For our purpose, there is no need to compute the exact
form of the mean value (\ref{dsk}), but only its dependence
on $\Delta \sigma$. This can be done in three steps:
\begin{enumerate}
        \item introduce the adimensional integration variable
	\beq
	    y ^{\mu \nu} \left [ C \right ]
            \equiv
            \frac{Y ^{\mu \nu} \left [ C \right ]}
                 {\Delta \sigma}
 	\quad ;
        \eeq
	\item shift the new integration variable as follows:
	\beq
	    y ^{\mu \nu} \left [ C \right ]
            \rightarrow
            \bar{y} ^{\mu \nu} \left [ C \right ]
            \equiv
	    y ^{\mu \nu} \left [ C \right ]
            -
            \frac{\ttt{\Delta A} K ^{\mu \nu}}
                 {2 m ^{2} \left( \Delta \sigma \right) ^{2}}
	    \quad ;
        \eeq
	\item rescale the integration variable as
	\begin{equation}
	\bar{y} ^{\mu \nu} \left [ C \right ]
        \rightarrow
	Z ^{\mu \nu} \left [ C \right ]
        \equiv
        \bar{y} ^{\mu\nu} \left [C \right ]
        \left[
            1
            +
            \frac{\left( \Delta A \right) ^2}
                 {4 m ^{4} \left( \Delta \sigma \right) ^4}
        \right] ^{1/2}
        \quad .
	\end{equation}
\end{enumerate}
Then, we obtain
\beq
    \left \langle \Delta S \right \rangle
    =
    \frac{\Delta A}
         {\sqrt{2} \Lambda _{DB} 2 m^2 \left( 2 \pi \right) ^{3/4}}
    \left [
        \funint{Z}
            \left(
	        \frac{\Lambda _{DB} Z ^{\mu \nu}}
                     {\left( \Delta \sigma \right)}
                \sqrt{1 + \beta ^{-2}}
                +
	        \Lambda _{DB} K ^{\mu \nu}
            \right) ^{2}
            e ^{- Z ^{\mu \nu} Z _{\mu \nu}/2}
    \right] ^{1/2}
    \quad ,
\label{sclfr}
\eeq
where
\bea
    \Lambda _{DB} ^{-1}
    & \equiv &
    \sqrt{\frac{1}{2} K ^{\mu \nu} K _{\mu \nu}}
\label{dbarea}
    \\
    \beta
    & \equiv &
    \frac{\Delta A}{2 m ^{2} \left(\Delta \sigma \right) ^{2}}
    \quad .
\eea
The parameter $\beta$ measures the ratio of the
``temporal'' to ``spatial''
uncertainty, while the area $\Lambda _{DB}$ sets the
scale of the surface variation at which the {\sl String} momentum is
$K _{\mu \nu}$. Therefore, with the particle analogy in mind,
we see that $\Lambda _{DB}$ can be assigned  the role of
{\sl loop De Broglie Area.}
Let us assume, for the moment, that $\Delta A$ is independent
of $\Delta\sigma$, so that either quantity can be
treated as a free parameter in the Theory. A notable exception to this
hypothesis will be discussed shortly. Presently, we note that
taking the limit
$ \left( \Delta \sigma \right) \rightarrow 0$,
affects only the first term of the
integral (\ref{sclfr}) and that its weight with respect
to the second term is measured by the ratio
$\Lambda _{DB} / \left( \Delta \sigma \right)$.
If the area resolution is much larger than the loop De Broglie area,
then the first term is negligible:
$\left \langle \Delta S \right \rangle$ is independent of $\Delta \sigma$
and $\left \langle S \right \rangle$ scales as
\beq
    \Lambda _{DB} \ll \left( \Delta \sigma \right)
    \quad
    :
    \quad
    \mathcal{S} _H
    \approx
    \left( \Delta \sigma \right)^{D _{H} - 2}
    \quad .
\label{dcl}
\eeq
In this case, independence of $\left( \Delta \sigma \right)$ is
achieved by assigning $D _{H} = 2$.
As one might have anticipated, the detecting apparatus is
unable to resolve the graininess of the {\sl String} stack, which therefore
appears as a smooth two dimensional surface. \\
The fractal, or quantum, behavior manifests itself below
$\Lambda _{DB}$, when the first term in (\ref{sclfr}) provides the
leading contribution
\beq
    \Lambda _{DB} \gg \left( \Delta \sigma \right)
    \quad
    :
    \quad
    \mathcal{S} _{H}
    \approx
    \frac{N \ttt{\Delta A}}{\Delta \sigma}
    \left( \Delta \sigma \right) ^{D _{H} - 2}
    \sqrt{
          1
          +
          \frac{4 m ^{4} \left( \Delta \sigma \right) ^{4}}
               {\left( \Delta A \right) ^{2}}
         }
    \quad .
\label{dhq}
\eeq
This expression is less transparent than the relation(\ref{dcl}),
as it involves also the $\Delta A$ resolution. However, one may now consider
two special  sub cases in which the Hausdorff dimension can be
assigned a definite value.\\
In the first case, we keep $\Delta A$ fixed and scale $\Delta \sigma$
down to zero. Then,
$$
    \left \langle \Delta S \right \rangle
    \propto
    \left( \Delta \sigma \right) ^{-1}
$$
diverges, because of larger and larger shape fluctuations, and
\beq
    \mathcal{S} _{H}
    \approx
    \frac{A}{\Delta \sigma}
    \left( \Delta \sigma \right) ^{D _{H} - 2}
\eeq
requires $D _{H} = 3$.\\
The same result can be obtained also in the second subcase, in
which both $\Delta \sigma$ and $\Delta A$ scale down to zero,
but in such a way that their ratio remains constant,
\beq
    \left [
        \frac{ 2 m ^{2} \left( \Delta \sigma \right) ^{2}}
             {\left( \Delta A \right)}
    \right \rceil _{\Delta \sigma \rightarrow 0}
    =
    \mathrm{const.}
    \equiv
    \frac{1}{b}
    \quad .
\eeq
The total area of the {\sl Parameter Space}
$A = N \ttt{\Delta A}$ is kept fixed.
Therefore, as
$\Delta A \sim \left( \Delta \sigma \right) ^{2} \rightarrow 0$,
then $N \rightarrow \infty$ in order to keep $A$ finite. Then,
\beq
    \left \langle \Delta S \right \rangle
    \propto
    \frac{\Delta A}{\Delta \sigma}
    \sqrt{1 + \frac{1}{b ^{2}}}
    \propto
    \Delta \sigma
\label{selfsim}
\eeq
and
\beq
    \mathcal{S} _H
    \propto
    A
    \left( \Delta \sigma \right) ^{D _{H} - 2}
    \frac{1}{\Delta \sigma}
    \sqrt{1 + \frac{1}{b ^{2}}}
    \quad ,
\eeq
which leads to $D _{H} = 3$ again. In the language of fractal
geometry, this interesting subcase corresponds to
{\it self--similarity}. Thus, the condition (\ref{selfsim})
defines a special class of {\it self--similar loops} characterized by
an average area variation which is proportional to $\Delta \sigma$ at any
scale.
\end{proof}


%% file: chap08.tex
\pageheader{}{The ``Double'' Classical Limit.}{}
\chapter{The ``Double'' Classical Limit}
\label{8.douclalimcha}

\begin{start}
``$\euf{I}$t's all right,\\
you can trust him.''\\
\end{start}

\section{Couplings and Limits}

\beq
   T
    =
    \frac{c }{\pi \alpha '}
    \quad .
\eeq
is the tension of a {\it classical} {\sl String}.
In the limit of vanishing Regge slope  $\alpha ' \to 0$ the {\sl String} tension
diverges $T \to \infty$. \\
At the quantum level the classical action is measured in $\hbar$ units. Thus,
one can introduce an {\it effective coupling constant}
\beq
   T^\prime
    \equiv
    \frac{c }{\hbar\pi \alpha '}
    \quad .
\eeq
Accordingly, the {\it Classical Limit} $\hbar \to 0$ is usually identified
with the infinite tension limit. However, we shall keep distinct the two
limiting procedures in what follows. To avoid any confusion,
we shall maintain the term ``classical limit'' in the original
form, i.e. $\hbar \to 0$ with $\alpha^\prime$ fixed, while in the infinite
tension, or {\it pointlike limit},  $\hbar$ is fixed while
$\alpha^\prime\to 0$.

\section{Classical Limit}

In this chapter we will see how, starting from the {\sl String
Functional Wave Equation}, it is possible to recover,
in what we will call the {\it Classical Limit}, a very interesting
Classical Field Theory of Extended Objects \cite{noi2}.
In this
formulation an extended object (we will concentrate on $1$-dimensional
ones, i.e. {\sl Strings}, but the same is true for membranes and
$p$-{\sl{}branes} in complete generality) is conveniently described
through a {\it current}, which is seen as a source of the object
itself and has support only over the {\sl World--Sheet} (or in general
the {\sl World--HyperTube}).

As a first step in taking this limiting procedure we see how the
{\sl Functional Schr\"o\-din\-ger Equation} can be derived starting
from a Lagrangian written in terms of the {\sl String Wave Functional},
$\Psi \qCq$.
\begin{props}[Quantum Lagrangian Density I]\spbcorr{}.\\
The Lagrangian density
\beq
    \Lag \ttt{\Psi , \Psi ^{\ast}}
    =
    \frac{2\pi\alpha^\prime}{4}
    \norme{\Gamma}
    \oint _{\Gamma} ds
        \sqrt{Y ^{\prime 2}}
        \frac{\delta \Psi ^{\ast}}
             {\delta Y ^{\mu \nu} \tst}
        \frac{\delta \Psi}
             {\delta Y _{\mu \nu} \tst}
    +\Psi ^{\ast}i\partial _{A}\Psi
    \label{8.qualagequ}
\eeq
is a Lagrangian Density associated with the
{\sl Functional Schr\"odinger Equation} {\rm (\ref{3.funlooschequ})}.
\end{props}
\begin{proof}
To get the desired result we have to make a variation of, say, the
$\Psi \qtq{C ; A}$ functional and compute the corresponding variation
of the action
$$
    S = \funint{C} \int dA \Lag \qtq{\Psi , \Psi ^{\ast}}
    \quad .
$$
After integration by parts this gives the desired ressult.
\end{proof}
Now we specialize the result (\ref{8.qualagequ})
above to the case of tension eigenstates
 \beq
     \Psi =\exp\left(-{i\over 2\pi\alpha^\prime}A\right)\Phi
 \quad .
 \eeq
Then, we obtain a Lagrangian Density for the ``stationary'', i.e.
$A$ independent, wave functional
\beq
    \Lag \ttt{\Phi , \Phi ^{\ast}}
    =
    \frac{1}{4}
    \norme{\Gamma}
    \oint _{\Gamma} ds
        \sqrt{Y ^{\prime 2}}
        \frac{\delta \Phi ^{\ast}}
             {\delta Y ^{\mu \nu} \tst}
        \frac{\delta \Phi}
             {\delta Y _{\mu \nu} \tst}
    +
    \left(
        \frac{1}{2 \pi \alpha '}
    \right) ^{2}
    \Phi ^{\ast}
    \Phi
    \quad ,
    \label{8.staqualagequ}
\eeq
where
$\Phi \equiv \Phi \qCq$ is a functional of the loop $C$,
$
     C : Y ^{\mu} = Y ^{\mu} \tst
$.
To carry on the limit procedure we introduce
the following {\sl Parametrization} of the loop $C$
\beq
    Y ^{\mu} \tst
    =
    y ^{\mu}
    +
    \sqrt{2\pi\alpha '} \kappa ^{\mu} \tst
    \quad .
    \label{8.embfunexp}
\eeq
where, $y$ is the loop {\it center of mass}, or ``{\it{}zero mode}'' defined as:
\beq
 y ^{\mu}=\norme{\Gamma}\oint _{\Gamma} ds \sqrt{\ttt{\By '}^{2}}Y^{\mu}(s)
 \eeq

Accordingly,  we can write
\beq
    \Phi \left [ C \right ]
    =
    \Phi
    \left [
        y ^{\mu}
        +
        \sqrt{\alpha '} \kappa ^{\mu} \tst
    \right ]
    \quad ,
\eeq
which is an appropriate form to expand the {\sl String} field in powers of
$ \sqrt{\alpha '}$.

On the other hand, any complex  functional can be written in terms of
modulus and a phase. The novelty is that we can express the phase as a
contour integral  and  write the {\sl Loop Functional} as
$$
    \Psi \qCq
    =
    \sqrt{P \qCq}
    \exp
        \left\{
            \frac{i}{\hbar}
            \oint _{C}
                A _{\mu} d Y ^{\mu}
        \right\}
    \quad ,
$$
where
$$
P \qCq = \vert \Psi \qCq \vert^2
$$
and
$$
     \exp
        \left\{
            \frac{i}{\hbar}
            \oint _{C}
                A _{\mu} d Y ^{\mu}
        \right\}
$$
is the {\it Abelian Wilson Loop} associated to a fictitious point charge
traveling along the closed contour $C$. The Stokes Theorem allows one
to express the Wilson factor in term of the flux of the field strength of $A$
across any surface bounded by $C$
\begin{eqnarray}
    \Phi \qCq
    & = &
    \sqrt{P \qCq}
    \exp
        \left\{
            \frac{i}{\hbar}
            \oint _{C= \partial \mathcal{W} }
                A _{\mu} d Y ^{\mu}
        \right\}
    \nonumber \\
    & = &
    \sqrt{P \qCq}
    \exp
        \left\{
            \frac{i}{2 \hbar}
            \int _{\mathcal{W}}
                \partial _{[ \mu } A _{\nu ]}
                d X ^{\mu} \wedge d X ^{\nu}
        \right\}
    \nonumber \\
    & = &
    \sqrt{P \qCq}
    \exp
        \left\{
            \frac{i}{2 \hbar}
            \int _{\mathcal{W}}
                B _{\mu \nu}
                d X ^{\mu} \wedge d X ^{\nu}
        \right\}
    \quad ,
    \label{8.stawavfunwilloo}
\end{eqnarray}
where we defined the field strength of $A_\mu$ as
$
    B _{\mu \nu}
    =
    \partial _{[ \mu} A _{\nu ]}
    \quad .
$

Using expression (\ref{8.stawavfunwilloo}) for the
{\sl Wave Functional} we can re express
the Lagrangian (\ref{8.staqualagequ}) in terms of $P \qCq$ and the
 field strength $B _{\mu \nu}$.
\begin{props}[Quantum Lagrangian Density II]\spbcorr{}.\\
The expression of the Lagrangian Density {\rm (\ref{8.staqualagequ})}
in terms of $P \qCq$ and $B _{\mu \nu}$ is
\beq
    L
    =
    -
    \frac{\hbar ^{2}}{4}
    \norme{\Gamma}
    \! \! \!
    \oint _{\Gamma} ds
        \sqrt{Y ^{\prime 2}}
        \left [
            \frac{\delta \sqrt{P \qCq}}
                 {\delta Y ^{\mu \nu} \tst}
            \frac{\delta \sqrt{P \qCq}}
                 {\delta Y _{\mu \nu} \tst}
            +
            \frac{P \qCq}
                 {4 \hbar ^{2}}
            B ^{\mu \nu}
            B _{\mu \nu}
        \right ]
    +
    \left(
        \frac{1}{2 \pi \alpha '}
    \right) ^{2}
    \! \! \!
    P \qCq
    \quad .
\label{8.profielag}
\eeq
\end{props}
\begin{proof}
We first compute the {\sl Holographic Derivative} of $\Phi$
\bea
    \frac{\delta \Phi \qCq}
         {\delta Y ^{\mu \nu} \tst}
    & = &
    \left [
        \frac{\delta \sqrt{P \qCq}}
             {\delta Y ^{\mu \nu} \tst}
        +
        \frac{i}{2 \hbar}
        \sqrt{P \qCq}
        B _{\mu \nu} \tst
    \right ]
\label{8.wavfunder}
\eea
where, $B _{\mu \nu} \tst$ is a shorthand notation for the $B _{\mu \nu}(x)$
field evaluated at the point $x^\mu=Y^(s)$ along the loop.
The modulus square of (\ref{8.wavfunder}) is
\beq
    \frac{\delta \Phi ^{\ast} \qCq}
         {\delta Y ^{\mu \nu} \tst}
    \frac{\delta \Phi \qCq}
         {\delta Y _{\mu \nu} \tst}
    =
    \left [
        \left(
            \frac{\delta \sqrt{P \qCq}}
                 {\delta Y ^{\mu \nu} \tst}
        \right) ^{2}
        +
        \left(
            \frac{\sqrt{P \qCq} B _{\mu \nu}}
                 {2 \hbar}
        \right) ^{2}
    \right ]
    \quad .
\eeq
Then we can substitute this results into expression
(\ref{8.staqualagequ}) together with equation (\ref{8.stawavfunwilloo})
for $\Phi \qCq$ to get the desired result.
\end{proof}
As a last step towards the main result of this chapter,
we give the following definition.
\begin{defs}[Functional Current]\spbcorr{}.\\
    The \underbar{Functional Current}, which is the ``Probability
    Current'', associated with the Theory described by the
    Lagrangian Density {\rm (\ref{8.qualagequ})} is
    \bea
        J _{\mu \nu} \qCq
        & = &
        \frac{\hbar}{2 i}
        \norme{\Gamma}
        \oint _{\Gamma}
        ds \sqrt{Y ^{\prime 2}}
        \left [
            \Psi ^{\ast} \qCq
            \frac{\delta \Phi \qCq}
                 {\delta Y ^{\mu \nu} \tst}
            -
            \Psi \qCq
            \frac{\delta \Phi ^{\ast} \qCq}
                 {\delta Y ^{\mu \nu} \tst}
        \right ]
        \nonumber \\
        & = &
        \frac{\hbar}{2 i}
        \norme{\Gamma}
        \oint _{\Gamma}
        ds \sqrt{Y ^{\prime 2}}
        \left [
            \Psi ^{\ast} \qCq
            \frac{\delta \Psi \qCq}
                 {\delta Y ^{\mu \nu} \tst}
            -
            \Psi \qCq
            \frac{\delta \Psi ^{\ast} \qCq}
                 {\delta Y ^{\mu \nu} \tst}
        \right ]
        \quad .
    \label{8.funcur}
    \eea
\end{defs}
Now we are ready  to compute the {\it Classical Limit}
and see what is  the resulting Theory. As a first step
we give the following proposition.
\begin{props}[Classical Lagrangian Density and Current]\spbcorr{}.\\
    In what we will call the classical limit, i.e. the limit
    where we  neglect $\mathcal{O}(\hbar)$ terms and keep only first order
    quantities in the inverse {\sl String} tension,
    the Lagrangian Density {\rm (\ref{8.profielag})} and  the current
    {\rm (\ref{8.funcur})}
    turn into the following {\it local} quantities:
    \bea
        L
        \quad
        & \longrightarrow &
        \quad
        L _{\mathrm{cl.}}
        =
        -
        \frac{1}{4}
        P \ttt{y}
        \partial _{[ \mu} A _{\nu ]}
        \partial ^{[ \mu} A ^{\nu ]}
        +
        \left(
            \frac{1}{2 \pi \alpha '}
        \right) ^{2}
        P \ttt{y}
        \nonumber \\
        J _{\mu \nu}
        \quad
        & \longrightarrow &
        \quad
        P \ttt{y} B _{\mu \nu} \ttt{y}
        \quad .
    \eea
\end{props}
\begin{proof}
To obtain the desired results we first see the consequences
of taking the classical limit in the two quantities that
appear in the expression (\ref{8.stawavfunwilloo})
for the wave functional, $\Phi \qCq$
In particular the expansion \ref{8.embfunexp} in $P \qCq$ gives
\beq
    P \qCq
    =
    P
    \left [
        y ^{\mu} + \sqrt{2 \pi \alpha '} \kappa ^{\mu} \tst
    \right ]
    \longrightarrow
    P \left( y ^{\mu} \right)
    \label{8.proexp} \\
\eeq
and in
$
 B _{\mu \nu}
 =
 B _{\mu \nu} \tst
 =
 B _{\mu \nu} \left( Y ^{\rho} \tst \right)
$
results in
\beq
    B _{\mu \nu}(s)
    =
    B _{\mu \nu}
    \left(
        y ^{\mu} + \sqrt{2 \pi \alpha '} \kappa ^{\mu} \tst
    \right)
    \longrightarrow
    B _{\mu \nu}
    \left( y  \right)
    =
    \partial _{[ \mu} A _{\nu ]} \left( y \right)
    \quad .
    \label{8.fieexp}
\eeq
Note how taking the limit of large {\sl String} tension amounts to
{\it squeezing} the {\sl String} to a single point, which is the
{\sl String} center of mass. At larger scales its extension
is no more detectable and we recover from functionals, ordinary
Quantum Fields.
Substituting the last results in the expression (\ref{8.profielag})
for the Lagrangian, we get in the limit $\sqrt{2 \pi \alpha '}  \to 0$, 

\beq
    L
    \longrightarrow
    -
    \frac{1}{4}
    P \ttt{y}
    \partial _{[ \mu} A _{\nu ]}
    \partial ^{[ \mu} A ^{\nu ]}
    +
    \left(
        \frac{1}{2 \pi \alpha '}
    \right) ^{2}
    P \ttt{y}
    +
    \mathcal{O} \left( \hbar ^{2} \right)
    \quad .
\eeq
Moreover if we let also $\hbar \to 0$, we remain with the
Lagrangian Density
\beq
    \Lag _{\mathrm{cl.}}
    =
    -
    \frac{1}{4}
    P \ttt{y}
    \partial _{[ \mu} A _{\nu ]}
    \partial ^{[ \mu} A ^{\nu ]}
    +
    \left(
        \frac{1}{2 \pi \alpha '}
    \right) ^{2}
    P \ttt{y}
\eeq
or, equivalently, with the action
\beq
    S _{\mathrm{cl.}}
    =
    \int d ^{4} y
    P \left( y \right)
    \left(
        -
        \frac{1}{4}
        \partial _{[ \mu} A _{\nu ]}
        \partial ^{[ \mu} A ^{\nu ]}
        +
        \left(
            \frac{1}{2 \pi \alpha '}
        \right) ^{2}
    \right)
    \quad .
\label{8.zermodclaact}
\eeq
The action (\ref{8.zermodclaact}) can now be varied with respect to $P \ttt{y}$,
\beq
    \frac{\delta S _{\mathrm{cl.}}}
         {\delta P \ttt{y}}
    =
    0
    \quad \Rightarrow \qquad
    -
    \frac{1}{4}
    B _{\mu \nu} \left( y \right)
    B ^{\mu \nu} \left( y \right)
    +
    \frac{1}{\left( 2 \pi \alpha '\right) ^{2}}=0
    \quad ,
\eeq
and, with respect to $A _{\mu}$:
\beq
    \frac{\delta S _{\mathrm{cl.}}}
         {\delta \A _{\mu} \ttt{y}}
    =
    0
    \quad \Rightarrow \qquad
    \partial _{\nu}
    \left [P \ttt{y}
        \partial ^{[ \mu}
        A ^{\nu ]}
    \right ]
    =
    0
    \quad .
\eeq
We can now follow the same procedure for the functional current
Inserting again expression (\ref{8.wavfunder}) together with
expansions (\ref{8.proexp}-\ref{8.fieexp}) in (\ref{8.funcur})
we get
\bea
    J _{\mu \nu} \qCq
    \longrightarrow
    J _{\mu \nu} \ttt{y}
    & = &
    \frac{\hbar}{2 i}
    \norme{\Gamma}
    \oint _{\Gamma}
    ds \sqrt{Y ^{\prime 2}}
    \left [
        \frac{2 i}{\hbar}
        P \ttt{y}
        \partial _{[ \mu} A _{\nu ]} \ttt{y}
    \right ]
    \nonumber \\
    & = &
    P \ttt{y} B _{\mu \nu} \ttt{y}
    \label{8.curprofierel}
    \quad .
\eea
\end{proof}
Now we can turn to the final result, i.e. to recover the
Classical Field Formulation for a Closed {\sl String}
described in \cite{noi2}.
\begin{props}[Gauge Theory of the String Geodesic Field]\spbcorr{}.
    The classical limit of the Theory described by the Lagrangian
    {\rm (\ref{8.qualagequ})}
    is the Classical Formulation of the Gauge Theory for the
    {\sl String} Geodesic Field\footnote{Please, look at the work
    cited above for details on this formulation}.
\end{props}
\begin{proof}
Starting from equation (\ref{8.curprofierel}) we see that
\beq
    B ^{\mu \nu} \ttt{y}
    =
    \frac{J ^{\mu \nu} \ttt{y}}
         {P \ttt{y}}
\eeq
and we can use this result to rewrite the classical action
(\ref{8.zermodclaact}) in terms of the classical current; substituting we get
\beq
    I _{\mathrm{cl.eq.}}
    =
    \int d ^{4} y
        \left [
            -
            \frac{1}{4}
            \frac{
                  J ^{\mu \nu} \ttt{y}
                  J _{\mu \nu} \ttt{y}
                 }
                 {P \ttt{y}}
            +
            \left(
                \frac{1}{2 \pi \alpha '}
            \right) ^{2}
            P \ttt{y}
            +
            B _{\nu}
            \partial _{\mu} J ^{\mu \nu}
        \right ]
    \quad ,
\eeq
where the last term is necessary to assure that
the current $J ^{\mu \nu}$ is divergence less and
$B _{\mu}$ is a non-dynamical Lagrange multiplier
enforcing the previous condition.
Varying with respect to $P \left( x \right)$ we
obtain
\bea
    \frac{\delta I _{\mathrm{cl.eq.}}}
         {\delta P \ttt{y}}
    =
    0
    & \Rightarrow &
    \frac{1}{4}
    \frac{
          J ^{\mu \nu} \ttt{y}
          J _{\mu \nu} \ttt{y}
         }
         {P \ttt{y}}
         +
         \left(
             \frac{1}{2 \pi \alpha '}
         \right) ^{2}
    =
    0
    \nonumber \\
    & \Rightarrow &
    P \left( y \right) ^{2}
    =
    -
    \left( 2 \pi \alpha ' \right) ^{2}
    \frac{
          J ^{\mu \nu} \ttt{y}
          J _{\mu \nu} \ttt{y}
         }
         {4}
    \nonumber \\
    & \Rightarrow &
    P \ttt{y}
    =
    \pm
    \left( 2 \pi \alpha ' \right)
    \sqrt{
        \frac{
              J ^{\mu \nu} \ttt{y}
              J _{\mu \nu} \ttt{y}
             }
             {4}
         }
    \quad .
\eea
Substituting then this solution in the action we get
\beq
    I _{\mathrm{cl.eq.red.}}
    =
    \int d ^{4} x
        \left [
            \frac{1}{2 \pi \alpha '}
            \sqrt{
                  -
                  \frac{1}{4}
                  J ^{\mu \nu} \ttt{x}
                  J _{\mu \nu} \ttt{x}
                 }
            +
            A _{\mu}
            \partial _{\nu} J ^{\mu \nu}
        \right ]
    \quad ,
\eeq
which is the classical action for the Gauge Theory
of the {\sl String} geodesic field!
\end{proof}

%% file: chap09.tex
\pageheader{}{Boundary {\it versus} Bulk Relation.}{}
\chapter{Boundary {\it versus} Bulk Relation}
\label{9.connection}

\begin{start}
``$\euf{T}$his is totally different.''\\
``No! No different!\\
Only different in your mind.\\
You must unlearn\\
what you have learned.''\\
\end{start}

\section{Overview}

In this section
we recover a general representation for the quantum state of a relativistic
closed line ({\it loop}) in terms of {\sl String} degrees of freedom.
The general form of the loop functional splits into the product
of the {\it Eguchi functional}, encoding the {\it Holographic Quantum
Dynamics}, times the Polyakov path--integral, taking into account the
full {\it Bulk} Dynamics, times a {\it loop effective action}, which is
needed to renormalize {\it Boundary} ultraviolet divergences.
The Polyakov {\sl String} action is derived as an {\it effective action}
from the {\sl Covariant Schild Action} (\ref{2.covschhamact})
by functionally integrating
out the {\sl World--Sheet} coordinates.
The {\sl Holographic Coordinates} description
of the {\sl Boundary Shadow Dynamics},
is shown to be induced by the ``zero mode'' of the {\it Bulk} quantum
fluctuations. Finally, we briefly comment about a ``{\it{}unified,
fully covariant}'' description of points, loops and {\sl Strings} in terms
of {\it Matrix Coordinates}.

Non perturbative effects in Quantum Field Theory are usually difficult
to study because of the missing of appropriate mathematical tools.
In the few cases where some special invariance, like electric/magnetic
duality or supersymmetry, allows to open a window over the strong
coupling regime  of the Theory, one  usually is faced with a new
kind of solitonic excitations describing extended field configurations,
which are ``dual'' to the pointlike states of the perturbative regime.
Remarkable examples of extended structures of this sort ranges from the Dirac
magnetic monopoles \cite{dirac} up to the loop states of
{\sl Quantum Gravity}.\\
The Dynamics of these extended objects is usually formulated in terms
of an ``Effective Theory'' of Relativistic {\sl Strings}, i.e. one
switches from a Quantum Field theoretical framework to a
different description by taking for granted that there is some
relation between the two. Such a kind of relation
has  recently shown up in the context of Superstring Theory as a web of
dualities among different phases of different superstring models.
On the other hand,  there is still no explicit way to connect pointlike and
stringy phases of non--supersymmetric Field Theories like QCD and
Quantum Gravity.\\
The main purpose of this section is to recover  a general
{\it String Representation} of a loop wave functional in terms of
the {\it Bulk} and {\it Boundary} wave functional of a
{\sl Quantum String}.
The loop wave functional can describe the quantum state of
a $1$-dimensional, closed excitation of some Quantum Field Theory, while
the corresponding {\sl String} functional is induced by the quantum fluctuations
of an open {\sl World--Sheet}. We shall find a quite
general relation linking
these two different objects.
The matching between a quantum loop and a quantum {\sl String}
being provided by
a Functional Fourier transform over an abelian vector field.\\
As a byproduct of this new {\it Loops/Strings connection}, we shall clarify
the interplay  between the {\it Bulk}
{\sl Quantum Dynamics}, encoded into the Polyakov
path--integral \cite{polya} and the induced {\it Boundary} Dynamics,
which, as we already saw, is the {\it Holographic Description}
of the {\it Boundary} quantum fluctuations.

\section{Loop and String States}
\label{9.connecbou}

To accomplish the task, that we briefly outlined above,
we will use the {\sl Covariant Schild Action}
of equation (\ref{2.covschhamact}) that we already introduced
in section \ref{2.covschactsec} using the canonical
formalism described in section \ref{2.repcanforsec}.
The quantum state of the {\sl Closed Bosonic String} is described
as in chapters \ref{5.strfunsol} and \ref{7.fractal}
by $\Psi \qCq$, the complex functional of the {\sl Target Space}
{\it Boundary}
$C = \partial \mathcal{W}$ or, equivalently, of the {\sl Parameter Space}
{\it Boundary} $\Gamma = \partial \Sigma$. After the results we gained
in the previous chapters, the following proposition should not be
a surprise: 
\begin{props}[String Wave Functional: Covariant Formulation]\spbcorr{}.\\
The quantum state of a {\sl Closed Bosonic String} can be expressed
as a ``phase space'' path--integral, namely
\beq
\Psi \qCq =
\int _{\partial \Sigma = \Gamma}
    [\mathcal{D} X ^{\mu}][\mathcal{D} \PPdd{m}{\mu}][\mathcal{D} g_{ab}]
    \exp
    \left(
        i
        \int _{\Sigma} d ^{2} \Bsigma
        \sqrt{\dete{\Bg}}
        \Lag _{(\bs{g})} \ttt{\Bx , \Bp , g _{ab}}
    \right)
    \quad ,
\label{9.pathuno}
\eeq
    where $\Lag _{(\bs{g})}$, the Covariant Schild Lagrangian Density,
    can be read out of equation {\rm (\ref{2.covschhamact})} as
    $$
    \Lag _{(\bs{g})} \ttt{\By , \Bp , g _{ab}}
    =
    g ^{mn}
       \left [
           \partial _{m} X ^{\mu} \ttt{\Bsigma}
           \PPdd{n}{\mu} \ttt{\Bsigma}
           -
           \frac{1}{2 \cpu}
           \PPdu{m}{\mu} \ttt{\Bsigma}
           \PPdd{n}{\mu} \ttt{\Bsigma}
       \right ]
    $$
and in the path--integral {\rm (\ref{9.pathuno})} we sum over all the
the {\sl String} {\sl World--Sheets} having the closed curve $\Gamma$
in {\sl Parameter Space} (equivalently
$C$ in {\sl Target Space}) as the only {\it Boundary}.
\end{props}
\begin{proof}
We apply here the same concepts that we developed in chapter
\ref{3.strfunqua}. There we saw that the propagator for an
extended object (string) can be interpreted as the probability
amplitude to obatin a shape modification from an initial one,
described by a loop $C _{0}$, to a final one, described by
a loop $C$. We wrote the propagator
$
    K \qtq{Y ^{\mu} , Y ^{\mu} _{0} ; A}
$
to fully emphasize this situation. Now the quantity which we call
$\Psi \qtq{C}$ is equivalent to $K \qtq{Y ^{\mu} ,\emptyset ; A}$,
where with $\emptyset$ we denote a vanishing initial string configuration.
Note that, since we are using the {\sl Covariant Schild Action}, we have also
to functionally integrate over the {\sl World--Sheet} metric, $g ^{ab}$,
since now it is an additioinal dynamical variable.
\end{proof}
We note now that to a closed line and an open $2$-surface can be given
different geometrical characterizations.
In this case we are going to use a formulation slightly different
from the one used extensively before, which resembles more the
contents of chapter \ref{8.douclalimcha}:
we will use the associated currents
(the {\it Boundary} as well as the {\it Bulk} one) which physically
represent the sources of our extended objects.
\begin{defs}[Bulk Current]\spbcorr{}.\\
The  \underbar{\it Bulk Current} $J^{\mu\nu}(x ; \mathcal{W})$
is a rank $2$ antisymmetric tensor distribution having non vanishing
support over the two dimensional {\sl World--Sheet} $\mathcal{W}$,
parametrized by $X ^{\mu} \ttt{\Bsigma}$:
\beq
   J ^{\mu\nu}(x ; \mathcal{W})
   \dfn
   \int _{\Sigma} d ^{2} \Bsigma
       \dot{X} ^{\mu \nu} \ttt{\Bsigma}
       \delta ^{D} \qtq{x - \Bx \ttt{\Bsigma}}
   \quad .
   \label{9.jbulk}
\eeq
\end{defs}
\begin{defs}[Loop Current]\spbcorr{}.\\
The  Loop Current $J ^{\mu} \ttt{y ; L}$ is a vector distribution
with support
over a closed line $L$ which represents a loop and is parametrized by
$l ^{\mu} \tst$:
\beq
    J ^{\mu} \ttt{x ; L}
    \dfn
    \oint _{\Gamma} ds
        \frac{d l ^{\mu} \tst}{d s}
        \delta \qtq{y - \bs{l} \tst}
    \label{9.jloop}
    \quad .
\eeq
\end{defs}
\begin{defs}[Boundary Current]\spbcorr{}.\\
The \underbar{Boundary Current} is the divergence of the {\sl Bulk Current}:
\beq
    J ^{\mu} \ttt{x ; \partial \mathcal{W}}
    \dfn
    \partial _\lambda J ^{\lambda\mu} \ttt{x ; \mathcal{W}}
    \quad .
    \label{9.jbordo}
\eeq
\end{defs}
Please, note that at this stage the {\sl Loop Current} is not linked
in any way to the {\sl Bulk Current}, i.e. the loop can be considered
free: it is not the {\it Boundary} of the {\sl World--Sheet}. Of course
we need to implement such a constraint, because our {\sl String} is the
only free {\it Boundary} of its history, and
a natural way to link a closed line to a surface is by ``appending''
the surface to the assigned loop \cite{axion}.
This matching condition can be formally written by identifying the loop
current with the {\it Boundary} current:
\beq
J ^{\mu} \ttt{x ; L}
=
J ^{\mu} \ttt{x ; \partial \mathcal{W} = C}
=
\partial _{\nu} J ^{\nu\mu} \ttt{x ; \mathcal{W}}
\label{9.div}
\quad :
\eeq
this is the mathematical way of requiring
$L \equiv C = \partial \mathcal{W}$.
We remark again that equation (\ref{9.div})
defines $J ^{\mu} \ttt{x ; L \equiv C}$ as the current associated to the
{\it Boundary} $C$ of the  surface $\mathcal{W}$.
In the absence of (\ref{9.div})
$J ^{\mu} \ttt{x ; L}$ is a loop current with no reference to
any surface. In this sense, equation (\ref{9.div}) is
a formal description of the ``{\it{}gluing operation}'' between
the surface $\mathcal{W}$ and the closed line $C$.

Accordingly, one would relate loops states to {\sl String} states through
a functional relation of the type
\beq
    \bar{\Psi} \qtq{L}
    =
    \int \qtq{\mathcal{D} C}
    \bar{\delta} \qtq{L - C}
    \Psi \qtq{C}
    \quad ,
\label{9.lstate}
\eeq
which in a more explicit way can be written
\beq
    \bar{\Psi} \qtq{L}
    =
    \int \qtq{\mathcal{D} C}
    \bar{\delta} \qtq{L - \partial \mathcal{W}}
    \Psi \qtq{\partial \mathcal{W}}
    \quad ;
\label{9.lstateuno}
\eeq
here, we introduced a {\it Loop Dirac Functional} which picks up
the assigned loop configuration $L$ among the (infinite) family of all
the allowed {\sl String} {\it Boundary} configurations $\partial \mathcal{W} = C$:
the definition of such an object is definitely non trivial and one
of its possible representations, that we will take as definition,
is given below; in particular a more appropriate definition
of such a {\it Loop Dirac Functional}
can be offered by the
``current representation'' of extended objects (provided by (\ref{9.jbulk}),
(\ref{9.jloop}), (\ref{9.jbordo})). To make rigorous the previous
formal step, we thus give the following definition.
\begin{defs}[Loop Dirac Functional]\spbcorr{}.
\label{9.loodirfundef}\\
The \underbar{Loop Dirac Functional} (formally written in
equation {\rm (\ref{9.lstate})}) can be given a suitable ``Fourier'' form
as a functional integral over a vector field $A _{\mu} \ttt{x}$:
\bea
    \bar{\delta} \qtq{L - \partial \mathcal{W}}
    & = &
    \delta
    \left[
        J ^{\mu} \ttt{x ; L} - J ^{\mu} \ttt{x ; \partial \mathcal{W}}
    \right]
    \nonumber \\
    & = &
    \funint{A _{\mu} \ttt{x}}
    \exp
    \left\{
        -i
        \int d ^{D} x \,
        A _{\mu} \ttt{x}
        \left[
            J ^{\mu} \ttt{x ; L} - J ^{\mu} \ttt{x ; \partial \mathcal{W}}
        \right]
    \right\}
    \quad .
    \label{9.delta}
\eea
\end{defs}
This gives us the possibility of rewriting the {\sl Loop Functional}
in a clearer form.
\begin{props}[Loop Functional State]\spbcorr{}.\\
The loop functional {\rm (\ref{9.lstate})} can be written as
\bea
    & & \esci
    \bar{\Psi} \qtq{L}
    =
    \funint{C} [\mathcal{D} X ^{\mu}][\mathcal{D} \PPdd{m}{\mu}]
               [\mathcal{D} g_{ab}][\mathcal{D} A _{\mu}]
    \exp
    \left\{
    i
    \int _{\Sigma} d ^{2} \Bsigma
         \sqrt{\dete{\Bg}}
         \Lag _{(\bs{g})} \ttt{\Bx , \Bp , g _{ab}}
    \right\}
    \cdot
    \nonumber\\
    & & \qquad \qquad \qquad \cdot
    \exp
    \left\{
     -
     i
     \int d^{D} x \,
         A _{\mu} \ttt{x}
         \left[
             J ^{\mu} \ttt{x ; L}
             -
             J ^{\mu} \ttt{x ; \partial \mathcal{W}}
         \right]
    \right\}
    \quad .
    \label{9.pathdue}
    \eea
\end{props}
\begin{proof}
The result follows by substituting equation (\ref{9.pathuno}) and
applying definition \ref{9.loodirfundef} in equation
(\ref{9.lstateuno}).
\end{proof}
These seemingly harmless manipulations are definitely non trivial.
A proper implementation of the {\it Boundary} conditions introduces an
{\it abelian vector field} coupled both to the loop and
the {\it Boundary} currents. The first integral in (\ref{9.delta}) is
the {\it circulation} of $A$ along the loop $L$:
\beq
    \int d ^{D} y \,
        A _{\mu} \ttt{y}
        J ^{\mu} \ttt{y ; L}
    =
    \oint _{L} \form{d l} ^{\mu}
        A _{\mu} \ttt{l}
    \quad ;
\label{9.circ1}
\eeq
in the same way the second
represents the circulation along $\partial \mathcal{W} = C$:
\beq
    \int d ^{D} x \,
        A _{\mu} \ttt{x}
        J ^{\mu} \ttt{x ; \partial \mathcal{W}}
    =
    \oint _{\partial \mathcal{W}} \form{d Y} ^{\mu}
        A _{\mu} \ttt{Y}
    \quad .
\label{9.flux}
\eeq
Now, let us recall the usual definition of the Wilson factor.
\begin{defs}[Wilson Factor]\spbcorr{}.
\label{9.wisfacdef}\\
The \underbar{Wilson Factor} associated with a loop $L$ is
\beq
    W \qtq{A _{\mu} , L}
    \dfn
    \exp
    \left[
        -
        i
        \oint _{L} \form{d l} ^{\mu}
            A _{\mu} \ttt{l}
    \right]
    \quad .
    \label{9.wilson}
\eeq
\end{defs}
In terms of the Wilson Factor we can define the Loop Transform of a
{\sl Loop Functional} $\Psi [\tilde{L}]$ of a loop $\tilde{L}$:
\begin{defs}[Loop Transform]\spbcorr{}.\\
    The \underbar{Loop Transform} of the {\sl Loop Functional}
    $\Psi [\tilde{C}]$ is the functional integral over all the
    possible Loops of the product of the {\sl Wilson Factor}
    times the {\sl Loop Functional}:
    $$
        \psi \qtq{A _{\mu} \ttt{x}}
        \dfn
        \int [\mathcal{D} C]
            W [A _{\mu} , C]
            \Psi \qCq
        \quad .
    $$
\end{defs}
Note that, if $C$ is parametrized by $Y ^{\mu} \tst$,
we can more clearly express the equation above as
$$
        \psi \qtq{A _{\mu} \ttt{x}}
        \dfn
        \int [\mathcal{D} Y ^{\mu} \tst]
            W [A _{\mu} , C]
            \Psi \qCq
        \quad .
$$
\begin{nots}[Loop Transform and Inverse Loop Transform]\spbcorr{}.\\
    We will use the following shorthand for the {\sl Loop Transform}
    that exchanges the loop $\tilde{L}$ with the abelian vector
    field $A _{\mu} \ttt{x}$:
    $$
        \euf{L} _{\bs{A} , \tilde{L}}
        \quad : \qquad
        \psi \qtq{A _{\mu} \ttt{x}}
        =
        \euf{L} _{\bs{A} , \tilde{L}} \Psi [\tilde{L}]
        \quad ,
    $$
    as well as the following
    $$
        \euf{L} ^{-1}
        \quad : \qquad
        \Psi \qtq{L}
        =
        \euf{L} ^{-1} _{{L} , \bs{A}} \psi \qtq{A _{\mu} \ttt{x}}
        \quad ,
    $$
    for its inverse, which exchanges the abelian vector field
    $A _{\mu} \ttt{x}$ with its support $L$.
\end{nots}
We will give a natural name to $\psi \qtq{A _{\mu}}$:
\begin{nots}[Dual String Functional]\spbcorr{}.\\
    Let $\Psi \qCq$ be a {\sl String Functional}.
    We will call
    {\sl Loop Transformed String Functional}, $\psi \qtq{A _{\mu}}$,
    the \underbar{Dual Loop Functional}.
\end{nots}
We also note the {\sl Loop Dirac Functional} can be
expressed in terms of {\sl Wilson Loops}.
\begin{props}[Loop Dirac Functional and Wilson Loops]\spbcorr{}.\\
Using the {\rm definition \ref{9.wisfacdef}} for the {\sl Wilson Factor}
the {\sl Loop Dirac Functional} can be written as
\beq
    \bar{\delta} \left[ C - L \right]
    =
    \funint{A _{\rho}}
        W ^{-1} \qtq{A _{\mu} , C}
        W \qtq{A _{\mu} , L}
    \quad .
\eeq
\end{props}
\begin{proof}
We have
$$
    W ^{-1} \qtq{A _{\mu} , C}
    =
    \exp
    \left[
        i
        \oint _{C} \form{d Y} ^{\mu}
            A _{\mu} \ttt{\By}
    \right]
    =
    \exp
    \left[
        i
        \int d ^{D} x
            A _{\mu} \ttt{x}
            J ^{\mu} \ttt{x ; \partial \mathcal{W}}
    \right]
$$
and
$$
    W \qtq{A _{\mu} , L}
    =
    \exp
    \left[
        -
        i
        \oint _{L} \form{d l} ^{\mu}
            A _{\mu} \ttt{\bs{l}}
    \right]
    =
    \exp
    \left[
        -
        i
        \int d ^{D} x
            A _{\mu} \ttt{x}
            J ^{\mu} \ttt{x ; L}
    \right]
    \quad .
$$
Then
$$
    W ^{-1} \qtq{A _{\mu} , C}
    W \qtq{A _{\mu} , L}
    =
    \exp
    \left\{
        -
        i
        \int d ^{D} x
            A _{\mu} \ttt{x}
            \left[
                J ^{\mu} \ttt{x ; \partial L}
                -
                J ^{\mu} \ttt{x ; \partial \mathcal{W}}
            \right[
    \right\}
$$
so that
\bea
    \funint{A _{\rho}}
        W ^{-1} \qtq{A _{\mu} , C}
        W \qtq{A _{\mu} , L}
    & = &
    \funint{A _{\rho}}
        \exp
        \left\{
            -
            i
            \int d ^{D} x
                A _{\mu} \ttt{x}
                \left[
                    J ^{\mu} \ttt{x ; \partial L}
                    -
                    J ^{\mu} \ttt{x ; \partial \mathcal{W}}
                \right[
        \right\}
    \nonumber \\
    & = &
    \bar{\delta} \left[ C - L \right]
    \quad ,
\eea
where we used the equality $C = \partial \mathcal{W}$.
\end{proof}
In this way we can see that there is a natural relation between
a {\sl Loop Functional} and the corresponding {\sl String Functional}:
\begin{props}[Relation between Loop and String Functionals]\spbcorr{}.\\
Let $\Psi \qtq{C}$ be a {\sl String Functional}. The corresponding
Loop Functional $\bar{\Psi} \qtq{L}$ can be written as
\beq
\bar{\Psi} \qtq{L}
=
\funint{A _{\mu}}
    W^{-1} \qtq{A _{\mu} , L }
    \psi \qtq{A _\mu}
    \quad ,
\label{9.looptransf}
\eeq
where $\psi \qtq{A _{\mu}}$ is the {\sl Dual String Functional}.
\end{props}
\begin{proof}
We can use the representation of the {\sl Dirac Fuction}
in terms of Wilson Loop Factors. Then starting from
(\ref{9.lstateuno}) we get
\bea
    \bar{\psi} \qtq{L}
    & = &
    \funint{C}
        \bar{\delta} \qtq{L - \partial \mathcal{W}} \Psi \qtq{\partial \mathcal{W}}
    \nonumber \\
    & = &
    \funint{C} \qtq{\mathcal{D} A _{\rho}}
        W ^{-1} \qtq{A _{\mu} , L}
        W \qtq{A _{\mu} , C}
        \Psi \qtq{\partial \mathcal{W}}
    \nonumber \\
    & = &
    \funint{A _{\rho}}
        W ^{-1} \qtq{A _{\mu} , L}
        \funint{C}
        W \qtq{A _{\mu} , C}
        \Psi \qtq{\partial \mathcal{W}}
    \nonumber \\
    & = &
    \funint{A _{\rho}}
        W ^{-1} \qtq{A _{\mu} , L}
        \funint{C}
        \psi \qtq{A _{\mu}}
    \quad .
    \nonumber \\
\eea
Hence,
The vector field $A _{\mu} \ttt{x}$ is the Fourier conjugate variable to
the {\sl String} {\it Boundary} configuration
and the wanted result is obtained by projecting
$\phi[\, A_\mu(x)\,]$ along the loop $C$.
The whole procedure can be summarized as follows:
\bea
    \euf{L}  _{\bs{C} , \bs{A}} ^{-1}
    \: : \quad
    \Psi \qCq \: \hbox{\sl String Functional}
    &\longrightarrow&
    \psi \qtq{A _{\mu}} \: \hbox{\sl Dual String Functional}
    \nonumber \\
    \euf{L} _{\bs{A} , \bs{L}}
    \: : \quad
    \psi \qtq{A _{\mu}} \: \hbox{\sl Dual String Functional}
    &\longrightarrow&
    \bar{\Psi} \qtq{L} \: \hbox{\sl Loop Functional}
    \quad .
    \nonumber
\eea
The loop has thus been glued to the {\it Boundary} of the {\sl World--Sheet},
which in the formulation in terms of currents means that the {\sl Loop Current}
has been identified with the {\sl Boundary Current}, i.e. the divergence
of the {\sl Bulk Current}.
\end{proof}
Let us proceed by unraveling the information contained in the {\sl String}
functional $\Psi \qCq$. We already pointed out at the beginning
of this section that we
choose as the classical {\sl String} action the
{\sl ``covariant'' Schild action} of equation
(\ref{2.covschhamact}). We now would like to push forward the functional
integration and as a first result we get that:
\begin{props}[Bulk-Boundary Decoupling]\spbcorr{}.\label{9.finresfirste}\\
    In the {\sl Loop Functional} {\rm (\ref{9.pathuno})}
    the {\it Bulk} Dynamics
    decouples from the {\it Boundary} Dynamics.
\end{props}
\begin{proof}
To begin with, it is instrumental to extract
a pure {\it Boundary} term from the first integral
in (\ref{2.covschhamact})
\bea
    {1 \over 2}
    \int _{\mathcal{W}} \form{d X} ^{\mu} \wedge \form{d X} ^{\nu}
        P _{\mu \nu}
    & = &
    {1 \over 2}
    \int _{\mathcal{W}}
        \form{d} \left( Y ^{\mu}  \form{d Y} ^{\nu} P _{\mu \nu} \right)
    -
    {1 \over 2} \int _{\mathcal{W}} Y ^{\mu}
        \form{d P} _{\mu\nu} \wedge  \form{d Y} ^{\nu}
    \nonumber\\
    & = &
    {1\over 2}
    \oint _{C}
        Y ^{\mu} \form{d Y} ^{\nu} Q _{\mu \nu} \ttt{Y}
    -
    {1\over 2}
    \int _{\Sigma} d ^{2} \Bsigma
        Y ^{\mu} \ttt{\Bsigma}
        \epsilon ^{mn}
        \partial _{[m} \PPdd{n]}{\mu}
    \quad .
\label{9.phase}
\eea
Then, we recognize that $X ^{\mu} \ttt{\Bsigma}$ appears in the path--integral
only in the last term of (\ref{9.phase}) through a linear coupling to the
left hand side of the classical equation of motion  (\ref{9.uno}).
Accordingly, to integrate over the {\sl String} coordinates is tantamount to
integrate over a {\it Lagrange multiplier} enforcing the canonical
momentum to satisfy the classical equation of motion (\ref{9.uno}):
\beq
    \funint{X ^{\mu} \ttt{\Bsigma}}
        \exp
        \left[
            {1 \over 2}
            \int _{\Sigma} d ^{2} \Bsigma
            Y ^{\mu} \ttt{\Bsigma}
            \epsilon ^{mn} \partial_{[m} \PPdd{n]}{\mu}
        \right]
        =
        \delta
        \left[ \partial _{[m} \PPdd{n]}{\mu} \right]
        \quad .
\eeq
Once the {\sl String} coordinates have been integrated out, the resulting
path--integral reads
\bea
    & & \esci
    \Psi \qCq
    =
    \funint{g_{mn}}[\mathcal{D} \PPdd{m}{\mu}]
        \delta \left[ \partial _{[m} \PPdd{n]}{\mu} \right]
    \cdot
    \nonumber \\
    & & \qquad
    \cdot
    \exp
    \left(
        {i \over 2}
        \oint _{C} Y ^{\mu}  \form{d Y} ^{\nu} Q _{\mu \nu} \ttt{Y}
        -
        {i \over 2 \mu _{0}}
        \int _{\Sigma} d ^{2} \Bsigma \sqrt{\dete{\bs{g}}}
        g^{mn} \PPdd{m}{\mu} \PPdu{n}{\mu}
    \right)
    \quad .
    \label{9.pathcl}
\eea
Equation (\ref{9.pathcl}) shows that we have sum only over classical momentum
trajectories. Such a restricted integration measure spans the subset of
momentum  trajectories we found in proposition \ref{2.covschsolpro}.Thanks to
them we can now give a definite meaning to the integration measure over
the classical solutions, namely
\beq
    \funint{\PPdd{m}{\mu}}
        \delta \left[ \partial _{[m} \PPdd{n}{\mu} \right]
    =
    \int d ^{D} \bar{\eta}
    \int \qtq{d \bar{P} _{\mu\nu}}
    \funint{\tilde{\eta} _{\mu} \ttt{\Bsigma}}
    \quad ,
\eeq
where, we remark that the first two integrations are ``over numbers'' and not
over functions. We have to sum over all possible constant values of
$\bar{P} _{\mu \nu}$ and and $\bar{\eta} ^{\mu}$.
The constant mode of the {\sl Bulk Momentum} does not mix with the other modes
in the on--shell Hamiltonian because the cross term vanish identically
\bea
    \bar{P} _{[\mu\nu]} g ^{(mn)}
    \partial _{(m} Y ^{[\nu}
    \partial _{n)} \eta ^{\mu]}
    & \equiv &
    0
    \nonumber \\
    \delta ^{[mn]}
    \partial _{[m} Y ^{\mu}
    \partial _{n]} \eta _{\mu}
    & \equiv &
    0
    \quad .
\eea
Accordingly, {\it Boundary} Dynamics decouples from the {\it Bulk}
Dynamics\footnote{A boundary term
$$
    {1 \over 2 \sqrt{\mu_{0}}}
    \bar{P} _{\mu \nu}
    \oint _{\Gamma} dt ^{m}
        \eta^{[\mu}
        \partial _{m} X^{\nu]}
    +
    {1 \over 2}
    \oint _{\Gamma} d n ^{m}
        \eta ^{\mu}
        \partial _{m} \eta _{\mu}
$$
vanishes because of the {\it Boundary} conditions
(\ref{9.b1}), (\ref{9.b2}), (\ref{9.b3}).}:
\bea
    {1\over 2}
    \oint_\gamma\overline y^\mu  d  \overline y^\nu P_{\mu\nu}
    (\overline y)&=&{1\over 4}\overline P_{\mu\nu}
    \oint_\gamma d\sigma^m\,\overline y^{[\,\mu}\, \partial_m  \overline
y^{\nu\,]}
    \equiv {1\over 2}\overline P_{\mu\nu}\,\sigma^{\mu\nu}(\gamma)
    \label{9.bt}
    \\
    -
    {1 \over 2 \mu _{0}}
    \int _{\Sigma} d ^{2} \Bsigma
        \sqrt{\dete{\bs{g}}} g^{mn} \PPdd{m}{\mu} \PPdu{n}{\mu}
    & = &
    -
    {1 \over 4 \mu _{0}}
    \bar{P} _{\mu\nu} \bar{P} ^{\mu\nu}
    \int _{\Sigma} d ^{2} \Bsigma
    \sqrt{\dete{\bs{g}}}
    -
    {1 \over 2}
    \int _{\Sigma} d ^{2} \Bsigma
    \sqrt{\dete{\bs{g}}}
        \eta _{\mu}
        \Delta _{\bs{g}} \eta ^{\mu}
    \quad ,
    \label{9.honn}
\eea
where, $\Delta _{\bs{g}}$ is the covariant, {\sl World--Sheet}
D'Alembertian.
This is the desired result.
\end{proof}
The proof above shows as the dynamical areas,
we introduced in previous chapters as the relevant dynamical
variables to describe  what we called the {\sl Shadow Dynamics}
of the {\it Boundary}, are selected by the system itself.
We remark how the {\sl Holographic Coordinate}
appears as the canonical partner of the zero mode {\sl Bulk Area Momentum}
$\bar{P} _{\mu \nu}$. The zero mode {\sl Bulk Area Momentum}
transfers to the {\it Boundary} the {\sl World--Sheet} vibrations.
In this way we can
see how the {\sl Functional Schr\"odinger
Equation} for the {\sl String} is related to the
propagation of the classical modes: a {\it Quantum Dynamics} for the
{\it Boundary} turns out to be an {\it induced effect} of the
``{\it{}Classical}'' {\it Dynamics} of the {\it Bulk}.
Moreover the {\sl Area Time}
\beq
    \int _{\Sigma} d ^{2} \Bsigma
        \sqrt{\dete{\bs{g}}}
    \equiv
    A
    \label{9.area}
\eeq
provides an intrinsic evolution parameter for the system.
We take here the opportunity to develop a bit further this concept
at the Quatum Level. Indeed, a quantum {\sl World--Sheet} has not a definite area
in {\sl Parameter Space}, the metric in (\ref{9.area}) being itself
a quantum operator. Thus, the left hand side
of the definition (\ref{9.area}) has to be replaced by the
corresponding quantum expectation value.
Then, we can split the sum over the {\sl String} metrics into
a sum over metrics $h_{mn}$, with fixed quantum expectation value of the
proper area, times an ordinary integral over all the values of the area
quantum average:
\beq
    \funint{g _{mn} \ttt{\Bsigma}}
        \left( \dots \right)
    =
    \int _{0} ^{\infty} d A
        \exp \left( i \lambda A \right)
        \funint{h _{mn} \ttt{\Bsigma}}
            \exp \left(
                     -
                     i
                     \lambda
                     \int _{\Sigma} d ^{2} \Bsigma
                         \sqrt{\dete{\bs{h}}}
                 \right)
        \left( \dots \right)
    \quad .
\eeq
The $\lambda$ parameter enters the path--integral as a constant external
source enforcing the condition that the quantum average
of the proper area operator is $A$. From a physical point of view it
represents the {\sl World--Sheet} cosmological constant, or vacuum energy
density.

We thus arrive at the fundamental result of this section.
\begin{props}[Boundary {\it versus} Bulk Dynamics]\spbcorr{}.\\
    The {\it Boundary} and {\it Bulk} Dynamics
    encoded in the {\sl Loop Functional}
    can be factorized as
    \bea
        & & \esci \esci
        \bar{\Psi} \qtq{L}
        \equiv
        \Psi\qtq{Y ^{\mu} \tst , Y ^{\mu\nu} \qCq}
        \nonumber \\
        & = &
        \int d ^{D} \bar{\eta}
            \left[
                \exp \left(i S ^{\mathrm{eff}} \qCq \right)
            \right]
        \int _{0} ^{\infty} dA
            \exp \left( i \lambda A \right)
            \Psi \qtq{Y \qCq ; A}
            Z _{\mathrm{BULK}} ^{A}
        \quad .
        \label{9.main}
    \eea
    The {\it Bulk} Quantum Physics is encoded into the Polyakov
    partition function {\rm{}\cite{polya}}
    \beq
        Z _{\mathrm{BULK}} ^{A}
        =
        \funint{h _{mn}\ttt{\Bsigma}}
               [\mathcal{D} \tilde{\eta} _\mu \ttt{\Bsigma}]
            \exp
            \left[
                -
                {i \over 2}
                \int _{\Sigma} d ^{2} \Bsigma
                \sqrt{\dete{\bs{h}}}
                \tilde{\eta} _\mu
                \Delta _{h} \tilde{\eta} ^{\mu}
                -
                i
                \int _{\Sigma} d ^{2} \Bsigma
                \sqrt{\dete{\bs{h}}}
                \left( \kappa R + \lambda \right)
            \right]
        \label{9.polya}
    \eeq
    for a Scalar Field Theory covariantly coupled to $2D$ gravity
    on a disk$\,$\footnote{Our result refers to the disk
    topology. The extension to more
    complex {\sl World--Sheet} topology is straightforward: the single,
    {\it Bulk},
    partition functional has to be replaced by a sum over definite genus
    path--integrals,
    i.e.
    $
        Z_{\mathrm{BULK}} ^{A}
        \longrightarrow
        \sum _{\bs{g}} Z _{\mathrm{BULK}} ^{(\bs{g}),A}
    $.
    This is the starting point for introducing topology changing
    quantum processes in the framework of {\sl String} Theory.} whereas the
    {\it Boundary} Dynamics appears as a solution of the {\sl Functional
    Schr\"odinger Equation} {\rm (\ref{3.kerfunwavsigequ})}
    \beq
        \Psi \qtq{Y ^{\mu \nu} \qCq ; A}
        \equiv
        \int [d \bar{P} _{\mu \nu}]
            \exp
            \left[
                {i \over 2}
                \bar{P} _{\mu \nu}
                Y ^{\mu \nu} \qtq{C}
                -
                i
                \left(
                    {\bar{P} _{\mu \nu} \bar{P} ^{\mu \nu} \over 4 \mu _{0}}
                \right)
                A
            \right]
            \quad :
        \label{9.eguchi}
        \eeq
    this wave functional encodes the holographic quantum mechanics
    of the {\sl String} {\it Boundary} resulting from a superposition of the
    classical {\sl World--Sheet} solutions driven by the zero mode
    area momentum onto the {\it Boundary}.
    Finally, $S^{\mathrm{eff}} \qtq{C}$ is the effective
    action induced by the quantum fluctuation of the {\sl String}
    {\sl World--Sheet}.
    It is a local quantity written in terms of the
    ``counterterms'' needed to cancel
    the {\it Boundary} ultraviolet divergent terms$\,$\footnote{The required
    counterterms are
    proportional to the loop proper length and extrinsic curvature.}.
\end{props}
\begin{proof}
Starting from the result of proposition \ref{9.finresfirste}
we can write then the {\sl String} functional as
\bea
    & & \esci \esci \esci
    \Psi \qCq
    =
    \int d ^{D} \bar{\eta}
    \int _{0} ^{\infty} d A
        \exp
        \left(
            i \lambda A
        \right)
    \int [d \bPPdd{m}{\mu}]
    \funint{h _{mn} \ttt{\Bsigma}}
           [\mathcal{D} \tilde{\eta} ^{\mu} \ttt{\Bsigma}]
    \cdot
    \nonumber \\
    & & \qquad \qquad \qquad \cdot
    \exp
    \left(
        {i \over 2}
        \bar{P} _{\mu\nu} Y ^{\mu \nu} \qCq
        -
        {i \over 4 \mu _{0}}
        \bar{P} _{\mu \nu} \bar{P} ^{\mu \nu} A
    \right)
    \cdot
    \nonumber\\
    & & \qquad \qquad \qquad \cdot
    \exp
    \left(
        -
        {i \over 2}
        \int _{\Sigma} d ^{2} \Bsigma
            \sqrt{\dete{\bs{h}}}
            \tilde{\eta} _{\mu}
            \Delta _{\bs{g}} \tilde{\eta} ^{\mu}
        -
        i
        \lambda
        \int _{\Sigma} d ^{2} \Bsigma
            \sqrt{\dete{\bs{h}}}
    \right)
    \quad .
    \label{9.pathonn}
\eea
The {\sl World--Sheet} Scalar Field Theory contains geometry
dependent, ultraviolet divergent quantities. This $2$-dimensional
Quantum Field Theory on a Riemannian manifold can be
renormalized by introducing suitable {\it Bulk} and {\it Boundary}
``counter terms''.
These new terms absorb the ultraviolet divergencies, and
represent induced weight factors in the functional integration over
the metric, $[\mathcal{D} h_{mn}]$\footnote{This functional
integration measure has to be
factored out by the orbit of the diffeomorphism group, and in the
critical dimension by the Weyl group, as well.},
and the {\it Boundary} shape $[\mathcal{D} C]$.
\end{proof}
The most part of current investigations in Quantum {\sl String} Theory
starts from the Polyakov path--integral and elaborate {\sl String} Theory as
a Scalar Field Theory defined over a Riemann surface. {\sl String}
Perturbation
Theory come from this term as an {\it expansion in the genus} of the
Riemann surface. Against
this background,  we assumed {\it the phase space, covariant path--integral
for the Schild {\sl String} as the basic quantity} encoding the whole
information about {\sl String} Quantum behavior,
and we recovered the Polyakov
``partition functional''  as an {\it effective path--integral,} after
integrating out the {\sl String}
coordinates and factorizing out the {\sl Boundary Shadow
Dynamics}. It is worth to recall that the Conformal Anomaly and the
{\it critical dimension} are encoded into the Polyakov path--integral.
Accordingly, they are {\it Bulk} effects.
But, this is not the end of the story.
Our approach provides the {\sl Boundary Shadow Dynamics} as well.
The fluctuations of the  $\bar{\eta} ^{\mu}$  field {\it induce}
the non--local part of the effective action for $Y ^{\mu} \tst$ , while
the {\sl World--Sheet} vibrations induce the local, geometry dependent
terms.
Furthermore, the Eguchi {\sl String Wave
Functional} $\Psi[C ; A]$  encodes the {\sl Quantum Holographic Dynamics}
of the {\it Boundary} in terms of area coordinates.
$\Psi$ gives the  amplitude to
find a closed {\sl String}, with area tensor $Y ^{\mu \nu}$ as the only
{\it Boundary} of a {\sl World--Sheet} with {\sl Parameter Space} of
proper area $A$, as it was  originally
introduced  in the areal formulation of ``{\sl{}String} Quantum Mechanics''.
This overlooked approach implicitly broke the accepted ``dogma''
that {\sl String}
Theory is intrinsically a ``Second Quantized Field Theory'', which cannot be
given  a First Quantized, or Quantum Mechanical, formulation.
This claim is correct as far as it
is referred to the infinite vibration modes of the
{\sl World--Sheet} {\it Bulk}. On the
other hand,  {\it Boundary} vibrations are induced by the
$1$-dimensional field
living on it and by the constant zero mode of the area momentum
$\bar{P} _{\mu\nu}$. From this viewpoint, the Eguchi approach appears as a
sort of minisuperspace approximation of the full {\sl String} Dynamics in
momentum space: all the infinite {\it Bulk} modes, except the constant one,
has been frozen out. The ``Field Theory'' of this single mode
collapses into
a generalized ``Quantum Mechanics'',
where both spatial and timelike coordinates are
replaced by area tensor and scalar respectively:
\begin{equation}
\hbox{Second Quantized {\it Bulk} Dynamics}
\longrightarrow
\hbox{First Quantized {\it Boundary} Dynamics}
\quad .
\end{equation}
In the absence of external interactions, the quantum state of the  ``free
{\sl World--Sheet} {\it Boundary}''  is represented by a
{\sl Gaussian Wave Functional}
\beq
    \Psi [\sigma\ ; A]
    \propto
    \left(
        {m ^{2} \over A}
    \right) ^{(D-1)/2}
    \exp
    \left(
        -
        i
        \mu _{0}
        {Y ^{\mu\nu} \qCq Y _{\mu\nu} \qCq \over 4 A}
    \right)
\eeq
solving the corresponding ``Schr\"odinger equation'' \cite{noi3}:
\beq
    -
    {1\over 4 m ^{2}}
    \norme{\Gamma}
    \oint _{\Gamma} ds \sqrt{\ttt{\By ' \tst} ^{2}}
        {\delta ^{2} \Psi[C ; A]
         \over
         \delta Y ^{\mu \nu} \tst
         \delta Y _{\mu \nu} \tst
        }
    =
    i
    {\partial \over \partial A }
    \Psi[C ; A]
    \quad .
\label{9.schrod}
\eeq
Our formulation enlightens the
complementary role played by the {\it Bulk} and {\it Boundary}
formulation of
{\sl String} Quantum Dynamics and the subtle interplay between the two.
The wave equation (\ref{9.schrod}) displays one of the most intriguing
aspects of Eguchi formulation: the role of spatial coordinates is
played by the area tensor while the area $A$ is the evolution parameter.
Such an unusual Dynamics is now explained as an induced effect due
to the {\it Bulk} zero mode $\bar{P} ^{\mu \nu}$. In this way
{\it a bare Quantum Loop is dressed with the degrees of freedom carried
by a Quantum} {\sl String}.

%% file: chap10.tex
\pageheader{}{Nonstandard \& Speculative.}{}
\chapter{Nonstandard \& Speculative}
\label{10.nonstacha}

\begin{start}
$\euf{W}$e ``[\dots] must have the deepest commitment,\\
the most serious mind.''\\
\end{start}

\section{Nonstandard Functional Quantization}
\label{10.nonstafunquasec}

As we have seen in chapter \ref{2.hamjac},
section \ref{2.areequide} it is possible to express the fundamental
equation of motion for the {\it Boundary} of a domain $\Sigma$
in terms of {\it Area Dynamics}
as the {\it second derivative} of the loop {\sl Holographic Coordinates}
with respect to the area $A$ in {\sl Parameter Space}
(cf. equation (\ref{2.newareequ}), which we report here for convenience):
\beq
    \frac{d ^{2} Y ^{\mu \nu} \qtq{C ; A}}{d A ^{2}} = 0
    \quad .
    \label{10.newareequ}
\eeq
Moreover, the functional equation of continuity can be
written as (cf. equation (\ref{2.areconequ}))
\beq
     \frac{d P \qtq{C ; A}}{d A}
     +
     \frac{1}{4 m ^{2}}
     \norme{\Gamma}
     \oint _{\Gamma} ds \sqrt{{\By '} ^{2}}
         \frac{\delta}
              {\delta Y ^{\mu \nu} \tst}
         \left [
            P \qtq{C ; A}
            \frac{\delta S \qtq{C ; A}}
                 {\delta Y _{\mu \nu} \tst}
         \right ]
     =
     0
\quad .
\eeq

Now we observe that all the results presented in
appendix \ref{C.nonstopro} can be generalized cases more general
than that of a single scalar field. In particular we are going to
use those results for a multiplet of scalar fields. The classical
quantities are exactly the {\sl Holographic Coordinates}
$Y ^{\mu \nu}$, so that the corresponding {\sl NonStandard Stochastic
Process} is now denoted by $\bs{\hat{Y}} ^{\mu \nu}$.
Now, we state a central\\[5mm]
{\bf Axiom (Stochastic Area Evolution)}. {\it The areal evolution of a
Quantum  {\sl String} is a stochastic process in
the variable $A$}.

According to this axiom we can now substitute
in the {\sl Area Newton Equation} above (\ref{10.newareequ})
the second derivative with what we call the
{\it NonStandard Stochastic Acceleration Operator},
as follows
\beq
    \frac{d ^{2} Y ^{\mu \nu} \qtq{C ; A}}{d A ^{2}}
    \qquad
    \longrightarrow
    \qquad
    \frac{1}{2}
    \left(
        \mfd \mbd
        +
        \mbd \mfd
    \right)
    \bs{\hat{Y}} ^{\mu \nu} \left( A \right)
\quad .
\eeq
The {\sl NonStandard Mean Forward/Backward Derivative} is defined in
appendix \ref{C.nonstopro} as well.
Hence, the \underbar{{\it Stochastic Equation of Motion}}
for the {\sl String} turns out to be
\beq
    \frac{1}{2}
    \left(
        \mfd \mbd
        +
        \mbd \mfd
    \right)
    \bs{\hat{Y}} ^{\mu \nu} \left( A \right)
    =
    0
    \label{10.stostrequmot}
    \quad ,
\eeq
where
$\bs{\hx} ^{\mu \nu}$ is a tensorial {\sl Stochastic Process}
on $\kawspace$ defined in a way analogous to that
of definition \ref{C.nonstoprodef}.
We first give the following definition, since this will be central in
the conclusions of this chapter:
\begin{defs}[Non Standard Wave Functional]\spbcorr{}.\\
    The \underbar{Functional Nonstandard Loop Wavefunctional} is
    \beq
        \bs{\hLam} \left [ C , A \right ]
        \dfn
        \sqrt{\sh{P} \left [ C , A \right ]}
        e ^{i \sh{S} \left [ C , A \right ]}
        =
        e ^{
            \sh{R} \left [ C , A \right ]
            +
            i \sh{S} \left [ C , A \right ]
           }
    \quad ,
    \label{10.nonstawavfun}
    \eeq
    where we rename the probability density in the following way
    $$
        \sqrt{\sh{P} \qtq{C , A}}
        \dfn
        e ^{\sh{R} \qtq{C , A}}
    $$
    i.e.
    \beq
        \frac{1}{2} \log \sh{P} \qtq{C , A}
        =
        \sh{R} \qtq{C , A}
    \quad .
    \label{10.nonstaredproden}
    \eeq
\end{defs}
Starting from the previous results, we can get some useful relations
to be used later on.
\begin{props}[Derivatives of the Non Standard
              Stochastic Process]\spbcorr{}.\\
    The {\sl Area Derivative} of $\bs{\hLam}$ is given by
    \beq
        \frac{\partial \bs{\hLam} \left [ C , A \right ]}
             {\partial A}
        =
        \left(
            \frac{\partial \sh{R} \left [ C , A \right ]}
                 {\partial A}
            +
            i
            \frac{\partial \sh{S} \left [ C , A \right ]}
                 {\partial A}
        \right)
        \bs{\hLam} \left [ C , A \right ]
    \quad ;
    \label{10.areder}
    \eeq
    moreover the first and second functional derivatives with respect to
    the {\sl Nonstandard Holographic Coordinates}
    $\bs{\hx} ^{\mu \nu} \tst$, respectively, are
    \bea
        \frac{\delta \bs{\hLam} \left[ C , A \right]}
             {\delta \bs{\hx} ^{\mu \nu} \left( u \right)}
        & = &
        \left [
            \frac{\delta}
                 {\delta \bs{\hx} ^{\mu \nu} \left( u \right)}
            \left(
                \sh{R}
                +
                i
                \sh{S}
            \right)
        \right ]
        \bs{\hLam} \left[ C , A \right]
        \label{10.firfunder}
        \\
        \frac{\delta ^{2} \bs{\hLam} \left[ C , A \right]}
             {
              \delta \bs{\hx} ^{\mu \nu} \left( u \right)
              \delta \bs{\hx} _{\mu \nu} \left( u \right)
             }
        & = &
        \left [
            \frac{\delta ^{2}}
                 {
                  \delta \bs{\hx} ^{\mu \nu} \left( u \right)
                  \delta \bs{\hx} _{\mu \nu} \left( u \right)
                 }
            \left(
                \sh{R}
                +
                i
                \sh{S}
            \right)
        \right ]
        \bs{\hLam} \left[ C , A \right]
        +
        \nonumber \\
        & & \qquad \qquad
        +
        \left [
            \frac{\delta}
                 {\delta \bs{\hx} ^{\mu \nu} \left( u \right)}
            \left(
                \sh{R}
                +
                i
                \sh{S}
            \right)
        \right ] ^{2}
        \bs{\hLam} \left[ C , A \right]
        \quad .
        \label{10.secfunder}
    \eea
\end{props}
\begin{proof}
The first result is just the application of the chain rule
to the exponential and is thus evident. The same is true for the second
one.
\bea
    \frac{\delta ^{2} \bs{\hLam} \left[ C , A \right]}
         {
          \delta \bs{\hx} ^{\mu \nu} \left( u \right)
          \delta \bs{\hx} _{\mu \nu} \left( u \right)
         }
    & = &
    \frac{\delta}
         {\delta \bs{\hx} ^{\mu \nu} \left( u \right)}
    \left\{
        \left [
            \frac{\delta}
                 {\delta \bs{\hx} _{\mu \nu} \left( u \right)}
            \left(
                \sh{R}
                +
                i
                \sh{S}
            \right)
        \right ]
        \bs{\hLam} \left[ C , A \right]
    \right\}
    \nonumber \\
    & = &
    \left [
        \frac{\delta ^{2}}
             {
              \delta \bs{\hx} ^{\mu \nu} \left( u \right)
              \delta \bs{\hx} _{\mu \nu} \left( u \right)
             }
        \left(
            \sh{R}
            +
            i
            \sh{S}
        \right)
    \right ]
    \bs{\hLam} \left[ C , A \right]
    +
    \nonumber \\
    & & \qquad \qquad
    +
    \left [
        \frac{\delta}
             {\delta \bs{\hx} ^{\mu \nu} \left( u \right)}
        \left(
            \sh{R}
            +
            i
            \sh{S}
        \right)
    \right ] ^{2}
    \bs{\hLam} \left[ C , A \right]
    \quad .
\eea
\end{proof}

Now we can turn to the stochastic process in Kawabata space
that we defined for a scalar field in section \ref{C.stoprokawspasec}.
In particular we can analyse more closely how the {\sl Forward
and Backward Drift Operators} are
related with classical dynamical quantities. This is done in the
following way.
\begin{props}[Non Standard Drift Velocity]\spbcorr{}.\\
    The average of the {\sl Forward
    and Backward Drift Operators} is proportional to the first
    functional derivative of the classical action, i.e.
    \beq
        \hbar
        \frac{\delta \sh{S} \left [ C , A \right ]}
             {\delta \bs{\hx} ^{\mu \nu} \left( u \right)}
        =
        \frac{1}{2}
        \left(
            \bs{\A} ^{+} _{A}
            +
            \bs{\A} ^{-}_{A}
        \right)
        \bs{\hx} ^{\mu \nu} \left( u \right)
    \quad .
    \label{10.nonstadrivel}
    \eeq
\end{props}
\begin{proof}
We start from the forward and backward Fokker Planck equations
that can be written by analogy with those presented in appendix
\ref{C.nonstopro} in the case of a single scalar field;
then we take their
average. This gives the following result:
\beq
    \frac{\partial \sh{P} \qtq{C , A}}{\partial A}
    +
    \int d s
        \frac{\delta}{\delta \bs{\hx} ^{\mu \nu} \tst}
        \left [
            \left(
                \frac{1}{2}
                \left(
                    \bs{\A} ^{+} _{A}
                    +
                    \bs{\A} ^{-} _{A}
                \right)
                \bs{\hx} ^{\mu \nu} \tst
            \right)
            \sh{P} \qtq{C , A}
        \right ]
    =
    0
    \label{10.fokplaequave}
\eeq
Thus, we see that the final result has exactly the same form of the
continuity equation
(\ref{8.conequ}), i.e. we can identify
\beq
    \frac{1}{2}
    \left(
        \A ^{+} _{A}
        +
        \A ^{-} _{A}
    \right)
    \hx ^{\mu \nu} \tst
\eeq
with the drift velocity. So, this expression is a
functional gradient, because thanks to equation (\ref{2.corfunder})
we have
$$
    \hbar
    \frac{\delta \sh{S} \left [ C , A \right ]}
         {\delta \bs{\hx} ^{\mu \nu} \left( u \right)}
    =
    Q _{\mu \nu}
    =
    \frac{1}{2}
    \left(
        \bs{\A} ^{+} _{A}
        +
        \bs{\A} ^{-}_{A}
    \right)
    \bs{\hx} _{\mu \nu} \left( u \right)
$$
where $\sh{S}$ is the classical action!
\end{proof}

After identifying the immaginary part of the exponent in
expression (\ref{10.nonstawavfun}) with the classical action, we can also
express the difference of the {\sl Forward
and Backward Drift Operators} in terms of the redefined form
(\ref{10.nonstaredproden}) for the probability density.
This gives the following result:
\begin{props}[Nonstandard Osmotic Velocity]\spbcorr{}.\\
    The difference of the {\sl Forward
    and Backward Drift Operators}, applied on the {\sl NonStandard
    Tensorial Stochastic Process} in
    Kawabata Space, can be expressed as the following gradient,
    \beq
        \frac{1}{2}
        \left(
            \bs{\A} ^{+} _{A}
            -
            \bs{\A} ^{-}_{A}
        \right)
        \bs{\hx} ^{\mu \nu} \left( u \right)
        =
        \frac{\hbar}{2}
        \frac{\delta \sh{R} \left [ C , A \right ]}
             {\delta \bs{\hx} ^{\mu \nu} \left( u \right)}
        \quad .
    \label{10.nonstaosmvel}
    \eeq
\end{props}
\begin{proof}
Again we start from the forward and backward Fokker Planck equations
multiplying by $1/2$ their difference to get:
\beq
    \oint _{\Gamma} ds
        \frac{\delta}{\delta \bs{\hx} ^{\mu \nu} \tst}
        \left [
            \left(
                \frac{1}{2}
                \left(
                    \bs{\A} ^{+} _{A}
                    -
                    \bs{\A} ^{-} _{A}
                \right)
                \bs{\hx} ^{\mu \nu} \tst
            \right)
            \sh{P} \qtq{\bs{\hx} , A}
        \right ]
    =
    \frac{\hbar}{2}
    \oint _{\Gamma} ds
        \frac{\delta \sh{P} \qtq{\bs{\hx} , A}}
             {\delta \bs{\hx} ^{\mu \nu} \tst ^{2}}
    \quad ,
\eeq
which we can rewrite
\beq
    \oint _{\Gamma} ds
        \frac{\delta}{\delta \bs{\hx} _{\mu \nu} \tst}
        \left\{
            \left [
                \left(
                    \frac{1}{2}
                    \left(
                        \bs{\A} ^{+} _{A}
                        -
                        \bs{\A} ^{-} _{A}
                    \right)
                    \bs{\hx} ^{\mu \nu} \tst
                \right)
                \sh{P} \qtq{\bs{\hx} , A}
            \right ]
            - \frac{\hbar}{2}
            \frac{\delta \sh{P} \qtq{\bs{\hx} , A}}
                 {\delta \bs{\hx} ^{\mu \nu} \tst}
        \right\}
    =
    0
    \quad .
\eeq
Thanks to the redefinition (\ref{10.nonstaredproden})
the previous equation becomes
\beq
    \oint _{\Gamma} ds
        \frac{\delta}{\delta \bs{\hx} _{\mu \nu} \tst}
        \left\{
            \left [
                \left [
                    \left(
                        \frac{1}{2}
                        \left(
                            \bs{\A} ^{+} _{A}
                            -
                            \bs{\A} ^{-} _{A}
                        \right)
                        \bs{\hx} ^{\mu \nu} \tst
                    \right)
                \right ]
            - \frac{\hbar}{2}
                \frac{\delta \sh{R} \qtq{\bs{\hx} , A}}
                     {\delta \bs{\hx} ^{\mu \nu} \tst}
            \right ]
            \sh{P} \qtq{\bs{\hx} , A}
        \right\}
    =
    0
    \quad .
\eeq
A sufficient condition to  satisfy  the above is that
$$
    \frac{1}{2}
    \left(
        \bs{\A} ^{+} _{A}
        -
        \bs{\A} ^{-} _{A}
    \right)
    \bs{\hx} ^{\mu \nu} \tst
    =
    - \frac{\hbar}{2}
    \frac{\delta \sh{R} \qtq{\bs{\hx} , A}}
         {\delta \bs{\hx} ^{\mu \nu} \tst}
    \quad ,
$$
which is the desired result.
\end{proof}

We can now turn to the central result of this chapter:
\begin{props}[Non Standard Functional
              Schr\"odinger Wave Equation]\spbcorr{}.\\
    The {\sl NonStandard Loop WaveFunctional} in Kawabata Space,
    $\bs{\hLam}$, satisfies the functional wave equation
    \beq
        \frac{\hbar ^{2}}{2}
        \oint ds
            \frac{\delta ^{2} \bs{\hLam} \left [ C , A \right]}
                 {
                  \delta \bs{\hx} ^{\mu \nu} \tst
                  \delta \bs{\hx} _{\mu \nu} \tst
                 }
        =
        i
        \hbar
        \frac{\partial \bs{\hLam} \left [ C , A \right]}
             {\partial A}
    \quad .
    \eeq
\end{props}
\begin{proof}
We start the proof from the forward and backward Fokker--Planck equations
using them to express equation (\ref{10.stostrequmot}):
\bea
    & & \esci \esci
    -
    \frac{\partial}{\partial A}
        \left(
            \frac{1}{2}
            \left(
                \bs{\A} ^{+} _{A}
                +
                \bs{\A} ^{-} _{A}
            \right)
            \bs{\hx} ^{\mu \nu} \ttt{t}
        \right)
    =
    -
    \hbar
    \int ds
        \frac{\delta ^{2}}
             {
              \delta \bs{\hx} ^{\alpha \beta} \tst
              \delta \bs{\hx} _{\alpha \beta} \tst
             }
        \left [
            \left(
                \bs{\A} ^{+} _{A}
                -
                \bs{\A} ^{-} _{A}
            \right)
            \bs{\hx} ^{\mu \nu} \ttt{t}
        \right ]
    \nonumber \\
    & &
    -
    \frac{1}{4}
    \int d s
    \left\{
        \left [
            \left(
                \bs{\A} ^{+} _{A}
                -
                \bs{\A} ^{-} _{A}
            \right)
            \bs{\hx} ^{\alpha \beta} \tst
        \right ]
    \frac{\delta}{\delta \bs{\hx} ^{\alpha \beta} \tst}
        \left [
            \left(
                \bs{\A} ^{+} _{A}
                -
                \bs{\A} ^{-} _{A}
            \right)
            \bs{\hx} ^{\mu \nu} \ttt{t}
        \right ]
    +
    \right.
    \nonumber \\
    & & \qquad \qquad
    \left .
    -
        \left [
            \left(
                \bs{\A} ^{+} _{A}
                +
                \bs{\A} ^{-} _{A}
            \right)
            \bs{\hx} ^{\alpha \beta} \tst
        \right ]
    \frac{\delta}{\delta \bs{\hx} ^{\alpha \beta} \tst}
        \left [
            \left(
                \bs{\A} ^{+} _{A}
                +
                \bs{\A} ^{-} _{A}
            \right)
            \bs{\hx} ^{\mu \nu} \ttt{t}
        \right ]
    \right\}
    \quad ,
\eea
which we can express
in terms of the quantities $\sh{S}$ and $\sh{R}$
using results
\ref{10.nonstadrivel} and \ref{10.nonstaosmvel} as,
\bea
    & & \esci \esci
    -
    \hbar
    \frac{\partial}{\partial A}
    \left(
        \frac{\delta \sh{S}}
             {\delta \bs{\hx} ^{\mu \nu} \ttt{t}}
    \right)
    =
    -
    \frac{\hbar ^{2}}{2}
    \int ds
        \frac{\delta ^{2}}{\delta \bs{\hx} ^{\alpha \beta} \tst ^{2}}
        \ttt{\frac{\delta \sh{R}}{\delta \bs{\hx} ^{\mu \nu} \ttt{t}}}
    +
    \nonumber \\
    & &
    -
    \hbar ^{2}
    \int ds
        \left\{
            \frac{\delta \sh{R}}{\delta \bs{\hx} ^{\alpha \beta} \tst}
            \frac{\delta}{\delta \bs{\hx} _{\alpha \beta} \tst}
            \ttt{\frac{\delta \sh{R}}{\delta \bs{\hx} ^{\mu \nu} \ttt{t}}}
            -
            \frac{\delta \sh{S}}{\delta \bs{\hx} ^{\alpha \beta} \tst}
            \frac{\delta}{\delta \bs{\hx} _{\alpha \beta} \tst}
            \ttt{\frac{\delta \sh{S}}{\delta \bs{\hx} ^{\mu \nu} \ttt{t}}}
        \right\}
    \quad .
\label{10.repart}
\eea
We will use these result in a while. Indeed
at the same time we have also relation (\ref{10.fokplaequave}),
which using equation (\ref{10.nonstadrivel}) we can rewrite as,
\beq
    \frac{\partial \sh{P} \left [ C , A \right ]}
         {\partial A}
    =
    -
    \hbar
    \oint _{\Gamma} ds
        \frac{\delta}{\delta \bs{\hx} ^{\alpha \beta} \tst}
        \left[
            \frac{\delta \sh{S}}
                 {\delta \bs{\hx} _{\alpha \beta} \tst}
            \sh{P}
        \right]
\eeq
We can then substitute the definition of $\sh{P} \left [ C , A \right ]$
given in equation (\ref{10.nonstaredproden}),
and performing the functional derivatives,
we get\footnote{Without problems we can simplify by
$2 \sh{P} \left [ C , A \right ]$.}
\beq
    \frac{\partial \sh{R}}{\partial A}
    =
    -
    \frac{\hbar}{2}
    \int ds
        \frac{\delta ^{2} \sh{S}}
             {\delta \bs{\hx} ^{\alpha \beta} \tst ^{2}}
    -
    \hbar
    \int ds
        \frac{\delta \sh{S}}
             {\delta \bs{\hx} ^{\alpha \beta} \tst}
        \frac{\delta \sh{R}}
             {\delta \bs{\hx} _{\nu} \tst}
\quad .
\eeq
Multiplying now by $\hbar$ and taking the functional derivative
with respect to $\bs{\hx} ^{\mu \nu} \ttt{t}$, results in
\bea
    & & \esci \esci
    \hbar
    \frac{\partial}{\partial A}
    \left(
        \frac{\delta \sh{R}}{\delta \bs{\hx} ^{\mu \nu} \ttt{t}}
    \right)
    =
    -
    \frac{\hbar ^{2}}{2}
    \int ds
        \frac{\delta}{\delta \bs{\hx} ^{\alpha \beta} \tst ^{2}}
        \frac{\delta \sh{S}}{\delta \bs{\hx} ^{\mu \nu} \ttt{t}}
    +
    \nonumber \\
    & &
    -
    \hbar ^{2}
    \int ds
        \left\{
            \frac{\delta \sh{S}}{\delta \bs{\hx} ^{\alpha \beta} \tst}
            \frac{\delta}{\delta \bs{\hx} _{\mu \nu} \ttt{t}}
            \frac{\delta \sh{R}}{\delta \bs{\hx} ^{\alpha \beta} \tst}
            -
            \frac{\delta \sh{R}}{\delta \bs{\hx} ^{\alpha \beta} \tst}
            \frac{\delta}{\delta \bs{\hx} _{\mu \nu} \ttt{t}}
            \frac{\delta \sh{S}}{\delta \bs{\hx} ^{\alpha \beta} \tst}
        \right\}
    \quad .
\label{10.impart}
\eea
We can now add equation (\ref{10.repart}) to $i$ times
equation (\ref{10.impart}) to get
\bea
    & & \esci \esci
    i
    \hbar
    \frac{\partial}{\partial A}
    \left(
        \frac{\delta \sh{R}}{\delta \bs{\hx} ^{\mu \nu} \ttt{t}}
    \right)
    -
    \hbar
    \frac{\partial}{\partial A}
    \left(
        \frac{\delta \sh{S}}
             {\delta \bs{\hx} ^{\mu \nu} \ttt{t}}
    \right)
    =
    \nonumber \\
    & & \esci
    =
    -
    i
    \frac{\hbar ^{2}}{2}
    \int ds
        \frac{\delta}{\delta \bs{\hx} ^{\alpha \beta} \tst ^{2}}
        \frac{\delta \sh{S}}{\delta \bs{\hx} ^{\mu \nu} \ttt{t}}
    -
    \frac{\hbar ^{2}}{2}
    \int ds
        \frac{\delta ^{2}}{\delta \bs{\hx} ^{\alpha \beta} \tst ^{2}}
        \frac{\delta \sh{R}}{\delta \bs{\hx} ^{\mu \nu} \ttt{t}}
    +
    \nonumber \\
    & &
    -
    i
    \hbar ^{2}
    \int ds
        \left \{
            \frac{\delta \sh{S}}{\delta \bs{\hx} ^{\alpha \beta} \tst}
            \frac{\delta}{\delta \bs{\hx} _{\mu \nu} \ttt{t}}
            \frac{\delta \sh{R}}{\delta \bs{\hx} ^{\alpha \beta} \tst}
            -
            \frac{\delta \sh{R}}{\delta \bs{\hx} ^{\alpha \beta} \tst}
            \frac{\delta}{\delta \bs{\hx} _{\mu \nu} \ttt{t}}
            \frac{\delta \sh{S}}{\delta \bs{\hx} ^{\alpha \beta} \tst}
        \right\}
    +
    \nonumber \\
    & & \quad
    -
    \hbar ^{2}
    \int ds
        \left\{
            \frac{\delta \sh{R}}{\delta \bs{\hx} ^{\alpha \beta} \tst}
            \frac{\delta}{\delta \bs{\hx} _{\alpha \beta} \tst}
            \frac{\delta \sh{R}}{\delta \bs{\hx} ^{\mu \nu} \ttt{t}}
            -
            \frac{\delta \sh{S}}{\delta \bs{\hx} ^{\alpha \beta} \tst}
            \frac{\delta}{\delta \bs{\hx} _{\alpha \beta} \tst}
            \frac{\delta \sh{S}}{\delta \bs{\hx} ^{\mu \nu} \ttt{t}}
        \right\}
    \quad .
\eea
Rearranging this result we then get
\bea
    & & \esci \esci
    i
    \hbar
    \frac{\partial}{\partial A}
    \left\{
        \frac{\delta \sh{R}}
             {\delta \bs{\hx} ^{\mu \nu} \ttt{t}}
        +
        i
        \frac{\delta \sh{S}}
             {\delta \bs{\hx} ^{\mu \nu} \ttt{t}}
    \right\}
    =
    \nonumber \\
    & & \esci
    =
    -
    \frac{\hbar ^{2}}{2}
    \int ds
        \frac{\delta}{\delta \bs{\hx} ^{\alpha \beta} \tst ^{2}}
        \left\{
            \frac{\delta \sh{R}}{\delta \bs{\hx} ^{\mu \nu} \ttt{t}}
            +
            i
            \frac{\delta \sh{S}}{\delta \bs{\hx} ^{\mu \nu} \ttt{t}}
        \right\}
    +
    \nonumber \\
    & &
    -
    i
    \hbar ^{2}
    \int ds
        \left [
            \frac{\delta \sh{R}}{\delta \bs{\hx} ^{\alpha \beta} \tst}
            +
            i
            \frac{\delta \sh{S}}{\delta \bs{\hx} ^{\alpha \beta} \tst}
        \right]
        \frac{\delta}{\delta \bs{\hx} ^{\mu \nu} \ttt{t}}
        \left [
            \frac{\delta \sh{R}}{\delta \bs{\hx} _{\alpha \beta} \tst}
            +
            i
            \frac{\delta \sh{S}}{\delta \bs{\hx} _{\alpha \beta} \tst}
        \right]
\eea
and at the end
\bea
    & & \esci \esci
    i
    \hbar
    \frac{\delta}
         {\delta \bs{\hx} ^{\mu \nu} \ttt{t}}
    \left\{
        \frac{\partial}{\partial A}
        \left [
            \sh{R}
            +
            i
            \sh{S}
        \right ]
    \right\}
    =
    \nonumber \\
    & & \esci
    =
    -
    \frac{\hbar ^{2}}{2}
    \frac{\delta}
         {\delta \bs{\hx} ^{\mu \nu} \ttt{t}}
    \int ds
        \frac{\delta}{\delta \bs{\hx} ^{\alpha \beta} \tst ^{2}}
        \left\{
            \sh{R}
            +
            i
            \sh{S}
        \right\}
    +
    \nonumber \\
    & &
    -
    i
    \frac{\hbar ^{2}}{2}
        \frac{\delta}{\delta \bs{\hx} ^{\mu \nu} \ttt{t}}
    \int ds
        \left\{
            \frac{\delta}{\delta \bs{\hx} ^{\alpha \beta} \tst}
            \left [
                \sh{R}
                +
                i
                \sh{S}
            \right]
        \right\} ^{2}
    \quad .
\eea
We thus see that a total functional gradient can be {\it integrated
away} and after multiplying by $\bs{\hLam}$ the result we get
\bea
    & & \esci \esci
    i
    \hbar
    \frac{\partial}{\partial A}
    \left [
        \sh{R}
        +
        i
        \sh{S}
    \right ]
    \bs{\hLam}
    =
    \nonumber \\
    & & =
    \frac{\hbar ^{2}}{2}
    \int ds
        \left\{
            -
            \frac{\delta ^{2} \sh{R}}
                 {\delta \bs{\hx} ^{\mu \nu} \tst ^{2}}
            +
            i
            \frac{\delta ^{2} \sh{S}}
                 {\delta \bs{\hx} ^{\mu \nu} \tst ^{2}}
            +
            \left [
                \frac{\delta}
                     {\delta \bs{\hx} ^{\mu \nu} \tst}
                \left(
                    \sh{R}
                    +
                    i
                    \sh{S}
                \right)
            \right]
        \right\}
        \bs{\hLam}
\eea
which, remembering results (\ref{10.areder}-\ref{10.secfunder}),
gives exactly
\beq
    \frac{\hbar ^{2}}{2}
    \oint ds
        \frac{\delta ^{2} \bs{\hLam} \left [ C , A \right]}
             {
              \delta \bs{\hx} ^{\mu \nu} \tst
              \delta \bs{\hx} _{\mu \nu} \tst
             }
    =
    i
    \hbar
    \frac{\partial \bs{\hLam} \left [ C , A \right]}
         {\partial A}
\eeq
\end{proof}
This result is very similar to the {\sl String Functional Wave
Equation} that we derived in chapter \ref{3.strfunqua}
and, choosing in a proper way a weight for the NonStandard Stochastic
Quantities, can be cast in exactly the same form. Thus we have a completely
rigorous derivation of that equation in terms of NonStandard Analysis.

\section{Holographic Coordinates and $M$-Theory}
\label{10.Mthconsec}

\subsection{Boundary Shadow Dynamics and $M$-Theory}

One of the most enlightening features of the Eguchi approach is
the formal correspondence it establishes between the Quantum
Mechanics of point particles and {\sl String} loops. Such a relationship
is summarized in the {\it translation code} displayed in the
Table \ref{dictionary}. Instrumental to this correspondence
is the replacement of the canonical
{\sl String} coordinates $Y ^{\mu} \tst$ with the reparametrization
invariant {\sl Holographic Coordinates}
$Y ^{\mu \nu} \left[ C \right]$.
We observed in previous chapters that, surprising as it may appear
at first sight, the new
coordinates $Y ^{\mu\nu} \left[ C \right]$ are just as ``natural'' as
the old $Y ^{\mu} \tst$ for the purpose of  defining the {\sl String}
``position''.
As a matter of fact, a Classical Gauge Field Theory
of Relativistic {\sl Strings} was proposed several
years ago\footnote{We already saw in chapter \ref{8.douclalimcha}
the relation between our present ``{\it{}quantum}'' proposal and
that classical framework.} \cite{no},
but only recently it was extended to
generic $p$-{\sl{}branes} including their coupling to $(p+1)$-forms
and Gravity \cite{noi2}.\\
Now that we have established the connection between {\sl Holographic}
Quantization and the path--integral formulation of Quantum
{\sl Strings}, it seems almost compelling to ponder about the
relationship, if any, between the
$Y ^{\mu \nu} \left[ C \right]$ matrix
coordinates and the matrix coordinates which, presumably, lie
at the heart of the M-Theory formulation of superstrings.
Since the general framework of M-Theory is yet to be
discovered, it seems reasonable to focus on a specific matrix
model recently proposed for {\it Type IIB}
superstrings\footnote{A similar matrix action for the Type
IIA model has been conjectured in \cite{banks}.} \cite{ikkt}.
The Dynamics of this model is encoded into a simple Yang-Mills type action
\beq
    S _{\mathrm{IKKT}}
    =
    -
    \frac{\alpha}{4}
    \Tr{\left[ A ^{\mu} , A ^{\nu} \right] ^{2}}
    +
    \beta
    \Tr{\mathbb{I}}
    +
    \mathrm{fermionic\ part}
    \quad ,
\label{jap}
\eeq
where the $A ^{\mu}$  variables are represented by $N \times N$
hermitian matrices and $\mathbb{I}$ is the unit matrix.
{\it The novelty of
this formulation is that it identifies the ordinary spacetime
coordinates with the eigenvalues of the non-commuting Yang-Mills
matrices}.
In such a framework, the emergence of  {\it Classical spacetime}
occurs in the {\it large-$N$ limit},
i.e., when the commutator goes into a Poisson
bracket\footnote{This is not the canonical Poisson bracket
which is replaced by the Quantum Mechanical commutator.
Rather, it is the {\sl World--Sheet} symplectic structure which is
replaced by the (classical) matrix commutator for finite $N$.}
and the matrix trace operation turns into a double continuous
sum over the row and column indices, which amounts to a two
dimensional invariant integration. Put briefly,
\bea
    \lim _{N \rightarrow \infty} \hbox{``Tr''}
    & \rightarrow &
    -
    i
    \int d ^{2} \Bsigma
        \sqrt{\dete{\Bg}}
    \\
    -
    i
    \lim _{N \rightarrow \infty}
        \comm{A ^{\mu}}{A ^{\nu}}
    & \rightarrow &
    \pois{Y ^{\mu}}{Y ^{\nu}}
    \\
    \sqrt{\dete{\Bg}}
    \pois{Y ^{\mu}}{Y ^{\nu}}
    & \equiv &
    \dot{Y} ^{\mu \nu}
    \quad .
\eea
What interests us is that, {\it in such a  classical limit, the
{\rm IKKT} action}
(\ref{jap}) {\it turns into the Schild action in Equation}
(\ref{prop})
{\it once we make the identifications}
\bea
    \alpha
    & \longleftrightarrow &
    -
    m ^{2}
    \nonumber \\
    \beta
    & \longleftrightarrow &
    {\cal E}
    \\
   N \left( \Bsigma \right)
   & \longleftrightarrow &
   \sqrt{\dete{\Bg}}
   \quad ,
\eea
while the trace of the Yang--Mills commutator turns into the
oriented surface element
\beq
    lim _{N \to \infty}
        \Tr{\comm{A ^{\mu}}{A ^{\nu}}}
    \longrightarrow
    \int _{\Sigma} d ^{2} \Bsigma
        \dot{X} ^{\mu \nu} \ttt{\Bsigma}
    =
    Y ^{\mu \nu} \left( C \right)
    \quad ,
\eeq
$C$ being the image of $\Gamma = \partial \Sigma$ in {\sl Target Space},
as usual: $C = \By \ttt{\Gamma}$.
This formal relationship can be clarified by considering a
definite case. As an example let us consider a static $D$-{\sl{}String}
configuration both in the classical Schild formulation
and in the corresponding matrix description. It is
straightforward to prove that a length $L$ static {\sl String}
stretched along the $x ^{1}$ direction, i.e.
\bea
    X ^{\mu}
    & = &
    \sigma ^{0}
    T
    \delta ^{\mu 0}
    +
    \frac{L \sigma ^{1}}{2 \pi}
    \delta ^{\mu 1}
    \quad , \quad
    0 \le \sigma ^{0} \le 1
    \quad , \quad
    0 \le \sigma ^{1} \le 2 \pi
    \\
    X ^{\mu}
    & = &
    0
    \quad ,\quad
    \mu \ne 0 , 1
\eea
solves the classical equations of motion
\beq
    \pois{X _{\mu}}{\pois{X ^{\mu}}{X ^{\nu}}}
    =
    0
    \quad .
\eeq
During a time lapse $T$ the {\sl String} sweeps a timelike {\sl World--Sheet}
in the $0-1$ plane characterized by the {\sl Holographic Coordinates}
\beq
    Y ^{\mu\nu} \left( L , T \right)
    =
    \int _{0} ^{1} d \sigma ^{0}
    \int _0 ^{2 \pi} d \sigma ^{1}
    \sqrt{\dete{\Bg}}
    \quad , \qquad
    \pois{X ^{\mu}}{X ^{\nu}}
    =
    T
    L
    \delta ^{0 [ \mu}
    \delta ^{\nu ] 1}
    \quad .
\label{area}
\eeq
Equation (\ref{area}) gives both the area and the orientation of
the rectangular loop  which is the {\it Boundary} of the {\sl String
World--Sheet}. The
corresponding matrix solution, on the other hand, must satisfy the equation
\beq
    \comm{A _{\mu}}{\comm{A ^{\mu}}{A ^{\nu}}}
    =
    0
    \quad .
\label{ym}
\eeq
Consider, then, two hermitian, $N \times N$ matrices $\hat{q}$,
$\hat{p}$ with an approximate $c$-number commutation relation
$\displaystyle \comm{\hat{q}}{\hat{p}} = i$, when $ N \gg 1$.
Then, a solution of the Classical Equation of Motion (\ref{ym}),
corresponding to a {\it solitonic} state in {\sl String
Theory}, can be written as
\beq
    A ^{\mu}
    =
    T
    \delta ^{\mu 0}
    \hat{q}
    +
    \frac{L}{2 \pi}
    \delta ^{\mu 1}
    \hat{p}
    \quad .
\eeq
In the large-$N$ limit we find
\beq
    -
    i
    \comm{A ^{\mu}}{A ^{\nu}}
    =
    -
    i
    \frac{L T}{2 \pi}
    \delta ^{0 [ \mu}
    \delta ^{\nu ] 1}
    \comm{\hat{q}}{\hat{p}}
    \approx
    \frac{LT}{2 \pi}
    \delta ^{0 [ \mu}
    \delta ^{\nu ] 1}
\eeq
and
\beq
    -
    i
    \Tr{\comm{A ^{\mu}}{A ^{\nu}}}
    \approx
    \int _{0} ^{1} d \sigma ^{0}
    \int _{0} ^{2\pi} d \sigma ^{1}
        \sqrt{\dete{\Bg}}
        \pois{X ^{\mu}}{X ^{\nu}}
    =
    Y ^{\mu \nu} \left( L , T \right)
    \quad .
\eeq
These results, specific as they are, point to a deeper connection between the
loop space description of {\sl String} Dynamics and matrix models of superstrings
which, in our opinion, deserves a more detailed investigation.

\subsection{Covariant ``Speculation''}

We complete the observations of the previous section
by speculating about a ``fully covariant''
formulation of the {\sl Boundary Wave Equation} (\ref{9.schrod}) and its
physical consequences.\\
The spacelike and timelike
character of the {\sl Holographic Coordinates} $Y ^{\mu\nu}$ and
of the {\sl Areal Time} $A$ shows
up in the Schr\"odinger, ``non--relativistic'' form, of
the wave equation (\ref{9.schrod}) which is second order in
th {\sl Holographic Functional Derivatives} $\delta/
\delta Y^{\mu\nu} \tst$, and first order in $\partial/\partial A$.
Therefore, it is intriguing to ponder about the form and the meaning of the
corresponding {\it Klein--Gordon} equation. The first step towards the
``relativistic'' form of (\ref{9.schrod}) is to introduce an appropriate
coordinate system where $Y ^{\mu\nu}$ and $A$ can play a physically
equivalent role. Let us introduce a {\it matrix coordinate}
${\mathbb{X}}^{MN}$,
where certain components are $Y ^{\mu\nu}$ and $A$. There is a
great freedom in the choice of ${\mathbb{X}}^{MN}$. However, an interesting
possibility would be
\beq
{\mathbb{X}}^{MN}=\left(\begin{array}{cc}
              m\,Y^{\mu\nu} &
              \delta^{\mu(\nu)}Y_{(\nu)}^{C.M.}\\
              \delta^{\mu(\nu)}Y_{(\nu)}^{C.M.} & m\,A_{\mu\nu}
              \end{array}\right)
              \quad ,
\label{9.matrix}
\eeq
where
\beq
A_{\mu\nu}=\left(\begin{array}{cc}
                 \mathbb{O} & \mathbb{A}\\
                 -\mathbb{A}     & \mathbb{O}
                 \end{array}
            \right)
\eeq
with
\beq
    \mathbb{A} = \left(\begin{array}{cc} 0 & A \\ 0 & 0 \end{array} \right)
    \quad :
\eeq
here we arranged the {\sl String} center of mass coordinate $ Y_\nu^{C.M.}$
inside off--diagonal sub--matrices and build--up an antisymmetric
proper area tensor in oder to endow $A$ with the same tensorial
character as $Y^{\mu\nu}$. The {\sl String} {\it length
scale}, $1/m$, has been introduced to provide the block
diagonal area sub-matrices  the coordinate canonical dimension
of a {\it length}, in natural units. The most remarkable feature for
this choice of ${\mathbb{X}}^{MN}$ is that if we let $\mu , \nu$ to range
over four values, then ${\mathbb{X}}^{MN}$ is an $8 \times 8$ antisymmetric
matrix with {\it eleven} independent entries. Inspired by the
recent progress in non--commutative geometry, where
point coordinates are described by non--commuting matrices,  we
associate  to each matrix  (\ref{9.matrix}) a representative point
in an eleven dimensional space which is the product of the
$4$-dimensional spacetime $\times$ the $6$-dimensional {\sl Holographic
Loop Space} $\times$ the  $1$-dimensional {\sl Area Time} axis. Eleven
dimensional spacetime is the proper arena of $M$-Theory,
and we do not believe this is a mere coincidence.\\
According with the assignment of the eleven entries in ${\mathbb{X}}^{MN}$ the
 corresponding ``point'' can represent different physical objects:
\begin{enumerate}
    \item a pointlike particle:
    $$
        \{ Y^{\mu\nu}=0\ , A=0\ , Y_\nu^{C.M.}\} \quad ;
    $$
    \item a loop with center of mass in $Y_\nu^{C.M.}$ and {\sl Holographic
    coordinates} $Y ^{\mu\nu}$:
    $$
        \{ Y^{\mu\nu}(\gamma)\ , A=0\ ,Y_\nu^{C.M.}\} \quad ;
    $$
    \item an open surface of proper area $A$,
    {\it Boundary} {\sl Holographic Coordinates} $Y^{\mu\nu}$ and
    center of mass in $Y_\nu^{C.M.}$,i.e. a {\it real} {\sl String}:
    $$
        \{ Y^{\mu\nu}(\gamma)\ , A\ ,Y_\nu^{C.M.}\} \quad;
    $$
    \item a closed surface with {\sl Parameter Space} of proper area $A$,
    i.e. a virtual {\sl String},
    $$
        \{ Y^{\mu\nu}=0\ , A\ , Y_\nu^{C.M.}\} \quad .
    $$
\end{enumerate}

It is appealing to conjecture that ``Special Relativity'' in this
enlarged space will transform one of the above objects into another
by a reference frame transformation! From this vantage viewpoint
particles, loops, real and virtual {\sl Strings} would appear as the same
basic object as viewed from different reference frames. Accordingly,
a quantum field $\Phi({\mathbb{X}})$ would create and destroy the
objects listed above, or, a more basic object encompassing all of them.
A {\it unified Quantum Field Theory of points, loops and}
{\sl Srings}, and its relation, if any,  with $M$-Theory or
non--commutative geometry, is an issue which
deserves a throughly, future, investigation: of particolar
relevance are the possible consequences about the nature of spacetime.
This is a very deep and fundamental problem, that needs a careful
investigation and a better comprehension of the empirical meaning
of these higly theoretical formulation. Neverthless we give, in the next
section, a proposal, without pretend to be exhaustive, but only to
see how it could be possible to set up a convenient framework
to address this topis starting from the concepts we presented so far.

\section{Superconductivity and Quantum Spacetime}

To set up a correct environment, it is worth to recall that
Quantum {\sl Strings}, or more generally
?-{\sl{}branes} of various kind, are currently
viewed as the fundamental constituents of everything: not
only every matter particle or gauge boson must be
derived from the {\sl String} vibration spectrum, but spacetime
itself is built out of them.\\
If spacetime is no longer preassigned, then logical
consistency demands that a matrix representation of $p$-{\sl{}brane}
Dynamics cannot be formulated in any given background
spacetime. The exact mechanism by which $M$-Theory is supposed
to break this circularity is not known at present, but {\sl Loop
Quantum Mechanics} points to a possible resolution of that
paradox. If one wishes to discuss Quantum {\sl Strings} on the same
footing with point particles and other $p$-{\sl{}branes}, then their
Dynamics is best formulated in {\sl Loop Space} rather than in
physical spacetime. As we have repeatedly stated throughout
this paper, our emphasis on {\sl String} {\it shapes} rather than
on the {\sl String} constituent points, represents a departure from
the canonical formulation and requires an appropriate choice
of dynamical variables, namely the {\sl String Holographic
Coordinates} and the {\sl Areal Time}. Then, at the {\sl Loop Space}
level, where each ``point'' is representative of a particular loop
configuration, our formulation is purely Quantum Mechanical,
and there is no reference to the background spacetime. At the
same time, the functional approach leads to a precise
interpretation of the fuzziness of the underlying Quantum
spacetime in the following sense: when the resolution of the
detecting apparatus is smaller than a particle De Broglie
wavelength, then the particle Quantum trajectory behaves as a
Fractal curve of Hausdorff dimension two. Similarly we have
concluded on the basis of the {\sl Shape Uncertainty Principle}
that the Hausdorff dimension of a Quantum {\sl String World--Sheet}
is three, and that two distinct phases (smooth and fractal
phase) exist above and below the loop De Broglie area. Now,
if particle world--lines and {\sl String World--Sheets} behave as
fractal objects at small scales of distance, so does the
{\sl World--HyperTube} of a generic $p$-{\sl{}brane} including spacetime
itself \cite{nott}, and we are led to the general expectation
that a new kind of {\it Fractal Geometry} may provide an
effective dynamical arena for physical phenomena near the
{\sl{}String} or Planck scale in the same way that a smooth
Riemannian geometry provides an effective dynamical arena for
physical phenomena at large distance scales.\\
Once committed to that point of view, one may naturally ask,
``what kind of  physical mechanism can be invoked in the
framework of {\sl Loop Quantum Mechanics} to account for the
transition from the fractal to the smooth geometric phase of
spacetime?''. A possible answer consists in the phenomenon of
{\it $p$-{\sl{}brane} condensation}. In order to illustrate its
meaning, let us focus, once again, on {\sl String} loops. Then, we
suggest that what we call ``classical spacetime'' emerges as a
condensate, or {\sl String} vacuum similar to the ground state of a
superconductor. The large scale properties of such a state
are described by an ``effective Riemannian geometry''. At a
distance scale of order $(\alpha') ^{1/2}$, the condensate
``evaporates'', and with it, the very notion of Riemannian
spacetime. What is left behind, is the fuzzy stuff of Quantum
spacetime. \\
Clearly, the above scenario is rooted in the Functional
Quantum Mechanics of {\sl String} loops discussed in the previous
sections, but is best understood in terms of a model which
mimics the Ginzburg-Landau Theory of superconductivity. Let
us recall once again that one of the main results of the
Functional
approach to Quantum
{\sl Strings} is that {\it it is possible to describe the evolution
of the system without giving up
reparametrization invariance}. In that approach,
the clock that times the evolution of a closed bosonic {\sl String} is
the {\sl Area} of the {\sl Parameter Space} associated with the
{\sl World--Sheet}, which we rewrite here:
$$
    A
    =
    \frac{1}{2}
    \epsilon _{ab}
    \int _{\Xi} d \xi ^{a} \wedge d \xi ^{b}
   \quad .
$$
The choice of
such a {\it fiducial} surface is arbitrary,
corresponding to the freedom of choosing the
initial instant of time, i.e., a fiducial reference area.
Then one can take advantage of this arbitrariness in
the following way. In Particle Field Theory an arbitrary lapse of euclidean,
or Wick rotated, imaginary time between initial and final
field configurations is usually given the meaning of {\it inverse temperature}
$$
    i
    \Delta t
    \longrightarrow
    \tau
    \equiv \frac{1}{\kappa _{B} T}
$$
and the resulting Euclidean Field Theory provides a {\it finite temperature}
statistical description of vacuum fluctuations. \\
Following the same procedure, we suggest to analytically
extend  $A$ to imaginary values,
$i A\longrightarrow a$, on the assumption that the resulting
{\it finite area Loop Quantum Mechanics} will provide a
statistical description of the {\sl String} vacuum
fluctuations. This leads us to consider the following
effective (euclidean) lagrangian of the Ginzburg--Landau type,
\bea
    & &
    L \left( \Psi , \Psi^{\ast} \right)
    =
    \Psi ^{\ast}
    \frac{\partial}{\partial a}
    \Psi
    -
    \frac{1}{4 m ^{2}}
    \norm
    \oint _{C} d l \tst
    \left \vert
        \left(
            \frac{\delta}{\delta \sigma ^{\mu\nu} \tst}
            -
            i
            g
            A _{\mu \nu} \left( x \right)
        \right)
        \Psi
    \right \vert ^{2}
    +
    \nonumber \\
    & & \qquad \qquad \qquad \qquad
    \phantom{\left( \Psi , \Psi ^{\ast} \right)}
    -
    V \left( \left \vert \Psi \right \vert ^{2} \right)
    -
    \frac{1}{2 \cdot 3!}
    H ^{\lambda \mu \nu}
    H _{\lambda \mu \nu}
\label{model}
    \\
    & &
    V \left( \left \vert \Psi \right \vert ^{2} \right)
    \equiv
    \mu _{0} ^{2}
    \left( \frac{a _{c}}{a} - 1 \right)
    \left \vert \Psi \right \vert ^{2}
    +
    \frac{\lambda}{4}
    \left \vert \Psi \right \vert ^{4}\\
    & &
    H _{\lambda \mu \nu}
    =
    \partial _{[ \lambda} A _{\mu \nu ]}
\quad .
\eea
Here, the {\sl String} field is coupled to a Kalb--Ramond gauge potential
$A _{\mu\nu} \left( x \right)$ and $a _{c}$ represents a
{\it critical area} such that, when $ a \le a _{c}$ the potential energy
is minimized by the {\it ordinary vacuum} $\Psi \left[ C \right] = 0$,
while for $a > a _{c}$ {\sl Strings} condense into a {\it superconducting vacuum}.
In other words,
\beq
    \vert \Psi[C] \vert ^{2}
    =
    \cases{
           0
           &
           if
           $a \le a _{c}$
           \cr
           \cr
           \displaystyle
           \frac{\mu _{0} ^{2}}{\lambda}
           \left(1 - a _{c} / a \right)
           &
           if
           $a > a _{c}$
          }
\eeq
Evidently, we are thinking of the {\sl String} condensate as the
large scale, background  spacetime. On the other hand, as one
approaches distances of the order
$\left( \alpha' \right)^{1/2}$ {\sl Strings} undergo
more and more shape--shifting, transitions which destroy the
long range correlation of the {\sl String} condensate. As we have
discussed earlier, this signals the transition from the
smooth to the fractal phase of the {\sl String World--Sheet}.
On the other hand, the Quantum Mechanical approach
discussed in this paper is in no way restricted to {\sl{}String}--like
objects.
In principle\footnote{We have already seen in chapter
\ref{4.pbrcha} and in section \ref{6.minpbrequsec} how the results
given for the {\sl String} can be generalized without further
technical problems to the more general case of $p$-{\sl{}branes}.},
it can be extended to any quantum $p$-{\sl{}brane}
and the limiting value of the corresponding
fractal dimension is $D _{H} = p + 2$. Then, if the above over all
picture is correct, spacetime fuzziness acquires a well defined meaning.
Far from being a smooth, $4$-dimensional manifold
assigned ``ab initio'', spacetime is, rather, a ``process in the
making'', showing an ever changing fractal structure which responds
dynamically to the resolving power of the detecting apparatus.
At a distance scale of the order of Planck's length, i.e., when
\beq
    a _{c} = G _{N}
\label{newt}
\eeq
then the {\it whole of spacetime boils over} and no trace is
left of the large scale condensate of either {\sl Strings} or
$p$-{\sl{}branes}.\\
As a final remark, it is interesting to note that since the
original paper by A.D.Sakarov about gravity as spacetime elasticity,
$G _{N}$ has been interpreted as a {\it phenomenological parameter}
describing the large scale properties of the gravitational vacuum.
Eq.(\ref{newt}) provides a deeper insight into the meaning of $G_N$ as the
{\it critical value} corresponding to the transition point between an
``elastic'' Riemannian--type condensate of extended objects and a
Quantum phase which is just a Planckian foam of Fractal objects.

%% file: concl.tex
\pageheader{}{Conclusions.}{}
\makechaptermyhead{Conclusion}{I$\!$V}

\noindent{}
\begin{start}
``$\euf{Y}$ou know,\\
I did feel something.\\
I could almost see the remote.''\\
``That's good.\\
You have taken your first step\\
into a larger world.''\\
\end{start}

\noindent{}In the  ten chapters before this conclusion
we have touched a lot of arguments, some of them are very far away
one from  another. Even if the main subject has been a particular
formulation of the  Classical and  Quantum Dynamics, of Relativistic
Extended Objects, the main idea was to show how to develop,
using quite natural concepts,  an organic Theory without
assuming to much preexisting structures behind. Thus it seems
important to us  to summarize the results we obtained, to remark
the motivations which support our work, and the future
developments of this line of research.
Indeed the long road that brought after years of work and reflection
to the results we gave here is not at the end: perhaps at the
beginning, since it would be important to apply in a more concrete
way the concepts developed so far.

In particular, we have in mind the difficulties that have been found
so far in trying to approach the problem of quantizing Gravity: this
is a problem that refused any proposed solution; the
purely geometrical approach of General Relativity was not able to
describe short distance spacetime structure in a consistent way;
no better results came from the purely Quantum Mechanical approach,
or from Quantum Field Theory. The problem of {\it Quantizing Geometry}
remains still with us and there is still no satisfactory answer.

In our opinion all the difficulties are very deeply related to the
fact that, until now, Gravity Theory and Quantum Theory have been
viewed and interpreted as pertaining to very different physical realms:
in connection to large scale problems, the first, and related to
short range effects the second one. Thus a Quantum Theory of Gravity
should be something that pertains to both fields: it should be
something able to relate the physics at very different order of
magnitude in length scale, or energy scale, which is the same.
If we assume that, after all, purely quantum gravitational
effects are really part of our world, we see
only three possible roads toward the goal of finding a Theory able
to describe them, i.e. to give some phenomenological prediction
capable of being experimentally tested:
\begin{enumerate}
    \item in first instance it is possible to investigate
    quantum effects on scales macroscopic with respect to,
    say, the atomic scale: far from being a new research field
    we think we are simply speaking of the {\it Quantum Theory
    of Measurement}, where Quantum Effects are brought to our
    everyday scale by a detecting apparatus;
    \item in second instance we could also think to find on very
    small scales, say the scales of elementary particles, effects
    related to the large scale structure of spacetime: this topics
    are strongly related with research in {\it AstroParticle Physics};
    \item as a third possibility, we can start more or less by scratch,
    trying to see if there is some theoretical formulation capable to
    give a consistent framework to short as well as long range effects.
\end{enumerate}
By the way, it is clear that we adhered in this work to the third
possibility, which is the one developed in {\sl String} Theory,
but trying to give it a more Geometrical flavor, instead than
a Quantum Field Theoretical one. In our point of view the Theory
of extended objects has a main advantage in tackling problems associated
with the presence of different scales because an extended object
is composed by the {\sl Bulk}, its history\footnote{The {\sl World--Sheet}
for a {\sl String}.}, but also by the {\it Boundary}. Effects on the {\sl Bulk}
can be in some sense {\it local}, but can have some {\it global}
influence on the {\it Boundary} too. This is the reason why we introduced the
{\sl Boundary Shadow Dynamics} of the {\sl String} as a way to keep
track of both effects. Moreover, the formulation we gave in terms
of Hamilton--Jacobi equation is motivated by the fact that this is a
key tool in deriving a connection between the Classical and the Quantum
Theory, already at the point particle level.

The main result of our formulation is that it is the same Theory that
singles out, given the {\sl Parametrization} for the {\sl Bulk}, the
most appropriate canonical variables for the {\it Boundary} Quantum Dynamics
which is encoded in our {\sl Functional
Wave Equation}\footnote{Incidentally
we observe the  striking similarity between our equation and the
Wheeler-deWitt equation for the {\it Wave Function of the Universe}.}:
these are the {\sl Holographic Coordinates}. Moreover, also the
evolution parameter is given as a product of our description of
the {\it Boundary} Dynamics and is again an {\it areal quantity},
namely the area of the {\sl Parameter Space}, the fiducial manifold
on which the {\sl String} and {\sl World--Sheet Embedding Functions}
are defined.

Now, it is not difficult to imagine that a complete Quantum
Theory of Gravity would probably have relevant effects on our
view of the world, which for a physicist, probably, means spacetime.
Let us explain in more detail our ideas about this issue:
\begin{enumerate}
    \item the main revolution of General Relativity has been to recognize that
    the presence of mass and energy can influence the
    causal structure of our world and, {\it at the same time}, be
    driven from it: we are {\it backreacted} in some way;
    \item on the other hand, Quantum Theory gave us the very deep
    result that, thanks to the Uncertainty Principle, not all situations, that
    we could imagine starting from our classical background, are practically
    realizable: we are {\it constrained} in some other way.
\end{enumerate}
Thus, in particular,
it is a logical conclusion that in a {\it would be} Theory
of Quantum Gravity the distribution of mass and energy should
give a constrained backreacting quantum interaction that only in
the average, taken at some particular scale, reproduces our
classical world. Moreover, not all the possible configurations, or, which
is the same, spacetime structures\footnote{Or, again, mass and energy
distribution, i.e. what we perceive as motion, flow, and so on.},
{\it empirically detectable as true}
in this classical average, are realizable in the Quantum Gravitational
Realm.

We like to think that as happens in Atomic Physics, the Uncertainty
Principle could be advocated to stabilize systems that, otherwise,
would undergo collapse. That is to say that we like to think
the Uncertainty Principle as the {\it physical reason} for the existence
of extension at small scales, because pointlik{\it{}ness} explicitly
violates such a fundamental principle. Thus in a Quantum Theory
of Gravity to assume the {\it point} as the basic constituent
of spacetime is not consistent: the rules of Quantum Mechanics forbids
pointlike situations, because we are uncertainty bounded in our
empirical knowledge. In this way the reader can see how we have reach by
a completely different point of view the same conclusion already
present in contemporary  {\sl String} Theory: the elementary constituents
of a Quantum Gravitational spacetime must be extended objects, {\it not}
point at all. This is the ground of our proposal of describing the
Quantum Dynamics of Extended Objects with a first quantized functional
formulation: if points becomes extended, let us say {\sl Strings}, but way
not membranes or other extendons, then a Quantum Theory of spacetime
is a Theory defined in a space of loops, membranes, \dots ,
$p$-{\sl{}branes} in general. Even if
Classical {\sl String Dynamics} is based on the simple and intuitive notion
that the {\sl World--Sheet} of a Relativistic {\sl String} consists of a
{\it smooth, $2$-dimensional} manifold embedded in a preexisting spacetime,
switching from
Classical to Quantum Dynamics changes this picture in a fundamental way.
Indeed, already in the path--integral approach to
particle Quantum Mechanics, Feynman and
Hibbs were first to point out that the trajectory of a particle is
continuous but nowhere differentiable \cite{fh}. This is again
the Uncertainty Principle at work\footnote{In this sense, we stress that
again, trying to
combine Quantum Theory with a Theory of spacetime, points
directly toward a formulation in terms of extended objects.}:
when a particle is more and more
precisely located in space, its trajectory becomes more and more
erratic. Abbott and Wise were next to point out that a particle
trajectory, appearing as a smooth line of topological
dimension one, turns into a {\it fractal} object of
{\it Hausdorff dimension} two, when the
resolution of the detecting apparatus is smaller than the particle
De Broglie wavelength \cite{abbw}.
Now in our point of view extended objects, not points, are
the constituents of spacetime at very short scales.
This is the reason why we extended the path--integral
approach to the {\sl String} case: in this case one is taking into
account the coherent contributions from all
the {\sl World--Sheets} satisfying some preassigned {\it Boundary}
conditions and one finds that a {\sl String Quantum World--Sheet}
is a fractal, non--differentiable surface. The way we explained
in this Thesis to give a {\it quantitative} support to this expectation
takes advantage by working with a first
quantized formulation of {\sl String}
Dynamics: in this way we are quantizing the
{\sl String} motion not through the displacement of each point on
the {\sl String}, but through the {\sl String} {\it shape}. The outcome
of this novel approach is  a {\it Quantum Mechanics of Loops},
describing {\it quantum shape shifting transitions}, i.e.,
the {\sl Shadow Dynamics} of the {\sl String}.
The emphasis on {\sl String} shapes, rather than points,
meets the interpretation we gave a few lines above of
{\sl Strings} as constituents of spacetime. We think worth to
note that this departure from the canonical formulation which requires
an appropriate choice of dynamical variables, namely, the {\sl String}
{\sl Holographic Coordinates} $Y \left [ C \right ]$
and the areal time $A$, immediately present itself with a very
striking and relevant property. In particular, the {\sl Shadow
Dynamics} is not able to completely single out a precise loop
in the Quantum {\sl String} First Quantized Dynamics. From the information
encoded in the {\sl Holographic Coordinates} we can not say that our
{\sl String}, i.e. extended spacetime point, is a square, or a circle, \dots{},
but only that {\it there is only one parallelogram in space that
has on the coordinate planes the projections such that their areas
coincide with the components of the {\sl Holographic Coordinates} Tensor}.
Thus our functional approach enables us to extend the {\it Quantum
Mechanical} discussion to spacetime itself: but we have to
adhere to the following principle: {\it events cannot be localized
with infinite precision, because spacetime is no more made of points;
its elementary constituents are extended object (say {\sl Strings})
and we thus can only say that we are able to find a ``plaquette'',
with the form of a parallelogram, of which we know the projection
onto those object that classically are the coordinate planes and in
which our event takes place}.

The Dynamics of such extended spacetime elements could be a bit
confusing at first sight, if not contradictory:
following all the procedure the reader could object that
we deal with a relativistic
system in a Quantum Mechanical, i.e., non relativistic framework, as opposed
to a quantum field theoretical framework \cite{mr}. Even if this is not
new in theoretical physics \cite{franchi}, in any case, one has to
realize that there are two distinct levels of discussion in our approach.
The {\it spacetime level}, where the actual deformations
of the {\sl String}, thought as a constituent element, take place,
and where the formulation is fully relativistic, as
witnessed by the covariant structure of our equations with
respect to the Lorentzian indices. However, at the {\it loop space level},
where each ``point'' is representative of a particular loop
configuration, our formulation is Quantum Mechanical, in the sense that the
{\sl String} coordinates $Y ^{\mu \nu}$ and $A$
are not treated equally, as it is manifest,
for instance, in the loop Schr\"odinger equation (\ref{3.funlooschequ}).
As a matter of fact, this is the very reason for referring to that
equation as the ``Schr\"odinger equation'' of {\sl String} Dynamics:
the timelike
variable $A$ enters the equation through a first order partial
derivative, as opposed to the functional ``laplacian'', which is of second
order with respect to the spacelike variables $Y ^{\mu \nu}$. Far from being
an artifact of our formulation, we emphasize that this spacetime
covariant, Quantum Mechanics of loops, is a direct consequence of the
Hamilton--Jacobi formulation of {\sl Classical String Dynamics}.
Moreover, since with our present quantum mechanical formulation
we have given a concrete meaning to the fractalization of a
{\sl String} in terms of the {\sl Shape Uncertainty
Principle}, taking it as a fundamental constituent of spacetime transfers
this property to it as well.
We have concluded that the Hausdorff dimension of
a quantum {\sl String} is three, and that two distinct
geometric phases exist above and below the loop De Broglie area.
As a matter of fact, the {\sl Quantum String} is literally ``fuzzy''
to a degree which depends critically on a well defined parameter,
the ratio between temporal and spatial resolution.
Then, if the above over all picture is correct, {\sl String} as well as
$p$-{\sl{}brane} fuzziness not only acquires a well defined meaning,
but points to a fundamental change in our perception of physical
spacetime. It is also fractal, since its constituent are fractal objects.
Far from being a smooth, four--dimensional manifold assigned ``ab initio'',
spacetime is, rather, a ``process in the making'', showing an ever changing
fractal structure which responds dynamically to the resolving power of the
detecting apparatus. It is important to point out that
{\sl Classical String Theory} can be recovered, as we have shown in chapter
\ref{8.douclalimcha}. Of course we want that at traditional energy scales the world appears
exactly like we know it. But as shown in chapter \ref{9.connection} at very high energy
{\sl Bulk} and {\it Boundary} effects take place: enlightening is in particular
the relation between our formulation and the traditional way of looking at
{\sl String} Theory as the Polyakov {\sl Bulk Theory}.

Let us conclude our work with just a curious observation.
With our interpretation it is evident that the Quantum Gravitational spacetime
has a Fractal Structure since, as we already pointed out, its constituents
elements are extended and fractal objects. It is thus very appealing to see
that our functional equation can be recovered using an
interesting  mathematical tool,
namely {\sl NonStandard Analysis}: the problem of infinities and
infinitesimals disappears, thanks to it, from the mathematical computations
and thanks to our {\sl Functional Wave Equation}, that can be rigorously
defined using it, the Fractal Structure of Quantum Gravitational spacetime
looses all the divergence problems related to pointlike localization.
Moreover, {\sl NonStandard Analysis} looks a good tool in dealing problems
associated to Fractals, where the usual notion of differentiability
breaks down.
We believe that {\sl NonStandard Analysis} will be the natural framework to
solve other problems, and attack more fundamental questions, in both
Quantum Gravity and other different sectors of theoretical Physics,
even if we cannot give it for granted.\\
Finally, it is very tempting  to conjecture a relation between our
{\sl NonStandard} formulation and  NonCommutative Geometry that is, nowadays,
perhaps the best candidate to describe the short distance structure of
quantum spacetime: but \dots this is another story.

%% file: appA.tex
\pageheader{}{Detailed Calculations.}{}
\chapter{Detailed Calculations}
\label{A.detcal}


\section{Lagrangian, Hamiltonian and {\bf Coefficients}!}
\label{A.HamDer}

We devote this section to the standard derivation of the Hamiltonian
density from the Lagrangian density
in the case of a $p$-{\sl{}brane}, in order to motivate the values of the
numerical factors that appear as coefficients. With $\rp$ we
indicate the mass in the particle case, $m ^{2}$ in the {\sl String} case
and in general for the $p$-{\sl{}brane} we will have $\rp = m ^{p+1}$,
where $m$ is a constant with dimension of mass. The
factorial in front of the {\it Schild} Lagrangian density avoids
the overcounting due to the $p+1$ saturated antisymmetric indices
of the kinetic term.
\begin{props}[$p$-brane Schild Hamiltonian Density]\spbcorr{}.\\
    The {\it Schild Hamiltonian density} $\Ham \ttt{\Bp}$ is given by
    \beq
        \Ham \ttt{\Bp}
        =
        \frac{1}{2 \rp \fact{p+1}}
        P _{\multind{\beta}{1}{p+1}}
        P ^{\multind{\beta}{1}{p+1}}
    \eeq
    where,
    $P _{\multind{\mu}{1}{p+1}}$ is the momentum canonically conjugated
    to the {\sl World--HyperTube Volume Velocity}
    $\dot{X} ^{\multind{\nu}{1}{p+1}}$.
\end{props}
\begin{proof}
We start from the Schild Lagrangian density
\beq
    \Lag \ttt{\Bx  , \dot{\Bx}}
    =
    \frac{\rp}{2 \fact{p+1}}
    \dot{X} ^{\multind{\mu}{1}{p+1}}
    \dot{X} _{\multind{\mu}{1}{p+1}}
\quad ;
\eeq
moreover we defined the  {\sl $p$-{\sl{}brane} Bulk Area Momentum} as  the variable canonically
conjugated to
$\dot{X} ^{\multind{\beta}{1}{p+1}}$, so that we get as usual:
\bea
    P _{\multind{\alpha}{1}{p+1}}
    & = &
    \frac{\partial \Lag}{\partial \dot{X} ^{\multind{\alpha}{1}{p+1}}}
    \nonumber \\
    & = &
    \frac{\rp}{2 \fact{p+1}}
    \frac{
          \partial \left(
                       \dot{X} ^{\multind{\mu}{1}{p+1}}
                       \dot{X} _{\multind{\mu}{1}{p+1}}
                    \right)
         }
         {\partial \dot{X} ^{\multind{\alpha}{1}{p+1}}}
    \nonumber \\
    & = &
    \frac{\rp}{\fact{p+1}}
    \fact{p+1}
    \dot{X} ^{\multind{\alpha}{1}{p+1}}
    \nonumber \\
    & = &
    \rp \dot{X} ^{\multind{\alpha}{1}{p+1}}
    \quad .
\eea
Then, Legendre transforming, we get for the Hamiltonian the result:
\bea
    \Ham
    & = &
    \frac{1}{\fact{p+1}}
    P _{\multind{\mu}{1}{p+1}}
    \dot{X} ^{\multind{\mu}{1}{p+1}}
    -
    \Lag
    \nonumber \\
    & = &
    \frac{1}{\rp \fact{p+1}}
    P _{\multind{\mu}{1}{p+1}}
    P ^{\multind{\mu}{1}{p+1}}
    -
    \frac{1}{2 \rp \fact{p+1}}
    P _{\multind{\mu}{1}{p+1}}
    P ^{\multind{\mu}{1}{p+1}}
    \nonumber \\
    & = &
    \frac{1}{2 \rp \fact{p+1}}
    P _{\multind{\mu}{1}{p+1}}
    P ^{\multind{\mu}{1}{p+1}}
    \quad .
\eea
\end{proof}

We observe that this general result
correctly reproduces the
{\sl Bulk Area Momentum} in the
{\sl String} case (since $p=1$ gives $\rp = m ^{2}$)
and also the momentum of a non relativistic particle,
which is a $p=0$-{\sl{}brane} (so that\footnote{Of course we have
also to {\it formally}
understand under the ``dot~$\dot{\ }$~{}'' a time derivative.})
$\rp = m$).

\section{$\Bxi$ and $\Bpi$ Variation of the Action for
         a $p$-brane}
\label{A.EbalDer}

We will derive, the {\it energy balance equation} both
for {\sl Strings} and $p$-{\sl{}branes}. This  is the equation of motion
associated with a variation of the $\Bxi \ttt{\Bsigma}$
fields in the actions
(\ref{2.schhamfun}) and
(\ref{4.pbrrepact}) respectively. The interested reader can compare this
derivation with the procedure used in proposition \ref{2.fullrepequpro} to
relate the two methods.
To clear the way, we start from the case with less floating indices, i.e.
the {\it {\sl String} case}.
\begin{props}[Standard $\Bxi$ and $\Bpi$ Variations]\spbcorr{}.\\
    The equations of motion induced by variations of the $\Bxi$ and
    $\Bpi$ fields in the action {\rm (\ref{2.schhamfun})} are given by
    \bea
        \epsilon ^{mn}
        \partial _{m} \pi _{AB}
        \partial _{n} \xi ^{B}
        & = &
        0
        \label{A.enebalequ}
        \\
        \pi _{AB}
        & = &
        \epsilon _{AB} \, \Ham \ttt{\Bp}
        \quad .
        \label{A.momcom}
    \eea
\end{props}
\begin{proof}
The first result (equation (\ref{A.enebalequ}))
requires a longer procedure.
The only term contributing is the second one
in equation \ref{2.fulrepact};
we thus get:
\bea
    & & \esci \esci \esci
    \delta _{\Bxi}
        \left(
            \frac{1}{2}
            \int _{\Xi}
            \form{d \xi} ^{A} \wedge \form{d \xi} ^{B} \form{\pi} _{AB}
        \right) =
    \nonumber \\
    & {\scriptstyle 1.} & \qquad =
        \frac{1}{2}
        \int _{\Xi}
        \delta _{\Bxi}
        \left( \form{d \xi} ^{A} \wedge \form{d \xi} ^{B} \right)
        \pi _{AB}
    \nonumber \\
    & {\scriptstyle 2.} & \qquad =
        \frac{1}{2}
        \int _{\Xi}
        \form{d} \left ( \delta \xi ^{A}  \right) \wedge \form{d \xi} ^{B}
        \pi _{AB}
        +
        \frac{1}{2}
        \int _{\Xi}
        \form{d \xi} ^{A} \wedge \form{d} \left ( \delta \xi ^{B} \right)
        \pi _{AB}
    \nonumber \\
    & {\scriptstyle 3.} & \qquad =
        \frac{1}{2}
        \int _{\Xi}
        \form{d} \left ( \delta \xi ^{A}  \right) \wedge \form{d \xi} ^{B}
        \pi _{AB}
        -
        \frac{1}{2}
        \int _{\Xi}
        \form{d} \left ( \delta \xi ^{B} \right) \wedge \form{d \xi} ^{A}
        \pi _{AB}
    \nonumber \\
    & {\scriptstyle 4.} & \qquad =
        \int _{\Xi}
        \form{d} \left ( \delta \xi ^{A}  \right) \wedge \form{d \xi} ^{B}
        \pi _{AB}
    \nonumber \\
    & {\scriptstyle 5.} & \qquad =
        \int _{\Xi}
        \form{d} \left [
                         \delta \xi ^{A}  \form{d \xi} ^{B}
                         \pi _{AB}
                 \right]
        -
        \int _{\Xi}
        \left\{  \form{d \xi} ^{B} \wedge \form{d \pi _{AB}} \right\}
        \delta \xi ^{A}
    \nonumber \\
    & {\scriptstyle 6.} & \qquad =
        0
        -
        \int _{\Sigma}
        \left\{
                \form{d \sigma} ^{m} \wedge \form{d \sigma} ^{n}
                \partial _{m} \xi ^{B}
                \partial _{n} \pi _{AB}
        \right\}
        \delta \xi ^{A}
    \label{A.StrEbalVar}
\eea
We used the following properties in the various steps:
\begin{enumerate}
    \item the variation affects only the $\Xi$ volume form, since
    in the reparametrization invariant action we have
    $\Bp = \Bp ( \Bsigma )$;
    \item distributive property of the $\wedge$--product with respect to
    the variation $\delta$;
    \item symmetry property of the $\wedge$--product;
    \item symmetry property of $\pi _{AB}$;
    \item integration by parts and nihilpotency of the exterior derivative;
    \item the first contribution vanishes, because if we go to the {\it Boundary},
    there the $\Bxi$ variations vanish; then we have
    \beq
        \form{d \xi} ^{B} = \frac{\partial \xi ^{B}}{\partial \sigma ^{m}}
                            \form{d \sigma} ^{m}
                          \equiv \partial _{m} \xi ^{b}
                            \form{d \sigma} ^{m}
    \eeq
    and
    \beq
        \form{d \Ham} ( \Bp ) = \frac{\partial \Ham ( \Bp )}
                                     {\partial \sigma ^{n}}
                                     \form{d \sigma} ^{n}
                              \equiv \partial _{n} \Ham ( \Bp )
                                \form{d \sigma} ^{n}
    \quad .
    \eeq
\end{enumerate}
Remembering now that $\form{d \sigma} ^{m} \wedge \form{d \sigma} ^{n}$
is a $2$-form in $2$ dimensions we can write
$
 \form{d \sigma} ^{m} \wedge \form{d \sigma} ^{n} =
 \epsilon ^{mn} d ^{2} \Bsigma
$,
so that from equation (\ref{A.StrEbalVar}) we obtain
\beq
    \epsilon ^{mn}
    \partial _{m} \xi ^{B}
    \partial _{n} \pi _{AB}
    =
    0 \quad .
\eeq
The second equation has a simpler derivation since no integration by
parts is needed. The following chain of equalities leads to the result:
\bea
    & &
\eea
\end{proof}

Now, it is a routine to follow the same steps for the case of the $p$-{\sl{}brane},
because the generalizations of the previous properties work in the same way.
\begin{props}[$p$-brane Standard $\Bxi$ and $\Bpi$ Variation]\spbcorr{}.\\
    The equations of motion induced by variations of the $\Bxi$ and
    $\Bpi$ fields in the action {\rm (\ref{4.pbrrepact})} are given by
    \bea
    \frac{1}{p!}
    \epsilon ^{\multind{m}{1}{p}}{}^{m}
    \ttt{\partial _{m _{1}} \xi ^{A _{1}}}
    \cdot \dots \cdot
    \ttt{\partial _{m _{p}} \xi ^{a _{p}}}
    \partial _{m} \pi _{\multind{A}{0}{p}}
    & = &
    0
    \label{A.Ebaltwo}
    \\
    \pi _{\multind{A}{0}{p}}
    & = &
    \epsilon _{\multind{A}{0}{p}} \Ham \ttt{\Bp}
    \quad .
    \eea
\end{props}
\begin{proof}
Of course the first result needs again a longer computation:
\bea
    & & \esci \esci \esci
    \delta _{\Bxi}
        \left(
            \frac{1}{\fact{p+1}}
            \int _{\Xip}
            \multint{\xi}{A}{0}{p}
            \pi _{\multind{A}{0}{p}}
        \right) =
    \nonumber \\
    & {\scriptstyle 1.} & \qquad =
        \frac{1}{\fact{p+1}}
        \int _{\Xip}
        \delta _{\Bxi} \left( \multint{\xi}{A}{0}{p} \right)
        \pi _{\multind{A}{0}{p}}
    \nonumber \\
    & {\scriptstyle 2.} & \qquad =
        \frac{1}{\fact{p+1}}
        \sum _{i} ^{1, p+1}
        \int _{\Xip}
            \form{d \xi} ^{A _{0}}
            \wedge \dots \wedge
            \form{d} \left( \delta \xi ^{A _{i}} \right)
            \wedge \dots \wedge
            \form{d \xi} ^{A _{p}}
            \pi _{A_{0} \dots A _{i} \dots A _{p}}
    \nonumber \\
    & {\scriptstyle 3.} & \qquad =
        \frac{p+1}{\fact{p+1}}
        \int _{\Xip}
            \form{d} \left ( \delta \xi ^{A _{0}}  \right)
            \wedge
            \multint{\xi}{A}{1}{p}
        \pi _{\multind{A}{0}{p}}
    \nonumber \\
    & {\scriptstyle 4.} & \qquad =
        \frac{1}{p!}
        \int _{\Xip}
        \form{d} \left [
            \multint{\xi}{A}{1}{p}
            \ttt{\delta \xi ^{A _{0}}}
            \pi _{\multind{A}{0}{p}}
        \right ]
        +
      \nonumber \\
      & & \qquad \qquad \qquad \qquad
        -
        \frac{1}{p!}
        \int _{\Xip}
        \left [
            \multint{\xi}{A}{1}{p} \wedge
            \pi _{\multind{A}{0}{p}}
        \right ]
        \delta \xi ^{A _{0}}
    \nonumber \\
    & {\scriptstyle 5.} & \qquad =
        0
        -
        \int _{\Sigmap}
        \left\{
        \frac{1}{p!}
            \multint{\sigma}{m}{2}{p+1} \wedge \form{d \sigma} ^{m} \cdot
        \right .
        \nonumber \\
        & & \qquad \qquad \qquad \qquad \qquad
        \left . \cdot
                \ttt{\partial _{m _{2}} \xi ^{A _{2}}}
                \cdot \dots \cdot
                \ttt{\partial _{m _{p+1}} \xi ^{A _{p+1}}}
            \partial _{m} \pi _{\multind{A}{1}{p+1}}
        \right\}
        \delta \xi ^{A _{1}}
        \quad .
        \label{A.PbrEbalVar}
\eea
The main computational steps are:
\begin{enumerate}
    \item as before the variation affects only the integration volume
    in $\Xip$ space, which is now $(p+1)$-dimensional;
    \item by distributing the variation to the single basis $1$-forms, we
    obtain a sum of $(p+1)$-terms, which \dots
    \item \dots can be summed over by means of the antisymmetry property
    of the $\wedge$--product and of the $(p+1)$-dimensional
    $\boldsymbol{\epsilon}$ tensor;
    \item integrating  by parts (using of course $\bs{d} ^{2} = 0$) and
    \item taking into account that the $\Bxi$ fields are fixed on the {\it Boundary},
    $\partial \Xip$, gives the vanishing of the first term; for
    the second one we can then proceed expressing the differentials in terms
    of the new $\Bsigma $ variables.
\end{enumerate}
Again we remember that a $(p+1)$-form in $(p+1)$-dimensions is proportional to
the $\boldsymbol{\epsilon}$ tensor to conclude from equation
(\ref{A.PbrEbalVar})
that \beq
     \frac{1}{p!}
     \epsilon _{\multind{A}{1}{p+1}}
     \epsilon ^{\multind{m}{2}{p+1}}{}^{m}
    \ttt{\partial _{m _{2}} \xi ^{A _{2}}}
    \cdot \dots \cdot
    \ttt{\partial _{m _{p+1}} \xi ^{A _{p+1}}}
    \partial _{m} \pi _{\multind{A}{1}{p+1}}
    =
    0 \quad .
\eeq
\end{proof}

\section{Momentum Variation in Reparametrized Hamiltonian Formalism}
\label{A.LMulDer}

Let us  perform the variation of the action for the full reparametrized
Theory (\ref{4.pbrfulrepact}) with respect to $\Bpi$ to get
equation (\ref{4.LMulEqo}). The computation  is shown
in the general case. The {\sl String} case can be derived by setting
$p = 1$.
\begin{props}[$p$-brane $\Bpi$ Variation of the Action]\spbcorr{}.\\
    The equation of motion obtained by the $\Bpi$ variation
    of the action {\rm (\ref{4.pbrfulrepact})}
    (respectively {\rm (\ref{2.fulrepact})} for the {\sl String}) are given
    by
    \beq
        N ^{\multind{A}{1}{p+1}} = \dot{\xi} ^{\multind{A}{1}{p+1}}
        \quad .
    \eeq
\end{props}
\begin{proof}
There are no problems due to integrations by parts here. Then
\bea
    \delta _{\Bpi} S
    & = &
    \frac{1}{\fact{p+1}}
    \delta \left\{
               \int _{\Xip}
                   \pi _{\multind{a}{1}{p+1}}
                   \multint{\xi}{a}{1}{p+1}
           \right\}
    +
    \nonumber \\
    & & \qquad \qquad -
    \frac{1}{\fact{p+1}}
    \delta \left\{
               \int _{\Sigmap}
                   d ^{p+1} \Bsigma
        N ^{\multind{A}{1}{p+1}}
        \left [
            \pi _{\multind{A}{1}{p+1}}
            -
            \epsilon _{\multind{A}{1}{p+1}}
            \Ham ( \Bp )
        \right ]
           \right\}
    \nonumber \\
    & = &
    \frac{1}{\fact{p+1}}
    \delta \left\{
               \int _{\Sigmap}
                   d ^{p+1} \Bsigma
                   \left [
                       \pi _{\multind{A}{1}{p+1}}
                       \left(
                           \dot{\xi } ^{\multind{A}{1}{p+1}}
                           -
        	           N ^{\multind{A}{1}{p+1}}
                       \right)
        	       -
        	       \epsilon _{\multind{A}{1}{p+1}}
        	       \Ham ( \Bp )
        	   \right ]
           \right\}
    \nonumber \\
    & = &
    \frac{1}{\fact{p+1}}
    \int _{\Sigmap}
        d ^{p+1} \Bsigma
        \left [
               \dot{\xi } ^{\multind{A}{1}{p+1}}
               -
               N ^{\multind{A}{1}{p+1}}
        \right ]
        \delta  \pi _{\multind{A}{1}{p+1}}
    \quad ,
\eea
where,
\beq
    \dot{\xi } ^{\multind{A}{1}{p+1}}
    =
    \epsilon ^{\multind{m}{1}{p+1}}
    \ttt{\partial _{m _{1}} \xi ^{A _{1}}}
    \cdot \dots \cdot
    \ttt{\partial _{m _{p+1}} \xi ^{A _{p+1}}}
    \quad .
\eeq
The result of equation (\ref{4.LMulEqo}) (respectively \ref{2.fulrepeq5})
is then obtained from the stationarity condition $\delta _{\Bpi} S = 0$
$$
    \Rightarrow \quad
    \dot{\xi} ^{\multind{A}{1}{p+1}}
    =
    N ^{\multind{A}{1}{p+1}}
    \quad .
$$
\end{proof}

\section{Hamilton Principle with one Free Boundary}
\label{A.FutVarDer}

In this section we perform the variation of the action letting the
{\it future Boundary} vary freely. As we pointed out in subsection
\ref{2.hamjacrepsec} and in section \ref{4.pbrhamjacsec}, this is an important
kind of variational procedure, since it gives the opportunity to identify the
canonical momentum conjugated to the {\it Boundary} and define the
energy associated to the {\it Boundary} Dynamics.
It turns out that the energy
is now conjugated to the proper area of the {\sl String}
{\sl Parameter Space}
or, equivalently, to the hypervolume associated with the $p$-{\sl{}brane}
{\sl Parameter Space} when we consider higher dimensional objects
in $D$-dimensional spacetime. In this more general framework the
{\it Boundary}
is a $(p-1)$-{\sl{}brane}. Now, we have to
consider variations of the fields,
which satisfy the classical equations of
motion on the {\it Bulk}, i.e. on the
{\it World--HyperTube} swept by the evolution of the object itself.

As in the previous section, we start from the simpler {\sl String} case.
\begin{props}[Free Boundary Variation of the String Action]\spbcorr{}.
    \label{A.futvarderstrpro}\\
    The free {\it Boundary} variation of the action {\rm (\ref{2.redrepact})}
    with respect to the fields $\By$ is given by:
    \beq
        \delta _{\By \! \! _{(f)}} S
        =
        \int _{ \partial \Sigma}
                  d s \,
                  Q _{\mu \nu}
                  Y ^{\prime \nu} \delta Y _{(f)} ^{\mu}
        -
        \Ham \delta A
    \quad .
    \eeq
\end{props}
\begin{proof}
The computation is of standard type; we have only to perform in the
proper way the passage from the {\it Bulk} to the {\it Boundary}, i.e.
from the {\sl Parameter Space} $\Sigma$
to the {\sl Boundary Space} $\Gamma = \partial \Sigma$.
\bea
    \delta _{\By} S
    & = &
    \frac{1}{2}
    \int _{\mathcal{W}}
    \left (
        \delta P _{\mu \nu}
        \form{d X} ^{\mu} \wedge \form{d X} ^{\nu}
        +
        P _{\mu \nu}
        \form{d} \left ( \delta X ^{\mu} \right ) \wedge \form{d X} ^{\nu}
        +
        P _{\mu \nu}
        \form{d X} ^{\mu} \wedge \form{d} \left ( \delta X ^{\nu} \right )
    \right )
    +
    \nonumber \\
    & & \qquad
    -
    \delta
    \left (
        \frac{1}{2}
        \epsilon _{AB}
        \int _{\Xi}
            \form{d \xi} ^{A} \wedge \form{d \xi} ^{B}
    \right )
    \Ham ( \Bp )
    -
    \frac{1}{2}
    \epsilon _{AB}
    \int _{\Xi}
        \form{d \xi} ^{A} \wedge \form{d \xi} ^{B}
        \frac{1}{2m ^{2}}
        P _{\mu \nu} \delta P ^{\mu \nu}
    \nonumber \\
    {\scriptstyle 1.} & = &
    \int _{\Sigma}
        d ^{2} \Bsigma
        \left(
            \frac{\dot{X} ^{\mu \nu}}{2}
            -
            \frac{
                  \epsilon _{AB}
                  \dot{\xi} ^{AB}
                 }
                 {4 m ^{2}}
            P ^{\mu \nu}
         \right)
           \delta P _{\mu \nu}
    -
    \Ham \delta A
    +
    \int _{\mathcal{W}}
        P _{\mu \nu}
        \form{d} \left( \delta X ^{\mu} \right) \wedge \form{d X} ^{\nu}
    \nonumber \\
    {\scriptstyle 2.} & = &
    \int _{\Sigma}
        d ^{2} \Bsigma
        \left(
            \frac{\dot{X} ^{\mu \nu}}{2}
            -
            \frac{
                  \epsilon _{AB}
                  \dot{\xi} ^{AB}
                 }
                 {4 m ^{2}}
            P ^{\mu \nu}
        \right)
            \delta P _{\mu \nu}
    -
    \Ham \delta A
    +
    \nonumber \\
    & & \qquad \qquad +
    \int _{\mathcal{W}}
        \form{d} \left [
              P _{\mu \nu}
              \form{d X} ^{\nu} \delta X ^{\mu}
          \right ]
    -
    \int _{\mathcal{W}}
        \form{d} \left [
              P _{\mu \nu}
              \form{d X} ^{\nu}
          \right ]
          \delta X ^{\mu}
    \nonumber \\
    {\scriptstyle 3.} & = &
    \int _{\Sigma}
        d ^{2} \Bsigma
        \left(
            \frac{\dot{X} ^{\mu \nu}}{2}
            -
            \frac{
                  \epsilon _{AB}
                  \dot{\xi} ^{AB}
                 }
                 {4 m ^{2}}
            P ^{\mu \nu}
        \right)
        \delta P _{\mu \nu}
    -
    \Ham \delta A
    +
    \nonumber \\
    & & \qquad \qquad +
    \int _{ \partial \mathcal{W} = C}
              P _{\mu \nu}
              \form{d Y} ^{\nu} \delta Y _{(f)} ^{\mu}
    -
    \int _{\mathcal{W}}
              \form{d P} _{\mu \nu}
              \wedge
              \form{d X} ^{\nu}
              \delta X ^{\mu}
    \nonumber \\
    {\scriptstyle 4.} & = &
    \int _{\Sigma}
        d ^{2} \Bsigma
        \left(
            \frac{\dot{X} ^{\mu \nu}}{2}
            -
            \frac{
                  \epsilon _{AB}
                  \dot{\xi} ^{AB}
                 }
                 {4 m ^{2}}
            P ^{\mu \nu}
        \right)
        \delta P _{\mu \nu}
    -
    \Ham \delta A
    +
    \nonumber \\
    & & \qquad +
    \int _{ \partial \Sigma = \Gamma}
              d s
              Q _{\mu \nu}
              Y ^{\prime \nu} \delta Y _{(f)} ^{\mu}
    -
    \int _{\Sigma}
              d ^{2} \Bsigma
              \epsilon ^{AB}
              \partial _{A} P _{\mu \nu}
              \partial _{B} X ^{\nu}
              \delta X ^{\mu}
    \nonumber \\
    {\scriptstyle 5.} & = &
    \int _{\Gamma}
              d s
              Q _{\mu \nu}
              Y ^{\prime \nu} \delta Y _{(f)} ^{\mu}
    -
    \Ham \delta A
\eea
\begin{enumerate}
    \item we put together the first and the last term expressing all the
    quantities in terms of the new variables, $\Bsigma $; then, taking into
    account the  antisymmetry of both $P _{\mu \nu}$ and the
    $\wedge$--product, we can add together the second and third terms;
    as far as the fourth term is concerned, we recall that the Hamiltonian is
    constant along a classical trajectory and so can be moved outside the
    integral;
    \item we integrate by parts the last term;
    \item we compute the first term on the second line  {\it on the Boundary,}
    and apply the properties of the exterior derivative to the second term on
    the second line;
    \item we express the first integral on the second line in terms of
    the variable parametrizing the loop ($s \in \Gamma$); the notation
    $\delta Y ^{\mu} _{(f)}$  wants to remark that we are using a variational
    principle with {\it free end loop}; the same procedure is applied to the
    last integral;
    \item we select only variations that satisfy the classical equations
    of motion on the {\sl Bulk} (actually, we did it already as we took
    $\Ham ( \Bp )$
    outside the integral as
    $ \Ham$); then the first and last term vanish.
\end{enumerate}
\end{proof}

Following the same line we can now perform the same computation
for the $p$-{\sl{}brane} in $D$-dimensions. The steps are somewhat more
tedious due to the increased number of dimensions, but the computation
is basically the same.
\begin{props}[Free Boundary Variation of the $p$-brane Action]\spbcorr{}.\\
    The one free {\it Boundary} variation of the action {\rm (\ref{4.pbrrepact})}
    with respect to the fields $\Bx$ is given by:
    \beq
        \delta _{\By _{(f)}} S
        =
        \frac{1}{p!}
        \int _{\Gamma ( \Bs )}
                  d ^{p} \Bs \,
                  Q _{\multind{\mu}{1}{p+1}}
                  Y ^{\prime \multind{\mu}{2}{p+1}}
                  \delta Y ^{\mu _{1}}
        -
        \Ham \delta V
    \eeq
\end{props}
\begin{proof}
The proof follows the same steps as the previous one, with just a little
bit of more troubles due to the proliferation of indices. We thus get
\bea
    \delta S
    & = &
    \frac{1}{\fact{p+1}}
    \int _{\Wsp}
    \Bigr[
    \left .
        \delta P _{\multind{\mu}{1}{p+1}}
        \multint{X}{\mu}{1}{p+1}
        +
    \right.
    \nonumber \\
    & & \qquad \qquad
    \left . +
        \sum _{i} ^{1,p+1}
        \left(
            P _{\mu _{1} \dots \mu _{i} \dots \mu _{p+1}}
            \form{d X} ^{\mu _{1}}
            \wedge \dots \wedge
            \delta \left( \form{d X} ^{\mu _{i}} \right)
            \wedge \dots \wedge
            \form{d X} ^{\mu _{p+1}}
        \right)
    \right ]
    +
    \nonumber \\
    & & \qquad
    -
    \delta
    \left (
        \frac{1}{\fact{p+1}}
        \epsilon _{\multind{a}{1}{p+1}}
        \int _{\Xip}
            \multint{\xi}{a}{1}{p+1}
    \right )
    \Ham \left( \Bp \right)
    +
    \nonumber \\
    & & \qquad
    -
    \frac{1}{\fact{p+1}}
    \epsilon _{\multind{a}{1}{p+1}}
    \int _{\Xip}
        \frac{\multint{\xi}{a}{1}{p+1}}{\rp}
        P _{\multind{\mu}{1}{p+1}}
        \delta P ^{\multind{\mu}{1}{p+1}}
    \nonumber \\
    {\scriptstyle 1.} & = &
    \int _{\Sigmap}
        d ^{p+1} \Bsigma
        \left(
            \frac{\dot{X} _{\multind{\mu}{1}{p+1}}}{\fact{p+1}}
            -
            \frac{1}{\rp \fact{p+1}}
            \epsilon _{\multind{a}{1}{p+1}}
            \dot{\xi} ^{\multind{a}{1}{p+1}}
            P _{\multind{\mu}{1}{p+1}}
         \right)
           \delta P ^{\multind{\mu}{1}{p+1}}
    +
    \nonumber \\
    & & \qquad
    -
    \Ham \delta V
    +
    \frac{1}{p!}
    \int _{\Wsp}
        P _{\multind{\mu}{1}{p+1}}
        \form{d} \left( \delta X ^{\mu _{1}} \right)
        \wedge
        \multint{X}{\mu}{2}{p+1}
    \nonumber \\
    {\scriptstyle 2.} & = &
    \int _{\Sigmap}
        d ^{p+1} \Bsigma
        \left(
            \frac{\dot{X} _{\multind{\mu}{1}{p+1}}}{\fact{p+1}}
            -
            \frac{1}{\rp \fact{p+1}}
            \epsilon _{\multind{a}{1}{p+1}}
            \dot{\xi} ^{\multind{a}{1}{p+1}}
            P _{\multind{\mu}{1}{p+1}}
         \right)
           \delta P ^{\multind{\mu}{1}{p+1}}
    +
    \nonumber \\
    & & \qquad
    -
    \Ham \delta V
    +
    \frac{1}{p!}
    \int _{\Wsp}
        \form{d}
          \left [
              P _{\multind{\mu}{1}{p+1}}
              \multint{X}{\mu}{2}{p+1}
              \delta X ^{\mu _{1}}
          \right ] +
    \nonumber \\
    & & \qquad
    -
    \frac{1}{p!}
    \int _{\Wsp}
        \form{d}
           \left [
              p _{\multind{\mu}{1}{p+1}}
              \multint{X}{\mu}{2}{p+1}
           \right ]
           \delta X ^{\mu _{1}}
    \nonumber \\
    {\scriptstyle 3.} & = &
    \int _{\Sigmap}
        d ^{p+1} \Bsigma
        \left(
            \frac{\dot{X} _{\multind{\mu}{1}{p+1}}}{\fact{p+1}}
            -
            \frac{1}{\rp \fact{p+1}}
            \epsilon _{\multind{a}{1}{p+1}}
            \dot{\xi} ^{\multind{a}{1}{p+1}}
            p _{\multind{\mu}{1}{p+1}}
         \right)
           \delta p ^{\multind{\mu}{1}{p+1}}
    +
    \nonumber \\
    & & \qquad
    -
    \Ham \delta V
    +
    \frac{1}{p!}
    \int _{\partial \Wsp = \Dp}
              P _{\multind{\mu}{1}{p+1}}
              \multint{Y}{\mu}{2}{p+1}
              \delta Y ^{\mu _{1}} +
    \nonumber \\
    & & \qquad
    -
    \frac{1}{p!}
    \int _{\Sigmap}
        d ^{p+1} \Bsigma
        \epsilon ^{\multind{a}{1}{p+1}}
        \partial _{a _{1}} P _{\multind{\mu}{1}{p+1}}
        \partial _{a _{2}} X ^{\mu _{2}}
        \dots
        \partial _{a _{p+1}} X ^{\mu _{p+1}}
        \delta X ^{\mu _{1}}
    \nonumber \\
    {\scriptstyle 4.} & = &
    \frac{1}{p!}
    \oint _{\Dp}
              Q _{\multind{\mu}{1}{p+1}}
              \multint{Y}{\mu}{2}{p+1}
              \delta Y ^{\mu _{1}}
    -
    \Ham \delta V
    \label{A.classvar}
    \\
    {\scriptstyle 5.} & = &
    \frac{1}{p!}
    \oint _{\BB}
              d ^{p} \Bs \,
              Q _{\multind{\mu}{1}{p+1}}
              Y ^{\prime \multind{\mu}{2}{p+1}}
              \delta Y ^{\mu _{1}}
    -
    \Ham \delta V
\eea
\begin{enumerate}
    \item we collect together the terms corresponding to the previous ({\sl String})
    case, i.e. those containing the variation of the momentum $\delta \Bp$,
    expressing them as functions of the new variables $\Bsigma $; the terms
    in the $\sum $ can be summed over, as usually, thanks to
    the properties of
    the $\wedge$--product and of the $\boldsymbol{\epsilon}$ tensor (third
    term), whereas the second term is obtained by implementing  the  constancy
    of the Hamiltonian along a classical trajectory;
    \item we integrate by parts the last integral obtaining the last two ones
    \dots
    \item \dots then, we go to the {\it Boundary}
    in the first of them and switch
    to the $\Bsigma$ variables in the last one;
    \item now, we restrict ourself to variations which solve the classical
    equations of motions on the {\sl Bulk} \dots
    \item \dots and turn to the new variables in the first integral.
\end{enumerate}
\end{proof}

We stress again that in the quantity
\beq
\label{A.xprimetwo}
    Y ^{\prime \multind{\mu}{1}{p}}
    =
    \epsilon ^{\multind{a}{1}{p}}
    \partial _{a _{1}} Y ^{\mu _{1}}
    \dots
    \partial _{a _{p}} Y ^{\mu _{p}}
    \quad ,
    \label{A.xprimedef}
\eeq
quoted also in definition \ref{4.pbrholcoo} and in equation
(\ref{4.xprimeone}), the derivation involves only the
$p$ {\it Boundary} variables $\Bs = (s _{1} , \dots , s _{p})$;
i.e., $\boldsymbol{Y}'$ is the tangent element to the $p$-{\sl{}brane}
{\it Boundary}.
So, one is led to the interpretation of
\beq
    q _{\mu}
    =
    Q _{\mu}{}_{\multind{\mu}{1}{p}}
    Y ^{\prime \multind{\mu}{1}{p}}
\eeq
as the projection of the momentum on the {\it Boundary} configuration,
in the case of the {\sl String} as well as of the $p$-{\sl{}brane}.

\section{Functional Integration of the $\Bxi ( \Bsigma )$ Fields}

In this section we perform the functional integration of the
$\Bxi$ fields in the case of the $p$-{\sl{}brane}.
The main reason to devote a section for this
computation is to point out the way in which, using a {\sl Functional
Fourier Transform}, it is possible to get a {\sl Functional Integral
Representation} of a {\sl Functional Dirac Delta} over the {\it classical
equation of motion}. Moreover we stress again the relevance of
the passage to the {\it Boundary} and we carefully choose the notation
to underline that.
\begin{props}[Functional Integration over $\Bxi$ Fields]\spbcorr{}.\\
    The functional integration over the $\Bxi$ fields singles
    out a functional Dirac delta over the classical equation
    of motion of the {\sl Bulk} $\Wsp$ times a term dependent
    only from the fields on the {\it Boundary}:
    \bea
        & & \esci \esci \esci \esci
        \int _{\Bzeta _{0} ( \Bs )} ^{\Bzeta ( \Bs)}
            [D \Bxi ( \Bsigma )]
            \exp
            \left\{
                \frac{i}{\fact{p+1}}
                \int _{\Xi ( \Bsigma )}
                   \pi _{\multind{A}{1}{p+1}}
                   \multint{\xi}{A}{1}{p+1}
            \right\}
        =
        \nonumber \\
        \qquad & = &
        \exp
        \left\{
            \frac{i}{\fact{p+1}}
            \int _{\partial \Xi ( \Bsigma )}
                \zeta ^{A _{p+1}}
                \pi _{\multind{A}{1}{p+1}}
                \multint{\zeta}{A}{1}{p}
        \right\}
        \cdot
        \nonumber \\
        & & \qquad \qquad \cdot
        \delta
        \left [
            \epsilon ^{\multind{m}{1}{p+1}}
            \partial _{m _{p+1}} \pi _{\multind{A}{1}{p+1}}
            \partial _{m _{1}} \xi ^{A _{1}}
            \dots
            \partial _{m _{p}} \xi ^{A _{p}}
        \right ]
    \quad .
    \eea
\end{props}
\begin{proof}
The first step to perform the integration is to use the following
identity
\bea
    & &
    \int _{\Xi}
        \pi _{\multind{A}{1}{p+1}}
        \multint{\xi}{A}{1}{p+1}
    =
    \nonumber \\
    & & \qquad
    =
    \int _{\Xi}
        \!
        \form{d}
          \left(
              \xi ^{A _{p+1}}
              \pi _{\multind{A}{1}{p+1}}
              \multint{\xi}{A}{1}{p}
          \right)
    +
    \nonumber \\
    & & \qquad \qquad
    -
    \int _{\Xi}
        \!
        \xi ^{A _{p+1}}
        \form{d}
          \left(
              \pi _{\multind{A}{1}{p+1}}
              \multint{\xi}{A}{1}{p}
          \right)
    \quad ,
\eea
which becomes, after an integration by parts of the first term,
\bea
    & &
    \int _{\Xi}
        \pi _{\multind{A}{1}{p+1}}
        \multint{\xi}{A}{1}{p+1}
    =
    \nonumber \\
    & & \qquad
    =
    \int _{\partial \Xi}
        \zeta ^{A _{p+1}}
        \pi _{\multind{A}{1}{p+1}}
        \multint{\zeta}{A}{1}{p}
    +
    \nonumber \\
    & & \qquad \qquad
    -
    \int _{\Xi}
        \xi ^{A _{p+1}}
        \form{d \pi} _{\multind{A}{1}{p+1}}
              \wedge
              \multint{\xi}{A}{1}{p}
\quad ,
\eea
where, the variable $\Bzeta$ indicates the fields
{\it on the Boundary}. Let us quickly recall the general setting where
$\Xi \ttt{\Bsigma}$ is a parametrization of the domain of variation of
the $\Bxi$ fields, defined on the {\sl Parameter Space} $\Sigma$, and
$\Sigma \ttt{\Bs}$ is a parametrization of the {\it Boundary} $\Gamma$ of $\Sigma$.
Then, the fields $\Bxi$, restricted to the {\it Boundary} $\Gamma$, will vary in the
{\it Boundary} $\partial \Xi$ of the domain $\Xi$, and will have the induced
parametrization
$$
    \Bzeta \ttt{\Bs} \dfn \Bxi \ttt{\Bsigma \ttt{\Bs}}
    \quad .
$$
Hence, we find
\bea
    & & \esci \esci
    \int _{\Bzeta _{0} ( \Bs )} ^{\Bzeta ( \Bs )}
        [D \Bxi ( \Bsigma )]
        \exp
        \left\{
            \frac{i}{\fact{p+1}}
            \int _{\Xi}
               \pi _{\multind{A}{1}{p+1}}
               \multint{\xi}{A}{1}{p+1}
        \right\}
    =
    \nonumber \\
    \qquad & = &
    \int _{\Bzeta _{0} ( \Bs )} ^{\Bzeta ( \Bs )}
        [D \Bxi ( \Bsigma )]
        \exp
        \left\{
            \frac{i}{\fact{p+1}}
            \int _{\partial \Xi}
                \zeta ^{A _{p+1}}
                \pi _{\multind{A}{1}{p+1}}
                \multint{\zeta}{A}{1}{p}
            +
        \right .
        \nonumber \\
        & &
        \qquad \qquad \qquad \qquad \qquad
        \left .
            - \frac{i}{\fact{p+1}}
            \int _{\Xi}
                \xi ^{A _{p+1}}
                \form{d \pi} _{\multind{A}{1}{p+1}}
                \multint{\xi}{A}{1}{p}
        \right\}
    \nonumber \\
    {\scriptstyle 1.} & = &
        \exp
        \left\{
            \frac{i}{\fact{p+1}}
            \int _{\partial \Xi}
                \zeta ^{A _{p+1}}
                \pi _{\multind{A}{1}{p+1}}
                \multint{\zeta}{A}{1}{p}
        \right\}
    \cdot
    \nonumber \\
    & & \qquad \cdot
    \int _{\Bzeta _{0} ( \Bs )} ^{\Bzeta ( \Bs )}
        [D \Bxi ( \Bsigma )]
        \exp
        \left\{
            - \frac{i}{\fact{p+1}}
            \int _{\Xi}
                \xi ^{A _{p+1}}
                \form{d \pi} _{\multind{A}{1}{p+1}}
                \wedge
                \multint{\xi}{A}{1}{p}
        \right\}
    \nonumber \\
    {\scriptstyle 2.} & = &
        \exp
        \left\{
            \frac{i}{\fact{p+1}}
            \int _{\partial \Xi}
                \zeta ^{A _{p+1}}
                \pi _{\multind{A}{1}{p+1}}
                \multint{\zeta}{A}{1}{p}
        \right\}
    \cdot
    \nonumber \\
    & & \qquad \cdot
    \int _{\Bzeta _{0} ( \Bs )} ^{\Bzeta ( \Bs )}
        [D \Bxi ( \Bsigma )]
        \exp
        \left\{
            - \frac{i}{\fact{p+1}}
            \int _{\Xi}
                \! \! \!
                d ^{p+1} \Bsigma
                \epsilon ^{\multind{m}{1}{p+1}}
                \cdot
        \right .
        \nonumber \\
        & & \qquad \qquad \qquad \qquad
        \left .
                \cdot
                \xi ^{A _{p+1}}
                \ttt{\partial _{m _{p+1}} \pi _{\multind{A}{1}{p+1}}}
                \ttt{\partial _{m _{1}} \xi ^{A _{1}}}
                \cdot \dots \cdot
                \ttt{\partial _{m _{p}} \xi ^{A _{p}}}
        \right\}
    \nonumber \\
    {\scriptstyle 3.} & = &
        \exp
        \left\{
            \frac{i}{\fact{p+1}}
            \int _{\partial \Xi}
                \zeta ^{A _{p+1}}
                \pi _{\multind{A}{1}{p+1}}
                \multint{\zeta}{A}{1}{p}
        \right\}
        \cdot
        \nonumber \\
        & & \qquad \qquad \cdot
        \delta
        \left [
            \epsilon ^{\multind{m}{1}{p+1}}
            \ttt{\partial _{m _{p+1}} \pi _{\multind{A}{1}{p+1}}}
            \ttt{\partial _{m _{1}} \xi ^{A _{1}}}
            \cdot \dots \cdot
            \ttt{\partial _{m _{p}} \xi ^{A _{p}}}
        \right ]
\eea
\begin{enumerate}
    \item the integral in the exponent, which is calculated on the {\it Boundary},
    is constant with respect to the path--integration and can
    be safely taken outside it;
    \item we change variables in the second integral introducing the
    $\Bsigma $ ones and \dots
    \item \dots we perform the path--integral taking into account the functional
    integral representation of the functional Dirac
    $\delta [ \dots ]$ function. 
\end{enumerate}
\end{proof}

\section{Functional Integration of the $\Bpi \ttt{\Bsigma}$ Fields}

This integration with respect to the momenta has always to be performed
when we start with a {\it Hamiltonian path--integral}. In this case the
process is non ambiguous, because we have just to compute a
functional integral which, restricted by a {\it Functional Dirac Delta},
collapses into an ordinary integral of one real variable.
\begin{props}[Functional Integration over the $\Bpi$ Fields]\spbcorr{}.\\
    The functional integration over the momentum conjugated to the
    $\Bxi$ fields, $\Bpi$, collapses into an ordinary integral over all the
    possible values of the classical energy of the system, i.e.
    we have
    \bea
    & & \esci \esci
    \int
        [ \mathcal{D} \pi _{\multind{A}{1}{p+1}} ]
        \delta
        \left [
            \epsilon ^{\multind{m}{1}{p+1}}
            \partial _{m _{p+1}} \pi _{\multind{A}{1}{p+1}}
            \ttt{\partial _{m _{1}} \xi ^{A _{1}}}
            \cdot \dots \cdot
            \ttt{\partial _{m _{p}} \xi ^{A _{p}}}
        \right ] \cdot
        \nonumber \\
        \qquad & & \cdot
        \exp
        \left\{
            \frac{i}{\fact{p+1}}
            \left [
                \int _{\Xi ( \Bsigma )}
                \form{d}
                  \left(
                      \xi ^{A _{p+1}}
                      \pi _{\multind{A}{1}{p+1}}
                      \multint{\xi}{A}{1}{p}
                  \right)
                +
            \right .
        \right .
        \nonumber \\
        & & \qquad \qquad \qquad \qquad \qquad \qquad
        \left .
            \left .
                -
                \int _{\Sigma}
                d ^{p+1} \Bsigma
                N ^{\multind{A}{1}{p+1}}
                \pi _{\multind{A}{1}{p+1}}
            \right ]
        \right\}
    =
    \nonumber \\
    & & \qquad =
    \int _{0} ^{\infty}
        d E
        e ^{iEV}
        \exp
        \left\{
            -
            \frac{i E}{\fact{p+1}}
            \left(
                \int _{\Sigma}
                    d ^{p+1} \Bsigma
                    N ^{\multind{A}{1}{p+1}}
                    \epsilon _{\multind{A}{1}{p+1}}
            \right)
        \right\}
    \eea
\end{props}
\begin{proof}
As we already briefly pointed out a few lines above
we can use the fact that the functional Dirac delta in front of the
integrand, requires $\pi _{\multind{a}{1}{p+1}}$ to satisfy the classical
equation  of motion; but at the classical level the energy balance equation
implies that the Hamiltonian $ \Ham ( \Bsigma ) $ is independent of the
$\Bsigma$'s, i.e. $H ( \Bsigma ) \equiv E \equiv \mathrm{const.}$; thus, we
can conclude
\beq
    \pi _{\multind{A}{1}{p+1}}
    =
    \epsilon _{\multind{A}{1}{p+1}}
    \Ham ( \Bsigma )
    =
    E
    \epsilon _{\multind{A}{1}{p+1}}
\quad ,
\eeq
and the functional integration turns into  an ordinary integration
over all the possible values of $E$. Accordingly, our functional integral
reduces to the following intermediate expression
\bea
    & &
    \int _{0} ^{\infty}
        d E
        \exp
        \left\{
            \frac{iE}{\fact{p+1}}
            \left [
                \int _{\Xi}
                \epsilon _{\multind{a}{1}{p+1}}
                d \left(
                      \xi ^{a _{p+1}}
                      \multint{\xi}{a}{1}{p}
                  \right)
                +
            \right .
        \right .
        \nonumber \\
        & & \qquad \qquad \qquad \qquad \qquad
            \qquad \qquad \qquad \qquad \qquad
        \left .
            \left .
                -
                \int _{\Sigma}
                d ^{p+1} \Bsigma
                N ^{\multind{a}{1}{p+1}}
                \epsilon _{\multind{a}{1}{p+1}}
            \right ]
        \right\}
        \quad ,
    \label{A.piintpar}
\eea
where,we recognize the first term as the definition of the differential
element associated to the {\sl World--HyperTube}
swept by the $p$-{\sl{}brane}; indeed,
we have
\beq
    V
    =
    \frac{\epsilon _{\multind{a}{1}{p+1}}}{\fact{p+1}}
    \int _{\Xi}
    \multint{\xi}{a}{1}{p+1}
    =
    \frac{\epsilon _{\multind{a}{1}{p+1}}}{\fact{p+1}}
    \int _{\Xi}
    \form{d}
      \left(
          \xi ^{a_{p+1}}
          \multint{\xi}{a}{1}{p}
      \right)
  \quad ,
\eeq
which, inserted in formula (\ref{A.piintpar}), gives the final result
\beq
    \int _{0} ^{\infty}
        d E
        e ^{iEV}
        \exp
        \left\{
            -
            \frac{i E}{\fact{p+1}}
            \left(
                \int _{\Sigma}
                    d ^{p+1} \Bsigma
                    N ^{\multind{a}{1}{p+1}}
                    \epsilon _{\multind{a}{1}{p+1}}
            \right)
        \right\}
\quad .
\eeq
\end{proof}

\section{Functional Integration of the $\Bp \ttt{\Bsigma}$ Fields}

Since the functional integration over the $\Bp$ fields is gaussian
it can be performed exactly; we will not care about the
determinant from the integration, because it will be re-adsorbed into an
overall normalization constant. Then, we find the following result:
\begin{props}[Functional Integration over the $\Bp$ Fields]\spbcorr{}.\\
    If we set $N = \epsilon _{\multind{A}{1}{p+1}} N ^{\multind{A}{1}{p+1}}$
    then, the functional integration over the momentum $[\mathcal{D} \Bp]$
    gives the following result:
    \bea
        & & \esci
        \funint{P ^{\alpha \beta}}
            \exp
            \left\{
                \frac{i}{\fact{p+1}}
                \int d ^{p+1} \Bsigma
                    \left [
                        P _{\multind{\mu}{1}{p+1}}
                        \dot{X} ^{\multind{\mu}{1}{p+1}}
                        +
                    \right .
            \right .
        \nonumber \\
        & & \qquad \qquad \qquad \qquad
            \left .
                    \left .
                        +
                        \frac{
                              \epsilon _{\multind{A}{1}{p+1}}
                              N ^{\multind{A}{1}{p+1}}
                             }
                             {2 \rp \fact{p+1}}
                        P _{\multind{\mu}{1}{p+1}}
                        P ^{\multind{\mu}{1}{p+1}}
                    \right ]
            \right\}
        \simeq
        \nonumber \\
        & & \qquad \qquad \qquad \qquad \qquad \qquad \qquad \simeq
        \exp
        \left\{
            -
            i
            \int d ^{p+1} \Bsigma
                \left [
                    -
                    \frac{\rp}{2 N \fact{p+1}}
                    \dot{X} ^{2}
                \right ]
        \right\}
    \quad .
    \eea
\end{props}
\begin{proof}
This is a gaussian integration which can be performed exactly;
nevertheless we avoid the computation of the functional determinant
that will be adsorbed in an overall normalization constant and this
is the meaning of the {\it proportionality} symbol in the last two
steps. Moreover, we use the {\it Minkowski--space result}; a more
appropriate mathematical procedure would require
an analytic continuation to Euclidean space, where the resulting integral
would be convergent, and a {\it  counter Wick rotation} at the end
of the calculations to get the final result. Apart from these technicalities,
the calculation goes as follows:
\bea
    & & \esci
    \funint{P}
        \exp
        \left\{
            \frac{i}{\fact{p+1}}
            \int d ^{p+1} \Bsigma
                \left [
                    P _{\multind{\mu}{1}{p+1}}
                    \dot{X} ^{\multind{\mu}{1}{p+1}}
                    +
                \right .
        \right .
    \nonumber \\
    & & \qquad \qquad \qquad \qquad +
        \left .
                \left .
                    \frac{
                          \epsilon _{\multind{a}{1}{p+1}}
                          N ^{\multind{a}{1}{p+1}}
                         }
                         {2 \rp \fact{p+1}}
                    P _{\multind{\mu}{1}{p+1}}
                    P ^{\multind{\mu}{1}{p+1}}
                \right ]
        \right\}
    =
    \nonumber \\
    {\scriptstyle 1.} & & \qquad
    =
    \funint{P}
        \exp
        \left\{
            \int d ^{p+1} \Bsigma
                \left [
                    i N
                    \frac{P _{\multind{\mu}{1}{p+1}} P ^{\multind{\mu}{1}{p+1}}}
                         {2 \rp \fact{p+1}}
                    +
                    \frac{i \dot{X} ^{\multind{\mu}{1}{p+1}}}
                         {\fact{p+1}}
                    P _{\multind{\mu}{1}{p+1}}
                \right ]
        \right\}
    \nonumber \\
    {\scriptstyle 2.} & & \qquad
    \simeq
    \exp
    \left\{
        \int d ^{p+1} \Bsigma
            \left [
                - i
                \frac{\rp \fact{p+1}}{2 N}
                \frac{i \dot{X} ^{\multind{\mu}{1}{p+1}}}
                     {\fact{p+1}}
                \frac{i \dot{X} _{\multind{\mu}{1}{p+1}}}
                     {\fact{p+1}}
            \right ]
    \right\}
    \nonumber \\
    {\scriptstyle 3.} & & \qquad
    \simeq
    \exp
    \left\{
        -
        i
        \int d ^{p+1} \Bsigma
            \left [
                -
                \frac{\rp}{2 N \fact{p+1}}
                \dot{\Bx} ^{2}
            \right ]
    \right\}
    \quad .
\eea
\begin{enumerate}
    \item we adopted the convention written in the proposition dots
    \item \dots using the {\it Minkowski--space} result for the Gaussian
    integration and
    \item symplifing the result.
\end{enumerate}
\end{proof}

\section{Saddle Point Evaluation}

To understand the physical meaning of equation (\ref{4.befsadpoieva})
we have to eliminate the {\it Lagrange multiplier}
$N ^{\multind{A}{1}{p+1}}$ which appears in the combination
$N = \epsilon _{\multind{A}{1}{p+1}} N ^{\multind{A}{1}{p+1}}$.
\begin{props}[Saddle Point Evaluation]\spbcorr{}.\\
    The exponent in equation {\rm (\ref{4.befsadpoieva})}
    is proportional to the Nambu--Goto Lagrangian density when estimated
    at the {\it saddle point}, i.e. we have
    $$
        -
        \frac{\rp}{2 N \fact{p+1}}
        \dot{\Bx} ^{2}
        +
        N
        E
        =
        \left( 2 E \rp \right) ^{\frac{1}{2}}
        \left [
            -
            \frac{\dot{X} ^{2}}{\fact{p+1}}
        \right ] ^{\frac{1}{2}}
        \quad .
    $$
\end{props}
\begin{proof}
The phase stationarity point is characterized by the vanishing of the first
derivative with respect to $N$, i.e. we have to solve
\beq
    \frac{d}{d N}
    \left [
        -
        \frac{\rp}{2 N \fact{p+1}}
        \dot{\Bx} ^{2}
        +
        N
        E
    \right \rceil _{N = \hat{N}}
    =
    0
    \quad ,
\eeq
where $\hat{N}$ is the stationarity point. So, we find
\beq
    \frac{\rp}{2 \hat{N} ^{2} \fact{p+1}}
    \dot{\Bx} ^{2}
    +
    E
    =
    0
    \quad \Rightarrow \qquad
    \hat{N} ^{2}
    =
    -
    \frac{\rp}{2 E \fact{p+1}}
    \dot{\Bx} ^{2}\ .
\eeq
Thus the exponent turns out to be
\bea
    \left .
    -
    \frac{\rp}{2 N \fact{p+1}}
    \dot{\Bx} ^{2}
    +
    N
    E
    \right \rceil _{N = \hat{N}}
    & = &
    \frac{1}{\hat{N}}
    \left [
        -
        \frac{\rp}{2 \fact{p+1}}
        \dot{\Bx} ^{2}
        -
        \hat{N} ^{2}
        E
    \right ]
    \nonumber \\
    & = &
    \frac{\left( 2 E \right) ^{\frac{1}{2}}}
         {
          \left(
              -
              \frac{\rp}{\fact{p+1}}
              \dot{\Bx} ^{2}
          \right) ^{\frac{1}{2}}
         }
    \left [
        -
        \frac{\rp}{2 \fact{p+1}}
        \dot{\Bx} ^{2}
        -
        \frac{\rp}{2 \fact{p+1}}
        \dot{\Bx} ^{2}
    \right ]
    \nonumber \\
    & = &
    \left( 2 E \rp \right) ^{\frac{1}{2}}
    \left [
        -
        \frac{\dot{\Bx} ^{2}}{\fact{p+1}}
    \right ] ^{\frac{1}{2}}
    \quad .
\eea
\end{proof}

\section{Equations Satisfied by the Kernel}
\label{A.kerwavder}

This section is devoted to the derivation of the equations for the
kernel
$$K \qtq{\BB ( \Bsigma) , \BB ( \Bsigma _{0}) ; V} \quad .$$
\begin{props}[Functional Differential Equations for the Kernel]\spbcorr{}.\\
    The Propagation Kernel of a $p$-{\sl{}brane} satisfies the following
    equations:
    \bea
        \frac{
              \partial K
                  \left [
                      \BB ( \Bsigma ),
                      \BB _{0} ( \Bsigma );
                      V
                  \right ]
             }
             {\partial V}
        & = &
        -
        \frac{i E}{\hbar}
        K
            \left [
                \By 
                \By _{0} 
                V
            \right ]
        \\
        \frac{
              \delta K
                  \left [
                      \By 
                      \By _{0} 
                      V
                  \right ]
             }
             {\delta Y ^{\mu} ( \Bs )}
        & = &
        \frac{i}{\hbar}
        \int _{\By _{0} ( \Bs )} ^{\By ( \Bs )}
        \int _{\zeta _{0} ( \Bs )} ^{\zeta ( \Bs )}
            \left [ D \mu ( \Bsigma ) \right ]
            Q _{ \mu \multind{\mu}{1}{p}}
            Y ^{\prime \multind{\mu}{1}{p}}
            \exp \left( \frac{i S}{\hbar} \right)
    \: .
    \eea
\end{props}
\begin{proof}
The result can be obtained starting from equations
(\ref{KernelVar}--\ref{ActionVar}). The left hand side of equation
(\ref{KernelVar}), i.e. the total variation of the Kernel with respect to the
physical coordinates, which are the ones singled out by Hamilton--Jacobi Theory,
i.e.
$V$ and $\By$, can be rewritten as
\beq
    \delta K
    \left [
        \By ,
        \By _{0} ;
        V
    \right ]
    =
    \frac{
          \partial K
                   \left [
                       \By ,
                       \By _{0} ;
                       V
                   \right ]
         }
         {\partial V}
    d V
    +
    \oint _{\Sigma}
        \frac{d ^{p} \Bs}{\sqrt{\By ^{\prime 2}}}
        \frac{
              \delta K
              \left [
                  \By ,
                  \By _{0} ; 
                  V
              \right ]
             }
             {\delta Y ^{\mu} ( \Bs )}
             \delta Y ^{\mu} ( \Bs )
    \quad ;
\eeq
then, inserting this equality, as well as formula (\ref{ActionVar})
into equation (\ref{KernelVar}) we obtain
\bea
    & & \esci
    \frac{\partial K
                   \left [
                       \By ,
                       \By _{0} ;
                       V
                   \right ]
         }
         {\partial V}
    d V
    +
    \oint _{\Sigma}
        \frac{d ^{p} \Bs }{\sqrt{\By ^{\prime 2}}}
        \frac{
              \delta K
              \left [
                  \By ,
                  \By _{0} ;
                  V
              \right ]
             }
             {\delta Y ^{\mu} ( \Bs )}
             \delta Y ^{\mu} ( \Bs )
    =
    \nonumber \\
    & & \qquad \qquad
    =
    i
    \int _{\By ( \Bs _{0})} ^{\By ( \Bs )}
    \int _{\Bzeta ( \Bs _{0})} ^{\Bzeta ( \Bs )}
        [D \mu ( \Bsigma )]
        \oint _{C}
            \frac{1}{p!}
            Q _{\multind{\mu}{1}{p+1}}
            \multint{Y}{\mu}{2}{p+1}
            \delta Y ^{\mu _{1}}
        \cdot
        \nonumber \\
        & & \qquad \qquad \qquad \qquad \qquad
            \qquad \qquad \qquad \qquad \qquad
        \cdot
        \exp
        \left\{
            i
            S \left [
                  \Bx ,
                  \Bp ,
                  \Bxi ,
                  \BN ;
                  V
              \right ]
        \right\}
    +
    \nonumber \\
    & & \qquad \qquad
    -
    i
    E
    \int _{\By ( \Bs _{0})} ^{\By ( \Bs )}
    \int _{\Bzeta ( \Bs _{0})} ^{\Bzeta ( \Bs )}
        [D \mu ( \Bsigma )]
        \exp
        \left\{
            i
            S \left [
                  \Bx ,
                  \Bp ,
                  \Bxi ,
                  \BN ;
                  V
              \right ]
        \right\}
    d V
    \nonumber \\
    & &
    =
    i
    \int _{\By ( \Bs _{0})} ^{\By ( \Bs )}
    \int _{\Bzeta ( \Bs _{0})} ^{\Bzeta ( \Bs )}
        [D \mu ( \Bsigma )]
        \oint _{\Sigma}
            \frac{d ^{p} \Bs}{\sqrt{\By ^{\prime 2}}}
            \frac{1}{p!}
                    Q _{\multind{\mu}{1}{p+1}}
                    Y ^{\prime \multind{\mu}{2}{p+1}}
                    \delta Y ^{\mu _{1}}
        \cdot
        \nonumber \\
        & & \qquad \qquad \qquad \qquad \qquad
            \qquad \qquad \qquad \qquad \qquad
        \cdot
        \exp
        \left\{
            i
            S \left [
                  \Bx ,
                  \Bp ,
                  \Bxi ,
                  \BN ;
                  V
              \right ]
        \right\}
    +
    \nonumber \\
    & & \qquad \qquad
    -
    i
    E
    K
    \left [
    	\By ,
    	\By _{0} ;
    	V
    \right ]
    d V
    \quad ;
\eea
comparing the first and the last quantities,
equating the terms corresponding to $dV$ and $\delta X ^{\mu _{1}}$
yields the desired result, i.e.
\bea
    \frac{
          \partial K
              \left [
                  \By 
                  \By _{0} 
                  V
              \right ]
         }
         {\partial V}
    & = &
    -
    \frac{i E}{\hbar}
    K
        \left [
            \By ,
            \By _{0} ;
            V
        \right ]
    \\
    \frac{
          \delta K
              \left [
                  \By , 
                  \By _{0} ;
                  V
              \right ]
         }
         {\delta Y ^{\mu} ( \Bs )}
    & = &
    \frac{i}{\hbar}
    \int _{\By _{0} ( \Bs )} ^{\By ( \Bs )}
    \int _{\Bzeta _{0} ( \Bs )} ^{\Bzeta ( \Bs )}
        \left [ D \mu ( \Bsigma ) \right ]
        Q _{ \mu \multind{\mu}{1}{p}}
        Y ^{\prime \multind{\mu}{1}{p}}
        \exp \left( \frac{i S}{\hbar} \right)
    \quad .
\eea
\end{proof}

\section{Kernel Functional Wave Equation}
\label{A.kerfunwavder}

As already pointed out in the text, the Kernel for an extended object
considered as the only dynamical {\it Boundary} of its history
can be obtained by solving a Schr\"o{}dinger-like functional
equation. This is the main
 result we get by starting from the Hamilton--Jacobi
description of the {\it Quantum Dynamics of the Boundary}. We already
derived this result according with the procedure of section
\ref{3.strfunwvequ}. Now, we will apply the same procedure for a $p$-{\sl{}brane}.
The result is not surprising, since, apart from some slightly different
notation, we get exactly the same formula as (\ref{A.kerfunwavequ}).
\begin{props}[$p$-brane Functional Schr\"oedinger Equation]\spbcorr{}.\\
    The propagation Kernel of the $p$-{\sl{}brane} satisfies the following
    Schr\"odinger--like functional equation:
    \bea
        & & \esci \esci \esci \esci
        -
        \frac{\hbar}{2 \rp p !}
        \norbe{\BB}
        \oint _{\BB} \frac{d ^{p} \Bs}{\sqrt{Y ^{\prime 2}}}
        \frac{\delta ^{2}}
             {\delta Y ^{\mu} \ttt{\Bs} \delta Y _{\mu} \ttt{\Bs}}
            K \left [
                  \By \ttt{\Bs}
                  ,
                  \By _{0} \ttt{\Bs}
                  ;
                  V
              \right ]
        =
        \nonumber \\
        & & \qquad \qquad \qquad \qquad =
        i
        \hbar
        \frac{\partial}{\partial V}
        K \left [
              \By \ttt{\Bs}
              ,
              \By _{0} \ttt{\Bs}
              ;
              V
          \right ]
    \eea
\end{props}
\begin{proof}
The proof reproduces exactly the one we gave for the {\sl String}. Thus, we just
briefly recall the various steps; firstly
\bea
    \frac{\delta ^{2}
          K \left [
                \By \left( \Bs \right)
                ,
                \By _{0} \left( \Bs \right)
                ;
                V
            \right ]
         }
         {
          \delta Y ^{\mu} \left( \Bs \right)
          \delta Y _{\mu} \left( \Bs \right)
         }
    & = &
    -
    \frac{1}{\hbar ^{2}}
    \int _{\By _{0} \left( \Bs \right)} ^{\By \left( \Bs \right)}
    \funinte{\mu \left( \Bsigma \right)}
            {\Bzeta _{0} \left( \Bs \right)}
            {\Bzeta \left( \Bs \right)}
        q _{\mu} q ^{\mu}
        e ^{\frac{i}{\hbar} S}
    \label{A.pbrsecfundermomave}
    \\
    \frac{\partial}{\partial A}
    K \left [
          \By \left( \Bs \right)
          ,
          \By _{0} \left( \Bs \right)
          ;
          V
      \right ]
    & = &
    -
    \frac{i E}{\hbar}
    K \left [
          \By \left( \Bs \right)
          ,
          \By _{0} \left( \Bs \right)
          ;
          V
      \right ]
      \quad ,
    \label{A.pbrkervolder}
\eea
which are the relations equivalent to (\ref{A.secfundermomave}) and
(\ref{A.kerareder}). Now, of course $q _{\mu}$ is the momentum projected on the
 $p$-dimensional {\it Boundary}, i.e.
\beq
    q _{\mu}
    =
    Q _{\mu \multind{\mu}{1}{p}}
    Y ^{\prime \multind{\mu}{1}{p}}
    \quad .
\eeq
The final step requires to identify equation (\ref{A.pbrsecfundermomave}) with
the expectation value of the square of the {\sl Boundary Momentum}, and  by
substituting (\ref{A.pbrsecfundermomave},\ref{A.pbrkervolder}) in the
Hamilton--Jacobi equation. Thus, we get the final result
\beq
    \frac{\hbar ^{2}}{2 \rp p!}
    \norbe{\BB}
    \oint _{\BB} \frac{d ^{p} \Bs}{\sqrt{\By ^{\prime 2}}}
    \frac{
          \delta ^{2}
          K \left [
                \By 
                ,
                \By _{0} 
                ;
                V
            \right ]
         }
         {\delta Y ^{\mu} \left( \Bs \right) \delta Y _{\mu} \left( \Bs \right)}
    =
    i \hbar
    \frac{\partial
          K \left [
                \By 
                ,
                \By _{0} 
                ;
                V
                \right ]
         }
         {\partial V}
    \quad .
\eeq
\end{proof}

\section{Holographic Coordinates: Functional Derivatives}
\label{A.funcdersig}

In this section we compute the first and second functional derivatives
of the {\sl Holographic coordinates},
which are the most appropriate coordinates
of the Quantized Theory for a $p$-{\sl{}brane}.
\begin{props}[Functional Derivatives of the Holographic Coordinates]\spbcorr{}.
\label{A.funcdersigpro}\\
    If we consider the coordinates of the quantized $p$-{\sl{}brane},
    \beq
        Y ^{\multind{\mu}{1}{p+1}} [ \BB ]
        \equiv
        \oint _{\BB}
            Y ^{\mu _{1}} \multint{Y}{\mu}{2}{p+1}
        =
        \oint _{\BB}
            d \Bs
            Y ^{\mu _{1}} \ttt{\Bs}
            Y ^{\prime \multind{\mu}{2}{p+1}} \ttt{\Bs}
    \eeq
    where, for the reader's convenience, we recall  the expression for the
    {\sl Tangent Element to a $p$-brane} {\rm (\ref{hypel})},
    $$
        Y ^{\prime \multind{\mu}{2}{p+1}}
        =
        \epsilon ^{\multind{a}{2}{p+1}}
        \frac{\partial Y ^{\mu _{2}}}{\partial \sigma _{a _{2}}}
        \cdot \dots \cdot
        \frac{\partial Y ^{\mu _{p+1}}}{\partial \sigma _{a _{p+1}}}
    \quad ,
    $$
    Accordingly, the first and second functional derivatives are given by
    \bea
        \frac{\delta Y ^{\multind{\mu}{1}{p+1}} [ \BB ]}
             {\delta Y ^{\alpha} ( \bar{\Bsigma} )}
        & = &
        \delta ^{\mu _{1}} _{\alpha}
        Y ^{\prime \multind{\mu}{2}{p+1}} ( \bar{\Bs} )
        -
        \sum _{i} ^{2,p+1}
             \delta ^{\mu _{i}} _{\alpha}
             Y ^{ \prime
                  \multind{\mu}{2}{i-1}
                  \check{\mu} _{i}
                  \mu _{1}
                  \multind{\mu}{i+1}{p+1}
                  }
        \label{A.fundersig1}
        \\
        \frac{
              \delta ^{2}
              Y ^{\multind{\mu}{1}{p+1} [ \BB ]}
             }
             {
              \delta Y ^{\alpha} ( \bar{\Bsigma} )
              \delta Y ^{\beta} ( \tilde{\Bsigma} )
             }
        & = &
        \sum _{j} ^{2,p+1}
            \delta ^{\mu _{1} \mu _{j}} _{\alpha \beta}
            \epsilon ^{\multind{a}{2}{p+1}}
            \partial _{a _{2}} Y ^{\mu _{2}}
            \cdot \dots \cdot
            \partial _{a _{i}} \delta \ttt{\bar{\Bs} - \tilde{\Bs}}
            \cdot \dots \cdot
            \partial _{a _{p+1}} Y ^{\mu _{p+1}}
        +
        \nonumber \\
        & & \qquad
        -
        \sum _{i \neq j \atop i,j} ^{2,p+1}
            \delta ^{\mu _{i}} _{\alpha}
            \delta ^{\mu _{j}} _{\beta}
            \epsilon ^{\multind{a}{2}{i} \dots \multind{a}{j}{p+1}}
            \partial _{a _{2}} Y ^{\mu _{2}}
            \cdot \dots \cdot
            \partial _{a _{i}} Y ^{\mu _{1}}
        \cdot
        \nonumber \\
        & & \qquad \qquad \qquad \qquad
        \cdot
            \dots \cdot
            \partial _{a _{j}} \delta \ttt{\bar{\Bs} - \tilde{\Bs}}
            \cdot \dots \cdot
            \partial _{a _{p+1}} Y ^{\mu _{p+1}}\ .
        \label{A.fundersig2}
    \eea
\end{props}
\begin{proof}
We have to be very careful with indices in this computation, because
in contrast to the {\sl String} case, in the general case we do not want
to explicitly write down the chain rule for all term. Hence, we proceed
as follows:
\bea
    \frac{\delta \sigma ^{\multind{\mu}{1}{p+1}} [ \BB ]}
         {\delta Y ^{\alpha} ( \bar{\Bs} )}
    & = &
    \oint _{\BB}
        d ^{p} \Bs
        \frac{\delta Y ^{\mu _{1}} ( \Bs )}
             {\delta Y ^{\alpha} ( \bar{\Bs} )}
         Y ^{\prime \multind{\mu}{2}{p+1}} ( \Bs )
    +
    \oint _{\BB}
        d ^{p} \Bs
        Y ^{\mu _{1}} ( \Bs )
        \frac{\delta Y ^{\prime \multind{\mu}{2}{p+1}} ( \Bs )}
             {\delta Y ^{\alpha} ( \bar{\Bs} )}
     \nonumber \\
    {\scriptstyle 1.} & = &
    \oint _{\BB}
        d ^{p} \Bs
        \delta ^{\mu _{1}} _{\alpha}
        \delta \left( \Bs  - \bar{\Bs} \right)
        Y ^{\prime \multind{\mu}{2}{p+1}} ( \Bs )
    +
    \nonumber \\
    & & \qquad
    +
    \oint _{\BB}
        d ^{p} \Bs
        Y ^{\mu _{1}} ( \Bs )
        \sum _{i} ^{2,p+1}
            \epsilon ^{\multind{a}{2}{p+1}}
    \cdot
    \nonumber \\
    & & \qquad \qquad
    \cdot
            \frac{\partial Y ^{\mu _{2}} ( \Bs )}
                 {\partial s _{a _{2}}}
            \cdot \dots \cdot
            \frac{\delta}
                 {\delta Y ^{\alpha} ( \bar{\Bs} )}
            \left(
                \frac{\partial Y ^{\mu _{i}} ( \Bs )}
                     {\partial s _{a _{i}}}
            \right)
            \cdot \dots \cdot
            \frac{\partial Y ^{\mu _{p+1}} ( \Bs )}
                 {\partial s _{a _{p+1}}}
     \nonumber \\
    {\scriptstyle 2.} & = &
    \delta ^{\mu _{1}} _{\alpha}
    Y ^{\prime \multind{\mu}{2}{p+1}} ( \bar{\Bs} )
    +
    \nonumber \\
    & & \qquad
    +
    \oint _{\BB}
        d ^{p} \Bs
        Y ^{\mu _{1}} ( \Bs )
        \sum _{i} ^{2,p+1}
            \epsilon ^{\multind{a}{2}{p+1}}
    \cdot
    \nonumber \\
    & & \qquad \qquad
    \cdot
            \frac{\partial Y ^{\mu _{2}} ( \Bs )}
                 {\partial s _{a _{2}}}
            \cdot \dots \cdot
            \partial _{s _{a _{i}}}
            \left[
                  \delta ^{\mu _{i}} _{\alpha}
                  \delta \left( \Bs - \bar{\Bs} \right)
            \right]
            \cdot \dots \cdot
            \frac{\partial Y ^{\mu _{p+1}} ( \Bs )}
                 {\partial s _{a _{p+1}}}
    \nonumber \\
    {\scriptstyle 3.} & = &
    \delta ^{\mu _{1}} _{\alpha}
    Y ^{\prime \multind{\mu}{2}{p+1}} ( \bar{\Bs} )
    +
    \nonumber \\
    & & \qquad
    +
    \oint _{\BB}
        d ^{p} \Bs
        \sum _{i} ^{2,p+1}
            \partial _{s _{a _{i}}}
            \left[
                Y ^{\mu _{1}} ( \Bs )
                    \epsilon ^{\multind{a}{2}{p+1}}
                    \frac{\partial Y ^{\mu _{2}} ( \Bs )}
                         {\partial s _{a _{2}}}
            \right .
    \cdot
    \nonumber \\
    & & \qquad \qquad \qquad \qquad \qquad \qquad
    \cdot
            \left .
                   \dots \cdot
                          \delta ^{\mu _{i}} _{\alpha}
                          \delta \left( \Bs - \bar{\Bs} \right)
                    \cdot \dots \cdot
                    \frac{\partial Y ^{\mu _{p+1}} ( \Bs )}
                         {\partial s _{a _{p+1}}}
            \right]
    +
    \nonumber \\
    & & \qquad
    -
    \oint _{\BB}
        d ^{p} \Bs
        \delta \left( \Bs - \bar{\Bs} \right)
        \sum _{i} ^{2,p+1}
            \delta ^{\mu _{i}} _{\alpha}
            \partial _{s _{a _{i}}}
            \left[
                Y ^{\mu _{1}} ( \Bs )
                    \epsilon ^{\multind{a}{2}{p+1}}
            \right .
    \cdot
    \nonumber \\
    & & \qquad \qquad \qquad \qquad
    \cdot
            \left .
                    \frac{\partial Y ^{\mu _{2}} ( \Bs )}
                         {\partial s _{a _{2}}}
                    \dots \cdot
                          \check{
                                 \frac{\partial Y ^{\mu _{i}} ( \Bs )}
                                      {\partial s _{a _{i}}}
                                }
                    \cdot \dots \cdot
                    \frac{\partial Y ^{\mu _{p+1}} ( \Bs )}
                         {\partial \sigma _{a _{p+1}}}
            \right]
    \nonumber \\
    {\scriptstyle 4.} & = &
    \delta ^{\mu _{1}} _{\alpha}
    Y ^{\prime \multind{\mu}{2}{p+1}} ( \bar{\Bs} )
    +
    \nonumber \\
    & & \qquad
    -
    \sum _{i} ^{2,p+1}
         \! \delta ^{\mu _{i}} _{\alpha}
         \left[
             \epsilon ^{\multind{a}{2}{p+1}}
             \frac{\partial Y ^{\mu _{2}} ( \bar{\Bs} )}
                  {\partial s _{a _{2}}}
             \dots
                   \check{
                          \frac{\partial Y ^{\mu _{i}} ( \bar{\Bs} )}
                               {\partial s _{a _{i}}}
                         }
             \frac{\partial Y ^{\mu _{1}} ( \bar{\Bs} )}
                  {\partial _{s _{a _{i}}}}
             \dots
             \frac{\partial Y ^{\mu _{p+1}} ( \bar{\Bs} )}
                  {\partial s _{a _{p+1}}}
         \right]
    \nonumber \\
    {\scriptstyle 5.} & = &
    \delta ^{\mu _{1}} _{\alpha}
    Y ^{\prime \multind{\mu}{2}{p+1}} ( \bar{\Bs} )
    -
    \sum _{i} ^{2,p+1}
         \delta ^{\mu _{i}} _{\alpha}
         Y ^{ \prime
              \multind{\mu}{2}{i-1}
              \check{\mu} _{i}
              \mu _{1}
              \multind{\mu}{i+1}{p+1}
              }
\eea
The main steps goes as follows;  we first apply the {\it chain} rule,
letting the functional derivative act on $\By$ and $\By '$ respectively;
\begin{enumerate}
    \item the action on $\By$ singles out a Dirac delta
    on the {\it {\sl{}brane} parameters} times a Kronecker delta over
    field components, whereas, we must be more careful acting
    on $\By '$ since the result is still differentiated once with respect
    to the {\sl{}brane} parameters\footnote{We assume that we can exchange
    functional and ordinary derivatives.}; in this way \dots
    \item \dots we can use the Dirac delta in the first term to
    {\it kill} the integration over $d ^{p} \Bs$ and integration
    by parts in the second term; this procedure \dots
    \item \dots singles out a {\it Boundary} term ({\it the second one})
    and a term where we find again a Dirac delta over the parameters;
    now \dots
    \item \dots
    the first term vanishes because $\Gamma$ has no {\it Boundary};
    integration in the second term is
    {\it killed} again by the Dirac delta; moreover we observe that applying
    the {\it chain rule} to the partial derivative, only acting on
    $Y ^{\mu _{1}}$ we get a non--vanishing result: in the other cases
    we have only second derivatives which, being symmetric, give a
    vanishing result when the indices are saturated with the
    totally anti symmetric Levi--Civita tensor;
    \item using the {\it suppressed} ``$\:\check{\ }\:$'' {\it notation}
    we can then write the result in more compact form.
\end{enumerate}
Along the same lines one  computes the second functional
derivative; but, it is better to derive, as a preliminary result,
the first functional derivative of the {\sl $p$-brane Tangent Element}:
\bea
    & & \esci \esci \esci
    \frac{
          \delta Y ^{\prime
                     \multind{\mu}{2}{i-1}
                     \check{\mu} _{i}
                     \mu _{1}
                     \multind{\mu}{i+1}{p+1}
                      }
          \ttt{\bar{\Bs}}
         }
         {\delta Y ^{\beta} \ttt{\tilde{\Bs}}}
    =
    \nonumber \\
    & & =
    \epsilon ^{\multind{a}{2}{p+1}}
    \frac{
          \delta
          \left(
              \partial _{a _{2}} Y ^{\mu _{2}} \ttt{\bar{\Bs}}
              \cdot \dots \cdot
              \partial _{a _{i}} Y ^{\mu _{1}} \ttt{\bar{\Bs}}
              \cdot \dots \cdot
              \partial _{a _{p+1}} Y ^{\mu _{p+1}} \ttt{\bar{\Bs}}
          \right)
         }
         {\delta Y ^{\beta} \ttt{\tilde{\Bs}}}
    \nonumber \\
    & & =
    \sum _{j \neq i \atop j} ^{2,p+1}
        \epsilon ^{\multind{a}{2}{i} \dots \multind{a}{j}{p+1}}
        \partial _{a _{2}} Y ^{\mu _{2}} \ttt{\bar{\Bs}}
        \cdot \dots \cdot
        \partial _{a _{i}} Y ^{\mu _{1}} \ttt{\bar{\Bs}}
    \cdot
    \nonumber \\
    & & \qquad \qquad
    \cdot
        \dots \cdot
        \partial _{a _{j}} \left[
                               \delta ^{\mu _{j}} _{\beta}
                               \delta \ttt{\bar{\Bs} - \tilde{\Bs}}
                           \right]
        \cdot \dots \cdot
        \partial _{a _{p+1}} Y ^{\mu _{p+1}} \ttt{\bar{\Bs}}
    +
    \nonumber \\
    & & \qquad
    +
    \epsilon ^{\multind{a}{2}{p+1}}
    \cdot
    \partial _{a _{2}} Y ^{\mu _{2}} \ttt{\bar{\Bs}}
    \cdot \dots \cdot
    \nonumber \\
    & & \qquad \qquad
    \cdot
    \partial _{a _{i}} \left [
    			   \delta ^{\mu _{j}} _{\beta}
    		           \delta \ttt{\bar{\Bs} - \tilde{\Bsigma}}
    		       \right]
    \cdot \dots \cdot
    \partial _{a _{p+1}} Y ^{\mu _{p+1}} \left( \bar{\Bsigma} \right)
    \quad .
\eea
By neglecting  the complication of the suppressed/inserted indices, the
formula reads
\bea
    & & \esci
    \frac{
          \delta Y ^{\prime \multind{\mu}{2}{p+1}}
          \left( \bar{\Bsigma} \right)
         }
         {\delta Y ^{\beta} \left( \tilde{\Bsigma} \right)}
    =
    \nonumber \\
    & &
    \sum _{j} ^{2,p+1}
        \epsilon ^{\multind{a}{2}{p+1}}
        \partial _{a _{2}} Y ^{\mu _{2}} \ttt{\bar{\Bs}}
        \cdot \dots \cdot
        \partial _{a _{j}} \left[
                               \delta ^{\mu _{j}} _{\beta}
                               \delta \ttt{\bar{\Bs} - \tilde{\Bs}}
                           \right]
        \cdot \dots \cdot
        \partial _{a _{p+1}} Y ^{\mu _{p+1}} \ttt{\bar{\Bs}}
    \label{A.holcorfirfunder}
\quad .
\eea
Then, applying the {\it chain rule} to result (\ref{A.holcorfirfunder}),we can
write
\bea
    \frac{
          \delta ^{2}
          Y ^{\multind{\mu}{1}{p+1}} \qtq{\BB}
         }
         {
          \delta Y ^{\alpha} ( \bar{\Bs} )
          \delta Y ^{\beta} ( \tilde{\Bs} )
         }
    & = &
    \delta ^{\mu _{1}} _{\alpha}
    \frac{\delta Y ^{\prime \multind{\mu}{1}{p+1} \ttt{\bar{\Bs}}}}
         {\delta Y ^{\beta} \ttt{\tilde{\Bs}}}
    +
    \nonumber \\
    & & \qquad \qquad
    -
    \sum _{i} ^{2,p+1}
        \delta ^{\mu _{i}} _{\alpha}
        \frac{
              \delta
              Y ^{\prime
                  \multind{\mu}{1}{i-1}
                  \check{\mu} _{i}
                  \mu _{1}
                  \multind{\mu}{i+1}{p+1}
                 }
              \ttt{\bar{\Bs}}
             }
             {\delta Y ^{\beta} \ttt{\tilde{\Bs}}}
\eea
and the second functional derivative results to be
\bea
    \frac{
          \delta ^{2}
          Y ^{\multind{\mu}{1}{p+1}} \qtq{\BB}
         }
         {
          \delta Y ^{\alpha} \ttt{\bar{\Bs}}
          \delta Y ^{\beta} \ttt{\tilde{\Bs}}
         }
    \! \! \! & = & \! \! \!
    \delta ^{\mu _{1}} _{\alpha}
    \sum _{j} ^{2,p+1}
       \epsilon ^{\multind{a}{2}{p+1}}
       \partial _{a _{2}} Y ^{\mu _{2}}
       \cdot \dots \cdot
       \delta ^{\mu _{j}} _{\beta}
       \partial _{a _{j}}
       \delta \ttt{\bar{\Bs} - \tilde{\Bs}}
       \cdot \dots \cdot
       \partial _{a _{p+1}} Y ^{\mu _{p+1}}
    +
    \nonumber \\
    & & \qquad
    -
    \sum _{i} ^{2,p+1}
        \delta ^{\mu _{i}} _{\alpha}
        \epsilon ^{\multind{a}{2}{p+1}}
        \sum _{j \neq i \atop j} ^{2,p+1}
            \partial _{a _{2}} Y ^{\mu _{2}}
            \cdot \dots \cdot
            \partial _{a _{i}} Y ^{\mu _{1}}
    \cdot
    \nonumber \\
    & & \qquad \qquad \qquad \qquad
    \cdot
            \dots \cdot
            \delta ^{\mu _{j}} _{\beta}
            \partial _{a _{j}} \delta \ttt{\bar{\Bs} - \tilde{\Bs}}
            \cdot \dots \cdot
            \partial _{a _{p+1}} Y ^{\mu _{p+1}}
    +
    \nonumber \\
    & & \qquad
    -
    \sum _{i} ^{2,p+1}
        \delta ^{\mu _{i}} _{\alpha}
        \delta ^{\mu _{1}} _{\beta}
        \epsilon ^{\multind{a}{2}{p+1}}
        \partial _{a _{2}} Y ^{\mu _{2}}
        \cdot \dots \cdot
        \partial _{a _{i}} \delta \ttt{\bar{\Bs} - \tilde{\Bs}}
        \cdot \dots \cdot
        \partial _{a _{p+1}} Y ^{\mu _{p+1}}
    \nonumber \\
    \! \! \! & = & \! \! \!
    \sum _{j} ^{2,p+1} \! \!
        \left (
            \delta ^{\mu _{1}} _{\alpha}
            \delta ^{\mu _{i}} _{\beta}
            -
            \delta ^{\mu _{i}} _{\alpha}
            \delta ^{\mu _{1}} _{\beta}
        \right)
        \epsilon ^{\multind{a}{2}{p+1}}
        \partial _{a _{2}} Y ^{\mu _{2}}
        \cdot \dots \cdot
        \partial _{a _{i}} \delta \ttt{\bar{\Bs} - \tilde{\Bs}}
        \cdot \dots \cdot
        \partial _{a _{p+1}} Y ^{\mu _{p+1}}
    +
    \nonumber \\
    & & \qquad
    -
    \sum _{i \neq j \atop i,j} ^{2,p+1}
        \delta ^{\mu _{i}} _{\alpha}
        \delta ^{\mu _{j}} _{\beta}
        \epsilon ^{\multind{a}{2}{i} \dots \multind{a}{j}{p+1}}
        \partial _{a _{2}} Y ^{\mu _{2}}
        \cdot \dots \cdot
        \partial _{a _{i}} Y ^{\mu _{1}}
    \cdot
    \nonumber \\
    & & \qquad \qquad \qquad \qquad
    \cdot
        \dots \cdot
        \partial _{a _{j}} \delta \ttt{\bar{\Bs} - \tilde{\Bs}}
        \cdot \dots \cdot
        \partial _{a _{p+1}} Y ^{\mu _{p+1}}
    \nonumber \\
    \! \! \! & = & \! \! \!
    \sum _{j} ^{2,p+1}
        \delta ^{\mu _{1} \mu _{j}} _{\alpha \beta}
        \epsilon ^{\multind{a}{2}{p+1}}
        \partial _{a _{2}} Y ^{\mu _{2}}
        \cdot \dots \cdot
        \partial _{a _{i}} \delta \ttt{\bar{\Bs} - \tilde{\Bs}}
        \cdot \dots \cdot
        \partial _{a _{p+1}} Y ^{\mu _{p+1}}
    +
    \nonumber \\
    & & \qquad
    -
    \sum _{i \neq j \atop i,j} ^{2,p+1}
        \delta ^{\mu _{i}} _{\alpha}
        \delta ^{\mu _{j}} _{\beta}
        \epsilon ^{\multind{a}{2}{i} \dots \multind{a}{j}{p+1}}
        \partial _{a _{2}} Y ^{\mu _{2}}
        \cdot \dots \cdot
        \partial _{a _{i}} Y ^{\mu _{1}}
    \cdot
    \nonumber \\
    & & \qquad \qquad \qquad \qquad
    \cdot
        \dots \cdot
        \partial _{a _{j}} \delta \ttt{\bar{\Bs} - \tilde{\Bs}}
        \cdot \dots \cdot
        \partial _{a _{p+1}} Y ^{\mu _{p+1}}\ .
\eea
\end{proof}

We can check the last result by setting $p=1$,
the {\sl String} case:
\beq
    \frac{\delta ^{2} Y ^{\mu _{1} \mu _{2}}}
         {\delta Y ^{\alpha} (s) \delta Y ^{\beta} (s)}
    =
    \left(
        \delta ^{\mu _{1}} _{\alpha}
        \delta ^{\mu _{2}} _{\beta}
        -
        \delta ^{\mu _{2}} _{\alpha}
        \delta ^{\mu _{1}} _{\beta}
    \right)
    \frac{d}{ds} \delta \left( s - \bar{s} \right)
    =
    \delta ^{\mu _{1} \mu _{2}} _{\alpha \beta}
    \frac{d}{ds} \delta \left( s - \bar{s} \right)
    \quad ,
\label{A.fundersig2app}
\eeq
where we took into account $\Bs \equiv s$ and
$\bar{\Bs} \equiv \bar{s}$, because a {\sl String} is one dimensional object!
In the membrane case, $p=2$, we have
\bea
    \frac{\delta ^{2} \sigma ^{\mu _{1} \mu _{2} \mu _{3}}}
         {\delta Y ^{\alpha} (\Bs) \delta Y ^{\beta} (\bar{\Bs})}
    & = &
    \left(
        \delta ^{\mu _{1}} _{\alpha}
        \delta ^{\mu _{2}} _{\beta}
        -
        \delta ^{\mu _{2}} _{\alpha}
        \delta ^{\mu _{1}} _{\beta}
    \right)
    \epsilon ^{a _{2} a _{3}}
    \left( \partial _{a _{3}} x ^{\mu _{3}} \right)
    \partial _{a _{2}} \delta \left( \Bs - \bar{\Bs} \right)
    +
    \nonumber \\
    & &
    +
    \left(
        \delta ^{\mu _{1}} _{\alpha}
        \delta ^{\mu _{3}} _{\beta}
        -
        \delta ^{\mu _{3}} _{\alpha}
        \delta ^{\mu _{1}} _{\beta}
    \right)
    \epsilon ^{a _{2} a _{3}}
    \left( \partial _{a _{2}} Y ^{\mu _{2}} \right)
    \partial _{a _{3}} \delta \ttt{\Bs - \bar{\Bs}}
    +
    \nonumber \\
    & &
    +
    \left(
        \delta ^{\mu _{2}} _{\alpha}
        \delta ^{\mu _{3}} _{\beta}
        -
        \delta ^{\mu _{3}} _{\alpha}
        \delta ^{\mu _{2}} _{\beta}
    \right)
    \epsilon ^{a _{2} a _{3}}
    \left( \partial _{a _{2}} Y ^{\mu _{1}} \right)
    \partial _{a _{2}} \delta \left( \Bs - \bar{\Bs} \right)
    \nonumber \\
    & = &
    \epsilon ^{a _{2} a _{3}}
    \partial _{a _{2}} \delta \ttt{\Bs - \bar{\Bs}}
    \cdot
    \nonumber \\
    & & \qquad \qquad \cdot
    \left [
        \delta ^{\mu _{1} \mu _{2}} _{\alpha \beta}
        \partial _{a _{3}} Y ^{\mu _{3}}
        +
        \delta ^{\mu _{3} \mu _{1}} _{\alpha \beta}
        \partial _{a _{3}} Y ^{\mu _{2}}
        +
        \delta ^{\mu _{1} \mu _{3}} _{\alpha \beta}
        \partial _{a _{3}} Y ^{\mu _{1}}
    \right ]
    \quad .
\eea

\section{Functional Derivatives of the Classical Action}
\label{A.funderclaact}

In this appendix we compute the functional derivatives of the classical action
(\ref{4.claactgue}) in order to find a solution of the equation
(\ref{kerfunwav}).
\begin{props}[Functional Derivatives of the $p$-brane Classical Action]\spbcorr{}.\\
    The first and second functional derivatives of the classical action
    $$
    S _{\mathrm{cl.}} \left [ \BB (\Bsigma ), \BB _{0} ( \Bsigma) ; V \right ]
    =
    \frac{\beta}{2 \left( p + 1 \right) V}
    \Sigma ^{\multind{\mu}{1}{p+1}} [ \BB - \BB _{0}]
    \Sigma _{\multind{\mu}{1}{p+1}} [ \BB - \BB _{0}]
    $$
    are
    \bea
        \frac{\delta S _{\mathrm{cl.}}}{\delta Y ^{\mu _{1}}\ttt{\Bs}}
        & = &
        \frac{\beta}{V}
        \Sigma ^{\multind{\mu}{1}{p+1}} \qtq{\BB - \BB _{0}}
        Y ' _{\multind{\mu}{2}{p+1}}
        \label{A.funderact1}
        \\
        \frac{\delta ^{2} S _{\mathrm{cl.}}}
             {\delta Y ^{\mu _{1}} \ttt{\Bs} \delta Y _{\mu _{1}}\ttt{\Bs}}
        & = &
        \left( D - p + 1 \right)
        \frac{\beta}{V}
        \By ^{\prime 2} \ttt{\Bs}
        \label{A.funderact2}
    \quad ,
    \label{A.triclaactfirfunder}
    \eea
 where $\Sigma^{\multind{\mu}{1}{p+1}} [ \BB - \BB _{0}]$
 is given by {\rm (\ref{4.sigdef})}.
\end{props}
\begin{proof}
To get the desired results we refer
to equations (\ref{A.fundersig1}--\ref{A.fundersig2})
and we find
\bea
    \frac{\delta S _{\mathrm{cl.}}}{\delta Y ^{\alpha}\ttt{\Bs}}
    & = &
    \frac{\beta}{2 V \ttt{p + 1}}
    2
    \Sigma _{\multind{\mu}{1}{p+1}} \qtq{\BB - \BB _{0}}
    \frac{\delta Y ^{\multind{\mu}{1}{p+1}}\qtq{\BB}}
         {\delta Y ^{\alpha} \ttt{\Bs}}
    \nonumber \\
    & = &
    \frac{\beta}{V \ttt{p + 1}}
    \left [
        \Sigma ^{\multind{\mu}{1}{p+1}} \qtq{\BB - \BB _{0}}
        \delta ^{\mu _{1}} _{\alpha}
        Y ^{\prime \multind{\mu}{2}{p+1}\ttt{\bar{\Bs}}}
        +
    \right .
    \nonumber \\
    & & \qquad \qquad
    -
    \left .
        \Sigma _{\multind{\mu}{1}{p+1}} \qtq{\BB - \BB _{0}}
        \sum _{i} ^{2,p+1}
            \delta ^{\mu _{i}} _{\alpha}
            Y ^{\prime
                \multind{\mu}{2}{i-1}
                \check{\mu} _{i}
                \mu _{1}
                \multind{\mu}{i+1}{p+1}
               }
    \right ]
    \nonumber \\
    & = &
    \frac{\beta}{V \left( p + 1 \right)}
    \left [
        \Sigma _{\multind{\mu}{1}{p+1}} \qtq{\BB - \BB _{0}}
        \delta ^{\mu _{1}} _{\alpha}
        Y ^{\prime \multind{\mu}{2}{p+1}\ttt{\bar{\Bs}}}
        +
    \right .
    \nonumber \\
    & & \qquad \qquad
    +
    \left .
        p \Sigma _{\multind{\mu}{1}{p+1}} \qtq{\BB - \BB _{0}}
        \delta ^{\mu _{1}} _{\alpha}
        Y ^{\prime
            \multind{\mu}{2}{i-1}
            \check{\mu} _{i}
            \mu _{i}
            \multind{\mu}{i+1}{p+1}
           }
    \right ]
    \nonumber \\
    & = &
    \frac{\beta}{V}
        \Sigma _{\alpha \multind{\mu}{2}{p+1}} \qtq{\BB - \BB _{0}}
        Y ^{\prime \multind{\mu}{2}{p+1}\ttt{\bar{\Bs}}}
    \quad .
\eea
From the first line of the set of equalities above we get
\begin{eqnarray}
    \frac{\delta ^{2} S _{\mathrm{cl.}}}
         {\delta Y ^{\alpha} \ttt{\Bs} \delta Y ^{\beta} \ttt{\bar{\Bs}}}
    & = &
    \frac{\beta}{V \ttt{p + 1}}
    \frac{\delta}{\delta Y ^{\beta}\ttt{\Bs}}
    \left [
        \Sigma _{\multind{\mu}{1}{p+1}} \qtq{\BB - \BB _{0}}
        \frac{\delta Y ^{\multind{\mu}{1}{p+1}} \qtq{\BB}}
             {\delta Y ^{\alpha} \ttt{\Bs}}
    \right ]
    \nonumber \\
    & = &
    \frac{\beta}{V \ttt{p+1}}
    \left [
        \left(
            \frac{\delta Y ^{\multind{\mu}{1}{p+1}} \qtq{\BB}}
                 {\delta Y ^{\alpha} \ttt{\Bs}}
        \right) ^{2}
        +
        \frac{\delta ^{2} Y ^{\multind{\mu}{1}{p+1}} \qtq{\BB}}
             {\delta Y ^{\alpha} \ttt{\Bs} \delta Y ^{\beta} \ttt{\bar{\Bs}}}
        \Sigma _{\multind{\mu}{1}{p+1}}
    \right]
\label{A.intermediate}
\end{eqnarray}
Now, we saturate both the discrete indices $\alpha$ and $beta$ and the
continuous
ones, i.e $\Bs$ and $\bar{\Bs}$. From equation
(\ref{A.fundersig2app}) follows
that the second term in the sum has zero trace since it is skew symmetric in
$\alpha$ and $\beta$. Then, we remain with the first term in the square brackets
of (\ref{A.intermediate}):
\begin{eqnarray}
    \frac{\delta ^{2} S _{\mathrm{cl.}}}
         {\delta Y ^{\alpha}\ttt{\Bs} \delta Y _{\alpha} \ttt{\Bs}}
    & = &
    \frac{\beta}{V \ttt{p+1}}
        \frac{\delta Y ^{\multind{\mu}{1}{p+1}} \qtq{\BB}}
             {\delta Y ^{\alpha} \ttt{\Bs}}
        \frac{\delta Y _{\multind{\mu}{1}{p+1}} \qtq{\BB}}
             {\delta Y ^{\beta} \ttt{\bar{\Bs}}}
    \nonumber \\
    & = &
    \frac{\beta}{V \ttt{p+1}}
    \left(
        \delta ^{\mu _{1}} _{\alpha}
        Y ^{\prime \multind{\mu}{2}{p+1}}
        -
        \sum _{i} ^{2,p+1}
             \delta ^{\mu _{i}} _{\alpha}
             Y ^{\prime
                 \multind{\mu}{2}{i-1}
                 \check{\mu} _{i}
                 \mu _{1}
                 \multind{\mu}{i+1}{p+1}
                }
    \right)
    \cdot
    \nonumber \\
    & & \qquad \qquad \qquad
    \cdot
    \left(
        \delta _{\alpha \mu _{1}}
        Y ' _{\multind{\mu}{2}{p+1}}
        -
        \sum _{j} ^{2,p+1}
             \delta _{\mu _{j} \alpha}
             Y ^{\prime
                 \multind{\mu}{2}{j-1}
                 \check{\mu} _{j}
                 \mu _{1}
                 \multind{\mu}{j+1}{p+1}
                }
    \right)
    \nonumber \\
    & = &
    \frac{\beta}{V \ttt{p+1}}
    \left[
        \delta ^{\alpha} _{\alpha}
        Y ^{\prime \multind{\mu}{2}{p+1}}
        Y ' _{\multind{\mu}{2}{p+1}}
    +
    \right .
    \nonumber \\
    & & \qquad \qquad
    +
        \sum _{i} ^{2,p+1}
             \delta ^{\mu _{i}} _{\alpha}
             Y ^{\prime
                 \multind{\mu}{2}{i-1}
                 \check{\mu} _{i}
                 \mu _{1}
                 \multind{\mu}{i+1}{p+1}
                }
    \cdot
    \nonumber \\
    & & \qquad \qquad \qquad \cdot
        \sum _{j} ^{2,p+1}
             \delta _{\mu _{j} \alpha}
             Y ^{\prime
                 \multind{\mu}{2}{j-1}
                 \check{\mu} _{j}
                 \mu _{1}
                 \multind{\mu}{j+1}{p+1}
                }
    +
    \nonumber \\
    & & \qquad \qquad
    -
        \delta ^{\mu _{1}} _{ \alpha}
        Y ^{\prime \multind{\mu}{2}{p+1}}
        \sum _{j} ^{2,p+1}
             \delta _{\mu _{j} \alpha}
             Y ^{\prime
                 \multind{\mu}{2}{j-1}
                 \check{\mu} _{j}
                 \mu _{1}
                 \multind{\mu}{j+1}{p+1}
                }
    +
    \nonumber \\
    & & \qquad \qquad
    \left .
    -
        \delta _{\alpha \mu _{1}}
        Y ' _{\multind{\mu}{2}{p+1}}
        \sum _{i} ^{2,p+1}
             \delta ^{\mu _{i}} _{\alpha}
             Y ^{\prime
                 \multind{\mu}{2}{i-1}
                 \check{\mu} _{i}
                 \mu _{1}
                 \multind{\mu}{i+1}{p+1}
                }
    \right]
    \nonumber \\
    & = &
    \frac{\beta}{V \left( p + 1 \right)}
    \left[
        \delta ^{\alpha} _{\alpha}
        Y ^{\prime \multind{\mu}{2}{p+1}}
        Y ' _{\multind{\mu}{2}{p+1}}
    +
    \right .
    \nonumber \\
    & & \qquad \qquad
    +
        \sum _{i,j} ^{2,p+1}
             \delta ^{\mu _{i}} _{\mu _{j}}
             Y ^{\prime
                 \multind{\mu}{2}{i-1}
                 \check{\mu _{i}}
                 \mu _{1}
                 \multind{\mu}{i+1}{p+1}
                }
             Y ^{\prime
                 \multind{\mu}{2}{j-1}
                 \check{\mu _{j}}
                 \mu _{1}
                 \multind{\mu}{j+1}{p+1}
                }
    +
    \nonumber \\
    & & \qquad \qquad
    -
        \sum _{j} ^{2,p+1}
             \delta _{\mu _{j}} ^{\mu _{1}}
             Y ^{\prime
                 \multind{\mu}{2}{j-1}
                 \check{\mu} _{j}
                 \mu _{1}
                 \multind{\mu}{j+1}{p+1}
                }
             Y ^{\prime \multind{\mu}{2}{p+1}}
    +
    \nonumber \\
    & & \qquad \qquad
    \left .
    -
        \sum _{i} ^{2,p+1}
             \delta ^{\mu _{i}} _{\mu _{1}}
             Y ^{\prime
                 \multind{\mu}{2}{i-1}
                 \check{\mu} _{i}
                 \mu _{1}
                 \multind{\mu}{i+1}{p+1}
                }
             Y ' _{\multind{\mu}{2}{p+1}}
    \right]
    \nonumber \\
    & = &
    \frac{\beta}{V \ttt{p+1}}
    \left[
        \delta ^{\alpha} _{\alpha}
        \left( \By ' \right) ^{2}
    +
    \right .
    \nonumber \\
    & & \qquad \qquad
    +
        \sum _{i=j=k} ^{2,p+1}
             \delta ^{\mu _{k}} _{\mu _{k}}
             \left ( \By ' \right) ^2
    +
    \nonumber \\
    & & \qquad \qquad
    +
        \sum _{i \neq j \atop i,j}
             Y ^{\prime
                 \multind{\mu}{2}{i-1}
                 \check{\mu} _{i}
                 \mu _{1}
                 \multind{\mu}{i+1}{j}
                 \dots
                 \mu _{p+1}
                }
             Y ^{\prime
                 \multind{\mu}{2}{i}
                 \dots
                 \mu _{j-1}
                 \check{\mu} _{j}
                 \mu _{1}
                 \multind{\mu}{j+1}{p+1}
                }
    +
    \nonumber \\
    & & \qquad \qquad
    -
        \sum _{j} ^{2,p+1}
             Y ^{\prime
                 \multind{\mu}{2}{j-1}
                 \check{\mu} _{j}
                 \mu _{j}
                 \multind{\mu}{j+1}{p+1}
                }
             Y ^{\prime \multind{\mu}{2}{p+1}}
    +
    \nonumber \\
    & & \qquad \qquad
    \left .
    -
        \sum _{i} ^{2,p+1}
             Y ^{\prime
                 \multind{\mu}{2}{i-1}
                 \check{\mu} _{i}
                 \mu _{i}
                 \multind{\mu}{i+1}{p+1}
                }
             Y _{\prime \multind{\mu}{2}{p+1}}
    \right]
    \nonumber \\
    & = &
    \frac{\beta}{V \ttt{p+1}}
    \left[
        \delta ^{\alpha} _{\alpha}
        \ttt{\By '} ^{2}
    +
    \right .
    \nonumber \\
    & & \qquad \qquad
    +
        \delta ^{\alpha} _{\alpha} p \ttt{\By '} ^{2}
        -
        \ttt{ p ^{2} - p } \ttt{\By '} ^{2}
    +
    \nonumber \\
    & & \qquad \qquad
    -
        p \ttt{\By ' } ^{2}
    +
    \nonumber \\
    & & \qquad \qquad
    \left .
    -
        p \ttt{\By ' } ^{2}
    \right]
    \nonumber \\
    & = &
    \frac{\beta}{V \ttt{ p + 1 }}
    \left [
        D +
        D p -
        p \ttt{ p - 1 } -
        2 p
    \right ]
    \ttt{\By ' } ^{2}
    \nonumber \\
    & = &
    \frac{\beta}{V \ttt{ p + 1 }}
    \ttt{ p + 1 } \ttt{ D - p}
    \ttt{\By ' } ^{2}
    \nonumber \\
    & = &
    \frac{\beta}{V}
    \ttt{ D - p}
    \ttt{\By ' } ^{2}
    \label{A.triclaactsecfunder}
\end{eqnarray}
\end{proof}

\section{Solutions for the $\alpha$ and $\beta$ Kernel Ansatz Parameters}
\label{A.alphabeta}

To determine the parameters $\alpha$, $\beta$, we insert
(\ref{A.triclaactfirfunder}), (\ref{A.triclaactsecfunder})
into (\ref{trial1}) and (\ref{trial2}). The results are:
\bea
    & &
    \left\{
        \matrix{
        \displaystyle
        2
        \rp
        \frac{\alpha}{{\red V}}
        p!
        =
        -
        \norm
        {\green \oint \frac{d ^{p} \Bs}{\sqrt{\By ^{\prime 2}}}}
            \left( D - p + 1 \right)
            \frac{\beta}{{\red V}}
            {\green \By ^{\prime 2}}
        \hfill
        \cr
        \cr
        \displaystyle
        {\blue 2}
        \rp
        \frac{{\cyan \beta} p!}{{\blue 2} \left( p + 1 \right) {\cyan V ^{2}}}
        (-)
        \Sigma ^{2}
        =
        -
        \norm
        \cdot
        \cr
        \qquad \qquad \qquad \qquad
        \cdot
        \oint \frac{d ^{p} \Bsigma}{\sqrt{\By ^{\prime 2}}}
            \frac{\beta ^{{\cyan 2}}}{{\cyan V ^{2}}}
            \Sigma ^{\mu _{1} \multind{\mu}{2}{p+1}}
            Y ' _{\multind{\mu}{2}{p+1}}
            \Sigma _{\mu _{1} \multind{\nu}{2}{p+1}}
            Y ^{\prime \multind{\nu}{2}{p+1}}
        }
    \right .
\nonumber \\
    & &
    \left\{
        \matrix{
        \displaystyle
        \alpha
        =
        -
        \frac{\left( D - p + 1 \right) \beta}{2 \rp \fact{p}}
        \hfill
        \cr
        \cr
        \frac{\rp {\green \fact{p}}}{{\red \left( p + 1 \right)}}
        {\gray \Sigma ^{2}}
        =
        -
        \norbe{\BB}
        {\magenta \oint \frac{d ^{p} \Bsigma}{\sqrt{\By ^{\prime 2}}}}
            \frac{{\green \fact{p}}}{{\red \fact{p+1}}}
            {\gray \Sigma ^{2}}
            {\magenta \By ^{\prime 2}}
            {\cyan \beta}
    }
    \right .
\nonumber \\
    & &
    \left\{
        \matrix{
        \displaystyle
        \alpha
        =
        -
        \frac{\left( D - p + 1 \right) \beta}{2 \rp p!}
        \hfill
        \cr
        \cr
        \beta
        =
        \rp
        p!
        \hfill
        }
    \right .
\nonumber \\
    & &
    \left\{
        \matrix{
        \displaystyle
        \alpha
        =
        -
        \frac{\left( D - p + 1 \right)}{2}
        \hfill
        \cr
        \cr
        \beta
        =
        \rp
        \fact{p}
        \hfill
    }
    \right .
\eea


%% file: appB.tex
\pageheader{}{Nonstandard Analysis.}{}
\chapter{NonStandard Analysis}
\label{B.nonstanda}

\section{Short Introduction}

In view of the developments that we describe in chapter \ref{10.nonstacha}
we think it is important to give here a brief account of an
interesting branch of mathematics, namely {\it NonStandard Analysis},
which could play a deep role (still to be discovered!) in Relativistic
Quantum Field Theory and Quantum Gravity.
We used this technical apparatus to derive the
{\sl String Functional Wave Equation}
as an independent  check of the path--integral derivation
in chapter \ref{3.strfunqua}. {\it NonStandard
Analysis} can be applied to many different type of problems, but  it
is not yet very popular among theoretical physicists.
We don't subscribe to this attitude
because {\it NonStandard Analysis} provides a powerful tool to give a
{\it rigorous} treatment of {\it infinitesimal} and {\it infinities},
which are
ubiquitous especially in High Energy Theoretical Physics.
We will give in this appendix more information than the ones strictly needed
to follow the material in appendix \ref{C.nonstopro} and chapter
\ref{10.nonstacha}. We hope to excite the reader's interest  on this subject.

\subsection{Infinitesimals and Infinities}

In our opinion the main reason why physicist should be interested
in the study of {\it Nonstandard Analysis} is the key observation
that most of the more difficult problems in Physics are in a
more or less direct way related to the existence of singularities, i.e.
infinities, of various kind. {\it Infinity} is a strange concept that
need a careful handling in mathematical proofs.
For example, when the concept of  {\it limit to  Infinity} is introduced in
a course of calculus, one must face the failure of the usual
$\epsilon$- and $\delta$-neighborhood procedure.
The same problem shows up when dealing with infinitesimal quantities:
the product of a quantity that, in some limit, is {\it divergent},
times one that, in the same limit, is {\it infinitesimal} is not well defined
in general. The usual way of dealing these problems is {\it very far
away} from the original concept of {\it infinitesimal} and {\it infinity}
as introduced by Leibniz. {\it NonStandard Analysis} recovers
a more intuitive point of view and gives to it a proper foundation.
Approaching {\it NonStandard Analysis} from a (beautyfull) purely
mathematical
side would require  a long digression about {\it logic}, {\it propositions},
{\it quantifiers}, \dots{}.
On the contrary, we would like to give the reader only the more immediate,
and useful, definitions and results pertaining to the so called the
{\it NonStandard Reals}.

{\it NonStandard Entities} represent an extension of the corresponding
ones in standard analysis. This means that {\it NonStandard
Entities} include the {\it standard ones} but there is something more.
Such a  ``{\it{}something more}'' is obtained employing a single new word,
namely {\it standard}, as a {\it predicate} so that every mathematical entity
about which we can speak can be {\it standard} or can be {\it non standard}.
\begin{nots}[Standard Entity]\spbcorr{}.\\
    To assert that the entity $x$ is standard we will say
    \begin{center}
        $x \quad$IS$\quad$STANDARD
    \end{center}
    or in symbols
    $$
        \stand{x}
        \quad .
    $$
\end{nots}
This new predicate will not change the numbers we already know, but will give
us a suitable definition of expressions like {\it infinitely large} and
{\it infinitely small}.
This is achieved through the following procedure.
\begin{defs}[Filter]\spbcorr{}.\\
    Let $\mathbb{K}$ be a countable set. Let
    $\mathcal{P} \ttt{\mathbb{K}}$ be the power set of
    $\mathbb{K}$, i.e. the set of all subsets of
    $\mathbb{K}$; consider $\mathcal{U}$,
    $\emptyset \subset \mathcal{U} \subset \mathcal{P} \ttt{\mathbb{K}}$
    a set of subsets of $\mathbb{K}$, which satisfies the following
    properties:
    \begin{enumerate}
        \item $\emptyset \notin \mathcal{U}$;
        \item $
               A \in \mathcal{U}
               \quad \wedge \quad
               B \in \mathcal{U}
               \quad \Longrightarrow \quad
               A \cup B \in \mathcal{U}$;
        \item $
               A \in \mathcal{U}
               \quad \wedge \quad
               B \in \mathcal{P} \ttt{\mathbb{K}}
               \quad \wedge \quad
               A \subseteq B
               \quad \Longrightarrow \quad
               B \in \mathcal{U}
              $
    \end{enumerate}
    $\mathcal{U}$ is called a \underbar{Filter}.
\end{defs}
\begin{defs}[Ultrafilter]\spbcorr{}.\\
    Let $\mathbb{K}$ be a countable set and $\mathcal{U}$ a
    {\sl Filter} on $\mathbb{K}$. If in addition $\mathcal{U}$ satisfies
    \begin{enumerate}
        \item[4.] $
              B \in \mathcal{P} \ttt{\mathbb{K}}
              \quad \Longrightarrow \quad
              B \in \mathcal{U}
              \quad \vee \quad
              \mathbb{K} \backslash B \in \mathcal{U}
              $
    \end{enumerate}
    it is called an \underbar{Ultrafilter} on $\mathbb{K}$.
\end{defs}
\begin{defs}[Free Ultrafilter]\spbcorr{}.\\
    Let $\mathbb{K}$ be a countable set and $\mathcal{U}$ an
    {\sl Ultrafilter} on $\mathbb{K}$. If in addition to the properties of
    an {\sl Ultrafilter} $\mathcal{U}$ satisfies
    \begin{enumerate}
        \item[5.] $
              \forall U , U \subset \mathbb{K}
              \quad \wedge \quad
              \exists n \in \N | \# U = n
              \quad \Longrightarrow \quad
              U \notin \mathcal{U}
              $
    \end{enumerate}
    then it is called a \underbar{Free Ultrafilter}.
\end{defs}
{\bf Notes:}
\begin{itemize}
    \item property $1.$ states that $\mathcal{U}$ is a
    {\it proper} {\sl Filter};
    \item property $2.$ is called the {\it finite intersection property};
    \item property $3.$ is called the {\it superset property};
    \item property $4.$ requires the {\it maximality} for $\mathcal{U}$.
\end{itemize}
Even if these definitions makes perfectly sense, it is by no means obvious
that a {\sl Free Ultrafilter} exists. It is thus easier to {\it assume} this
property:\\[2mm]
{\bf Axiom (Ultrafilter)}: {\it Let $\mathcal{F}$ be a filter on
$\mathbb{K}$; there is an ultrafilter $\mathcal{U}$ on $\mathbb{K}$
which contains}\footnote{It is possible to prove that
this axiom follows from the {\it axiom of choice}, i.e. Zorn's lemma.}
$\mathcal{F}$.

Now we have all the basic elements to build an extension of $\R$ in which
{\it infinitesimal} and {\it infinities} will be properly defined.
Of course we will loose something in the process, namely the new
$\it field$\footnote{Beware, we are not speaking of physical fields!}
will be non Archimedean.  To proceed toward this result we first set up some
notations.
\begin{nots}[Sequence]\spbcorr{}.\\
    We will indicate a sequence by
    $$
        \seqn{a}{n}
        \quad .
    $$
\end{nots}
Moreover, we will indicate with $\mathcal{U}$ a {\sl Free Ultrafilter}
on $\N$ and with $\R ^{\N}$ the set of all sequence of real numbers.
\begin{props}[Operations in $\R ^{\N}$]\spbcorr{}.\\
    Let us consider $\seqn{a}{i} \in \R ^{\N}$ and
    $\seqn{b}{j} \in \R ^{\N}$. Defining
    $$
        \matrix{
            \boxplus : & \R ^{\N} \times \R ^{\N}
                       & \longrightarrow
                       & \R ^{\N}
                       \cr
                       & \ttt{\seqn{a}{i} , \seqn{b}{j}}
                       & \longrightarrow
                       & \seqn{s}{k} \dfn \left\{ a _{k} + b _{k} \right\} _{k \in \N}
               }
    $$
    and
    $$
        \matrix{
            \boxtimes : & \R ^{\N} \times \R ^{\N}
                        & \longrightarrow
                        & \R ^{\N}
                        \cr
                        & \ttt{\seqn{a}{i} , \seqn{b}{j}}
                        & \longrightarrow
                        & \seqn{p}{k} \dfn \left\{ a _{k} \times b _{k} \right\} _{k \in \N}
               }
    $$
    $\ttt{\R ^{\N}, \boxplus , \boxtimes}$ becomes a {\it commutative ring} with
    {\it unit element} (namely the sequence
    $\seqn{u}{i} | \forall i \in \N \, , \: u _{i} = 1$) and
    {\it zero element} (namely the sequence
    $\seqn{z}{j} | \forall j \in \N \, , \: z _{j} = 0$).
Furthermore, this {\it ring} is not a {\it field} since it has zero divisors.
\end{props}

We thus see that this naive way of trying to construct an extension
of the real numbers fails; there are too many elements in $\R ^{\N}$.
A standard way to cut out some unwanted presence is to
construct an equivalence relation and consider the quotient of our structure
modulo this relation with, the hope that the quotient structure turns out
to have some richer structure. In our case we can define the following
relation:
\begin{defs}[Equivalence Modulo Ultrafilter]\spbcorr{}.\\
    Consider $\N$ and a {\sl Free Ultrafilter} $\mathcal{U}$ on $\N$.
    Moreover let us have $\seqn{a}{i} \in \R ^{\N}$ and
    $\seqn{b}{j} \in \R ^{\N}$. We will say that the two sequences
    are equal \underbar{Almost Everywhere} with respect to $\mathcal{U}$
    and we will write
    $$
        \seqn{a}{i} \equiv _{\mathcal{U}} \seqn{b}{j}
    $$
    if and only if
    $$
        \left\{ n \in \N \, | \, a _{n} = b _{n} \right\} \in \mathcal{U}
    \quad .
    $$
\end{defs}
It can be now proved, using the properties which define an {\sl Ultrafilter},
that
\begin{props}[Equivalence Relation Modulo Ultrafilter]\spbcorr{}.\\
    $\equiv _{\mathcal{U}}$ is an {\it equivalence relation} on $\R ^{\N}$.
\end{props}

Thus, the following definition makes completely sense now.
\begin{defs}[Nonstandard Reals]\spbcorr{}.\\
    We consider all the equivalence classes
    $\cl{\seqn{a}{i}}$ of elements of $\Rns$
    generated by the relation
    $\equiv _{\mathcal{U}}$
    $$
        \cl{\seqn{a}{i}}
        \dfn
        \left\{
            \seqn{x}{m} \in \R ^{\N} \, | \, \seqn{x}{m} \equiv _{\mathcal{U}}
\seqn{a}{i}
        \right\}
    $$
    and define $\Rns$ as
    $$
        \Rns
        \dfn
        \left\{
            \cl{\seqn{a}{i}} \, | \, \seqn{a}{i} \in \R ^{\N}
        \right\}
        =
        \ttt{\R ^{\N}} \slash \equiv _{\mathcal{U}}
    \quad .
    $$
    This is the \underbar{Ultrapower} generated by $\mathcal{U}$
    and its elements $\cl{\seqn{a}{i}} \in \Rns$ are called
    \underbar{Hyperreals} or \underbar{NonStandard Reals}.
\end{defs}
\begin{props}[Linearly Ordered Structure on $\Rns$]\spbcorr{}.\\
    Consider the structure
    $\ttt{\Rns , + _{\mathcal{U}} , \times _{\mathcal{U}} , < _{\mathcal{U}}}$
    defined as:
    $$
        \matrix{
            + _{\mathcal{U}} : & \Rns \times \Rns
                               & \longrightarrow
                               & \Rns
                               \cr
                               & \ttt{\cl{\seqn{a}{i}} , \cl{\seqn{b}{j}}}
                               & \longrightarrow
                               & \cl{\seqn{s}{k}} \dfn \cl{\seqn{a}{i} \boxplus \seqn{b}{j}}
               }
    $$
    $$
        \matrix{
            \times _{\mathcal{U}} : & \Rns \times \Rns
                                    & \longrightarrow
                                    & \Rns
                                    \cr
                                    & \ttt{\cl{\seqn{a}{i}} , \cl{\seqn{b}{j}}}
                                    & \longrightarrow
                                    & \cl{\seqn{p}{k}} \dfn \cl{\seqn{a}{i} \boxtimes \seqn{b}{j}}
               }
    $$
    $$
        \matrix{
            < _{\mathcal{U}} : & \cl{\seqn{a}{i} < _{\mathcal{U}} \cl{\seqn{b}{j}}}
                                    & \Longleftrightarrow
                                    & {n \in \N | a _{n} < b _{n}} \in \mathcal{U}
                                    \cr
            \leq _{\mathcal{U}} : & \cl{\seqn{a}{i} \leq _{\mathcal{U}} \cl{\seqn{b}{j}}}
                                  & \Longleftrightarrow
                                  & \cl{\seqn{a}{i} < _{\mathcal{U}} \cl{\seqn{b}{j}}}
                                    \vee
                                    \cl{\seqn{a}{i} = _{\mathcal{U}} \cl{\seqn{b}{j}}}
               }
    $$
    It is a {\it linearly ordered field}.
\end{props}

In such a way, one constructs a field with the same {\it operations}
as the reals $\R$. It would be very nice if, as the name $\Rns$ already
anticipate, we could also prove that it is an extension of the reals,
i.e. that it is possible to find an {\it order preserving isomorphism}
of $\R$ into $\Rns$. After the following definition:
\begin{defs}[Non Standard Isomorphism]\spbcorr{}.\\
    Let us consider $x \in \R$. We define the map $\ast$ as follows:
    $$
        \matrix{
            \ast : & \R
                   & \longrightarrow
                   & \Rns
                   \cr
                   & x
                   & \longrightarrow
                   & \cl{\seqn{\chi}{i}}
                   \cr
                   \cr
                   &
                   & \hfill \mathrm{where}
                   & \chi _{i} = x \quad , \forall i \in \N
                }
    \quad .
    $$
    The map $\ast$ is called the \underbar{NonStandard Isomorphism}.
\end{defs}
It is possible to prove that the map defined in the above definition
has the property we would like to require in order to embed
$\R$ in $\Rns{}\,$.
\begin{props}[Order Preserving Isomorphism]\spbcorr{}.\\
    The {\sl NonStandard Isomorphism} $\ast$
    is an Order Preserving Isomorphism of fields.
\end{props}

We described the construction of $\Rns$ because it is a standard
procedure we will use in what follow to define some other spaces,
which we are going to use in our application to the derivation
of the functional wave equation for {\sl String} Dynamics. Moreover,
with a few examples we will see in the next subsection that
this extension of the reals, $\R$, to the {\sl NonStandard Reals},
$\Rns$, is not just a mathematical exercise, but can be really useful
in treating in a more handy, but nevertheless perfectly rigorous way,
{\it infinitesimals} and {\it infinities}


\subsection{Some Examples}

A key step in the outlined direction can be done by observing that
$\Rns$ contains elements, which are bigger than any real number,
as well as positive elements, which are smaller than any real
positive number.
\begin{props}[An Infinite Number and an Infinitesimal Number]\spbcorr{}.\\
    The element $\omega = \clue{n}{n} \in \Rns$ is bigger than any
    real number.\\
    The element $\varepsilon = \clue{\frac{1}{n}}{n}$ is positive and
    smaller than any positive number.
\end{props}

Then, the following definitions will introduce some useful names for
some interesting subsets of $\Rns$.
\begin{defs}[Infinite, Infinitesimal, Limited,
             Appreciable Numbers]\spbcorr{}.\\
    We define four subsets of elements in $\Rns$.
    \begin{itemize}
        \item A number $\omega = \cl{\seqn{o}{k}} \in \Rns$
        is \underbar{Infinitely Large} or \underbar{Infinite}
        or \underbar{Unlimited} if and only if
        $$
            \forall {}^{\ast} \,x \,| \, \stand{{}^{\ast} x}
            \: Longrightarrow \:
            {}^{\ast} x < _{\mathcal{U}} \absv{\omega}
            \quad .
        $$
        \item A number $\varepsilon = \cl{\seqn{i}{k}} \in \Rns$
        is \underbar{Infinitely Small} or \underbar{Infinitesimal}
        if and only if
        $$
            \forall \, n \in N
            \: \Longrightarrow \:
            \absv{\varepsilon} < _{\mathcal{U}} {}^{\ast} \ttt{\frac{1}{n}}
            \quad .
        $$
        \item A number $\rho = \cl{\seqn{r}{k}} \in \Rns$
        is \underbar{Limited} or \underbar{Finite} if and only if it is
        not {\sl Infinite}.
        \item A number $\rho = \cl{\seqn{r}{k}} \in \Rns$
        is \underbar{Appreciable} if and only if it is
        not {\sl Infinite} nor {\sl Infinitesimal}.
    \end{itemize}
\end{defs}
Moreover, for easier writing, it turns useful to
introduce the following notation:
\begin{nots}[$\mho$, $@$, $\pounds$, $\infty$]\spbcorr{}.\\
    The following symbols are used to characterize {\sl Infinitesimal},
    {\sl Appreciable}, {\sl Limited} and {\sl Infinite} numbers:
    \begin{itemize}
        \item $\mho$ represents an {\sl Infinitesimal} number;
        \item $@$ represents an {\sl Appreciable} number;
        \item $\pounds$ represents a {\sl Limited} number;
        \item $\infty$ represents an {\sl Infinite} number.
    \end{itemize}
\end{nots}
Then, it is possible to use the tables \ref{B.nonstacomrul},
for example to perform operations, where in standard calculus,
longer limit procedures should be performed.

\begin{table}
\noncenteredtables
\vbox{
    \hbox to \textwidth {\hfil
\begintable
    $+$  \|    $\mho$    |     $@$     |  $\pounds$  |   $\infty$   \crthick
  $\mho$ \| $\mho$ | $@$ | $\pounds$ | $\infty$ \cr
  $@$    \| | $\pounds$ | $\pounds$ | $\infty$ \cr
$\pounds$\| | | $\pounds$ | $\infty$ \cr
$\infty$ \| | | | ?
\endtable
\hfil
\begintable
    $-$  \|    $\mho$    |     $@$     |  $\pounds$  |   $\infty$   \crthick
  $\mho$ \| $\mho$ | $@$ | $\pounds$ | $\infty$ \cr
  $@$    \| | $\pounds$ | $\pounds$ | $\infty$ \cr
$\pounds$\| | | $\pounds$ | $\infty$ \cr
$\infty$ \| | | | ?
\endtable
\hfil
         }
    \hbox to \textwidth {\hfil(a) \hskip 1 cm \hfil(b)\hfil}
    }
\vskip 1 cm
\vbox{
    \hbox to \textwidth{\hfil
\begintable
$\times$ \|    $\mho$    |     $@$     |  $\pounds$  |   $\infty$   \crthick
  $\mho$ \| $\mho$ | $\mho$ | $\mho$ | ? \cr
  $@$    \| | $@$ | $\pounds$ | $\infty$ \cr
$\pounds$\| | | $\pounds$ | ? \cr
$\infty$ \| | | | $\infty$
\endtable
\hfil
\begintable
  $/$    \|    $\mho$    |     $@$     |  $\pounds$  |   $\infty$   \crthick
  $\mho$ \| ? | $\infty$ | ? | $\infty$ \cr
  $@$    \| $\mho$ | $@$ | $\pounds$ | $\infty$ \cr
$\pounds$\| ? | ? | ? | $\infty$ \cr
$\infty$ \| $\mho $| $\mho$ | $\mho$ | ?
\endtable
\hfil
         }
    \hbox to \textwidth {\hfil(c) \hskip 1 cm \hfil(d)\hfil}
    }
\centeredtables
\caption{Computation tables with Infinity, Infinitesimal, Limited and
Appreciable numbers: (a) sum;
(b) subtraction , $\mathrm{row} - \mathrm{column}$;
({}c{}) product;
(d) quotient, $\mathrm{row} / \mathrm{column}$.}
\label{B.nonstacomrul}
\end{table}

This example explains, although in a very simple and
{\it poor} way, why\footnote{If not {\it how}; indeed the subject
is much more deep and vast, and the account we gave here is
just to explain the flavor of what we think is a really interesting
branch of mathematics.} {\sl NonStandard Analysis} can be helpful
in tackling with the problem of {\sl Infinitesimals} and {\sl Infinities}
and this is the main reason that leads us, in this thesis work,
to find a {\it NonStandard procedure} to motivate on a rigorous way
the results obtained with the formal manipulations of chapter
\ref{3.strfunqua}.

\section{Ultra Euclidean Space}
\label{B.ulteucspasec}

It is possible to perform the {\sl UltraPower construction} in the
previous section
in a slightly more involved case. We remember that in all our discussion
$\mathcal{U}$ denotes a {\sl Free Ultrafilter}
over the natural numbers $\N$.

We then proceed along the same line that we have employed in the previous
section, namely we define an equivalence relation,
denoted with $\simeq _{\mathcal{U}}$, thanks to the
{\sl Free Ultrafilter} $\mathcal{U}$: this time the space we consider
is the following one
$$
    \prod _{n \in \N} \R ^{n}
$$
Thus, sequences are now sequences of vectors in spaces of increasing
dimension. In particular
\begin{nots}[Element of $\prod _{n \in \N} \R ^{n}$]\spbcorr{}.\\
    We will denote an element in the infinite sequence of products
    of $\R ^{n}$ as n ranges in $1 , \dots , \infty$ with German
    letters
    $$
    \euf{a}
    =
    \left\{
        \bs{a} ^{(n)}
    \right\} _{n \in \N}
    \in
    \prod _{n \in \N} \R ^{n}
    \quad .
    $$
\end{nots}
Then, the equivalence relation modulo the {\sl Free Ultrafilter}
identifies elements of $\prod _{n \in \N} \R ^{n}$, i.e. sequences
whose component vectors agree for a set of spaces indexed by an
element which is in the {\sl Free Ultrafilter}.
\begin{defs}[Equivalence Relation on
             $\prod _{n \in \N} \R ^{n}$]\spbcorr{}.\\
\label{B.RtNrel}
    On $\prod _{n \in \N} \R ^{n}$ the {\sl Free Ultrafilter}
    $\mathcal{U}$ defines the following relation
    $$
        \euf{a} \simeq _{\mathcal{U}} \euf{b}
        \quad
        \Longleftrightarrow
        \quad
        \left\{
            n \in N
            \, | \,
            \bs{a} ^{(n)}
            =
            \bs{b} ^{(n)}
        \right\}
        \in
        \mathcal{U}
        \quad ,
    $$
    where we assume
    \bea
        \euf{a}
        & = &
        \left\{
            \left( \udin{a}{1}{1} \right)
            ,
            \left( \udin{a}{1}{2} , \udin{a}{2}{2} \right)
            ,
            \dots
            ,
            \left(
                \udin{a}{1}{i}
                ,
                \udin{a}{2}{i}
                ,
                \dots
                ,
                \udin{a}{i}{i}
            \right)
            ,
            \dots
        \right\}
        \nonumber \\
        \euf{b}
        & = &
        \left\{
            \left( \udin{b}{1}{1} \right)
            ,
            \left( \udin{b}{1}{2} , \udin{a}{2}{2} \right)
            ,
            \dots
            ,
            \left(
                \udin{b}{1}{i}
                ,
                \udin{b}{2}{i}
                ,
                \dots
                ,
                \udin{b}{i}{i}
            \right)
            ,
            \dots
        \right\}
    \quad .
    \eea
\end{defs}
This relation is of course an equivalence relation so that
it is meaningful to define a partition of $\R ^{\N}$ into
disjoint equivalence classes of elements; this leads directly to
the following
\begin{defs}[Ultra Euclidean Space]\spbcorr{}.
\label{B.ulteucspa}\\
    We call \underbar{Ultra Euclidean Space} the quotient
    \beq
        \uespace
        =
        \left( \prod _{n \in \N} \R ^{n} \right)
        /
        \simeq _{\mathcal{U}}
    \quad .
    \label{B.RtNquo}
    \eeq
\end{defs}
We remark that  the basic idea of the above procedure is simply a
{\it chinese box construction} with some identifications. Indeed,
 we stress again that a generic element
in $\left( \prod _{n \in \N} \R ^{n} \right)$ is simply an
{\it infinite sequence of finite sequences of increasing length},
\beq
    \euf{a}
    =
    \left\{
        \left( \udin{a}{1}{1} \right)
        ,
        \left( \udin{a}{1}{2} , \udin{a}{2}{2} \right)
        ,
        \dots
        ,
        \left(
            \udin{a}{1}{i}
            ,
            \udin{a}{2}{i}
            ,
            \dots
            ,
            \udin{a}{i}{i}
        \right)
        ,
        \dots
    \right\}
    \in
    \prod _{n \in \N} \R ^{n}
\eeq
and
\begin{equation}
    \bs{a} ^{(n)}
    =
    \left( \udin{a}{1}{n} , \udin{a}{2}{n} , \dots , \udin{a}{n}{n} \right)
    \in
    \R ^{n}
\quad .
\end{equation}
An element of the {\sl Ultra Euclidean Space} is thus an equivalence class
in the relation defined by the {\sl Ultrafilter} $\mathcal{U}$.
\begin{nots}[Element of the Ultra Euclidean Space $\uespace$]\spbcorr{}.\\
    We will denote an element of the {\sl Ultra Euclidean Space} as
    $$
        \cl{\euf{a}}
        =
        \clue{\bs{a} ^{(n)}}{n}
        \in
        \uespace
        \quad .
    $$
\end{nots}
Some properties of the {\sl Ultra Euclidean Space} $\uespace$ can be
easily proved. The proofs can be an useful exercise in applying
simple properties of {\sl NonStandard Analysis}.
\begin{props}[Linearity of $\uespace$]\spbcorr{}.\\
    $\uespace$ is a \underbar{Linear Space} over $\Rns$.
\end{props}

Moreover,
\begin{props}[Euclideanity of $\uespace$]\spbcorr{}.\\
    $\uespace$ is an \underbar{Euclidean Space} with scalar
    product defined by
    \bea
        \scal{\euf{a}}
             {\euf{b}}
             {\uespace}
        & = &
        \scal{\clue{a ^{(n)}}{n}}
             {\clue{b ^{(n)}}{n}}
             {\uespace}
        \nonumber \\
        & \dfn &
        \clue{
              \scal{\bs{a} ^{(n)}}
                   {\bs{b} ^{(n)}}
                   {\R ^{n}}
             }
             {n}
        =
        \clue{
              \sum _{i} ^{1,n}
                  \udin{a}{i}{n}
                  \udin{b}{i}{n}
             }
             {n}
    \quad .
    \eea
\end{props}
Let us note that, in the definition of the scalar product,
\beq
    \scal{\bs{a} ^{(n)}}
         {\bs{b} ^{(n)}}
         {\R ^{n}}
\eeq
is the usual scalar product in $\R ^{n}$, so that actually
$\left\{ \scal{\bs{a} ^{(n)}}{\bs{b} ^{(n)}}{\R ^{n}} \right\} _{n \in \N}$
is a sequence, i.e. is in $\R ^{\N}$; then the
{\sl Ultrafilter Equivalence Relation} sets its equivalence class
in $\Rns$.

%% file: appC.tex
\pageheader{}{Nonstandard Stochastic Processes.}{}
\chapter{Nonstandard Stochastic Processes}
\label{C.nonstopro}

\section{Functional Spaces}

We quote the following usual definitions and notations:
\begin{nots}[Schwartz Space]\spbcorr{}.\\
    The Schwartz space over $\R ^{d}$,
    is denoted by
    $$
         \mathscr{S} \ttt{\R ^{d}}
    \quad .
    $$
\end{nots}
\begin{nots}[\protect $\mathcal{L} ^{2}$ Space]\spbcorr{}.\\
    The $\mathcal{L} ^{2}$ Space over $\R ^{d}$,
    is denoted by
    $$
         \mathcal{L} ^{2} \ttt{\R ^{d}}
    \quad .
    $$
\end{nots}
\begin{nots}[Complete Orthonormal Set]\spbcorr{}.\\
    A complete orthonormal set, belonging to
    $
         \mathscr{S} \ttt{\R ^{d}}
    $,
    which is complete orthonormal in
    $
         {\cal L} ^{2} \ttt{\R ^{d}}
    $,
    is denoted by
    $$
        \left\{ e _{n} \right\} _{n \in \N}
        \quad .
    $$
\label{C.ortset}
\end{nots}
From the results of section \ref{B.ulteucspasec}
we can now proceed to some central
definitions: the principal space for the procedure of section
\ref{10.nonstafunquasec} and
some other tool that are used therein.
\begin{defs}[Kawabata Space]\spbcorr{}.\\
    Consider
    $
        \left\{ e _{n} \right\} _{n \in \N}
    $
    as in {\rm notation \ref{C.ortset}}.
    The \underbar{Kawabata Space} is a set of $\Rns$
    valued functions and is defined as
    \bea
    & & \esci \esci
        \kawspace
        \dfn
        \left\{
            \bs{\hphi}
            \left | \right .
            \bs{\hphi} = \clue{\phi _{n}}{n}
        \right .
    \nonumber \\
    & & \qquad \qquad \qquad
        \left .
                       = \clue{\sum _{i} ^{1,n} \udin{a}{i}{n} e _{i}}{n}
            \, , \:
            \forall
            \euf{a} = \clue{\bs{a} ^{(n)}}{n} \in \uespace
        \right\}
    \qquad .
    \eea
\end{defs}
\begin{nots}[$n$th-approximation of an
             Elment of Kawabata Space]\spbcorr{}.\\
    We will call
    $\phi _{n}$,
    in the definition above,
    an \underbar{Approximation}\footnote{Or more specifically
    the \underbar{$n$th-Approximation}.} of $\bs{\hphi}$.
\end{nots}
Some explanations are worth to clarify the construction:
\begin{enumerate}
    \item it is important to remember that
    $e _{i} = e _{i} \ttt{-}$
    is a function on $\R ^{d}$, which assigns to each
    $\bs{v} \in \R ^{d}$ the number
    $e _{i} \ttt{\bs{v}} \in \R \,$;
    \item moreover for each $n$, $ \bs{a} ^{(n)}$ is a vector in $\R ^{n}$;
    we then use its components $\udin{a}{j}{n}$ as coefficients
    to write an $n$-th order expansion $\phi _{n} \ttt{-}$,
    in terms of the first $n$ elements of the base
    $\left\{ e _{j} \ttt{-} \right\} _{j \in \N}$;
    \item thus $\forall j \in \N$ we obtain an approximation of
    $\phi \ttt{-}$; suppose now to apply $\forall j \in \N$
    the approximation $\phi _{j} \ttt{-}$ to a vector
    $\bs{v} \in \R ^{d}$; this results is the sequence of real
    numbers
    $\left\{ \phi _{j} \ttt{\bs{v}} \right\} _{j \in \N} \in \R ^{\N}$,
    which can be uniquely associated throught the equivalence relation
    defined by the {\sl Ultrafilter} $\mathcal{U}$ to the {\sl NonStandard} real
    $\clue{\phi _{j} \ttt{\bs{v}}}{j} \in \Rns$: this is
    actually the value of $\bs{\hphi} \ttt{\bs{v}}$.
\end{enumerate}
We now define an appropriate structure on $\kawspace$.
\begin{defs}[Scalar Product on Kawabata Space]\spbcorr{}.\\
    Consider $\bs{\hphi}, \bs{\hpsi} \in \kawspace$. The
    scalar product of these two elements is defined as follows:
    \begin{eqnarray}
        \scal{\bs{\hphi}}{\bs{\hpsi}}{\kawspace}
        & = &
        \scal{\clue{\sum _{i} ^{1,n} \udin{a}{i}{n} e _{i}}{n}}
             {\clue{\sum _{i} ^{1,n} \udin{a}{i}{n} e _{i}}{n}}
             {\kawspace}
        \label{C.homone} \\
        & \dfn &
        \clue{
              \sum _{i,j} ^{1,n}
                  \udin{a}{i}{n} \udin{b}{j}{n}
                  \scal{e _{i}}
                       {e _{j}}
                       {\mathscr{S} \ttt{\R ^{d}}}
             }
             {n}
        \nonumber \\
        & = &
        \clue{
              \sum _{i,j} ^{1,n}
                  \udin{a}{i}{n} \udin{b}{i}{n}
             }
             {n}
        \nonumber \\
        & = &
        \scal{\clue{\bs{a} ^{(n)}}{n}}
             {\clue{\bs{b} ^{(n)}}{n}}
             {\uespace}
        \nonumber \\
        & = &
        \scal{\euf{a}}{\euf{b}}{\uespace}
    \label{C.homtwo}
    \quad ,
    \end{eqnarray}
    where of course $\euf{a} = \clue{\bs{a} ^{(n)}}{n}$
    and $\euf{b} = \clue{\bs{b} ^{(n)}}{n}$.
\end{defs}
Then two results follows.
\begin{props}[Homeomorphism between $\kawspace$ and $\uespace$]\spbcorr{}.\\
    $\ttt{\kawspace , \scal{-}{-}{\kawspace}}$
    is homeomorphic to
    $\ttt{\uespace , \scal{-}{-}{\uespace}}$.
\end{props}
\begin{proof}
This result follows immediatly comparing the left hand side of equation
(\ref{C.homone}) with (\ref{C.homtwo}):
$$
    \scal{\bs{\hphi}}{\bs{\hpsi}}{\kawspace}
    =
    \scal{\euf{a}}{\euf{b}}{\uespace}
    \quad .
$$
\end{proof}
\begin{props}[Local Properties of Elements in Kawabata Space]\spbcorr{}.\\
    $\forall \, \bs{\hphi} \in \kawspace \; , \: \bs{\hphi}$ is {\it locally
    integrable} and {\it locally differentiable}.
\end{props}
In particular we can define
\begin{defs}[Integral of an Element of Kawabata Space]\spbcorr{}.\\
    The integral of $\bs{\hphi}$ is defined ad
    $$
        \int _{\R ^{d}} d ^{d} x \, \bs{\hphi} \ttt{x}
        \dfn
        \clue{
              \sum _{i} ^{1,n}
                  \udin{a}{i}{n}
                  \int _{\R ^{d}}
                      d ^{d} x \,
                      e _{i} \ttt{x}
             }{n}
        \in \Rns
        \quad .
    $$
\end{defs}
Moreover we also have
\begin{defs}[Directional Derivative of an
             Element of Kawabata Space]\spbcorr{}.\\
    The directional derivative of $\bs{\hphi}$ in the direction of $x$ is
    $$
        \ttt{\partial _{j} \bs{\hphi}}
        \dfn
        \clue{
              \sum _{i} ^{1,n}
                  \udin{a}{i}{n}
                  \partial _{j} e _{i} \ttt{x}
             }
             {n}
        \in
        \Rns
        \quad .
    $$
\end{defs}
\begin{defs}[Functional on Kawabata Space]\spbcorr{}.\\
    Let
    $\bs{\hpsi}
     =
     \clue{\sum _{i} ^{1,n} \udin{a}{i}{n} e _{i}}{n}
     \in
     \kawspace
    $.
    Consider $\forall n \in \N$ the $n$-th approximation of $\bs{\hpsi}$,
    and the function
    $$
    \matrix{
        F ^{(n)} &
        : &
        \R ^{n} &
        \longrightarrow &
        \R \cr
        &
        &
        \bs{a} ^{(n)} &
        \longrightarrow &
        F ^{(n)} \ttt{a ^{(n)}}
           }
    \quad .
    $$
    Then the map
    $$
    \matrix{
        \shF &
        : &
        \kawspace &
        \longrightarrow &
        \Rns \cr
        &
        &
        \hpsi &
        \longrightarrow &
        \clue{F ^{(n)} \ttt{\bs{a} ^{(n)}}}{n}
           }
    $$
    is a \underbar{Functional} on $\kawspace$.
\end{defs}
\begin{nots}[Functional on Kawabata Space]\spbcorr{}.\\
    We will write a functional on $\kawspace$ with its argument
    enclosed in square brackets, i.e. as $\shF \left[ \hpsi \right]$.
\end{nots}
This notation has the following idea behind: since usually it is
possible to confuse themselves with notations, we use no {\it oversymbols}
for elements in $\Rns$. Then, since elements on $\kawspace$ have a
hat ``~$\hat{\ }$~'' over them and functionals on $\kawspace$ have a
check ``~$\check{\ }$~'' which is something like an {\it inverse hat},
we see that acting with a functional on $\kawspace$ on an element of
$\kawspace$ the check and the hat annihilates, so we can immediatly
understand that we just end in $\Rns$ with a {\sl NonStandard Real Number}.

To see that this procedure is consistent, we note that:
\begin{enumerate}
    \item $\bs{\hpsi}$ is completely defined by the coefficients
    of its expansion in terms of the complete orthonormal set
    $e _{i} \ttt{-}$; in terms of the definition of
    $\uespace$ we can thus consider the associated element
    $\euf{a} = \clue{\bs{a} ^{(n)}}{n}$;
    \item since $F ^{(n)}$ values are in $\R$, the sequence
    $\left\{ F ^{(n)} \ttt{\bs{a} ^{(n)}} \right\} _{n \in \N}$
    is a sequence of real numbers and we can apply to it the usual
    identification procedure {\it modulo} the {\sl Ultrafilter}
    $\mathcal{F}$, so
    that the result $\clue{F ^{(n)} \ttt{\bs{a} ^{(n)}}}{n}$
    is in $\Rns$;
    \item since a function $f$, even if we think at the usual
    situation, in which no {\sl NonStandard} elements appear, is completely
    determined by its coefficients in an expansion in terms of a
    suitably choosen functional basis, it is sensible that a functional
    of $f$ is completely defined in terms of its action on the
    coefficients that determine $f$.
\end{enumerate}
Keeping in mind these observations, we can easily understand the
forms of the {\sl NonStandard} definitions of {\it functional
derivative} and
{\it functional integral}. We stress again that an element of $\kawspace$
is a function from $\R ^{d}$ to $\R$, which is clearly indicated
by the notation of the following definitions.
\begin{defs}[Functional Derivative of
             a Functional on Kawabata Space]\spbcorr{}.
    \label{C.nonfunderdef}\\
    Let
    $\shF \left [ \bs{\hpsi} \right ]
     =
     \clue{F ^{(n)} \ttt{\bs{x} ^{(n)}}}{n}
    $
    be a functional of
    $\bs{\hpsi} \ttt{-}
     =
     \clue{\sum _{i} ^{1,n} \udin{x}{i}{n} e _{i} \ttt{-}}{n}
    $
    on $\kawspace$. The functional derivative of
    $\shF \left [ \hat{\ } \right ]$ with respect to
    $\bs{\hpsi} \ttt{-}$
    is
    $$
        \frac{\delta \shF \left[ \bs{\hpsi} \right]}
             {\delta \bs{\hpsi} \ttt{-}}
        =
        \clue{
              \sum _{j} ^{1,n}
                  \frac{\partial F \ttt{\bs{x} ^{(n)}}}
                       {\partial \udin{x}{j}{n}}
                  e _{j} \ttt{-}
             }{n}
        \in \kawspace
    \quad .
    $$
\end{defs}
Note that as is natural, if $x \in \R ^{d}$, then
\begin{equation}
    \frac{\delta \shF \left[ \bs{\hpsi} \right]}
         {\delta \bs{\hpsi} \ttt{x}}
    \in
    \Rns
\quad .
\end{equation}
Moreover we have
\begin{defs}[Functional Integral of
             a Functional on Kawabata Space]\spbcorr{}.
    \label{C.nonfunintdef}\\
    Let
    $\shF \left [ \bs{\hpsi} \right ]
     \dfn
     \clue{F ^{(n)} \ttt{\bs{x} ^{(n)}}}{n}
    $
    be a functional of
    $\bs{\hpsi} \ttt{-}
     =
     \clue{\sum _{i} ^{1,n} \udin{x}{i}{n} e _{i} \ttt{-}}{n}
    $
    on $\kawspace$. The functional integral of
    $\shF \left [ \hat{\ } \right ]$ with respect to
    $\bs{\hpsi} \ttt{-}$
    is
    $$
    \int _{\kawspace}
        \shF \left [ \bs{\hpsi} \right ]
    \dfn
    \clue{\int F ^{(n)} \ttt{\bs{x} ^{(n)}}}{n}
    \delta \bs{\hpsi}
    \in
    \R
    \quad .
    $$
\end{defs}
We end this section with the definition of operators on $\kawspace$
and some remarks.
\begin{defs}[Operator on Kawabata Space]\spbcorr{}.\\
    Consider
    $
    \left\{ \bs{f} ^{(n)} \ttt{-} \right\} _{n \in \N}
    $
    a sequence of functions
    $$
    \matrix{
        \bs{f} ^{(n)} \ttt{-} &
        : &
        \R ^{n} &
        \longrightarrow &
        \R ^{n} \cr
        &
        &
        x &
        \longrightarrow &
        \bs{f} ^{(n)} \ttt{x}
           }
    \quad .
    $$
    An operator on $\kawspace$ is an application $\F$ such that
    $$
    \matrix{
        \F \ttt{-} &
        : &
        \kawspace &
        \longrightarrow &
        \kawspace \cr
        &
        &
        \bs{\hpsi} = \clue{\sum _{i} ^{1,n} \udin{x}{i}{n} e _{i}}{n} &
        \longrightarrow &
        \F \bs{\hpsi} = \clue{
                              \sum _{j} ^{1,n}
                                  f ^{(n)} _{j} \ttt{\bs{x} ^{(n)}}
                                  e _{j}
                             }{n} &
           }
    \quad .
    $$
\end{defs}
This definition recalls the same ideas of the previous ones. Moreover
we observe that all the given definitions can be generalized adding
some parametric dependence on one or more continuos or discrete
parameters. Then we note that
\begin{props}[Structure of Operators in Kawabata Space]\spbcorr{}.
\label{C.nonstaopeperfunpro}\\
    Let
    $$
    \mathcal{A} \left [ \hat{\ } \right ]
    =
    \clue{
        \sum _{j} ^{1,n}
            \udin{a}{j}{n} \ttt{-}
            e _{j}
         }{n}
    $$
    be an element in
    the set of all operators on $\kawspace$; consider
    now a functional
    $$
    \shF \left [ \bs{\hpsi} \right ]
    =
    \clue{F ^{(n)} \ttt{\bs{x} ^{(n)}}}{n}
    $$
    on $\kawspace$\footnote{We remember that both, an operator
    as well as a functional, act on elements of $\kawspace$; the
    first associates to this elements another element of $\kawspace$
    whereas the secon associates a {\sl NonStandard} Real.}.
    We will give the following meaning to the composition between
    an operator and a functional: by definition it is again an operator
    $\tilde{\mathcal{F}}$ defined as
    \beq
    \tilde{\mathcal{F}} \left [ \hat{\ } \right ]
    \dfn
    \mathcal{A} \left [ \hat{\ } \right ]
    \sh{F} \left [ \hat{\ } \right ]
    \dfn
    \clue{
          \sum _{j} ^{1,n}
              \udin{a}{j}{n} \ttt{-}
              F ^{(n)} \ttt{-}
              e _{j}
         }{n}
    \quad .
    \label{C.nonfunopepro}
    \eeq
\end{props}
That this is a meaningful definition can be seen since
    $$
    \tilde{\mathcal{F}} \left [ \bs{\hpsi} \right ]
    =
    \clue{
          \sum _{j} ^{1,n}
              \udin{a}{j}{n} \ttt{\bs{x} ^{(n)}}
              F ^{(n)} \ttt{\bs{x} ^{(n)}}
              e _{j} \ttt{-}
         }{n}
    \in \kawspace
    \quad .
    $$
is again an elemen of $\kawspace$.

\section{Diffusion Process on \protect $\kawspace$}
\label{C.stoprokawspasec}

The procedure to properly define a diffusion process on $\kawspace$
is similar to the one used in the various definitions above. We pick
up a family of ordinary diffusion processes
\beq
    \left\{
        \bs{X} ^{(n)} \ttt{t}
    \right\} _{n} ^{1 , \infty}
\label{C.famordstopro}
\eeq
and construct with them an element of $\uespace$ for each $t$,
i.e. a diffusion process in $\uespace$. This is a sequence of
diffusion processes in euclidean spaces of increasing dimension.
More explicitly this means that we consider
$$
    \left\{
        \bs{X} ^{(1)} \ttt{t}
        ,
        \bs{X} ^{(2)} \ttt{t}
        ,
        \dots
        ,
        \bs{X} ^{(i)} \ttt{t}
        ,
        \dots
    \right\}
$$
where each $\bs{X} ^{(j)} \ttt{t}$ is a random variable
in $\R ^{j}$. Once we have this sequence, by means of the usual
equivalence relation defined thanks to the {\sl Ultrafilter}
$\mathcal{U}$, we
obtain the definition of the stochastic process in $\kawspace$.
\begin{defs}[Stochastic Process on Kawabata Space]\spbcorr{}.
    \label{C.nonstoprodef}\\
    Let
    $
        \left\{
            \bs{X} ^{(n)} \ttt{t}
        \right\} _{n} ^{1 , \infty}
    $
    be a family of stochastic processes, with
    $
        \bs{X} ^{(n)}
    $
    an $n$-dimensional random variable. Then $\bs{\hPsi} \ttt{t}$
    defined as
    $$
        \bs{\hPsi} \ttt{t}
        \dfn
        \clue{
              \sum _{j} ^{1,n}
                  \udin{X}{j}{n} \ttt{t}
                  e _{j}
             }{n}
        \in
        \kawspace
    $$
    is a \underbar{Stochastic Process} on $\kawspace$ or a
    \underbar{Nonstandard Stochastic Process}.
\end{defs}
We note that of course the {\sl Stochastic Process on Kawabata Space}
is a one parameter $\Rns$ valued family of functions on $\R ^{d}$.

Now we would be really glad if we were able to obtain a complete
generalization of the properties of stochastic processes in $\R ^{d}$
to Kawabata space. In particulare we need probability distributions
for our processes and a notion of volume is the first step toward
this goal. The following definition is thus natural, in the light
of all the work done so far. Indeed we need a definiton of volume
element which allows the standard definition of {\it probability}
in each component $\R ^{i}$ in which the $i$-dimensional
random process $\bs{X} ^{(i)}$ is living in.
Then it is natural to define the following sequence
\beq
    \left\{
        d ^{n} x ^{(n)}
    \right\} _{n \in \N}
    =
    \left\{
        d \udin{x}{1}{1}
        ,
        d \udin{x}{2}{1}
        d \udin{x}{2}{2}
        ,
        \dots
        ,
        d \udin{x}{i}{1}
        d \udin{x}{i}{2}
        \dots
        d \udin{x}{i}{i}
        ,
        \dots
    \right\}
\eeq
and use the {\sl Ultrafilter} $F$ to obtain the usual identification.
\begin{defs}[Elementary Volume Element on Kawabata Space]\spbcorr{}.
\label{C.nonelevoleledef}\\
    The elementary {\it volume element} on $\kawspace$ is
    $$
        \delta \bs{\hpsi}
        \dfn
        \clue{d ^{n} x ^{(n)}}{n}
        \quad .
    $$
\end{defs}
Then we can define the {\it NonStandard Probability Distribution}
associated with the {\sl NonStandard Stochastic Process}.
To motivate the following definition, we
stress again that the starting point is always the family of stochastic
processes \ref{C.famordstopro},
the first of them living in $\R$, the second in
$\R ^{2}$, the \dots in \dots, the $i$-th in $\R ^{i}$ and so on.
Each of this processes has an associated probability distribution
$p ^{(j)} \ttt{\bs{x} ^{(j)} , t}$, defined on $\R ^{j}$, from
which it is possible to compute the probability $P ^{(j)} \ttt{\bs{x} ^{(j)}}$
of finding a value of the random
variable $\bs{X} ^{(j)}$ in a small volume element $d ^{j} x ^{(j)}$
around $\bs{x} ^{(j)}$ at time $t$:
\beq
    P ^{(j)} = \prob{\bs{X} ^{(j)} \ttt{t} \in d ^{j} x ^{(j)}}
             = p ^{(j)} \ttt{\bs{x} ^{(j)} , t}
               d \udin{x}{1}{j} \cdot \cdot \cdot d \udin{x}{j}{j}
    \quad .
\eeq
We can generalize this definition to a {\it NonStandard--Real
Probability} on $\kawspace$ as follows:
\begin{defs}[Non Standard Probability]\spbcorr{}.
\label{C.nonprodef}\\
    Let $\bs{\hPsi}$ be a {\sl NonStandard Stochastic Process}
    on $\kawspace$
    with $i$-th approximation $\sum _{j} ^{1,i} \udin{\bs{X}}{j}{i} e _{i}$
    having probability $P ^{(i)} \ttt{\bs{x} ^{(i)}}$ of being
    found in a volume element
    $d ^{i} x ^{(i)}$ around $\bs{x} ^{(i)}$.
    The \underbar{NonStandard Probability} of finding the process
    $\bs{\hPsi}$ in a {\sl NonStandard Volume Element} $\delta \bs{\hpsi}$
    around
    $\bs{\hpsi} = \clue{\sum _{l} ^{1,k} x _{l} ^{(k)} e _{l}}{k}$
    is defined as
    \bea
        \nsprob{\bs{\hpsi} \ttt{t} \in \delta \bs{\hpsi}}
        & = &
        \nsprob{
                \clue{
                      \sum _{i} ^{1,n}
                          \udin{\bs{X}}{i}{n} \ttt{t}
                          e _{i}
                     }{n}
                \in
                \clue{d ^{n} x ^{(n)}}{n}
               }
        \nonumber \\
        & \dfn &
        \clue{P ^{(n)}}{n}
        \label{C.nonprodefequ}
        \quad .
    \eea
\end{defs}
Then we can prove the following result.
\begin{props}[Non Standard Probability Distribution]\spbcorr{}.\\
    The {\sl NonStandard Probability Distribution} of finding the
    {\sl NonStandard Stochastic Process} $\bs{\hPsi}$ in a
    {\sl NonStandard Volume Element}
    $\delta \bs{\hpsi}$ around $\bs{\hpsi}$ can be expressed as
    $$
        \nsprob{\bs{\hPsi} \ttt{t} \in \delta \bs{\hpsi}}
        =
        \sh{P} \left[ \bs{\hpsi} , t \right]
        \delta \bs{\hpsi}
    $$
    with
    $$
        \sh{P} \left[ \bs{\hpsi} , t \right]
        =
        \clue{p ^{(n)} \ttt{x ^{(n)} , t}}{n}
    \quad .
    $$
\end{props}
\begin{proof}
The result follows from defintions
\ref{C.nonelevoleledef}, \ref{C.nonprodef}
and from the definition of product
between numbers of $\Rns$. Indeed we have starting from
equation (\ref{C.nonprodefequ}) in definition \ref{C.nonprodef}
\bea
    \nsprob{\bs{\hPsi} \ttt{t} \in \delta \bs{\hpsi}}
    & = &
    \clue{P ^{(n)}}{n}
    \nonumber \\
    & = &
    \clue{
          \prob{X ^{(j)} \ttt{t} \in d ^{j} x ^{(j)}}
         }{j}
    \nonumber \\
    & = &
    \clue{
          p ^{(j)} \ttt{x ^{(j)} , t}
          d ^{j} x ^{(j)}
         }{j}
    \nonumber \\
    & = &
    \clue{
          p ^{(j)} \ttt{x ^{(j)} , t}
         }{j}
    \clue{
          d ^{j} x ^{(j)}
         }{j}
    \nonumber \\
    & = &
    \sh{P} \left [ \bs{\hpsi} , t \right ]
    \delta \bs{\hpsi}
    \quad .
    \nonumber
\eea
\end{proof}

Note that the {\sl NonStandard Probability Distribution}
$\sh{P} \left [ \bs{\hpsi} , t \right ]$ is a continuos family
of functionals on $\kawspace$.
Now each of the stochastic processes which is used to define the
$n$-approximation of the {\sl NonStandard Stochastic Process}
$\bs{\hPsi}$, being a process in $\R ^{n}$, has a probability density,
which we called $p ^{(n)} \ttt{\bs{x} ^{(n)} , t}$, that
satisfies the Fokker-Plank equation,
\beq
    \frac{\partial p ^{(n)}}{\partial t}
    =
    -
    \sum _{i} ^{1,n}
        \frac{\partial}{\partial x ^{(n)} _{i}}
        \left [
            \udin{a}{i}{n} p ^{(n)}
        \right]
    +
    \beta
    \sum _{i} ^{1,n}
        \frac{\partial ^{2} p ^{(n)}}
             {\partial \udin{x}{i}{n} {}^{2}}
\eeq
or equivalently the corresponding Langevin equation
\beq
    d \bs{X} ^{(n)} \ttt{t}
    =
    \bs{a} ^{(n)} \ttt{\bs{X} ^{(n)} \ttt{ t } , t} dt
    +
    \bs{B} ^{(n)} \ttt{dt}
\eeq
$\bs{a} ^{(n)} \ttt{\bs{x} ^{(n)} , t}$ being the {\it drift function}.
Define now the following family of operators on $\kawspace$.
\begin{defs}[Forward Drift Operator]\spbcorr{}.\\
    Let us have as usual
    $\bs{\hpsi} = \clue{\sum _{i} ^{1,n} \udin{x}{i}{n} e _{i}}{n} $.
    Then the \underbar{Forward Drift Operator} is
    $$
    \matrix{
        \bs{\A} ^{+} _{t} &
        : &
        \kawspace &
        \longrightarrow &
        \kawspace \cr
        &
        &
        \bs{\hpsi} &
        \longrightarrow &
        \bs{\A} ^{+} _{t} \bs{\hpsi}
        =
        \clue{
              \sum _{i} ^{1,n}
                  \udin{a}{i}{n}
                  \ttt{x ^{(n)} , t}
                  e _{i} \ttt{-}
             }{n}
           }
        \quad ,
    $$
    where we observe that
    $$
        \left[ \bs{\A} ^{+} _{t} \bs{\hpsi} \right ] \ttt{x}
        =
        \clue{
              \sum _{i} ^{1,n}
                  \udin{a}{i}{n}
                  \ttt{x ^{(n)} , t}
                  e _{i} \ttt{x}
             }{n}
        \in
        \Rns
    \quad .
    $$
\end{defs}
In the same way the {\it Backward Drift Operator} $\bs{\A} ^{-} _{t}$
can be defined and it turns out to be related to the {\sl Forward} one
by
\beq
    \left(
        \bs{\A} ^{+} _{t}
        -
        \bs{\A} ^{-} _{t}
    \right)
    \bs{\hpsi} \ttt{y}
    =
    \frac{\beta}{2}
    \frac{\delta \log \sh{P} \qtq{\bs{\hpsi} , t}}
         {\delta \bs{\hpsi} \ttt{y}}
\quad .
\eeq

Consider now:
\beq
    \left[
        \bs{\A} _{t} ^{+} \hat{-}
    \right] \ttt{\circ}
    \sh{P} \left [
               \hat{-} , t
           \right ]
    =
    \clue{
          \sum _{i} ^{1,n}
              \udin{a}{i}{n}
              \ttt{- , t}
              e _{i} \ttt{y}
         }{n}
    \clue{p ^{(m)} \ttt{- , t}}{m}
    \quad .
\eeq
This is a sensible expression, which we can easily read as the product
of an operator on $\kawspace$ times a functional on $\kawspace$,
which we gave a meaning in proposition \ref{C.nonstaopeperfunpro} to.
Note that we have indicated the possible arguments in this expression
with two different symbols: $\circ$ indicates that there is a dependence
from a point in $\R ^{d}$ and $-$ indicates the dependence from
the elements of $\kawspace$. If we just saturate the $-$ entry
we get an element of $\kawspace$ because the probability distribution
gives a {\sl NonStandard Real Number} and the {\sl Operator} an
element of $\kawspace$, which is a vector space; so
\bea
    & & \esci \esci
    \left[
        \bs{\A} _{t} ^{+} \bs{\hpsi}
    \right] \ttt{\circ}
    \sh{P} \left [
               \bs{\hpsi} , t
           \right ]
    =
    \clue{
          \sum _{i} ^{1,n}
              \udin{a}{i}{n}
              \ttt{\bs{x} ^{(n)} , t}
              e _{i} \ttt{y}
         }{n}
    \cdot
    \nonumber \\
    & & \qquad \qquad \cdot
    \clue{p ^{(m)} \ttt{\bs{x} ^{n} , t}}{m}
    \in
    \kawspace
\eea
where we assumed that $\bs{\hpsi}$ is as in the definition
above. On the contrary, if we saturate the $\circ$ entry, we get a functional
on $\kawspace$, because when we apply it to $\bs{\hpsi}$ the result
is in $\Rns$. Then, since
\beq
    \left[
        \bs{\A} _{t} ^{+} \hat{-}
    \right] \ttt{y}
    \sh{P} \left [
               \hat{-} , t
           \right ]
    =
    \clue{
          \sum _{i} ^{1,n}
              \udin{a}{i}{n}
              \ttt{- , t}
              e _{i} \ttt{y}
         }{n}
    \clue{p ^{(m)} \ttt{- , t}}{m}
\eeq
is a functional on $\kawspace$
it is a complete meaningful operation to apply to it the functional
derivative operation of definition \ref{C.nonfunderdef}.
If we first compute the product
of the functional $\sh{P}$ with the operator $\bs{\A}$,
following equation (\ref{C.nonfunopepro}),
we then get
\bea
    & & \esci \esci
    \frac{\delta}{\delta \bs{\hpsi} \ttt{y}}
    \left\{
        \left[
            \bs{\A} _{t} ^{+} \bs{\hpsi}
        \right] \ttt{y}
        \sh{P} \left [
                   \bs{\hpsi} , t
               \right ]
    \right\}
    \nonumber \\
    & &
    =
    \frac{\delta}{\delta \bs{\hpsi} \ttt{y}}
    \left\{
        \clue{
            \sum _{i} ^{1,n}
                \udin{a}{i}{n}
                \ttt{\bs{x} ^{(n)} , t}
                p ^{(m)} \ttt{\bs{x} ^{(m)} , t}
                e _{i} \ttt{y}
             }{n}
    \right\}
    \nonumber \\
    & &
    =
    \clue{
        \sum _{j} ^{1,n}
            \frac{\partial}{\partial \udin{x}{i}{n}}
            \left(
                \sum _{i} ^{1,n}
                \udin{a}{i}{n}
                \ttt{\bs{x} ^{(n)} , t}
                p ^{(n)} \ttt{\bs{x} ^{(n)} , t}
                e _{j} \ttt{y}
            \right)
            e _{j} \ttt{y}
         }{n}
    \nonumber \\
    & &
    =
    \clue{
        \sum _{j} ^{1,n}
        \sum _{i} ^{1,n}
            \frac{\partial}{\partial \udin{x}{i}{n}}
            \left(
                \udin{a}{i}{n}
                \ttt{\bs{x} ^{(n)} , t}
                p ^{(n)} \ttt{\bs{x} ^{(n)} , t}
            \right)
            e _{i} \ttt{y}
            e _{j} \ttt{y}
         }{n}
    \quad .
\eea
Now we can write
\bea
    & & \esci \esci
    \int d ^{d} y
        \frac{\delta}{\delta \bs{\hpsi} \ttt{y}}
        \left\{
            \left[
                \bs{\A} ^{+} _{t} \bs{\hpsi}
            \right] \ttt{y}
            \sh{P} \left [
                       \bs{\hpsi} , t
                   \right ]
        \right\}
    \nonumber \\
    & &
    =
    \int d ^{d} y
        \clue{
            \sum _{j} ^{1,n}
            \sum _{i} ^{1,n}
                \frac{\partial}{\partial \udin{x}{i}{n}}
                \left(
                    \udin{a}{i}{n}
                    \ttt{\bs{x} ^{(n)} , t}
                    p ^{(n)} \ttt{\bs{x} ^{(n)} , t}
                \right)
                e _{j} \ttt{y}
                e _{j} \ttt{y}
             }{n}
    \nonumber \\
    & &
    =
    \clue{
        \sum _{j} ^{1,n}
        \sum _{i} ^{1,n}
            \frac{\partial}{\partial \udin{x}{i}{n}}
            \left(
                \udin{a}{i}{n}
                \ttt{\bs{x} ^{(n)} , t}
                p ^{(n)} \ttt{\bs{x} ^{(n)} , t}
            \right)
        \scal{e _{j}}{e _{j}}{{\mathscr{S} \ttt{\R ^{d}}}}
         }{n}
    \nonumber \\
    & &
    =
    \clue{
        \sum _{j} ^{1,n}
        \sum _{i} ^{1,n}
            \frac{\partial}{\partial \udin{x}{i}{n}}
            \left(
                \udin{a}{i}{n}
                \ttt{\bs{x} ^{(n)} , t}
                p ^{(n)} \ttt{\bs{x} ^{(n)} , t}
            \right)
        \delta _{ij}
         }{n}
    \nonumber \\
    & &
    =
    \clue{
        \sum _{i} ^{1,n}
            \frac{\partial}{\partial \udin{x}{i}{n}}
            \left(
                \udin{a}{i}{n}
                \ttt{\bs{x} ^{(n)} , t}
                p ^{(n)} \ttt{\bs{x} ^{(n)} , t}
            \right)
         }{n}
    \label{C.nonfokplafirter}
\eea
and we also have
$$
    \frac{\delta \sh{P} \qtq{\bs{\hpsi} , t}}
         {\delta \bs{\hpsi} \ttt{x}}
    =
    \clue{
          \sum _{i} ^{1,n}
              \frac{\partial p ^{(n)} \ttt{\bs{x} ^{(n)} , t}}
                   {\partial \udin{x}{i}{n}}
              e _{i} \ttt{x}
         }
         {n}
$$
and
$$
    \frac{\delta ^{2} \sh{P} \qtq{\bs{\hpsi} , t}}
         {\delta \bs{\hpsi} \ttt{x} ^{2}}
    =
    \clue{
          \sum _{j} ^{1,n}
          \sum _{i} ^{1,n}
              \frac{\partial ^{2} p ^{(n)} \ttt{\bs{x} ^{(n)} , t}}
                   {\partial \udin{x}{j}{n} \partial \udin{x}{i}{n}}
              e _{j} \ttt{x}
              e _{i} \ttt{x}
         }
         {n}
    \quad .
$$
Thus, integrating over the $x$ variable, we get
\bea
    \int d ^{d} x
        \frac{\delta ^{2} \sh{P} \qtq{\bs{\hpsi} , t}}
             {\delta \bs{\hpsi} \ttt{x} ^{2}}
    & = &
    \clue{
          \sum _{j} ^{1,n}
          \sum _{i} ^{1,n}
              \frac{\partial ^{2} p ^{(n)} \ttt{\bs{x} ^{(n)} , t}}
                   {\partial \udin{x}{j}{n} \partial \udin{x}{i}{n}}
              \scal{e _{i}}{e _{j}}{\mathscr{S} \ttt{\R ^{d}}}
         }
         {n}
    \nonumber \\
    & = &
    \clue{
          \sum _{j} ^{1,n}
          \sum _{i} ^{1,n}
              \frac{\partial ^{2} p ^{(n)} \ttt{\bs{x} ^{(n)} , t}}
                   {\partial \udin{x}{j}{n} \partial \udin{x}{i}{n}}
              \delta _{ij}
         }
         {n}
    \nonumber \\
    & = &
    \clue{
          \sum _{i} ^{1,n}
              \frac{\partial ^{2} p ^{(n)} \ttt{\bs{x} ^{(n)} , t}}
                   {\partial x _{i} ^{(n) \, 2} {}^{2}}
         }
         {n}
    \quad .
    \label{C.nonfokplasecter}
\eea
Now generalising the procedure used so far to define
objects in $\kawspace$, we see that the {\it NonStandard
Fokker-Planck Equation} for the {\sl NonStandard Stochastic Process},
which proceeding as in the previous definitions is
\bea
    & & \esci \esci
    \frac{\partial \clue{p ^{(n)} \ttt{\bs{x} ^{(n)} , t}}{n}}
         {\partial t}
    =
    \nonumber \\
    & &
    =
    -
    \clue{
          \sum _{i} ^{1,n}
              \frac{\partial}{\partial \udin{x}{i}{n}}
              \left(
                  \udin{a}{i}{n} \ttt{\bs{x} ^{(n)} , t}
                  p ^{(n)} \ttt{\bs{x} ^{(n)} , t}
              \right)
         }{n}
    +
    \nonumber \\
    & & \qquad \qquad
    +
    \beta
    \clue{
          \sum _{i} ^{1,n}
              \frac{\partial ^{2} p ^{(n)} \ttt{\bs{x} ^{(n)} , t}}
                   {\partial x _{i} ^{(n) \, 2} {}^{2}}
         }{n}
    \quad ,
\eea
can be rewritten, using equations
(\ref{C.nonfokplafirter}-\ref{C.nonfokplasecter}), as
\beq
    \frac{\partial \sh{P} \qtq{\bs{\hpsi} , t}}
         {\partial t}
    =
    -
    \int d ^{d} y
        \frac{\delta}{\delta \bs{\hpsi} \ttt{y}}
        \left[
            \left(
                \bs{\A} ^{+} _{t} \bs{\hpsi} \ttt{y}
            \right)
            \sh{P} \qtq{\bs{\hpsi} , t}
        \right]
    +
    \beta
    \int d ^{d} y
        \frac{\delta ^{2} \sh{P} \qtq{\bs{\hpsi} , t}}
             {\delta \bs{\hpsi} \ttt{y} ^{2}}
    \quad .
\eeq
This is the forward Fokker-Planck equation. The backward Fokker-Planck
equation is also satisfied,
\beq
    \frac{\partial \sh{P} \qtq{\bs{\hpsi} , t}}
         {\partial t}
    =
    -
    \int d ^{d} y
        \frac{\delta}{\delta \bs{\hpsi} \ttt{y}}
        \left[
            \left(
                \bs{\A} ^{-} _{t} \bs{\hpsi} \ttt{y}
            \right)
            \sh{P} \qtq{\bs{\hpsi} , t}
        \right]
    -
    \beta
    \int d ^{d} y
        \frac{\delta ^{2} \sh{P} \qtq{\bs{\hpsi} , t}}
             {\delta \bs{\hpsi} \ttt{y} ^{2}}
\eeq

Consider now a stochastic process $\bs{\hPsi} \ttt{t}$
on $\kawspace$ space, defined as in (\ref{C.nonstoprodef}).
Thus $\bs{\hPsi}$
is a non-standard random variable, defined in terms of the
$1$, $2$, \dots{}, $i$, \dots{}, dimensional random variables
$\bs{X} ^{(1)}$, $\bs{X} ^{(n)}$, \dots{}, $\bs{X} ^{(i)}$, \dots{}.
A natural definition then arises:
\begin{defs}[Non Standard Expectation Value]\spbcorr{}.\\
    Let
    $$
        \bs{\hPsi} \ttt{t}
        =
        \clue{
              \sum _{j} ^{1,n}
                  \udin{X}{j}{n} \ttt{t}
                  e _{j}
             }{n}
    $$
    be a {\sl NonStandard Stochastic Process} and
    $$
        \bs{\hR} \left [ \bs{\hPsi} \ttt{t} , t \right ]
        =
        \clue{
              R ^{(n)} \ttt{\bs{X} ^{(n)} \ttt{ t } , t}
             }{n}
    $$
    a {\sl NonStandard Random Variable} defined starting from $\bs{\hpsi}$.\\
    The \underbar{NonStandard Expectation Value} of $\bs{\hR}$ is
    \bea
        \nexpect{\bs{\hR} \left [ \bs{\hPsi} \ttt{t} , t \right ]}
        & \dfn &
        \clue{
              \expect{R ^{(n)} \ttt{\bs{X} ^{(n)} \ttt{t} , t}}
             }{n}
        \nonumber \\
        & = &
        \int _{\kawspace}
            R \left [ \bs{\hpsi} \ttt{t} , t \right ]
            \sh{P} \qtq{\bs{\hpsi} \ttt{t} , t}
            \delta \bs{\hpsi}
    \quad .
    \eea
\end{defs}
After this definition we can at the end define the most important
quantities we are looking for.
\begin{defs}[Mean Non Standard Forward/Backward Derivative]\spbcorr{}.\\
    Let
    $$
        \bs{\hPsi} \ttt{t}
        =
        \clue{
              \sum _{j} ^{1,n}
                  \udin{X}{j}{n} \ttt{t}
                  e _{j}
             }{n}
    $$
    be a non-standard stochastic process.
    The \underbar{Mean NonStandard Forward Derivative}
    of the {\sl NonStandard Stochastic Variable}
    $$
        \bs{\hR} \left [ \bs{\hPsi} \ttt{t} , t \right ]
        =
        \clue{
              R ^{(n)} \ttt{X ^{(n)} \ttt{ t } , t}
             }{n}
    $$
    is given by
    $$
        \mfd
        \bs{\hR} \left [ \bs{\hPsi} \ttt{t} , t \right ]
        \dfn
        \lim _{\Delta t \to 0 ^{+}}
        \frac{
              \cnexpect{
                        \bs{\hR}
                        \left [
                            \bs{\hPsi} \ttt{t + \Delta t}
                            ,
                            t + \Delta t
                        \right ]
                        -
                        \bs{\hR} \left [ \bs{\hPsi} \ttt{t} , t \right ]
                       }
                       {\hPsi \ttt{t}}
             }
             {\Delta t}
    \quad .
    $$
    Similarly the \underbar{Mean NonStandard Backward Derivative} is
    $$
        \mbd
        \bs{\hR} \left [ \bs{\hPsi} \ttt{t} , t \right ]
        \dfn
        \lim _{\Delta t \to 0 ^{+}}
        \frac{
              \cnexpect{
                        \bs{\hR} \left [ \bs{\hPsi} \ttt{t} , t \right ]
                        -
                        \bs{\hR}
                        \left [
                            \bs{\hPsi} \ttt{t - \Delta t}
                            ,
                            t - \Delta t
                        \right ]
                       }
                       {\hPsi \ttt{t}}
             }
             {\Delta t}
    \quad .
    $$
\end{defs}
We can now prove the following.
\begin{props}[Mean NonStandard Forward/Backward Derivative]\spbcorr{}.\\
    Let
    $$
        \bs{\hR} \left [ \bs{\hPsi} \ttt{t} , t \right ]
        =
        \clue{
              R ^{(n)} \ttt{X ^{(n)} \ttt{ t } , t}
             }{n}
    $$
    be a {\sl NonStandard Stochastic Variable}
    of the {\sl NonStandard Stochastic Process}
    $$
        \bs{\hPsi} \ttt{t}
        =
        \clue{
              \sum _{j} ^{1,n}
                  \udin{X}{j}{n} \ttt{t}
                  e _{j}
             }{n}
    $$
    and
    $$
        \mfd
        \bs{\hR} \left [ \bs{\hPsi} \ttt{t} , t \right ]
    $$
    its {\sl Mean NonStandard Forward Derivative}.
    This last expression can
    be rewritten as
    $$
        \mfd
        \bs{\hR} \left [ \bs{\hPsi} \ttt{t} , t \right ]
        =
        \left(
            \frac{\partial}{\partial t}
            +
            \int d ^{d} x
                \bs{\A} ^{+} _{t}
                \bs{\hpsi} \ttt{x}
                \frac{\delta}{\delta \psi \ttt{x}}
            +
            \beta
            \int d ^{d} x
                \frac{\delta ^{2}}{\delta \bs{\hpsi} \ttt{x} ^{2}}
        \right)
        \bs{\hR} \left [ \bs{\hPsi} \ttt{t} , t \right ]
    \quad .
    $$
    Similarly, for the {\sl Mean NonStandard Backward Derivative} we get
    $$
        \mbd
        \bs{\hR} \left [ \bs{\hPsi} \ttt{t} , t \right ]
        =
        \left(
            \frac{\partial}{\partial t}
            +
            \int d ^{d} x
                \bs{\A} ^{-} _{t}
                \bs{\hpsi} \ttt{x}
                \frac{\delta}{\delta \bs{\hpsi} \ttt{x}}
            -
            \beta
            \int d ^{d} x
                \frac{\delta ^{2}}{\delta \bs{\hpsi} \ttt{x} ^{2}}
        \right)
        \bs{\hR} \left [ \bs{\hPsi} \ttt{t} , t \right ]
    \quad .
    $$
\end{props}
\begin{proof}
The result is a direct consequence of the definitions
of {\sl NonStandard Expectation Value}
of a {\sl NonStandard Random Variable},
and of the form of the {\sl NonStandard} Fokker-Planck equations.
Indeed we have
\bea
        \mfd
        \bs{\hR} \left [ \bs{\hPsi} \ttt{t} , t \right ]
        & = &
        \lim _{\Delta t \to 0 ^{+}}
        \frac{
              \cnexpect{
                        \bs{\hR}
                        \left [
                            \bs{\hPsi} \ttt{t + \Delta t}
                            ,
                            t + \Delta t
                        \right ]
                        -
                        \bs{\hR} \left [ \bs{\hPsi} \ttt{t} , t \right ]
                       }
                       {\hPsi \ttt{t}}
             }
             {\Delta t}
        \nonumber \\
        & = &
        \lim _{\Delta t \to 0 ^{+}}
        \frac{
              \clue{
                  \cexpect{
                           \bs{\hR}
                           \left [
                               \bs{X} ^{(n)} \ttt{t + \Delta t}
                               ,
                               t + \Delta t
                           \right ]
                           -
                           \bs{\hR} \left [ \bs{X} ^{(n)} \ttt{t} , t \right ]
                          }
                          {\bs{X} ^{(n)} \ttt{t}}
                   }{n}
             }
             {\Delta t}
        \nonumber \\
        & = &
        \clue{
            \left(
                 \frac{\partial}{\partial t}
                +
                \sum _{j} ^{1,n}
                    \udin{a}{j}{n}
                    \frac{\partial}{\partial \udin{x}{j}{n}}
                +
                \beta
                \sum _{j} ^{1,n}
                    \frac{\partial ^{2}}{\partial \udin{x}{j}{n} {}^{2}}
            \right)
            R ^{(n)} \left ( \bs{X} ^{(n)} \ttt{t} , t \right )
             }{n}
        \nonumber \\
        & = &
        \left(
            \frac{\partial}{\partial t}
            +
            \int d ^{d} x
                \bs{\A} ^{+} _{t}
                \bs{\hpsi} \ttt{x}
                \frac{\delta}{\delta \bs{\hpsi} \ttt{x}}
            +
            \beta
            \int d ^{d} x
                \frac{\delta ^{2}}{\delta \bs{\hpsi} \ttt{x} ^{2}}
        \right)
        \bs{\hR} \left [ \bs{\hPsi} \ttt{t} , t \right ]
    \quad .
    \nonumber
\eea
Of course the same procedure gives the result for the mean non-standard
backward derivative.
\end{proof}

%% file: appD.tex
\pageheader{}{Loop Derivatives.}{}
\chapter{Loop Derivatives.}
\label{D.looderapp}

\section{Overview}

It may be useful to give a brief review about
{\sl Holographic Functional Derivatives}, because they
are often confused with {\sl (Ordinary) Functional Derivatives}
in view of the formal relation
\beq
    \frac{\delta}{\delta Y ^{\mu} \left( s \right)}
    =
    Y ^{\prime \nu} \left( s \right)
    \frac{\delta}{\delta Y ^{\mu \nu} \left( s \right)}
    \quad.
    \label{D.funarederrel}
\eeq
Notwithstanding, there is a basic difference between these two types of
operations.\\
To begin with, an infinitesimal {\it shape variation} corresponds to
``cutting'' the loop $C$ at a particular point, say $y$, and then
joining the two end--points to  an infinitesimal
loop $\delta C _{y}$. Accordingly,
\beq
    \delta Y ^{\mu \nu} \left [C ; y \right ]
    =
    \oint _{\delta C _{y}}
    Y ^{\mu}
    d Y ^{\nu}
    \approx
    \left .
    d Y ^{\mu}
    \wedge
    d Y ^{\nu}
    \right \rceil _{y}
\label{D.darea}
\eeq
where $\left . d Y ^{\mu} \wedge dY ^{\nu} \right \rceil _{y}$
is the elementary oriented
area subtended by $\delta C _{y}$. A suggestive description of
this procedure, due to Migdal \cite{mig}, is that of adding a ``petal''
to the original loop. Then, we can introduce an ``intrinsic distance''
between the deformed and initial {\sl Strings}, as the infinitesimal,
oriented area variation,
$$
    \delta \sigma ^{\mu \nu} \left[ C ; y \right]
    \equiv
    \left .
    d Y ^{\mu}
    \wedge
    d Y ^{\nu}
    \right \rceil _{y}
    \quad .
$$
``Intrinsic'' means that the (spacelike) distance is invariant under
reparametrization and/or embedding transformations. Evidently, there is no
counterpart of this operation in the case of point--particles, because of
the lack of spatial extension. Note that, while $\delta Y^{\mu} \left( s
\right)$
represents a {\it smooth} deformation of the loop
$Y ^{\mu} = Y ^{\mu} \left( s \right)$, the addition of a petal introduces a
singular ``{\it cusp}'' at the contact point. Moreover, cusps produce infinities
in the
ordinary variational derivatives but not in the
{\sl Holographic Functional Derivatives} \cite{mig}.


\section{Functional and Holographic Derivatives}

Let us discuss the relation between {\sl Functional} and
{\sl Holographic Derivatives}.
A functional is a ``function of functions'', which we shall write
as $F \left[ f \right]$. We adopt the square brackets notation, in order
to avoid  confusion with the composition law of two functions
$F \left( f \left( x \right) \right)$.

In ordinary calculus, the derivative of a function $f(x)$ is a
map, which associates to each point the differential of the function
at the given point. The first derivative
of a functional is an application which associates to each ``{\it{}point}'',
represented in this case by a function $f$, a linear functional
$F' \left[ f \right]$, which acts on functions $v(x)$. Thus, we obtain
\beq
    F' \left[ f \right] \left( v \right) =
    \int dx \frac{\delta F \left[ f \right]}{\delta f ( x )} v(x)
\quad ,
\eeq
where
\beq
    \frac{\delta F \left[ f \right]}{\delta f ( x )} =
    \int dy \frac{\delta F \left( f(y) \right)}{\delta f(x)}
\quad .
\eeq

On the other side, the {\sl Holographic Derivative} of a loop, is given by
a different, more geometrical procedure. It amounts
to evaluate the linear increment of a functional over a closed loop $C$, when
 we deform $C$ by adding a small loop ({\it petal}) $\delta C$ in the way
 described above. To wash away any dependence from the attachment point,
i.e. to  implement reparametrization invariance, we ``average'' such a
``{\it petal addition}'' over the whole loop. Formally,
\beq
    \delta F \left[ C ; s \right ]
    =
    F \left [ C + \delta C ( s ) \right ] - F \left[ C \right ]
    =
    \int _{\delta C} \frac{\delta F \left[ C \right]}
                          {\delta Y ^{\mu \nu} (s)}
                     d Y ^{\mu} \wedge d Y ^{\nu}
\eeq
and
\beq
    \frac{\delta F}{\delta C ^{\mu \nu}}
    \equiv
    \frac{\delta F}{\delta Y ^{\mu \nu}}
    \equiv
    \left( \oint _{C} dl(s) \right) ^{-1}
    \oint _{C} dl(s) \frac{\delta F \left [ C \right ]}
                          {\delta Y ^{\mu \nu} (s)}
\quad ,
\eeq
where, one can use the following equivalent notations
$
    F \left [ C \right ]\ ,
$
or
$
    F \left [ \sigma ^{\mu \nu} \right ]\ ,
$
and
\beq
    F \left [ \sigma ^{\mu \nu} \right ] \equiv
    F \left [ \sigma ^{\mu \nu} \left ( s _{0} , s _{1} \right ) \right ]
\quad .
\eeq

In what follows,  we would like to clarify the relation between
{\sl Functional} and
{\sl Holographic Derivative}.
\begin{props}[Functional and Holographic Derivatives]\spbcorr{}.
\label{D.funholderrelpro}\\
    The {\sl Functional} and {\sl Holohgraphic Derivative} are
    related in the following way:
    \beq
        \frac{\delta}
             {\delta Y ^{\alpha} (s)}
        =
        Y ^{\prime \beta} (s)
        \frac{\delta}
             {\delta Y ^{\alpha \beta} (s)}
    \label{D.funholderrel}
    \eeq
\end{props}
\begin{proof}
To establish this relationship we recall the definition
$$
    Y _{\mu \nu}
    =
    \oint Y ^{\mu} d Y^{\nu}
    =
    \int _{\Gamma} ds Y ^{\mu} (s) Y ^{\prime \nu} (s)
$$
and derive the following properties of the derivative \cite{prop}.
The first {\sl Functional Derivative} of the {\sl Holographic
Coordinates} is
\bea
    \frac{\delta Y ^{\mu \nu} \left [ C \right ]}
         {\delta Y ^{\alpha} (t)}
    & = &
    \oint _{\Gamma} ds \frac{\delta Y ^{\mu \nu} \left[ x(s) \right]}
                            {\delta Y ^{\alpha} (t)}
    \nonumber \\
    & = &
    \oint _{\Gamma} ds \left [
                             \frac{\delta Y ^{\mu} (s)}
                                  {\delta Y ^{\alpha} (t)}
                             Y ^{\prime \nu} (s)
                             +
                             Y ^{\mu} (s)
                             \frac{d}{ds} \left(
                                                \frac{\delta Y ^{\nu} (s)}
                                                     {\delta Y ^{\alpha} (t)}
                                          \right)
                      \right ]
    \nonumber \\
    & = &
    \oint _{\Gamma} ds \left [
                            \delta _{\alpha} ^{\mu}
                            \delta ( s - t )
                            Y ^{\prime \nu} (s)
                            +
                            Y ^{\mu} (s)
                            \frac{d}{d s} \left(
                                                \delta _{\alpha} ^{\nu}
                                                \delta ( s - t )
                                          \right)
                      \right ]
    \nonumber \\
    & = &
    \delta _{\alpha} ^{\mu} Y ^{\prime \nu} (t)
    +
    \left [
           \delta ^{\nu} _{\alpha}
           Y ^{\mu} (s)
           \delta ( s - t )
    \right \rceil ^{\partial \Gamma = \partial \partial \Sigma = \emptyset}
    +
    \nonumber \\
    & & \qquad \qquad
    -
    \oint _{\Gamma} ds
        \delta _{\alpha} ^{\nu} Y ^{\prime \mu} (s) \delta ( s - t )
    \nonumber \\
    & = &
    \delta _{\alpha} ^{\mu} Y ^{\prime \nu} (t)
    -
    \delta _{\alpha} ^{\nu} Y ^{\prime \mu} (t)
    \nonumber \\
    & = &
    \delta _{\alpha \beta} ^{\mu \nu} Y ^{\prime \beta}
    \quad ;
\eea
then the second one turns out to be
\bea
    \frac{\delta ^{2} \sigma ^{\mu \nu} \left [ C \right ]}
         {\delta x ^{\alpha} (t) \delta Y ^{\beta} (u)}
    & = &
    \oint _{\Gamma} ds \frac{\delta ^{2} Y ^{\mu \nu} \left [ Y(s) \right ]}
                            {\delta Y ^{\alpha} (t) \delta Y ^{\beta} (u)}
    \nonumber \\
    & = &
    \oint ds \frac{\delta}{\delta Y ^{\alpha} (t)}
             \left [
                    \delta ^{\mu} _{\beta} \delta ( s - u ) Y ^{\prime \nu} (s)
                    +
                    Y ^{\mu} (s) \delta _{\beta} ^{\nu}
                    \frac{d}{d s}
                    \left(
                          \delta ( s - u )
                    \right)
             \right ]
    \nonumber \\
    & = &
    \oint ds \left [
                    \delta ^{\mu} _{\beta}
                    \delta ^{\nu} _{\alpha}
                    \delta ( s - u )
                    \frac{d}{d s}
                    \left(
                          \delta ( s - t )
                    \right)
                    +
             \right .
    \nonumber \\
    &   & \qquad \qquad \qquad +
             \left .
                    \delta ^{\mu} _{\alpha}
                    \delta ^{\nu} _{\beta}
                    \delta ( s - t )
                    \frac{d}{d s}
                    \left(
                          \delta ( s - u )
                    \right)
             \right ]
    \nonumber \\
    & = &
    \delta ^{\mu} _{\beta} \delta ^{\nu} _{\alpha}
    \frac{d}{d u} \left( \delta ( u - t ) \right)
    +
    \delta ^{\nu} _{\beta} \delta ^{\mu} _{\alpha}
    \frac{d}{d t} \left( \delta ( t - u ) \right)
    \nonumber \\
    & = &
    \delta ^{\mu} _{\beta} \delta ^{\nu} _{\alpha}
    \frac{d}{d u} \left( \delta ( u - t ) \right)
    -
    \delta ^{\nu} _{\beta} \delta ^{\mu} _{\alpha}
    \frac{d}{d u} \left( \delta ( t - u ) \right)
    \nonumber \\
    & = &
    \delta ^{\mu} _{\beta} \delta ^{\nu} _{\alpha}
    \frac{d}{d u} \left( \delta ( u - t ) \right)
    -
    \delta ^{\nu} _{\beta} \delta ^{\mu} _{\alpha}
    \frac{d}{d u} \left( \delta ( u - t ) \right)
    \nonumber \\
    & = &
    \delta ^{\mu \nu} _{\alpha \beta}
    \frac{d}{d u} \left( \delta ( u - t ) \right)
\quad .
\eea
This two result are in agreement with the more general case
computed in section \ref{A.funcdersig}, proposition \ref{A.funcdersigpro}.
We can now calculate the relation between {\sl Holographic} and
{\sl Functional Derivatives} using a generalization of the
{\it chain rule} to the functional case. We then get
\bea
    \frac{\delta}{\delta Y ^{\alpha} (s)}
    {\scriptstyle 1.} & = &
    \frac{1}{2}
    \frac{\delta Y ^{\mu \nu} \left [ C \right ]}
         {\delta Y ^{\alpha} (s)}
    \frac{\delta}
         {\delta Y ^{\mu \nu} (s)}
    \nonumber \\
    {\scriptstyle 2.} & = &
    \frac{1}{2}
    \delta ^{\mu \nu} _{\alpha \beta}
    Y ^{\prime \beta} (s)
    \frac{\delta}
         {\delta Y ^{\mu \nu} (s)}
    \nonumber \\
    {\scriptstyle 3.} & = &
    Y ^{\prime \beta} (s)
    \frac{\delta}
         {\delta Y ^{\alpha \beta} (s)}
    \quad ,
    \nonumber
\eea
which is the desired result.
\end{proof}


%% file: biblio.tex
\pageheader{}{Bibliography.}{}